\documentclass[11pt,a4paper,german,twoside ,bigheadings,titlepage]{scrreprt}

\usepackage{latexsym,alltt,amsmath,amssymb,textcomp,graphicx}
\usepackage[rflt]{floatflt}
\usepackage[latin1]{inputenc}
\usepackage[T1]{fontenc}
\usepackage[all]{xy}
\usepackage{bm}
\usepackage{ifpdf}
\usepackage[thref,hyperref,amsmath,thmmarks]{ntheorem}
\usepackage{natbib}
\usepackage[automark, nouppercase]{scrpage2}
\usepackage{geometry}
                                                            \let\SavedRightarrow=\Rightarrow
                                                                \usepackage{marvosym}
                                                            \let\Rightarrow=\SavedRightarrow

                                                            \usepackage[newenum]{paralist}
                                                            \setdefaultenum{a)}{1.)}{(i)}{A.}

                                                            \newcommand{\BLR}{[BLR]}
                                                            \newcommand{\BoschRG}{[Bosch]}
                                                            \newcommand{\Drago}{[Dra]}
                                                            \newcommand{\DraVol}{[ADFV]}
                                                            \newcommand{\EGAz}{[EGA II]}
                                                            \newcommand{\EGAv}{[EGA IV]}
                                                            \newcommand{\EGAvz}{[EGA IV$_2$]}
                                                            \newcommand{\EGAvv}{[EGA IV$_4$]}
                                                            \newcommand{\Hart}{[Har]}
                                                            \newcommand{\Haw}{[Haw]}
                                                            \newcommand{\Heath}{[Hea]}
                                                            \newcommand{\Liu}{[Liu]}
                                                            \newcommand{\Milne}{[Milne]}
                                                            \newcommand{\Mordell}{[Mor]}
                                                            \newcommand{\Neu}{[Neu]}
                                                            \newcommand{\pI}{[BF]}
                                                            \newcommand{\pII}{[Vol1]}
                                                            \newcommand{\pIII}{[ADFV]}
                                                            \newcommand{\pIV}{[Dra1]}
                                                            \newcommand{\pV}{[DDN1]}
                                                            \newcommand{\pVI}{[DDN2]}
                                                            \newcommand{\pVII}{[Dra2]}
                                                            \newcommand{\pVIII}{[DDNV]}
                                                            \newcommand{\pIX}{[Dra3]}
                                                            \newcommand{\pX}{[ADV]}
                                                            \newcommand{\pXI}{[Vla]}
                                                            \newcommand{\pXII}{[VV1]}
                                                            \newcommand{\pXIII}{[VV2]}
                                                            \newcommand{\pXIV}{[VV3]}
                                                            \newcommand{\pXV}{[Meu]}
                                                            \newcommand{\pXVI}{[Zel1]}
                                                            \newcommand{\pXVII}{[Zel2]}
                                                            
                                                            \newcommand{\pXIX}{[Dra5]}
                                                            \newcommand{\pXX}{[Wey]}
                                                            \newcommand{\pXXI}{[Dra6]}
                                                            \newcommand{\pXXII}{[Dra7]}
                                                            \newcommand{\pXXIII}{[VVZ]}
                                                            \newcommand{\pXXIV}{[DD]}
                                                            \newcommand{\pXXV}{[DFU]}
                                                            \newcommand{\pXXVI}{[DN1]}
                                                            \newcommand{\pXXVII}{[DN2]}
                                                            \newcommand{\pXXVIII}{[DN3]}
                                                            \newcommand{\pXXIX}{[DN4]}
                                                            \newcommand{\pXXX}{[Dra4]}

                                                            \newcommand{\SGAIII}{[SGA 3]}
                                                            \newcommand{\Schar}{[Schar]}
                                                            \newcommand{\SerreI}{[Serre2]}
                                                            \newcommand{\SerreII}{[Serre1]}
                                                            \newcommand{\SerreArit}{[Serre1]}

                                                            \newcommand{\rela}{of type (R) }
                                                            \newcommand{\relao}{of type (R)}
                                                            \newcommand{\GRo}{model of type (GR)}
                                                            
                                                            \newcommand{\GRos}{models of type (GR)}
                                                            \newcommand{\GR}{model of type (GR) }
                                                            \newcommand{\GRs}{models of type (GR) }
                                                            \newcommand{\SRo}{model of type (SR)}
                                                            \newcommand{\SRso}{models of type (SR)}
                                                            \newcommand{\SRos}{models of type (SR)}
                                                            \newcommand{\SR}{model of type (SR) }
                                                            \newcommand{\SRs}{models of type (SR) }
                                                            \DeclareMathOperator{\eins}{\textsc{i}}
                                                            
                                                            \DeclareMathOperator{\Verjungung}{\emph{rejuvenation}}

                                                            \DeclareMathOperator{\cJ}{\mathcal{J}}
                                                            \DeclareMathOperator{\Affin}{\mathbb{A}}
                                                            \DeclareMathOperator{\Proj}{\mathbb{P}}
                                                            \DeclareMathOperator{\Os}{{\cal O}_S}
                                                            \DeclareMathOperator{\Ox}{{\cal O}_X}

                                                            \DeclareMathOperator{\Oz}{{\cal O}_Z}
                                                            
                                                            \DeclareMathOperator{\maxi}{\mathfrak{m}}
                                                            \DeclareMathOperator{\prim}{\mathfrak{p}}
                                                            \DeclareMathOperator{\Prim}{\mathfrak{P}}
                                                            \DeclareMathOperator{\Liealg}{\mathfrak{g}}

                                                            \DeclareMathOperator{\Ccat}{\mathfrak{C}}

                                                            \DeclareMathOperator{\rat}{\mathbb{Q}}
                                                            \DeclareMathOperator{\ratb}{\overline{\mathbb{Q}}}
                                                            \DeclareMathOperator{\real}{\mathbb{R}}
                                                            \DeclareMathOperator{\complex}{\mathbb{C}}
                                                            \DeclareMathOperator{\Spec}{\mathrm{Spec}}
                                                            
                                                            \DeclareMathOperator{\Natural}{\mathbb{N}}
                                                            \DeclareMathOperator{\Ganz}{\mathbb{Z}}
                                                            
                                                            \DeclareMathOperator{\Hom}{\mathrm{Hom}}
                                                            \DeclareMathOperator{\Aut}{\mathrm{Aut}}
                                                            \DeclareMathOperator{\Ext}{\mathrm{Ext}}
                                                            \DeclareMathOperator{\Lie}{\mathrm{Lie}}
                                                            \DeclareMathOperator{\GHom}{\underline{\mathrm{Hom}}}
                                                            \DeclareMathOperator{\GDer}{\underline{\mathrm{Der}}}
                                                            \DeclareMathOperator{\GExt}{\underline{\mathrm{Ext}}}
                                                            \DeclareMathOperator{\Der}{\mathrm{Der}}

                                                            \newcommand{\vol}{v}

                                                            \newcommand{\st}{{space-time }}
                                                            \newcommand{\sto}{{space-time}}

                                                            \DeclareMathOperator{\m}{\mathfrak{m}}

                                                            \newcommand{\Ad}{\text{Ad}}
                                                            \newcommand{\ad}{\text{ad}}
                                                            \newcommand{\Int}{\text{Int}}
                                                            \newcommand{\intg}{\textit{Int(g)}}

                                                            \newcommand{\kil}{{\sigma}}
                                                            \newcommand{\con}{\omega}

                                                            \newcommand{\ca}{\mathcal{A}}

                                                            \newcommand{\cd}{\mathcal{D}}
                                                            \newcommand{\ce}{\mathcal{E}}
                                                            \newcommand{\cf}{\mathcal{F}}
                                                            \newcommand{\cg}{\mathcal{G}}
                                                            
                                                            \newcommand{\ci}{\mathcal{I}}
                                                            \newcommand{\cj}{\mathcal{J}}

                                                            \newcommand{\cm}{\mathcal{M}}
                                                            \newcommand{\cn}{\mathcal{N}}
                                                            \newcommand{\co}{\mathcal{O}}

                                                            \newcommand{\ct}{\mathcal{T}}

                                                            \newcommand{\ba}{\mathbb{A}}
                                                            
                                                            \newcommand{\bg}{\mathbb{G}}
                                                            \newcommand{\bn}{\mathbb{N}}
                                                            \newcommand{\bp}{\mathbb{P}}
                                                            \newcommand{\br}{\mathbb{R}}
                                                            \newcommand{\bv}{\mathbb{V}}
                                                            \newcommand{\bz}{\mathbb{Z}}

                                                            \newcommand{\ve}{\varepsilon}
                                                            \def\stackrellow#1#2{\mathrel{\mathop{#2}\limits_{#1}}}

                                                            \newcommand{\vect}{\mathfrak{t}}
                                                            \newcommand{\vecu}{\mathfrak{u}}
                                                            \newcommand{\vecv}{\mathfrak{v}}
                                                            \newcommand{\vecw}{\mathfrak{w}}

                                                            \newcommand{\vecz}{\mathfrak{z}}

                                                            \newcommand{\VECV}{\mathfrak{v}}
                                                            \newcommand{\VECW}{\mathfrak{w}}

                                                            \clearscrheadings 
                                                            \clearscrplain

                                                            \ohead{\pagemark} 
                                                            \ihead{\headmark} 
                                                            \setheadsepline{.4pt} 

                                                            \setheadwidth[0pt]{text}

                                                            {\theoremstyle{change}\theorembodyfont{\itshape}\newtheorem{Satz}{Proposition}[chapter]}
                                                            {\theoremstyle{change}\theorembodyfont{\itshape}\newtheorem{Theorem}[Satz]{Theorem}}
                                                            {\theoremstyle{change}\theorembodyfont{\itshape}\newtheorem{Lemma}[Satz]{Lemma}}
                                                            {\theoremstyle{change}\theorembodyfont{\itshape}}
                                                            {\theoremstyle{change}\theorembodyfont{\itshape}}
                                                            {\theoremstyle{change}\theorembodyfont{\rmfamily}}
                                                            {\theoremstyle{change}\theorembodyfont{\itshape}\newtheorem{Cor}[Satz]{Corollary}}
                                                            {\theoremstyle{change}\theorembodyfont{\rmfamily}\newtheorem{Rem}[Satz]{Remark}}
                                                            {\theoremstyle{change}\theorembodyfont{\rmfamily}\newtheorem{Ex}[Satz]{Example}}
                                                            {\theoremstyle{change}\theorembodyfont{\rmfamily}}
                                                            {\theoremstyle{change}\theorembodyfont{\itshape}}
                                                            {\theoremstyle{change}\theorembodyfont{\rmfamily}\newtheorem{Def}[Satz]{Definition}}
                                                            {\theoremstyle{change}\theorembodyfont{\rmfamily}}
                                                            {\theoremstyle{change}\theorembodyfont{\rmfamily}\newtheorem{Jac}[Satz]{Jacobi
                                                            Criterion}}
                                                            {\theoremstyle{change}\theorembodyfont{\rmfamily}}
                                                            {\theoremstyle{change}\theorembodyfont{\rmfamily}}
                                                            {\theoremstyle{change}\theorembodyfont{\rmfamily}}
                                                            {\theoremstyle{change}\theorembodyfont{\rmfamily}}
                                                            {\theoremstyle{change}\theorembodyfont{\rmfamily}}
                                                            {\theoremstyle{change}\theorembodyfont{\itshape}\newtheorem{HM}[Satz]{Hasse-Minkowski
                                                            theorem}}
                                                            {\theoremstyle{change}\theorembodyfont{\itshape}\newtheorem{Faltings}[Satz]{Faltings's
                                                            theorem}}
                                                            {\theoremstyle{change}\theorembodyfont{\itshape}}
                                                            {\theoremstyle{change}\theorembodyfont{\itshape}\newtheorem{YME}[Satz]{Yang-Mills
                                                            equation}}
                                                            {\theoremstyle{change}\theorembodyfont{\itshape}}
                                                            {\theoremstyle{change}\theorembodyfont{\itshape}\newtheorem{Bianchi}[Satz]{Bianchi-identity}}
                                                            {\theoremstyle{change}\theorembodyfont{\itshape}\newtheorem{structure}[Satz]{structure-equation}}
                                                            {\theoremstyle{change}\theorembodyfont{\rmfamily}\newtheorem{S-valued-point}[Satz]{Translations
                                                            by $S$-valued points on torsors.}}
                                                            {\theoremstyle{change}\theorembodyfont{\rmfamily}}
                                                            {\theoremstyle{change}\theorembodyfont{\rmfamily}\newtheorem{PropSpecModGrav}[Satz]
                                                            {Properties of \SRos.}}
                                                            {\theoremstyle{change}\theorembodyfont{\rmfamily}\newtheorem{EinsteinEq}[Satz]
                                                            {Einstein equations}}
                                                            {\theoremstyle{change}\theorembodyfont{\itshape}\newtheorem{Vac}[Satz]{Non-trivial
                                                            vacuum structure.}}
                                                            {\theoremstyle{break}\theorembodyfont{\rmfamily}\theoremprework{\hrule}\theorempostwork{\hrule}}
                                                            {\theoremheaderfont{\sc}\theorembodyfont{\upshape}\theoremstyle{nonumberplain}\theoremsymbol{\ensuremath{\square}}\newtheorem{proof}{proof}}
                                                            {\theoremheaderfont{\sc}\theorembodyfont{\upshape}\theoremstyle{nonumberplain}}

                                        \title{%
                                            {\large{Westf\"alische Wilhelms-Universit\"at M\"unster}} \vspace{-3mm}\\
                                            {\large{Fachbereich Mathematik und Informatik}} \vspace{8mm}\\
                                            \vspace{1cm} \hrule \vspace{5mm} \Huge  {\sffamily
                                            Arithmetic gravity and \qquad \qquad \qquad Yang-Mills theory:\qquad \qquad \qquad
                                            An approach to adelic physics via algebraic spaces}\\
                                            {\large\sffamily Dissertation} \vspace{5mm} \hrule \vspace{40mm}%
                                        }
                                        \author{\textbf{Ren\'e Schmidt}  \vspace{8mm} \\
                                        \textbf{- 2008 - \ \ }}
                                        \date{}

\begin{document}
%
%

\chapter*{\center{\Large{Arithmetic gravity and Yang-Mills theory:
    An approach to adelic physics via algebraic spaces}}}

\begin{center}
    {\Large{Ren\'e Schmidt}} \vspace{4mm} \\
    \textit{Mathematisches Institut der Westf\"alischen Wilhelms-Universit\"at \\
    Einsteinstraße 62, D-48419 M\"unster, Germany \\
    e-mail\footnote{alternative e-mail: rene.schmidt1@gmx.net}: \textsf{rene.schmidt@uni-muenster.de}} \vspace{25mm} \\
\end{center}
\centerline{\textbf{Abstract}} {This work is a dissertation thesis
written at the WWU M\"unster (Germany), supervised by Prof. Dr.
Raimar Wulkenhaar. We present an approach to adelic physics based
on the language of algebraic spaces. Relative algebraic spaces $X$
over a base $S$ are considered as fundamental objects which
describe space-time. This yields a formulation of general
relativity which is covariant with respect to changes of the
chosen domain of numbers $S$. With regard to adelic physics the
choice of $S$ as an excellent Dedekind scheme is of interest
(because this way also the finite prime spots, i.e. the
$\prim$-adic degrees of freedom are taken into account). In this
arithmetic case, it turns out that $X$ is a N\'eron model. This
enables us to make concrete statements concerning the structure of
the space-time described by $X$. Furthermore, some solutions of
the arithmetic Einstein equations are presented. In a next step,
Yang-Mills gauge fields are incorporated. }


\chapter*{}
\vspace{-28mm} \hrule \vspace{8mm} {\center{\textbf{\Huge
{\sffamily Arithmetic gravity and
 }} } \vspace{0mm}
  \center{\textbf{\Huge  {\sffamily Yang-Mills theory:
}}} \center{\textbf{\Huge {\sffamily An approach to
 adelic physics via}}}
 \center{\textbf{\Huge {\sffamily algebraic spaces}}}
 \vspace{6mm} \hrule
%
\ \\ \ \\ \ \\ \ \\ \ \\ \ \\
{\Large {\sffamily  Inaugural-Dissertation }} \\
    \vspace{3mm}
    \large {\sffamily  zur Erlangung des Doktorgrades }\\
    \large {\sffamily  der Naturwissenschaften im Fachbereich }\\
    \large {\sffamily  Mathematik und Informatik }\\
    \large {\sffamily  der Mathematisch-Naturwissenschaftlichen Fakult\"at }\\
    \large {\sffamily  der Westf\"alischen Wilhelms-Universit\"at M\"unster }
    \vspace{45mm}

\begin{center}
    \sffamily vorgelegt von  \\
    \textbf{\sffamily Dipl.-Math. Ren\'e Schmidt} \\
    \ \vspace{6mm} \\
    \sffamily - 2008 -
\end{center}

\newpage }

\ \\
{\textbf{\textsf{\huge{Danksagung}}}} \\ \\
Die vorliegende Dissertation ist das Ergebnis einer mehrj\"ahrigen
Forschungsarbeit am Fachbereich Mathematik und Informatik  der
Mathematisch-Naturwissenschaftlichen Fakult\"at der
Westf\"alischen Wilhelms-Universit\"at M\"unster. Die Arbeit wurde
erm\"oglicht durch die finanzielle Unterst\"utzung des
Graduiertenkollegs ``Analytische
Topologie und Metageometrie''. \\ \\
Mein besonderer Dank richtet sich an meinen Doktorvater Herrn
Prof. Dr. Raimar Wulkenhaar f\"ur die Betreuung und F\"orderung
meiner mathematischen und physikalischen Studien. Stets hatte er
ein offenes Ohr f\"ur die Fragestellungen, denen ich mich im Laufe
der Promotionsphase stellen musste. Ohne die von ihm erlernten
Techniken und Kenntnisse besonders im Bereich der Yang-Mills
Theorie w\"are das Zustandekommen dieser Dissertation nicht
m\"oglich gewesen.
\\ \\
Leider ist es mir an dieser Stelle nicht m\"oglich, all jenen
namentlich zu danken, die sowohl w\"ahrend meiner Schulzeit als
auch w\"ahrend meiner Zeit an der Universit\"at direkt oder
indirekt am Aufbau meines mathematischen und physikalischen
Fundamentes beteiligt waren.

Ganz speziell m\"ochte ich mich jedoch bei Herrn Prof. Dr.
Siegfried Bosch f\"ur die hervorragende Ausbildung bedanken, die
ich unter seiner Anleitung im algebraischen Bereich erfahren
durfte. In einem aufeinander aufbauenden Vorlesungszyklus, der
mich vom ersten Semester bis zum Diplom begleitete, gelang es ihm
meine Begeisterung f\"ur die algebraische Geometrie zu wecken. Die
bei Herrn Prof. Dr. Bosch erlernten Kenntnisse in relativer
algebraischer Geometrie einerseits und das von Herrn Prof. Dr.
Wulkenhaar vermittelte physikalische Wissen im Bereich der
Eichtheorie andererseits bilden das theoretische Fundament, das
dieser Dissertation zugrunde liegt.

Nicht unerw\"ahnt lassen m\"ochte ich ferner den Namen von Herrn
Dr. Matthias Strauch, der den Vorlesungszyklus von Herrn Prof. Dr.
Bosch im Rahmen von Seminaren begleitete. Stets stand er uns
Studenten f\"ur Fragen zur Verf\"ugung. Durch gemeinsames Rechnen
von \"Ubungsaufgaben, teilweise bis sp\"at in den Abend hinein,
hat er unsch\"atzbare Beitr\"age zum Aufbau auch meiner
mathematischen Kenntnisse geleistet.
\\ \\
Ein ganz besonders großer und pers\"onlicher Dank gilt schließlich
noch meinen Eltern. Sie schufen ein Umfeld, welches mir schon
w\"ahrend meines Studiums erm\"oglichte, mich vollst\"andig auf
die wissenschaftliche Arbeit zu konzentrieren.

\newpage

\tableofcontents

                                        \part*{{{Introduction}}}

                                        \section{Preface}\label{bbbb}

                                        Unless otherwise specified, let $K \subset \real$ be an algebraic
                                        number field (i.e. a finite algebraic extension of $\rat$), and
                                        let ${\footnotesize{\text{$\cal O$}}}_K$ be the ring of integral
                                        numbers of $K$ (i.e. the integral closure of $\Ganz$ in $K$). For
                                        example, think of ${\footnotesize{\text{$\cal O$}}}_K= \Ganz$ and
                                        $K=\rat$.

                                        In this thesis we present a reformulation of general relativity
                                        and (pure) Yang-Mills theory within the bounds of arithmetic
                                        algebraic geometry. Relative algebraic spaces $X \to S$ will be
                                        considered as fundamental objects which describe \sto. We will see
                                        that one is reduced to the well known theories in the special case
                                        $S= \Spec \complex$. But one may also make other choices for $S$.
                                        We are especially interested in the case $S= \Spec
                                        {\footnotesize{\text{$\cal O$}}}_K$.  This choice is motivated as
                                        follows.

                                        Since 1987, there have been many interesting applications of
                                        $p$-adic numbers in physics. In his influential paper  \pII, I.V.
                                        Volovich draws the vision of number theory as the ultimate
                                        physical theory, where numbers are proposed as the fundamental
                                        entities of the universe. It is argued that the development of
                                        physics over arbitrary (number) fields might be necessary. In
                                        particular, this implies the incorporation of $p$-adic numbers in
                                        physical theories. Since then, many $p$-adic models have been
                                        constructed. A very nice overall view with respect to the various
                                        applications of $p$-adic numbers in mathematic physics may be
                                        taken from the book \pXXIII.

                                        At first, let us mention that applications of $p$-adic numbers in
                                        quantum physics are of interest (\pXII, \pXIII, \pXIV, see also
                                        \pXV). Let us recall that there is a canonical way to generalize
                                        ordinary quantum mechanics to the $p$-adic world. As illustrated
                                        in \pXX \ and \pXIII, ordinary one-dimensional quantum mechanics
                                        can be given by a triple $(L_2(\real),$$W_{\infty}(z),$
                                        $U_{\infty}(t))$, where $L_2(\real)$ is the Hilbert space of
                                        complex-valued, square-integrable functions on $\real$, $z$ is a
                                        point of the real classical phase space, $W_{\infty}(z_{})$ is a
                                        unitary representation of the Heisenberg-Weyl group on
                                        $L_2(\real)$, and $U_{\infty}(t)$ is a unitary representation of
                                        the evolution operator on $L_2({\real})$. According to the
                                        Vladimirov-Volovich formulation, one-dimensional $p$-adic quantum
                                        mechanics is a triple
                                        \begin{align*}
                                            \Big(L_2(\rat_p),W_{p}(z),U_{p}(t) \Big)
                                        \end{align*}
                                        where $L_2(\rat_p)$ is the Hilbert space of complex-valued, square
                                        integrable functions  on $\rat_p$ (with respect to the Haar
                                        measure), and $W_{p}(z)$ is an unitary representation of the
                                        Heisenberg-Weyl group on $L_2(\rat_p)$. Furthermore $U_p(t)$ is an
                                        unitary evolution operator
                                        \begin{align*}
                                            U_p(t) \psi_p(x) = \int_{\rat_p} K_p(x,t;y,0) \psi_p(y) dy
                                        \end{align*}
                                        on $L_2(\rat_p)$, whose kernel $K_p(x,t;y,0)$ is defined in
                                        complete analogy to the real case by a path integral. The operator
                                        $U_p(t)$ and its kernel satisfy the canonical group relations
                                        $U_p(t+t')=U_p(t)U_p(t')$ and $K_p(x,t+t';y,0)= \int_{\rat_p}
                                        K_p(x,t;z,0) K_p(z,t';y,0) dz$. The $p$-adic Feynman path integral
                                        is investigated in \pVI, where an explicit formula for the kernel
                                        of quadratic Lagrangians is derived. This is the basis of a
                                        quantum mechanical treatment of the $p$-adic harmonic oscillator
                                        (\pIV). Like in the realm of real numbers, this is an exactly
                                        solvable model (see also \pXVI, \pXVII). However, these general
                                        techniques are not limited to non-relativistic quantum mechanics.
                                        For example, the free relativistic particle is considered in \pV.

                                        With regard to this thesis, gravity is of special interest. In
                                        \pIII, the $p$-adic version of general relativity in the setting
                                        of $p$-adic differential geometry is considered and cosmological
                                        models are studied (see also \pVII, \pIX, \pXXV, \pXXVI \ -
                                        \pXXIX). The starting point are the Einstein gravitational field
                                        equations
                                        \begin{align*}
                                            R_{\mu \nu} - \frac{1}{2} g_{\mu \nu} R + \Lambda g_{\mu \nu}
                                            = \kappa T_{\mu \nu}
                                        \end{align*}
                                        which make sense as well over $\real$ as over $\rat_p$ (for all
                                        $p$) if the constants $\Lambda$ and $\kappa$ are assumed to be
                                        rational numbers. As an application of $p$-adic quantum mechanics
                                        to the $p$-adic cosmological models, minisuperspace cosmological
                                        models are investigated (see e.g. \pVIII). For this purpose,
                                        Feynman's path integral method in the Hartle-Hawking approach \Haw
                                        \ was exploited (see \pXXI, \pXXII). Thereby, the wave function of
                                        the universe still takes complex values, but its argument is not
                                        necessarily real but also $p$-adic or even \textit{adelic}. Adeles
                                        enable us to regard real and $p$-adic numbers simultaneously. More
                                        precisely, an adele is an infinite tuple
                                        \begin{align*}
                                            x=(x_2, \ldots, x_p, \ldots, x_{\infty}),
                                        \end{align*}
                                        where $x_{\infty} \in \real$ and $x_p \in \rat_p$ with the
                                        restriction that one has $x_p \in \Ganz_p$ for all but a finite
                                        set of primes. In many papers, which were cited above, not only
                                        the $p$-adic models are formulated, but also the {adelic}
                                        generalization of these models are studied. These adelic models
                                        unify in a certain way the ordinary (i.e. $\real$-valued) and
                                        $p$-adic models. In this sense, adelic models may be considered as
                                        very canonical.

                                        Adelic physical models are the starting point of this thesis. But,
                                        instead of working directly with adeles and the respective adelic
                                        \st models as it is usually done, we will study a new, purely
                                        geometric approach to adelic physics based on relative algebraic
                                        spaces $X \to S$, $S= \Spec {\footnotesize{\text{$\cal O$}}}_K$.
                                        However, there are close relations between these two approaches as
                                        it may be seen in the following example.

                                        \begin{Ex}\label{5010}
                                            Let us choose $K=\rat$. Consequently,
                                            ${\footnotesize{\text{$\cal O$}}}_K= \Ganz$ and $S=\Spec \Ganz$.
                                            Furthermore
                                            assume that the relative algebraic space $X$ over $S$ is representable by a
                                            smooth, separated $S$-scheme, i.e. let us consider a smooth, separated morphism $\pi: X \to \Spec \Ganz$ of schemes.
                                            Set-theoretically, $\Spec \Ganz$ consists of infinitely many closed points
                                            (one point for each prime number $p$) plus one generic point
                                            which we will denote by $\infty$, and which corresponds to the zero ideal of $\Ganz$.
                                            Furthermore, $X$ may be viewed as union $\bigcup_p \pi^{-1}(p) \cup \pi^{-1}(\infty) $ of the fibres of
                                            $\pi$, and at least set-theoretically we obtain the following
                                            picture:
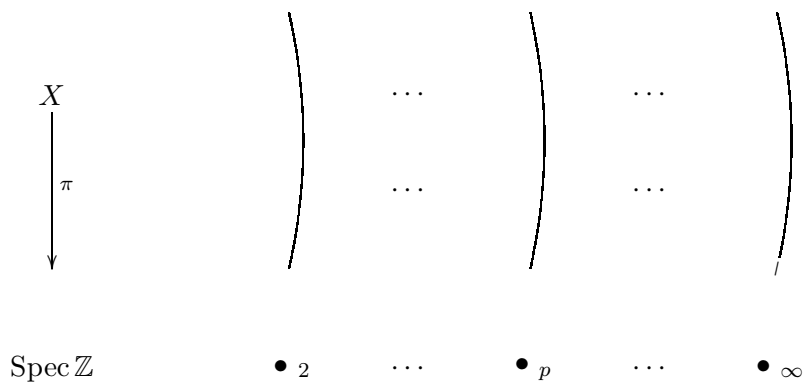
                                            \begin{equation*}
                                                \xymatrix{
                                                \  & &    \ar@{-}@/^/[ddd] &  & \ar@{-}@/^/[ddd] & & \ar@{{-}}@/^/[ddd] \\
                                                X \ar[dd]^{\pi} &  & \ & \cdots & \ & \cdots &  & & \\
                                                \  &   &   &  \cdots &   &  \cdots & & &  \\
                                                \ &  \ & \ & \ &  & & \\
                                                \Spec \Ganz &  & \ \bullet_{ \ 2} & \cdots & \ \bullet_{ \ p} & \cdots & \ \bullet_{ \ \infty } }
                                            \end{equation*}
                                            In our arithmetic setting (and in analogy to complex algebraic geometry), a ``physical point''
                                            $x$ is given by an $S$-valued point of $X$, i.e. by a section $s: \Spec \Ganz \hookrightarrow
                                            X$ of the structure morphism $\pi$ (i.e. $\pi \circ s = \text{id} $). More precisely, $x$ is
                                            given by the image of $s$ (this is a closed subscheme of $X$).
                                            However, set-theoretically, $s$ cuts out one closed
                                            point in each fibre. Thus, in analogy to the adelic situation,
                                            an $S$-valued point $x$ may be viewed as a set of points:
                                            \begin{align*}
                                                x= \{ x_2, \ldots, x_p, \ldots, x_{\infty} \}.
                                            \end{align*}
                                            Furthermore, according to the point of view of  adelic physics, each archimedean
                                            point (resp. each morphism over the archimedean prime spot at infinity) is
                                            only the archimedean component of an adelic point (resp.
                                            an adelic morphism). In short, everything in the archimedean world comes from the adelic level.
                                            If now $\varphi: Y \to \Spec \Ganz$ is an arbitrary
                                            smooth $S$-scheme and if we denote by $Y_K$ the pre-image $\varphi^{-1}(\infty)$
                                            of $\infty$ under $\varphi$, the above extension property
                                            (from the archimedean to the adelic level) reads
                                            as follows in algebraic geometry:
                                            \begin{align*}
                                                \emph{\text{For every $K$-morphism $f_K:Y_K \to X_K$, there is an }} \qquad \qquad \quad (\star) \\
                                                \emph{\text{$S$-morphism $f:Y \to X$ which extends $f_K$. \quad \qquad \qquad \quad \ \ \ \ }}
                                            \end{align*}
                                        \end{Ex}
                                        All in all, instead of adeles, the set $X(S)$ of $S$-valued points
                                        of an algebraic space $X \to S$ is the set of interest in our
                                        approach. The objective of this thesis is the investigation of a
                                        new approach to general relativity and (pure) Yang-Mills theory
                                        based on algebraic spaces. The condition $(\star)$ makes clear why
                                        N\'eron models will be of particular interest. We will illustrate in
                                        section \ref{5030} that our approach is naturally settled in the
                                        realm of adelic physics. But as the starting point are algebraic
                                        spaces, powerful algebraic geometric tools yield interesting, new
                                        insight.
                                        \newpage
                                        \section{Introduction}\label{baaa}

                                        According to the theory of general relativity, \st may be
                                        described  by means of a differentiable manifold. Thereby, gravity
                                        is encoded in a metrical tensor $g$ which satisfies the Einstein
                                        equations. More precisely, our starting point are the complex
                                        gravitational field equations. Then, any solution of the Einstein
                                        equations gives rise to a complex manifold. For technical reasons,
                                        we will once and for all assume that this classical \st manifold
                                        may be realized as a compact complex manifold $\mathfrak{X}$ which
                                        is \emph{Moishezon}. The latter condition means that
                                        \begin{align*}
                                            \text{transdeg}_{\complex} \Big( K(\mathfrak{X}) \Big) =
                                            \dim_{\complex} \mathfrak{X},
                                        \end{align*}
                                        where $K(\mathfrak{X})$ denotes the field of meromorphic functions
                                        on $\mathfrak{X}$. For example, all algebraic manifolds fulfill
                                        this equation. Therefore, following the ideas of \pIII, where it
                                        is among other things argued that one should restrict  to
                                        algebraic manifolds in quantum cosmology, our assumption is not
                                        too restrictive. However, let us at least mention that there are
                                        Moishezon manifolds which are not algebraic.

                                        The technical reason why we restrict attention to Moishezon
                                        manifolds is the following beautiful theorem due to Artin.

                                        \begin{Theorem}\label{5020}
                                            There is an equivalence of categories
                                            \begin{align*}
                                                \Big(  {\textrm{Moishezon manifolds}} \Big)
                                                \leftrightsquigarrow
                                                \Big(  {\text{smooth, proper algebraic spaces over }} \complex \Big)
                                            \end{align*}
                                        \end{Theorem}
                                        This theorem enables us to consider the ordinary complex
                                        space-time manifold $\mathfrak{X}$ as a complex algebraic space.
                                        Now the following observation is crucial. While, on the level of
                                        manifolds, the theory is essentially adapted to the complex
                                        numbers, the language of algebraic spaces offers to possibility to
                                        replace $\complex$ by any commutative ring.

                                        In 1987, I.V. Volovich suggested that a fundamental physical
                                        theory should be formulated in such a way that it is invariant
                                        under change of the underlying number field (see \pII). According
                                        to the 2006 paper \pXIX $ $ of B. Dragovich, such a number field
                                        invariant model has not yet been constructed. This motivates the
                                        following program which will be studied within the first part of
                                        this thesis:
                                        \begin{enumerate}
                                            \item
                                            Replace the pair $( \mathfrak{X}, g)$ consisting of a
                                            (complex) manifold $\mathfrak{X}$ and a metric $g$
                                            by a pair
                                            \begin{align*}
                                                \Big( X \to S , g \Big),
                                            \end{align*}
                                            where $X$ is a smooth, separated algebraic space over an arbitrary base $S$,
                                            and where $g$ is a metric over $X$ (see Definition \ref{5101}).

                                            \item
                                            Starting from exactly the same physical principles as in the realm of
                                            manifolds, deduce the equations of Einstein's theory of
                                            general relativity in the setting of algebraic spaces over
                                            an arbitrary base $S$ (thus realizing a number field invariant theory).
                                            Determine the pair $( X \to S , g )$ in such a way that
                                            Einstein's equations are fulfilled.

                                            \item
                                            Investigate properties of hypothetical \st models $( X \to S , g )$
                                            depending on the choice of the base $S$.
                                        \end{enumerate}
                                        \begin{Rem}\label{5021}
                                            Principally, there are many interesting possible choices for $S$.
                                            For example, there is the case of positive characteristic, i.e. $S$ might be chosen as the spectrum
                                            of a (finite) field or as function field of an algebraic curve over a finite field. However,
                                            in those models $( X \to S , g )$, which will be studied within the bounds of this
                                            thesis, we will often choose $S$ to be representable by an excellent
                                            Dedekind-scheme with field of fractions $K$ of characteristic
                                            zero (see Definition \ref{5022}). Then the following two cases are of interest:
                                            \begin{enumerate}
                                                \item
                                                $S$ is Zariski zero-dimensional and given by the spectrum
                                                $\Spec K$ of a field $K$ of characteristic zero.
                                                Especially in the case $K=\complex$, everything may be translated back into the
                                                language of manifolds (by Theorem \ref{5020}).

                                                \item
                                                $S$ is Zariski one-dimensional. In this case we are
                                                interested in the choice $S= \Spec {\footnotesize{\text{$\cal
                                                O$}}}_K$, where ${\footnotesize{\text{$\cal O$}}}_K \subset
                                                K$ is the ring of integral numbers of an algebraic number
                                                field $K$ (e.g. $K=\rat$ and ${\footnotesize{\text{$\cal O$}}}_K= \Ganz$).
                                            \end{enumerate}
                                        \end{Rem}
                                        But what is the physics behind the choice $S= \Spec
                                        {\footnotesize{\text{$\cal  O$}}}_K$? Why should we consider
                                        number fields instead of real or complex numbers? Following the
                                        ideas of B. Dragovich, V.S. Vladimirov, I.V. Volovich and many
                                        others (see also section \ref{bbbb}), let us state at least two
                                        arguments at this place. The first argument concerns the process
                                        of measurement. While it is not clear at all whether
                                        transcendental numbers can be the result of a measurement,
                                        integral (or rational) numbers can. Second, we know from Einstein
                                        that gravity is encoded in deformations of \st scales (described
                                        by means of the metrical tensor $g$). Looking at the energy scale
                                        that we experience, it is an empiric fact that we may assume that
                                        gravity is completely encoded in the archimedean scale and that
                                        non-archimedean, $\prim$-adic scales may be neglegted.
                                        Nevertheless, there is no reason why this should be true on all
                                        energy scales down to the Planck scale. It is an appealing project
                                        to study physical models where not only the ordinary, archimedean
                                        degrees of freedom are taken into consideration, but also the
                                        $\prim$-adic, non-archimedean degrees of freedom. This was one
                                        starting point for the adelic models which we cited in section
                                        \ref{bbbb}. Physically, the adelic approach means:
                                        \begin{align*}
                                            \textit{ \qquad \qquad There is one degree of freedom per primespot and dimension. \qquad \qquad} (*)
                                        \end{align*}
                                        As already indicated in Example \ref{5010}, the principle $(*)$ may as
                                        well be realized by considering algebraic spaces over
                                        ${\footnotesize{\text{$\cal O$}}}_K$. This motivates the following
                                        Definition \ref{5022} (whose physical motivation will be illustrated in
                                        Remark \ref{5023}). Recall that, given two relative algebraic spaces $X
                                        \to S$ and $Y \to S$, we denote by $X(Y)$ the set of $S$-morphisms
                                        $Y \to X$. Furthermore recall that for an algebraic space $\pi: X
                                        \to S$ we denotes the fibre of $\pi$ over the generic point of $S$
                                        by $X_K$ (physically this generic fibre represents the archimedean
                                        component of the algebraic space).

                                        \begin{Def}\label{5022}
                                            Let $S$ be an excellent Dedekind scheme with field of fractions $K$ of characteristic zero.
                                            Consider a pair $(X \to S, g)$ consisting of:
                                            \begin{enumerate}
                                                \item[$\bullet$]
                                                a smooth, separated algebraic space $\pi: X \to S$ over
                                                $S$

                                                \item[$\bullet$]
                                                a metric $g$ on $X$ (see Definition \ref{5101})
                                            \end{enumerate}
                                            such that the following conditions are fulfilled:
                                            \begin{enumerate}
                                                \item[(i)]
                                                $g$ satisfies the Einstein equations \ref{5121}.

                                                \item[(ii)]
                                                For each smooth algebraic space $Y \to S$ and each
                                                $K$-morphism $u_K:Y_K \to X_K$ there is an $S$-morphism $u:Y \to X$
                                                extending $u_K$.
                                            \end{enumerate}
                                            Then the pair $(X \to S, g)$ is called a \emph{\GRo}.
                                        \end{Def}
                                        \begin{Cor}\label{5022c}
                                            In the setting of Definition \ref{5022}, let us assume that the algebraic
                                            space $\pi: X \to S$ is representable by a smooth and separated
                                            $S$-scheme. Then, the morphism $u$ in Definition \ref{5022} b) is
                                            uniquely determined, i.e. $X \to S$ is the \emph{N\'eron model} of
                                            its generic fibre $X_K$ (see Definition \ref{0066}). In particular, the
                                            following statements hold:
                                            \begin{enumerate}
                                                \item
                                                If $u_K$ is an isomorphism so is $u$.

                                                \item
                                                For each \'etale $S$-scheme $S'$ with field of fractions
                                                $K'$ the canonical map $X(S') \to X_K(K')$ is bijective.
                                                %
                                            \end{enumerate}
                                        \end{Cor}\label{5022c}
                                        \begin{proof}
                                            In order to prove the uniqueness assertion let us choose two
                                            morphisms $u,v$ extending $u_K$. Using the separatedness of $X \to S$
                                            we conclude from \Liu, Prop. 3.3.11, that $u$ and $v$ are equal if they
                                            coincide on a dense subset of $Y$. Therefore, it suffices to
                                            show that the generic fibre $Y_K$ of $Y$ is dense in $Y$.
                                            This may be done as follows: Due to smoothness, the structure morphism $f:Y \to S$ is an
                                            open map of topological spaces (use \BLR, Prop. 2.4/8 and \EGAvz,
                                            $2.4.6$). The openness of $f$ implies that the pre-image $f^{-1}(D)$ of any
                                            dense subset $D$ of $S$ is dense in $Y$. Therefore, we are done, because the generic point
                                            of $S$ is dense in $S$. Consequently, $X \to S$ is
                                            the N\'eron model of its generic fibre.

                                            The statements a) and b) follow directly from the universal property of N\'eron
                                            models. For example, choose $Y=S'$ in order to see b).
                                        \end{proof}

                                        \begin{Rem}\label{5023}
                                            If $S=\Spec K$ is the spectrum of a field $K$,
                                            condition (ii) of Definition \ref{5022} is empty. If furthermore $K=\real$,
                                            any \GR induces a solution of Einstein's theory of general relativity (by evaluation at $\real$-valued points).
                                            This explains the label \emph{\GRo}, because (GR) shall remind of general
                                            relativity.
                                            However, in the case $S= \Spec {\footnotesize{\text{$\cal O$}}}_K$
                                            we arrive at the following physical interpretation:
                                            \begin{enumerate}
                                                \item[$\bullet$]
                                                Condition (ii) implements the ``adelic'' point of view.

                                                In order to see this, let us choose $S= \Spec \Ganz$ and therefore
                                                $K=\rat$. Recall that the generic fibre $X_K$ of $X$
                                                represents the archimedean component. Then condition (ii) says
                                                that the archimedean world is only the projection from the
                                                ``adelic'' level to the archimedean component.
                                                In truth, everything is defined over all prime spots,
                                                and there is one degree of freedom per prime spot.

                                                \item[$\bullet$]
                                                We saw in Corallary 0.5 that condition (ii) implies a canonical bijection $X_K(K)=X(S)$. Recall that
                                                $X_K(K)$ is the set of archimedean points, and that $X(S)$ is
                                                the set of ``adelic'' points.
                                                In the special case $K=\rat$, the  bijection $X_K(K) \cong X(S)$ means exactly that
                                                every archimedean point $x_{\infty} \in X_K(K)$ of
                                                $X$ is in truth only the archimedean element $x_{\infty}$ of an infinite
                                                set of points $ x = \{ x_2, \ldots, x_p, \ldots, x_{\infty} \}  \in X(S)$.
                                                Finally, Corallary 0.5 a) reflects the physically crucial statement
                                                that any ``deformation'' of the archimedean component by
                                                means of isomorphisms extends to the ``adelic'' level.
                                            \end{enumerate}
                                        \end{Rem}
                                        Furthermore, we  immediately obtain the interesting result that
                                        the pair $(X \to S, g)$ cannot be the flat Minkowski \st if we are
                                        in the ``adelic'' situation $S= \Spec {\footnotesize{\text{$\cal
                                        O$}}}_K$.

                                        \begin{proof}
                                            Let $S= \Spec {\footnotesize{\text{$\cal O$}}}_K$ and assume
                                            that $(X \to S, g)$ describes the flat, topologically trivial Minkowski \sto. Then
                                            \begin{enumerate}
                                                \item[$\bullet$]
                                                $g= \text{diag}( \pm 1, \pm 1, \pm 1, \pm 1)$ \qquad and

                                                \item[$\bullet$]
                                                $X = \Affin_S^n$ or $X= \mathbb{P}_S^n$ depending on
                                                whether we work projective or not.\footnote{Recall that
                                                the affine space $\Affin_S^n$ may be regarded as the algebraic geometric
                                                analogue of  flat space. In order to see this, let $S= \Spec R$
                                                be the spectrum of a commutative ring $R$. Then
                                                $\Affin_S^n = \Spec R[T_1, \ldots, T_n]$ is the spectrum
                                                of a polynomial ring in $n$ variables. Consequently, $\Affin_S^n(S) =
                                                \text{Hom}_{R}(R[T_1, \ldots, T_n],R)= R^n$.
                                                In the special case $R= \mathbb{K}$, $\mathbb{K}=\mathbb{R}, \mathbb{C}$,
                                                the \st induced by  $\Affin_S^n$ is the flat manifold $\Affin_S^n(S) =
                                                \mathbb{K}^n$.}
                                            \end{enumerate}
                                            But if $X = \Affin_S^n$, then $K^n \cong X_K(K) \neq X(S) \cong {\footnotesize{\text{$\cal O$}}}_K^n$, and if $X=
                                            \mathbb{P}_S^n$, not every morphism $u_K$ extends to an
                                            $S$-morphism $u$ (see Example \ref{2005}). Therefore, the flat, topologically trivial Minkowski \st is
                                            impossible.
                                        \end{proof}
                                        In chapter \ref{0080}, we will study a particularly simple class
                                        of \GRs which we call \SRso.

                                        \begin{Def}\label{5024}
                                            Let $(X \to S, g)$ be a \GR in the
                                            sense of Definition \ref{5022}, and let $X_K$ be the generic fibre of
                                            $X$. Then the pair $(X \to S, g)$ is called  a
                                            \emph{\SRo}, if in addition the
                                            following condition holds:
                                            \begin{enumerate}
                                                \item[(iii)]
                                                $X_K$ is a commutative $K$-group (see Definition \ref{1204}).
                                            \end{enumerate}
                                            More precisely, the $K$-group $X_K$ should be considered as
                                            a $K$-torsor under $X_K$ (see Definition \ref{1413}). The latter means that
                                            the special choice of a zero element of the group is forgotten
                                            as it should be for physical reasons.
                                        \end{Def}
                                        In order to generalize this notion slightly, one may also admit
                                        $K$-torsors $X_K$ under $K$-groups $G_K \neq X_K$. However, we
                                        will restrict attention to the case $G_K = X_K$.

                                        Definition \ref{5024} is motivated by \emph{special relativity}
                                        with {electromagnetism}: The Minkowski \st of special relativity
                                        naturally carries an additive, commutative group structure, and
                                        the gauge group of electromagnetism is commutative, too. This
                                        explains the label \emph{\SRo}, because (SR) shall remind of
                                        special relativity. Due to the additional condition (iii), \SRs
                                        are simpler than \GRos, and one can prove many properties of \SRs
                                        in an abstract manner without fixing a special model. More
                                        precisely, we will see in chapter \ref{0080} that the following
                                        statements are true for \emph{all} \SRso.

                                        \begin{PropSpecModGrav}\label{5025}
                                            Let $(X \to S, g)$ be a \SRo. Then
                                            the following statements are true:
                                            \begin{enumerate}
                                                \item
                                                $X \to S$ is \'etale-invariant. More precisely, this statement means the following:
                                                Let $\varphi:X \to X$ be an \'etale
                                                $S$-morhpism, and let $(X' \to
                                                S'$, $g')$ be the pair obtained from $(X \to S,g)$ by base
                                                change with an \'etale morphism $S' \to S$. Then, $(X \to
                                                S, \varphi^*g)$ and $(X' \to S'$, $g')$ are \SRos, too (see section \ref{0081}).

                                                \item
                                                $X$ cannot be the flat, topologically trivial Minkowski space
                                                (see section \ref{0102} for more details).

                                                \item
                                                The archimedean component $X_K(K)$ is bounded with respect to all $\prim$-adic
                                                norms.
                                                In the special case $K=\rat$ and under the assumption that
                                                there is a closed immersion $X_K
                                                \hookrightarrow \Affin_K^n$,  this is the following statement:
                                                For each prime number $p$, the $p$-adic
                                                manifold $X_{K}(\rat_p)$ is a bounded subset of some $\rat_p^n$
                                                with respect to the canonical $p$-adic norm $| \cdot |_p$.

                                                \begin{proof}
                                                    We know that $X_K$
                                                    possesses a global N\'eron model. Consequently, the local N\'eron
                                                    models exist (see Proposition \ref{0098}). Therefore, due to Proposition \ref{0105},
                                                    it is necessary that $X_K(K)$ or even the continuum
                                                    $X_K(\hat{K})$ is bounded
                                                    (see section \ref{0132} for details and the notion of bounded).
                                                \end{proof}

                                                \item
                                                The archimedean component $X_K(K)$ carries a discrete
                                                geometry. More precisely, Theorem \ref{2001a} tells us that
                                                $X_K(K)$ is a finitely generated abelian group, i.e.
                                                \begin{align*}
                                                    X_K(K) \cong \Ganz^d \oplus \Ganz/{(p_1^{\nu_1})} \oplus \cdots
                                                    \oplus \Ganz/{(p_s^{\nu_s})}
                                                \end{align*}
                                                for some prime numbers $p_i \in \Natural$ and integers $d,s,\nu_i
                                                \in \Natural$. In the special case $d=0$,
                                                $X_K(K)$ consists of only finitely many points.
                                                For a more detailed exposition we refer to chapter \ref{0300}.

                                                The finiteness of the set $X_K(K)$ would also be desirable in
                                                the case that $(X \to S,g)$ is a \GRo. However, it is
                                                unequally harder to see in how far this should be true.
                                                Nevertheless, motivated from what we already know about
                                                \SRos, it does not seem senseless to demand the finiteness
                                                of the set
                                                $X_K(K)$ in the case of \GRos.

                                                \item
                                                The privileged character of four-dimensional spaces: Let
                                                us consider pure gravity which is described by
                                                means of a euclidian metrical tensor $g_{\mu \nu}$. Over the
                                                field $\real$, the metric takes the form diag$(1, \ldots ,1)$
                                                in inertial systems.  Interpreting the metric as a
                                                quadratic form $g: \mathbb{R}^n \times \mathbb{R}^n \to \mathbb{R}^+$, we observe two fundamental
                                                properties:
                                                \begin{enumerate}
                                                    \item[$\bullet$]
                                                    anisotropy: $g(v,v)=0$ if and only if $v=0$.

                                                    \item[$\bullet$]
                                                    surjectivity: For all positive $r\in \real$ there exists a $v \in
                                                    \real^n$ such that $g(v,v)=r$. Physically, this means that
                                                    scales do not have holes.
                                                \end{enumerate}
                                                We will see in section \ref{0202b} that metrics over
                                                non-archimedean fields which fulfill the non-archimedean analogues of these two properties
                                                only exist in four dimensional spaces.
                                                Therefore, also in the case of number fields, we
                                                are naturally let to four dimensional spaces by
                                                means of the theorem of Hasse-Minkowski.

                                                Additionally, in the case of number fields, we will see in section \ref{0202b}
                                                that the archimedean tangent space
                                                $T_{X_K/K}(x_{\infty})$ at an archimedean point $x_{\infty} \in
                                                X_K(K)$ naturally carries the structure of a quarternion
                                                algebra. This implies that we may write pairs  $({\cal A}, \overline{{\cal A}})$
                                                consisting of a gauge field ${\cal A}$ and its anti-field $\overline{{\cal A}}$
                                                in the form
                                                $
                                                    ({\cal A}, \overline{{\cal A}}) = \sum_{\mu=0}^3 {\cal A}^{\mu}
                                                    \gamma_{\mu}
                                                $
                                                with gamma matrices fulfilling the relations
                                                \begin{align*}
                                                    \gamma_{\mu}\gamma_{\nu}+\gamma_{\nu}\gamma_{\mu} = 2
                                                    g_{\mu \nu}.
                                                \end{align*}

                                                \item
                                                If we do not demand the quasi-compactness of the
                                                archimedean component $X_K$ of $X$, one can prove that $X_K$
                                                possesses a N\'eron model if and only if there is an exact
                                                sequence
                                                \begin{equation*}
                                                    \xymatrix{
                                                    0 \ar[r] & T_K \ar[r] & X_K \ar[r] & A_K \ar[r] & 0}
                                                \end{equation*}
                                                over some algebraic closure of $K$, where $T_K$ is an algebraic torus and $A_K$ is an abelian
                                                variety (see section \ref{0120}). While $A_K$ is $\prim$-adically bounded, $T_K$ is
                                                not. This obtrudes the interpretation of $A_K$ as space part
                                                and the interpretation of the torus $T_K$ as an
                                                internal, gauge group part which we therefore associate with
                                                electromagnetism. Thus, $X_K$  should appear as
                                                $A_K$-torsor under $T_K$ (which is the algebraic geometric
                                                analogue of the differential geometric principal bundle of
                                                gauge theory).

                                                \item
                                                On the ``adelic'' level there is some kind of entanglement
                                                of dimensions. For example, it is in general not possible to
                                                diagonalize the metric at the ``adelic'' points of
                                                $X(S)$ (see also Remark \ref{0142rem}).
                                            \end{enumerate}
                                        \end{PropSpecModGrav}
                                        Let us remark that the statements a) and b) are also true for
                                        \GRos.

                                        %
                                        %
                                        %
                                        %
                                        %
                                        %
                                        %
                                        %
                                        %
                                        %
                                        %
                                        %
                                        %
                                        %
                                        %
                                        %
                                        %
                                        %
                                        %
                                        %
                                        %
                                        %
                                        %
                                        %
                                        %
                                        %
                                        %
                                        %
                                        %
                                        %
                                        %
                                        %
                                        %
                                        %
                                        %
                                        %
                                        %
                                        %
                                        %
                                        %
                                        %
                                        %
                                        %
                                        %
                                        %
                                        %
                                        %
                                        %
                                        %
                                        %
                                        %
                                        %
                                        %
                                        %
                                        %
                                        %
                                        \section{The continuum limit - Comparison with the adelic model}\label{5030}

                                        Within this section let $(X \to S,g)$ be a \emph{\GR} (see
                                        Definition \ref{5022}), and assume for simplicity that $S= \Spec \Ganz$ and
                                        that $X\to S$ is representable by a smooth $S$-scheme. Then, on
                                        the fundamental level, \st is given by the set $X(S)$. The
                                        metrical situation at a point $\alpha \in X(S)$ is described by
                                        the pull-back metric $\alpha^*g$.

                                        The purpose of this section is to answer the following questions:
                                        \begin{enumerate}
                                            \item[$\bullet$]
                                            What is the archimedean, real continuum limit $\mathfrak{X}_{\infty}$ of $(X \to S,g)$?

                                            \item[$\bullet$]
                                            What are the non-archimedean continuum limits $\mathfrak{X}_p$ of $(X \to S,g)$?

                                            \item[$\bullet$]
                                            Does the family $\mathfrak{X}_{\nu}$, $\nu=2, 3, 5, \ldots, p, \ldots, \infty$,
                                            give rise to an adelic continuum limit $\mathfrak{X}$ of $(X \to S,g)$ which
                                            is adelic in the sense of B. Dragovich, I.V. Volovich et. al.?
                                        \end{enumerate}
                                        We will see that the last question has a positive answer.
                                        Therefore, our new approach via algebraic spaces presented within
                                        the bounds of this thesis fits exactly into the beautiful context
                                        of adelic, general relativistic (cosmological) models.

                                        \subsection*{The archimedean continuum limit $\mathfrak{X}_{\infty}$}\label{5040}

                                        As usual, let $X_K$ denote the generic fibre of $X$, and denote by
                                        $g_K$ the pull-back of $g$ under the canonical inclusion $X_K
                                        \hookrightarrow X$. Then, the archimedean limit is obtained by
                                        evaluation at $K$-valued points:
                                        \begin{align*}
                                            \Big( X_K(K), g_K \Big)
                                        \end{align*}
                                        More precisely, the metric $g_K$ at an arbitrary point $x_{} \in
                                        X_K(K)$ is by construction the symmetric bilinear form
                                        $g_K(x):=x^*g_K: K \times K \to K$ (see the beginning of section
                                        \ref{0202b}). The continuum limit is obtained by applying the
                                        pull-back with $i:X_{\real} \to X_{K}$, where $i$ is induced by
                                        the canonical map $ K \hookrightarrow \real$. It is therefore the
                                        pair
                                        \begin{align*}
                                            (\mathfrak{X}_{\infty},g_{\infty}):=( X_K(\real), i^*g_{K}).
                                        \end{align*}
                                        By construction, $(\mathfrak{X}_{\infty},g_{\infty})$ may be
                                        endowed with the structure of a differentiable Riemannian
                                        manifold, and in addition it is a solution of Einstein's classical
                                        theory of general relativity.

                                        As already indicated in \ref{5025}, d), it is natural to assume
                                        that $X_K(K)$ is a finite set. Then, $X_K(K)$ is a discrete subset
                                        of the continuum $\mathfrak{X}_{\infty}$ and is invariant under
                                        arbitrary $K$-isomorphisms $\varphi_K:X_K \stackrel{\sim}\to X_K$.
                                        Furthermore, $\varphi_K^*g_K$ is a solution of the Einstein
                                        equations for all $K$-isomorphisms $\varphi_K$. Therefore,
                                        $(X_K(K),g_K)$ describes metrical states on smallest scales. At
                                        distances much larger than these smallest lengths, it is
                                        admissible to work with the continuum
                                        $(\mathfrak{X}_{\infty},g_{\infty})$ instead.

                                        \subsection*{The non-archimedean continuum limits $\mathfrak{X}_{p}$}\label{5040}

                                        Let $s \in S$ be the Zariski closed point corresponding to a prime
                                        number $p$ (i.e. more precisely corresponding to the prime ideal
                                        $(p) \subset \Ganz$). Let $\widehat{R}:={\widehat{{\cal
                                        O}}_{S,s}}$ denote the completion of $R:={{\cal O}_{S,s}}$ with
                                        respect to its maximal ideal $s$. In our special case,
                                        $\widehat{R}=\Ganz_p$ is the ring of $p$-adic integers with field
                                        of fractions $\widehat{K}=\rat_p$. By base change with the
                                        canonical morphism $\Spec \widehat{R} \to \Spec R \to S$, $X$
                                        induces the $\widehat{R}$-scheme $X_{\widehat{R}}:=X \otimes_S
                                        \widehat{R}$ which consists of a generic fibre and a special fibre
                                        (the latter is the fibre over the point $s$). In order to take
                                        into account all infinitesimal neighborhoods of the special fibre,
                                        let us onsider the formal completion $\widehat{X}_{\widehat{R}}$
                                        of $X_{\widehat{R}}$ along its special fibre. Let us point to the
                                        fact that $\widehat{X}_{\widehat{R}}$ coincides topologically with
                                        the special fibre of $X_{\widehat{R}}$. Furthermore, the metric
                                        $g$ induces a metric $g_p$ on $\widehat{X}_{\widehat{R}}$. Writing
                                        $\mathfrak{X}_{p}:=\widehat{X}_{\widehat{R}}$ for the formal
                                        completion, we define the \emph{non-archimedean continuum limit}
                                        of $(X \to S,g)$ {over $p$} as the pair
                                        \begin{align*}
                                            (\mathfrak{X}_{p},g_{p}).
                                        \end{align*}
                                        In order to illustrate the structure of the non-archimedean
                                        continuum limit, let us recall the notion of \emph{formal
                                        completion of a scheme} $Y$ \emph{along a closed subscheme} $Z
                                        \subset Y$. So $Z$ is defined by a quasi-coherent ideal $\ci
                                        \subset \co_Y$. Then consider the sheaf $\co_Z$ obtained from
                                        restricting the projective limit $\stackrellow{\longleftarrow
                                        n}{\lim} \co_Y/\ci^n$ to $Z$. It follows that $(Z, \co_Z)$ is a
                                        locally ringed topological space, the desired formal completion
                                        $\widehat{Y}$ of $Y$ along $Z$. Locally, the construction looks as
                                        follows: Let $Y=\Spec A$ and assume that $\ci$ is associated to
                                        the ideal $\mathfrak{i} \subset A$. Then, $(Z, \co_Z)= \text{Spf}
                                        \stackrellow{\longleftarrow n}{\lim} A/{\mathfrak{i}^n}=
                                        \text{Spf} \, \hat{A}$, where $\hat{A}$ denotes the
                                        $\mathfrak{i}$-adic completion of $A$.

                                        \subsection*{The adelic  continuum limit $\mathfrak{X}$}\label{5040}

                                        As illustrated above, the {\GR} $(X \to S,g)$ induces a family
                                        \begin{equation*}
                                                \xymatrix{
                                                &  &    \ar@{-}@/^/[ddd] &  & \ar@{-}@/^/[ddd] & & \ar@{{-}}@/^/[ddd] \\
                                                &   & \ & \cdots & \ & \cdots &  & & \\
                                                & &   &  \cdots &   &  \cdots & & &  \\
                                                &  \ & \ & \ &  & & \\
                                                &  &  (\mathfrak{X}_2, g_2) & \cdots & (\mathfrak{X}_p, g_p) & \cdots &
                                                (\mathfrak{X}_{ \ \infty }, g_{ \ \infty }) }
                                        \end{equation*}
                                        where $(\mathfrak{X}_{ \ \infty }, g_{ \ \infty })$ is a real
                                        manifold which is a solution of Einstein's ordinary theory of
                                        general relativity. Furthermore, each object $(\mathfrak{X}_{p},
                                        g_{p})$ lives (as topological space) over a finite prime spot and
                                        also solves the corresponding Einstein's equation.

                                        We will illustrate now that this family canonically induces an
                                        adelic object $\mathfrak{X}$ which is the adelic continuum limit
                                        of $(X \to S,g)$ in the sense of B. Dragovich, I.V. Volovich et.
                                        al. For this purpose, consider the ``generic fibre''
                                        $\mathfrak{X}_{p,
                                        \widehat{K}}:=\widehat{X}_{\widehat{K}}:=\widehat{X}_{\widehat{R}}
                                        \otimes_{\widehat{R}} \widehat{K}$ of
                                        $\mathfrak{X}_{p}=\widehat{X}_{\widehat{R}}$. This is a
                                        rigid-anaytic space, and the metric $g$ induces a metric $g_{p,
                                        \widehat{K}}$ on $\mathfrak{X}_{p, \widehat{K}}$ which is a
                                        solution of the  $p$-adic Einstein equations. The pair
                                        \begin{align*}
                                            (\mathfrak{X}_{p, \widehat{K}},g_{p, \widehat{K}}).
                                        \end{align*}
                                        will be called the \emph{$p$-adic continuum limit} of $(X \to
                                        S,g)$. Next, let us consider the direct product
                                        \begin{align*}
                                            \mathfrak{X}:=  \Big( \prod_p \mathfrak{X}_{p, \widehat{K}} \Big) \times \mathfrak{X}_{\infty}.
                                        \end{align*}
                                        We claim that $\mathfrak{X}$ is the desired adelic object. In
                                        order to illustrate this, let us fix some prime number $p$ and let
                                        us study the structure of the $p$-adic continuum limits
                                        $\mathfrak{X}_{p, \widehat{K}}$ in more detail. Thereby, we make
                                        use of the notation which was already introduced in the previous
                                        subsection ``The non-archimedean continuum limits
                                        $\mathfrak{X}_{p}$''.

                                        First of all, consider the polynomial ring $A=\widehat{R}[\zeta]$
                                        over $\widehat{R}$, where $\zeta$ denotes a set of variables
                                        $\zeta=(\zeta_1, \ldots, \zeta_n)$, and where $\mathfrak{i}=(p)$
                                        is the maximal ideal of $\widehat{R}$. So let $Y$ be the
                                        $n$-dimensional affine space $\Affin_{\widehat{R}}^n$ over
                                        ${\widehat{R}}$, and let $Z$ be its special fibre. Then the formal
                                        completion $\widehat{Y}$ of $Y$ along $Z$ yields the formal affine
                                        $n$-space $\text{Spf} \, \widehat{R} \langle \zeta \rangle$.
                                        Thereby, the $\widehat{R}$-algebra $\widehat{R} \langle \zeta
                                        \rangle$ of restricted power series in the variables $\zeta_1,
                                        \ldots, \zeta_n$ is defined as the subalgebra of the
                                        $\widehat{R}$-algebra $\widehat{R} [[\zeta]]$ of formal power
                                        series, consisting of all series $\sum_{\nu \in \Natural^n}
                                        c_{\nu} \zeta^n$ with coefficients $c_{\nu} \in \widehat{R}$
                                        constituting a zero sequence in $\widehat{R}$ (with respect to the
                                        $p$-adic topology on $\widehat{R}$). The ``generic fibre''
                                        $\widehat{Y}_{\widehat{K}}$ of $\widehat{Y}$ is the affinoid
                                        $\widehat{K}$-space $\text{Sp} (\widehat{R} \langle \zeta \rangle
                                        \otimes_{\widehat{R}} \widehat{K})= \text{Sp} \widehat{K} \langle
                                        \zeta \rangle$ which (set-theoretically) coincides with the set of
                                        maximal ideals of $\widehat{K} \langle \zeta \rangle$.
                                        Consequently, $\widehat{Y}_{\widehat{K}} \subset
                                        Y^{\text{an}}_{\widehat{K}} \cong Y(\widehat{K}) \cong
                                        {\widehat{K}}^n$ coincides with the unit ball
                                        $\mathbb{B}_{{\widehat{K}}}^n:=\{x \in {\widehat{K}}^n \mid |x|_p
                                        \leq 1 \}$ (where $Y^{\text{an}}$ denote the rigid-analytification
                                        of $Y_{\widehat{K}}:=Y \otimes_{\widehat{R}} {\widehat{K}}$). In
                                        the special case $n=1$, $\widehat{Y}_{\widehat{K}}$ may actually
                                        be identified with the $p$-adic integers $\Ganz_p$.

                                        Let us now return to our scheme $X_{\widehat{R}}$, but let us
                                        first assume that it is defined by an ideal $\mathfrak{a} \subset
                                        \widehat{R}[\zeta]$, where $\zeta$ denotes as above a set of
                                        variables $\zeta=(\zeta_1, \ldots, \zeta_n)$. Then one can show
                                        that $\widehat{X}_{\widehat{R}} = \text{Spf} \, \widehat{R}
                                        \langle \zeta \rangle / (\mathfrak{a})$. Thus, we find that
                                        $\mathfrak{X}_{p,{\widehat{K}}}= \text{Sp} \, \widehat{K} \langle
                                        \zeta \rangle / (\mathfrak{a}) \subset
                                        \mathbb{B}_{{\widehat{K}}}^n$, i.e. set theoretically we may
                                        write:
                                        \begin{align*}
                                            \mathfrak{X}_{p,{\widehat{K}}} = X_K(\widehat{K}) \cap
                                            \mathbb{B}_{{\widehat{K}}}^n \subset \Ganz_p^n.
                                        \end{align*}
                                        However, it is a general phenomenon that the generic fibre
                                        $\widehat{X}_{\widehat{K}}:=\widehat{X}_{\widehat{R}}
                                        \otimes_{\widehat{R}} \widehat{K}$ should be viewed as an open
                                        subspace of the rigid analytification
                                        ${X}_{\widehat{K}}^{\text{an}}$ of the scheme
                                        ${X}_{\widehat{K}}:={X}_{\widehat{R}} \otimes_{\widehat{R}}
                                        \widehat{K}$. Consequently, the direct product
                                        \begin{align*}
                                            \mathfrak{X}
                                            :=  \Big( \prod_p \mathfrak{X}_{p,{\widehat{K}}}  \Big) \times \mathfrak{X}_{\infty}
                                            \subset \Big( \prod_p \Ganz_p^n  \Big) \times \real^n \subset
                                            \Affin^n
                                        \end{align*}
                                        is a subset of some power of the ring
                                        $\Affin_{\Ganz}:=\widehat{\Ganz} \times \real \subset \Affin$ of
                                        adeles.

                                        Finally, in the general case, we can find a covering of $X$ by
                                        open subsets $U_i$, $i \in I$, such that there are $S$-immersions
                                        $U_i \hookrightarrow \Affin_S^n$ for all $i \in I$. Let
                                        $\mathfrak{X}_{\nu}$ (resp. $\mathfrak{U}_{i,\nu}$) be the
                                        respective continuum limits of $X$ (resp. $U_i$), where $\nu=2, 3,
                                        5, \ldots, p, \ldots, \infty$. Let $\mathfrak{X}=( \prod_p
                                        \mathfrak{X}_{p,{\widehat{K}}} ) \times \mathfrak{X}_{\infty}$ and
                                        $\mathfrak{U_i}=( \prod_p \mathfrak{U}_{i,p,{\widehat{K}}} )
                                        \times \mathfrak{U}_{i,\infty}$. By the above reasoning,
                                        $\mathfrak{U_i}$ is contained in an $n$-dimensional ring of adeles
                                        $\Affin^n$. Therefore, $\mathfrak{X}$ is obtained by gluing
                                        subsets of $n$-dimensional rings of adeles. Consequently, it is
                                        indeed some kind of ``adelic manifold''. In particular, we find
                                        that the adelic \st $\mathfrak{X}$ cannot be the flat,
                                        topologically trivial adele $(\Affin_{\Ganz})^n$. This follows
                                        from the fact that on the one hand $X$ cannot be the affine space
                                        $\Affin_S^n$ over $S$ (see the statement below Remark \ref{5023}), but
                                        that on the other hand $\mathfrak{X}=(\Affin_{\Ganz})^n$ if and
                                        only if $X=\Affin_S^n$.

                                        All in all, this shows that our arithmetic geometric approach to
                                        general relativity completely fits into the well established and
                                        beautiful setting of adelic physics.

\part{Arithmetic theory of general relativity}

\chapter{Einstein's equation in the setting of algebraic
spaces}\label{0010}

\section{Smoothness in algebraic geometry}\label{0011s}

In this section we will recapitulate some basic concepts of
differential calculus for schemes. We introduce notions like
unramified, \'etale and smooth morphisms and illustrate how these
concepts generalize structures known from differential geometry.
The aim is to perform general relativity in the setting of smooth
algebraic spaces instead of smooth manifolds. The algebraic
geometric Einstein equations will finally be derived in section
\ref{5100}. Using Grothendieck´s language of schemes, our
exposition on smoothness will follow \BLR, and we refer the reader
to chapter 2 of this book for more details.
\\ \\
In the following let $S$ be a base scheme and $X$ an $S$-scheme
(i.e. $X$ is a scheme together with a unique morphism $X \to S$).
Furthermore let $R$ be a commutative ring with neutral element
$1$, and let $A$ be an $R$-algebra. The reader, who is not so much
familiar with the notion of global schemes, may for simplicity
think of $X= \Spec A$ and $S= \Spec R$. As we are doing algebraic
geometry, we have to perform differential calculus with purely
algebraic methods. This is done by using so called derivations.

\begin{Def}\label{0012}
    \begin{enumerate}
        \item
        An $R$-derivation of $A$ into an $A$-module $M$ is an
        $R$-linear map $d:A \to M$ such that
        \begin{align*}
            d(fg)=f \cdot d(g)+g \cdot d(f) \quad \text{for all} \ f,g
            \in A
        \end{align*}
        The $A$-module of all $R$-derivations of $A$ into an $A$-module
        $M$ is denoted by $\Der_R(A,M)$.
        \item
        The module of \emph{relative differential forms (of degree
        one)} of $A$ over $R$ is an $A$-module $\Omega_{A/R}^1$ together
        with an $R$-derivation $d_{A/R}:A \to \Omega_{A/R}^1$
        which has the following universal property: For each
        $A$-module $M$, the canonical map
        \begin{equation*}
            \xymatrix{\Hom_A \left(\Omega_{A/R}^1,M \right) \ar[r]^{\thicksim \  \quad \quad}
            & \Der_R(A,M), \quad \varphi \mapsto \varphi \circ d_{A/R}}
        \end{equation*}
        is bijective.
    \end{enumerate}
\end{Def}
\begin{Rem}\label{0013} $\Omega_{A/R}^1$ exists and its universal property implies:
    \begin{enumerate}
        \item
        $\Omega_{A/R}^1$ is unique up to canonical isomorphism.
        \item
        Each morphism $\phi:A \to B$ of $R$-algebras induces a
        canonical $B$-linear map
        \begin{align*}
            \Omega_{A/R}^1 \otimes_A B \to \Omega_{B/R}^1, \quad f
            \cdot d_{A/R}(g) \otimes h & \mapsto \phi(f) \cdot h
            \cdot d_{B/R}(\phi(g))
        \end{align*}
        \item
        Considering $A$-derivations of $B$ as $R$-derivations
        yields a $B$-linear map
        \begin{align*}
            \Omega_{B/R}^1  \to \Omega_{B/A}^1, \quad d_{B/R}(g) & \mapsto
            d_{B/A}(g).
        \end{align*}
    \end{enumerate}
\end{Rem} \
\ \\
There is also a global notion of modules of differentials in terms
of sheaves over schemes which has similar functorial properties.
If $X$ is locally of finite type over $S$ there is a
quasi-coherent $\Ox$-module $\Omega_{X/S}^1$. It is called the
\emph{sheaf of relative differential forms (of degree 1)}.
Furthermore we have a canonical $\Ox$-linear map $d_{X/S}:\Ox \to
\Omega_{X/S}^1$, the \emph{exterior differential}.

Since $\Omega_{X/S}^1$ is quasi-coherent, $\left( \Omega_{X/S}^1,
d_{X/S}\right)$ can be described in local terms: For each open
affine subset $V= \Spec R$ of $S$ and for each open affine subset
$U= \Spec A$ of $X$ lying over $V$, the sheaf $\Omega_{X/S}^1
\mid_U$ is the quasi-coherent $\Ox \mid_U$-module associated to
the $A$-module $\Omega_{A/R}^1$, and the map $d_{X/S} \mid_U$ is
associated to the canonical map $d_{A/R}:A \to \Omega_{A/R}^1$. \\
\\
An important operation, which we will frequently make use of, is
the \emph{pull-back} of differential forms.

\begin{Def}\label{0014}
    \textbf{(pull-back of relative differential forms)} Let $f:X \to
    Y$ be an $S$-morphism. Then the map in Remark \ref{0013}, b) gives
    rise to a canonical $\Ox$-morphism
    \begin{align*}
        f^*\Omega_{Y/S}^1 \to \Omega_{X/S}^1.
    \end{align*}
    Each section $\omega$ of $\Omega_{Y/S}^1$ gives rise to a
    section $\omega'$ of $f^*\Omega_{Y/S}^1$ which is mapped to a
    section $f^*\omega$ of $\Omega_{X/S}^1$ under the
    above map. $f^*\omega$ is called the pull-back of $\omega$.
\end{Def}
Let us convince ourselves that this notion of pull-back reduces to
the one given in differential geometry. \\ \\
\underline{Physical interpretation:} Let $S= \Spec \real$, $X=
\Affin_{\real}^m$, $Y= \Affin_{\real}^n$ and let $f$ be the map
    \begin{align*}
        f:\Affin_{\real}^m \to \Affin_{\real}^n, \quad
        T:=\left(T_1, ..., T_m \right)^t \mapsto  p(T):=\left(p_1(T), \ldots , p_n(T)
        \right)^t,
    \end{align*}
where $T$ denotes a set of variables, and where $p_i, \ i=1,
\ldots , n$, are polynomials in $T$. Thus, on $\real$-valued
points $x \in X(\real)= \real^m$, $f$ is simply the map
    \begin{align*}
        f(\real):\real^m \to \real^n, \quad
        x:=(x_1, \ldots, x_m)^t \mapsto  p(x):=\left(p_1(x), \ldots , p_n(x)
        \right)^t.
    \end{align*}
On its ring of global sections, $f$ corresponds to the
$\real$-algebra homomorphism
\begin{align*}
    f^*: \real \left[S_1, \ldots, S_n \right] &\to \real \left[T_1, \ldots, T_m
    \right] = \real \left[T \right]. \\
    S_i &\mapsto p_i(T)
\end{align*}
Writing more compactly $S:= \left(S_1, \ldots, S_n \right)^t$ as a
set of variables $S_i$, we can write $f^*$ simply as:
\begin{align*}
    f^*(S)=p(T).
\end{align*}
This can be seen as follows: By Taylor expansion of $p$ around a
$\real$-valued point $x\in X(\real)=\real^m$ we find a matrix
$q(T) \in \mathrm{Mat}_{\real[T]}(n \times m) $ such that
\begin{align*}
    p(T)-p(x) &=q(T) \cdot (T-x).
\end{align*}
Therefore $ (f^*)^{-1}\left((T-x) \cdot \real[T] \right) =
(f^*)^{-1}\left((p(T)-p(x))\cdot \real[T]\right) = (S-p(x))\cdot
\real[S]$, i.e. the $\real$-valued point $x$ is mapped by $f$ to
the $\real$-valued point $p(x)$ as desired. Now going into the
local description  of the pull-back map, we find that we have the
following canonical commutative diagram.
\begin{equation*}
    \xymatrix{f^*\Omega_{Y/S}^1 = f^*\Omega_{\Affin_{\real}^n}^1 \quad \ar[rr] \ar@{=}[d] &&
    \Omega_{\Affin_{\real}^m}^1 = \Omega_{X/S}^1 \ar@{=}[d] \\
    \quad \Omega_{\Affin_{\real}^n}^1 \otimes_{\real[S]} \real[T] \quad \ar[rr] \ar@{=}[d] &&
    \quad \quad \Omega_{\Affin_{\real}^m}^1 \quad \quad \ar@{=}[d] \\
    \left( \bigoplus_{i=1}^n \real[S] \cdot dS_i \right) \otimes_{\real[S]} \real[T]  \ar[rr] \ar@{=}[d] &&
    \bigoplus_{i=1}^m \real[T] \cdot dT_i  \ar@{=}[d] \\
    \quad \bigoplus_{i=1}^n \real[T] \cdot dS_i \quad \ar[rr] && \bigoplus_{i=1}^m \real[T] \cdot
    dT_i \ ,  & dS_i \ar@{|->}[r] &  \sum_{j=1}^m \frac{\partial p_i}{\partial
    T_j} \cdot dT_j
    }
\end{equation*}
Thus the pull-back of a differential form $\omega = h(S) \cdot dS
\in \Omega_{Y/S}^1 = \Omega_{\Affin_{\real}^n}^1$ is
\begin{align*}
    f^*\omega &= h(f^*(S)) \cdot f^*(dS) = h(f^*(S)) \cdot
    d(f^*(S))= h(p(T)) \cdot d(p(T)) \\
    &= h(p(T)) \cdot \left[ \left( \frac{\partial p_i}{\partial
    T_j} \right) \cdot dT \right],
\end{align*}
and on $\real$-valued points $x \in X(\real)= \real^m$ we get
\begin{equation*}
    (f^*(h(S)dS))(x)= (h \circ p)(x) \cdot \left[ (Dp)(x) \cdot dx
    \right],
\end{equation*}
where $Dp$ denotes the Jacobian matrix of $p=f(\real): \real^m \to
\real^n$. This is just the notion of pull-back of differential
forms we are used to from differential geometry. \\ \\
There are some exact sequences which are induced by the maps
defined in Remark \ref{0013}. As we will make use of them later
let us state them here. For proofs we refer to \EGAvv, 16.4.
\begin{Satz}\label{0015}
    Let $f:X \rightarrow Y$ be an $S$-morphism. Then the canonical
    sequence of $\Ox$-modules
    \begin{equation*}
        \xymatrix{f^*\Omega_{Y/S}^1 \ar[r] &
        \Omega_{X/S}^1 \ar[r] &
        \Omega_{X/Y}^1 \ar[r] & 0}
    \end{equation*}
    is exact.
\end{Satz}
\begin{Satz}\label{0016}
    Let $j:Z \hookrightarrow X$ be an immersion of $S$-schemes. Let $\cJ$ be the sheaf of
    ideals defining $Z$ as subscheme
    of $X$. Then the canonical sequence of $\Oz$-modules
    \begin{equation*}
        \xymatrix{\cJ/\cJ^2 \ar[r]^{\delta} &
        j^*\Omega_{X/S}^1 \ar[r] &
        \Omega_{Z/S}^1 \ar[r] & 0}
    \end{equation*}
    is exact, where $\delta$ is locally given by
    the map $\overline{a} \mapsto d(a) \otimes 1, \ \overline{a} \equiv a \ \mathfrak{mod} \cJ^2$.
\end{Satz}
\begin{Satz}\label{0017}
    Let $X$ and $S'$ be $S$-schemes. Let $X':=X \times_S S'$ be the $S$-scheme obtained by base change, and let
    $p:X' \rightarrow X$ be the projection. Then the canonical map
    \begin{equation*}
        \xymatrix{p^*\Omega_{X/S}^1  \ar[r] & \Omega_{X'/S'}^1}
    \end{equation*}
    is an isomorphism.
\end{Satz}
\begin{Satz}\label{0018}
    Let $X_1$ and $X_2$ be $S$-schemes. If $ p_i:X_1 \times_S X_2 \rightarrow X_i$ are the projections for $i=1,2$,
    the canonical map
    \begin{equation*}
        \xymatrix{p_1^*\Omega_{X_1/S}^1 \oplus p_2^*\Omega_{X_2/S}^1  \ar[r]^{\quad \quad \sim} &
        \Omega_{X_1 \times_S X_2/S}^1}
    \end{equation*}
    is an isomorphism.
\end{Satz}
Let us now explain the concepts of unramified, \'etale and smooth
morphisms. These morphisms will provide us the analogues of
typical constructions from differential geometry like submersions
and immersions.
\begin{Def}\label{0019}
    A morphism of schemes $f:X \to S$ is called \emph{unramified} at a
    point $x\in X$ if there exist an open neighborhood $U$ of $x$
    and an $S$-immersion
    \begin{equation*}
        j:U \hookrightarrow \Affin_S^n
    \end{equation*}
    of $U$ into some linear space $\Affin_S^n$ over $S$ such that
    the following conditions are satisfied:
    \begin{enumerate}
        \item
        locally at $j(x)$ (i.e., in an open neighborhood of
        $j(x)$), the sheaf of ideals $\cJ$ defining $j(U)$  as a
        subscheme of $\Affin_S^n$ is generated by finitely many
        sections.
        \item
        the differential forms of type $dg$ with sections g of
        $\cJ$ generate $\Omega_{\Affin_S^n/S}^1$ at $j(x)$.
    \end{enumerate}
    The morphism $f:X \to S$ is called unramified if it is
    unramified at all points of $X$.
\end{Def}
\begin{Satz}\label{0020}
    Let $f:X \to S$ be locally of finite presentation, let $x \in X$ and $s:=f(x)$. Then the following conditions are
    equivalent:
    \begin{enumerate}
        \item
        $f$ is unramified at x.
        \item
        $\Omega_{X/S,x}^1 = 0$.
        \item
        The diagonal morphism $\Delta: X \to X \times_S X$ is a
        local isomorphism at $x$.
        \item
        The maximal ideal $\maxi_x$ of ${\mathcal O}_{X,x}$ is
        generated by the maximal ideal $\maxi_s$ of ${\mathcal
        O}_{S,s}$ and $k(x)$ is a finite separable extension of
        $k(s)$.
    \end{enumerate}
    The morphism $f:X \to S$ is called unramified if it is
    unramified at all points of $X$.
\end{Satz}
Before defining the central notion of smooth morphisms, let us
make some remarks on the notion of dimensionality in the realm of
schemes. Let $f:X \to S$ be a morphism of schemes. Then the
relative dimension $\dim_xf:= \dim_xf^{-1}(f(x)) \equiv \dim_x
X_{f(x)}$ of $f$ at a point $x \in X$ is in general defined via
lengths of chains of irreducible closed subsets of the fibre
$X_{f(x)}$. This is explained in detail in \Liu, Chapter $2.5$. In
the case of $S$-immersions $X \hookrightarrow \Affin_{S}^n$ (in
which we are interested), $X_{f(x)}$ is of finite type over the
field $k(f(x))$ and one can show that $\dim_xX_{f(x)}$ coincides
with the intuitively clear notion of relative dimension given in
the following definition of smooth morphisms.
\begin{Def}\label{0021}
    A morphism of schemes $f:X \to S$ is called \emph{smooth}  at a
    point $x\in X$ (of relative dimension $r$) if there exist an open neighborhood $U$ of $x$
    and an $S$-immersion
    \begin{equation*}
        j:U \hookrightarrow \Affin_S^n
    \end{equation*}
    of $U$ into some linear space $\Affin_S^n$ over $S$ such that
    the following conditions are satisfied:
    \begin{enumerate}
        \item
        locally at $y:=j(x)$ (i.e., in an open neighborhood of
        $j(x)$), the sheaf of ideals $\cJ$ defining $j(U)$  as a
        subscheme of $\Affin_S^n$ is generated by $(n-r)$ sections $g_{r+1}, \ldots, g_n$.
        \item
        the differential forms $dg_{r+1}(y), \ldots, dg_n(y)$ are linearly independent in
        $\Omega_{\Affin_S^n/S}^1 \otimes k(y)$.
    \end{enumerate}
    The morphism $f:X \to S$ is called smooth if it is
    smooth at all points of $X$.
\end{Def}
\underline{Physical interpretation:} Let $S= \Spec \real$ and let
$X= V(g_{r+1}, \ldots , g_n) \subset \Affin_{S}^n$ be the zero set
of $(n-r)$ polynomials $g_{r+1}, \ldots ,g_n \in \real[T_1, \ldots
,T_n]$ such that the canonical map $f:X \to
S$ is smooth of relative dimension $r$. Therefore \\
\begin{equation*}
    X(\real)= \left\{ (x_1, \ldots ,x_n)\in \real^n
    \mid g_i(x_1, \ldots ,x_n)=0 \ \text{for all} \ r+1 \leq i \leq n \right\}
\end{equation*}
is a $r$-dimensional differentiable manifold. $X(\real)
\hookrightarrow \Affin_{S}^n(\real)=\real^n$ is an immersion in
the sense of ordinary differential geometry.

Thus  the notion of smooth morphisms is a natural generalization
of immersions in the sense of ordinary differential geometry and
analogously unramified morphisms generalize submersions.
\begin{Def}\label{0022}
    A morphism of schemes $f:X \to S$ is called \emph{\'etale} (at a
    point) if it is smooth (at the point) of relative dimension
    $0$.
\end{Def}
\begin{Cor}\label{0023}
    An immersion $f:X \to S$ is \'etale if and only if $f$ is an
    open immersion.
\end{Cor}

\begin{proof}
    \BLR, Lemma 2.2/4
\end{proof}
\begin{Satz}\label{0024}
    Let $f:X \to S$ be a smooth morphism of schemes. Then:
    \begin{enumerate}
        \item
        $\Omega_{X/Y}$ is locally free. Its rank at $x\in X$ is
        equal to the relative dimension of $f$ at $x$.
        \item
        The canonical sequence of $\Ox$-modules
        \begin{equation*}
            \xymatrix{0 \ar[r]&
            f^*\Omega_{Y/S}^1 \ar[r] &
            \Omega_{X/S}^1 \ar[r] &
            \Omega_{X/Y}^1 \ar[r] & 0 }
        \end{equation*}
        is exact an locally split.
    \end{enumerate}
\end{Satz}
\begin{proof}
    \BLR, Prop. 2.2/5
\end{proof}

\begin{Satz}\label{0025}
    Let $f:X \to S$ be locally of finite presentation. Let $x \in
    X$ and set $s:=f(x)$. The following conditions are equivalent:
    \begin{enumerate}
        \item
        $f$ is smooth at $x$.
        \item
        $f$ is flat at $x$ and the fibre $X_s=X \times_S \Spec
        k(s)$ is smooth over $k(s)$ at $x$.
    \end{enumerate}
\end{Satz}

\begin{proof}
    \BLR, Prop. 2.4/8
\end{proof}
Like in differential geometry there is a Jacobi Criterion for
smoothness. The analogy to the classical Jacobi Criterion is
particularly evident in characterization $e)$ of the following
list of criterions for smoothness.
\begin{Jac}\label{0026}
    Let $X$ and $Z$ be $S$-schemes and let $j:X \hookrightarrow Z$ be
    a closed immersion which is locally of finite presentation.
    Let $\cJ \subset \Oz$ be the sheaf of ideals which defines $X$ as a subscheme of $Z$.
    Let $x \in X$ and $z:=j(x)$. Assume that, as an $S$-scheme, $Z$ is smooth  at $z$ of relative dimension $n$.
    Then the following conditions are equivalent:
    \begin{enumerate}
        \item
        As an  $S$-scheme, $X$ is smooth at $x$ of relative dimension
        $r$.
        \item
        The canonical sequence of $\Ox$-modules
        \begin{equation*}
            \xymatrix{0 \ar[r] &
            \cJ/\cJ^2 \ar[r] &
            j^*\Omega_{Z/S}^1 \ar[r] &
            \Omega_{X/S}^1 \ar[r] & 0}
        \end{equation*}
        is spilt exact at $x$, and $r=\mathrm{rank}(\Omega^1_{X/S} \otimes
        k(x))$.
        \item
        If $dz_1, \ldots ,dz_n$ is a basis of
        $(\Omega^1_{Z/S})_z$, and if $g_1, \ldots , g_N$ are local
        sections of $\Oz$ generating ${\cal J}_z$, there exists a
        re-indexing of the $z_1, \ldots , z_n$  and of the $g_1,
        \ldots ,g_N$ such that $g_{r+1}, \ldots, g_n$ generate $\cJ$ at
        $z$ and such that $dz_1, \ldots ,dz_r, dg_{r+1}, \ldots,
        dg_n$ generate $(\Omega^1_{Z/S})_z$.
        \item
        There exist local sections $g_{r+1}, \ldots, g_n$ of $\Oz$
        such that
        \begin{enumerate}
            \item[(i)]
            $g_{r+1}, \ldots, g_n$ generate ${\cal J}_z$.

            \item[(ii)]
            $dg_{r+1}(z), \ldots, dg_n(z)$ are linearly independent in $\Omega^1_{Z/S} \otimes _{\Oz}
            k(z)$, where $dg_i(z)$ is the image of $dg_i \in (\Omega^1_{Z/S})_z$
            in $\Omega^1_{Z/S} \otimes _{\Oz}
            k(z) = (\Omega^1_{Z/S})_z / {\maxi_z
            (\Omega^1_{Z/S})_z}$.
        \end{enumerate}
        \item
        There exist local sections $g_{r+1}, \ldots, g_n$ of $\Oz$
        such that
        \begin{enumerate}
            \item[(i)]
            $g_{r+1}, \ldots, g_n$ generate ${\cal J}_z$.

            \item[(ii)]
            considering a representation
            \begin{equation*}
            dg_j= \sum_{i=1}^n \frac{\partial g_j}{\partial z_i} dz_i
            \end{equation*}
            of the differential forms $g_{r+1}, \ldots, g_n$ with respect to a basis
            $dz_1, \ldots ,dz_n$ of $(\Omega^1_{Z/S})_z$ , there is a $(n-r)$-minor of the matrix
            $\left( \frac{\partial g_j}{\partial z_i} \right)_{}$
            which does not vanish at $z$.
        \end{enumerate}
    \end{enumerate}
\end{Jac}

\begin{proof}
    \BLR, Prop. 2.2/7
\end{proof}
\begin{Satz}\label{0027}
    Let $f:X \to Y$ be an $S$-morphism. Let $x\in X$ and set
    $y:=f(x)$. Assume that $X$ is smooth over $S$ at $x$ and that
    $Y$ is smooth over $S$ at y. Then the following conditions are
    equivalent:
    \begin{enumerate}
        \item
        $f$ is \'etale at x.
        \item
        The canonical homomorphism $(f^*\Omega_{Y/S}^1)_x \to
        (\Omega_{X/S}^1)_x$ is bijective.
    \end{enumerate}
\end{Satz}

\begin{proof}
    \BLR, Cor. 2.2/10
\end{proof}
\begin{Satz}\label{0028}
    Let $f:X \to S$ be a morphism and $x \in X$. Then the
    following conditions are equivalent:
    \begin{enumerate}
        \item
        $f$ is smooth of relative dimension $n$.
        \item
        There exists an open neighborhood $U$ of $x$ and a
        commutative diagram
        \begin{equation*}
            \xymatrix{U \ar[r]^g \ar[dr]_{f\mid_U} & \Affin_S^n \ar[d]^p \\
            & S}
        \end{equation*}
        where $g$ is \'etale and $p$ is the canonical projection.
    \end{enumerate}
\end{Satz}

\begin{proof}
    \BLR, Prop. 2.2/11
\end{proof}
\begin{Satz}\label{0029}
    If $X$ is a smooth scheme over a field $k$, the set of closed
    points $x$ of $X$ such that $k(x)$ is a separable extension of
    $k$ is dense in $X$.
\end{Satz}

\begin{proof}
    \BLR, Cor. 2.2/13
\end{proof}
\begin{Satz}\label{0030}
    Let $f:X \to S$ be a smooth morphism. Let $s$ be a point of
    $S$, and let $x$ be a closed point of the fibre $X_s=X \times_S \Spec
    k(s)$ such that $k(x)$ is a separable extension of $k(s)$.
    Then there exists an \'etale morphism $g:S' \to S$ and a point
    $s' \in S'$ above $s$ such that the morphism $f': X \times_S S' \to
    S'$ obtained from $f$ by the base change $S' \to S$ admits a
    section $h: S' \to X \times_S S'$, where $h(s')$ lies above
    $x$, and where $k(s')=k(x)$.
\end{Satz}

\begin{proof}
    \BLR, Prop. 2.2/14
\end{proof}

\begin{Satz}\label{0030a}
    Let $R$ be a local henselian ring with residue field $k$. Let
    $X$ be a smooth $R$-scheme. Then the canonical map $X(R) \to
    X(k)$ from the set of $R$-valued points of $X$ to the set of
    $k$-valued points of $X$ is surjective. In particular, if $R$
    is strictly henselian, the set of $k$-valued points of $X_k=X \otimes_R
    k$ which lift to $R$-valued points of $X$ is dense in $X_k$.
\end{Satz}

\begin{proof}
    \BLR, Prop. 2.3/5
\end{proof}
Let us finish with some remarks on flatness, because by
Proposition \ref{0025} smooth morphisms are in particular flat.

\begin{Satz}\label{0031}
    Let $f:X \to Y$ be locally of finite presentation. If $f$ is
    flat, then $f$ is open.
\end{Satz}

\begin{proof}
    \EGAvz, $2.4.6$
\end{proof}

\begin{Satz}\label{0032}
    Let $f:X \to Y$ be faithfully flat (i.e. flat and surjective) and quasi-compact.
    Then the topology  of $Y$ is the quotient topology of $X$ with
    respect to $f$, i.e. a subset $V \subset Y$ is open if and
    only if $f^{-1}(V)$ is open in $X$.
\end{Satz}

\begin{proof}
    \EGAvz, $2.3.12$
\end{proof}
Later, we will describe the physical models by means of  smooth
morphisms $f:X \to S$ of algebraic spaces, where $S$ is an Zariski
one-dimensional excellent Dedekind ring. This choice of $S$ and
the flatness of $f$ guarantee that the considered models have got
the properties that one would expect. Next to the fact that the
relative dimension of the universe $X$ is constant (see Lemma
\ref{0222}), $X$ may be considered as a ``continuous'' family of
fibres $X_s$, $s \in S$; i.e. if we remove one fibre $X_s \subset
X$ over a closed point $s\in S$, we can fill up this hole uniquely
from the data given on the open subscheme $X \setminus X_s \subset
X$.

\begin{Satz}\label{0033}
    Let $Y$ be a regular, integral scheme of dimension one, let
    $p \in Y$ be a closed point, and let $X \hookrightarrow
    \mathbb{P}_{Y-\{p\}}^n$ be a closed subscheme which is flat
    over $Y- \{p\}$. Then there exists a unique closed subscheme
    $\bar{X} \hookrightarrow \mathbb{P}_Y^n$, flat over $Y$, whose
    restriction to $\mathbb{P}_{Y-\{p\}}^n$ is $X$.
\end{Satz}

\begin{proof}
    \Hart,  Chap. III, Prop. 9.8
\end{proof}

\begin{Satz}\label{0034}
    Let $f:X \to Y$ be a morphism of schemes, with $Y$ integral
    and regular of dimension one. Then $f$ is flat if and only if
    every associated point $x \in X$ maps to the generic point of
    $Y$. In particular, if $X$ is reduced, this says that every
    irreducible component of $X$ dominates $Y$.
\end{Satz}

\begin{proof}
    \Hart,  Chap. III, Prop. 9.7
\end{proof}
Finally, let us conclude with the statements that illustrate the
very nice behavior of the relative dimension of smooth morphisms.
Thereby, the flatness of smooth morphisms is essential.

\begin{Lemma}\label{0222}
    Let $f:X \to S$ be locally of finite type and flat. Assume that $X$ is
    irreducible and that $Y$ is locally noetherian.
    Then the relative dimension of $f$ is constant on $X$.
\end{Lemma}

\begin{proof}
    \EGAvz, 14.2.2
\end{proof}

\begin{Lemma}\label{0223}
    Let $S$ be locally noetherian scheme and let $f:X \to S$ be a morphism of
    finite type smooth at a point $x$. Let $s:=f(x)$. Then
    $\Omega_{X/S}^1$ is free of rank $\dim_x X_s$ in a
    neighborhood of x.
\end{Lemma}

\begin{proof}
    \Liu, Prop. 6.2.5
\end{proof}
\section{The arithmetic Einstein equations}\label{5100}

Let $X \to S$ be a smooth, separated $S$-scheme of relative
dimension $n$. The purpose of this section is the derivation of
the fundamental equations of general relativity in our algebraic
geometric setting. As the ordinary differential geometric Einstein
equations are differential equation, we must expect that this
holds in algebraic geometry, too. The necessary techniques
concerning smoothness and differential calculus in algebraic
geometry were summarized in section \ref{0011s}.
Crucial are the following notions.
\begin{tabbing}
    \ \quad \= $\Omega_{X/S}^1$ \=  \quad
    \= sheaf of (relative) differential forms  \\
    \> ${\cal T}_{X/S}$ \> :=  ${\cal H}\text{om}_{{\cal O}_X}\left( \Omega_{X/S}^1 ,{\cal O}_X \right)$
     \quad \text{sheaf of (relative) vector fields}  (see Definition \ref{1128}) \\
    \> $T_{X/S}$ \> := $\mathbb{V}\left( \Omega_{X/S}^1 \right)$
    \quad
    (relative) tangent bundle (see Definition \ref{1129})
\end{tabbing}
One can prove (Remark \ref{1130}) that
\begin{align*}
    \Gamma(T_{X/S}/U):=\text{Hom}_X(U,T_{X/S}) \cong {\cal
    T}_{X/S}(U)
\end{align*}
for every Zariski open subset $U \subset X$. Therefore vector
fields correspond to sections of the tangent bundle.

\subsection*{The metric tensor}

Due to smoothness, the sheaves $\Omega_{X/S}^1$ and ${\cal
T}_{X/S}$ are locally free. Let us fix a local base $\{ \omega^i
\}$ of $\Omega_{X/S}^1$ which is dual to the local base  $\{
\partial_i \}$ of ${\cal T}_{X/S}$.

\begin{Def}\label{5101}
    Let $g:{T_{X/S} \times_X T_{X/S}}  \to  \Affin_X^1$ be an
    $X$-morphism which is bilinear (see section \ref{0200g}).
    Equivalently, $g$ may be interpreted as a
    global section of $\Omega_{X/S}^{\otimes 2}$.
    Locally, we may write
    \begin{align*}
        g= \sum\limits_{1 \leq i,j \leq n} g_{ij} \, \omega^i \otimes \omega^j \in \Omega_{X/S}^{\otimes
        2}, \quad \quad g_{ij}\in \Ox.
    \end{align*}
    Then $g$ is called a \emph{metric} if the following conditions
    hold for any sufficiently small open subset of $X$:
    \begin{enumerate}
        \item[(i)]
        The matrix $(g_{ij})$ is symmetric, i.e. $g_{ij}=g_{ji}$.

        \item[(ii)]
        The matrix $(g_{ij})$ is invertible, i.e. $\det(g_{ij}) \in
        {\cal O}_X^{ \, *}$.
    \end{enumerate}
\end{Def}

\subsection*{Covariant derivation}\label{5105}

\begin{Def}\label{5106}
    Let $\nabla:{T_{X/S} \times_X T_{X/S}}  \to  T_{X/S}$ be an
    $X$-morphism. Interpret $\nabla$ as a
    map
    \begin{align*}
        \nabla: {\cal T}_{X/S}(X) \times {\cal T}_{X/S}(X) \to {\cal
        T}_{X/S}(X), \quad ( \vecu , \vecv) \mapsto
        \nabla_{\vecu}\vecv.
    \end{align*}
    Let us assume that $\nabla$ is a $\co_S(S)$-bilinear map,
    where the $\co_X(X)$-module ${\cal T}_{X/S}(X)$ is viewed as
    $\co_S(S)$-module via the canonical morphism $\co_S(S) \to
    \co_X(X)$.
    Then $\nabla$ is called a \emph{covariant derivation} if the following conditions
    hold for all $f \in \Ox(X)$ and $\vecu, \vecv \in {\cal T}_{X/S}(X)$:
    \begin{enumerate}
        \item[(i)]
        $\nabla_{f \vecu}\vecv = f \nabla_{\vecu}\vecv$.

        \item[(ii)]
        $\nabla_{\vecu}(f \vecv)= (\vecu f) \vecv + f
        \nabla_{\vecu}\vecv$.
    \end{enumerate}
\end{Def}
Thereby, $\vecu f:= \vecu (f) = (d_{X/S}f)(\vecu)$ is the
canonical action of vector fields on functions (the differential
$d_{X/S}$ is introduced and explained directly above Definition
\ref{0014}).

\begin{Def}\label{5107}
    Let $\nabla$ be a covariant derivation, and let $\vecu,
    \vecv $ and $\vecw \in {\cal T}_{X/S}(X)$.
    \begin{enumerate}
        \item
        ${T}(\vecu,\vecv):= \nabla_{\vecu}\vecv- \nabla_{\vecv}\vecu- [{\vecu},\vecv]
        $ \quad is called the \emph{torsion} of $\nabla$.

        \item
        $\nabla$ is called \emph{torsion-free} if and only if ${T}(\vecu,\vecv) =0$ for all $\vecu,
        \vecv$.

        \item
        $\nabla$ is called \emph{metrical} if and only if $\vecu g(\vecv,\vecw)
        = g(\nabla_{\vecu}\vecv,\vecw)+g(\vecu,\nabla_{\vecu}\vecw)$ for all $\vecu,
        \vecv, \vecw$.

    \end{enumerate}
\end{Def}
In the same way as in differential geometry one proves that there
exists a uniquely determined covariant derivation $\nabla$ which
is metrical and torsion-free, the \emph{Levi-Civita connection}.
The Levi-Civita connection is completely determined by the
metrical tensor. More precisely, the Koszul formula holds.
\begin{align*}
    2 \, g(\nabla_{\vecu}\vecv,\vecw) = \vecu g(\vecv,\vecw)- \vecw g(\vecu,\vecv)+ \vecv g(\vecw,\vecu)
    + g([\vecu,\vecv],\vecw) + g([\vecw,\vecu],\vecv) - g([\vecv,\vecw],\vecu)
\end{align*}

\subsection*{Curvature}\label{5110}

From now on let $\nabla$ be the Levi-Civita connection. Then we
may introduce the curvature tensor
\begin{align*}
    {R}_{ \vecu \vecv }(\vecw):= \nabla_{\vecu}\nabla_{\vecv}\vecw-
        \nabla_{\vecv}\nabla_{\vecu}\vecw-\nabla_{[{\vecu},\vecv]}\vecw.
\end{align*}
Then the tensor
\begin{align*}
    {R}_{ \vecz \vecw \vecu \vecv }:= g({R}_{ \vecu \vecv }(\vecw),\vecz)
\end{align*}
is called the \emph{Riemannian curvature tensor}. The Riemannian
curvature tensor fulfills the following identities.

\begin{Satz}\label{5111}
    Let $\vecu, \vecv, \vecw, \vecz \in {\cal T}_{X/S}(X)$. Then:
    \begin{enumerate}
        \item
        ${R}_{ \vecu \vecv \vecw \vecz} = -{R}_{ \vecv \vecu \vecw
        \vecz}$

        \item
        ${R}_{ \vecu \vecv \vecw \vecz}=-{R}_{ \vecu \vecv \vecz
        \vecw}$

        \item
        ${R}_{ \vecu \vecv \vecw \vecz}={R}_{\vecw \vecz \vecu \vecv
        }$

        \item
        \emph{first Bianchi-identity:} \qquad \
        ${R}_{ \vecz \vecu \vecv \vecw}+{R}_{\vecz  \vecv \vecw \vecu
        }+{R}_{ \vecz \vecw \vecu \vecv }=0$

        \item
        \emph{second Bianchi-identity:} \quad
        $(\nabla_{\vecu} R)_{\vecv \vecw}+(\nabla_{\vecv} R)_{\vecw \vecu} + (\nabla_{\vecw} R)_{\vecu
        \vecv}=0$

        Thereby, $(\nabla_{\vecu} R)_{\vecv \vecw}(\vecz):= \nabla_{\vecu}(R_{\vecv\vecw}(\vecz))
        - R_{\nabla_{\vecu}\vecv,\vecw}(\vecz) -R_{\vecv,\nabla_{\vecu}\vecw}(\vecz)
        -R_{\vecv \vecw}(\nabla_{\vecu} \vecz)$.

    \end{enumerate}
\end{Satz}
More generally, the covariant derivation of arbitrary tensor
fields $S$ and $T$ with respect to a vector field $\vecv$ may
defined inductively as follows: $\nabla_{\vecv}(S \otimes T):=
\nabla_{\vecv}S \otimes T + \nabla_{\vecv}T \otimes S$.
\begin{Def}\label{5112}
    The bi-quadratic form
    \begin{align*}
        k(\vecu,\vecv):=R_{\vecu \vecv \vecu \vecv }
    \end{align*}
    is called intersection curvature.
\end{Def}

\begin{Satz}\label{5113}
    The Riemannian curvature tensor is completely determined by
    $k$. More precisely:
    \begin{enumerate}
        \item
        $4 \cdot R_{\vecu \vecv \vecv \vecw}= k(\vecu+\vecw,\vecv)-
        k(\vecu-\vecw,\vecv)$

        \item
        $6 \cdot R_{\vecu \vecv \vecw \vecz}
        = R_{\vecu, \vecv + \vecw, \vecv + \vecw, \vecz}
        - R_{\vecu, \vecv - \vecw, \vecv - \vecw, \vecz}
        - R_{\vecv, \vecu + \vecw, \vecu + \vecw, \vecz}
        + R_{\vecv, \vecu - \vecw, \vecu - \vecw, \vecz}$
    \end{enumerate}
\end{Satz}

\begin{Cor}\label{5114}
    Let $X \to S$ be a smooth $S$-scheme with metric $g$.
    Then $X$  is
    \emph{flat}, i.e. $k(\vecu, \vecv)=0$ for all $\vecu, \vecv \in {\cal
    T}_{X/S}(X)$, if and only if the Riemannian curvature tensor vanishes, i.e. $R =0$.
\end{Cor}

\begin{Def}\label{5115}
    Let $X \to S$ be a smooth $S$-scheme with metric $g$
    and consider a local base $\{ \partial_i \}$ of ${\cal T}_{X/S}$ and
    a local base $\{ \omega^i \}$ of $\Omega_{X/S}^1$. The
    curvature tensor $R_{ \vecu \vecv}(\vecw)$ is trilinear in
    $\vecu, \vecv, \vecw$ and therefore induces linear maps $R_{ \bullet
    \vecv}(\vecw)$. Taking the trace finally yields
    the symmetric bilinear form ric
    \begin{align*}
        \text{ric}(\vecv,\vecw)
        := \text{Tr}(R_{ \bullet \vecv}(\vecw))
    \end{align*}
    which is called the \emph{Ricci-form} of $(X \to S,g)$.
    Now consider the uniquely determined tensor Ric which is
    given by $g(\text{Ric}(\vecu),\vecv)=\text{ric}(\vecu,\vecv)$
    for all vector-fields $\vecu,\vecv$.
    The \emph{scalar curvature} sc is by definition the trace
    \begin{align*}
        \text{sc}
        := \text{Tr}(\text{Ric}(\bullet))
    \end{align*}
    of the linear map $\text{Ric}(\bullet)$.
    Furthermore, the \emph{divergence} $\text{div}(T)$ of any symmetric $(0,q)$-tensor
    $T:= \sum T_{i_1 \ldots i_q} \omega^{i_1} \otimes \ldots \otimes
    \omega^{i_q}$ is defined as follows: The covariant derivation
    $\nabla T$ of $T$ is a $(0,q+1)$-tensor
    $\nabla T:= \sum T_{i_1 \ldots i_q;j} \omega^{i_1} \otimes \ldots \otimes
    \omega^{i_q} \otimes \omega^j$. Then $\text{div}(T)$ is the
    $(0,q-1)$-tensor which is obtained by lifting the new
    variable and contracting it:
    \begin{align*}
        (\text{div}(T))_{i_1 \ldots i_{q-1}}=g^{i_q j} T_{i_1 \ldots
        i_q;j}.
    \end{align*}
\end{Def}

\subsection*{Einstein's equation}\label{5120}

Let $X \to S$ be a smooth $S$-scheme with metric $g$, and let
$\nabla$ be the Levi-Civita connection on $X$. Furthermore, let
$T$ denote the energy-stress tensor. This is a symmetric
$(0,2)$-tensor on $X$ with $\text{div}(T)=0$. Then the equations
of general relativity in our arithmetic setting are given by the
following system of equations:

\begin{EinsteinEq}\label{5121}
    \begin{align*}
        \text{ric}- \frac{1}{2} \, \text{sc} \cdot g = \kappa T
    \end{align*}
\end{EinsteinEq}
where $\kappa \in {\cal O}_S(S)$ is a constant. Now, having
written down the equations of general relativity in the setting of
arithmetic algebraic geometry, one can ask for solutions. Choosing
$S= \Spec \real$ and assuming that there exists a solution of the
corresponding algebraic geometric Einstein equations, it follows
that this solution gives rise to a differential geometric solution
of the ordinary, differential geometric Einstein equations. This
follows from the purely algebraic nature of the notions metric,
covariant derivation and curvature (see also section \ref{0200g}
for more details). However, we are interested in ``adelic'' \st
models. So, let us now make the choice $S= \Spec
{\footnotesize{\text{$\cal O$}}}_K$, where
${\footnotesize{\text{$\cal O$}}}_K$ denotes as usual the ring of
integral numbers of an algebraic number field $K$. Recall that the
generic fibre $X_K$ of $X \to S$ represents the archimedean limit.

In order to solve the Einstein equations, it is most convenient to
perform all computations locally and to glue the local solutions
in a second step. We will see that these local computations may be
performed in essentially the same way as in differential geometry.
On the one hand, this is due to the fact that the local ring
${\cal O}_{X,x}$ at a point $x \in X_K(K)$ may be embedded into a
ring of formal power series.

\begin{Satz}\label{0351}
    Let $X \to S$ be a smooth morphism of locally Noetherian
    schemes. Let $s \in S$ and $x \in X_s$ be a $k(s)$-rational
    point. Then there exists an isomorphism of ${\widehat{\cal
    O}}_{S,s}$-algebras
    \begin{align*}
        {\widehat{\cal O}}_{X,x} = {\widehat{\cal O}}_{S,s} [[x_1, \ldots, x_n]]
    \end{align*}
    where $(x_1, \ldots, x_n)$ is a set of variables and $n=\dim {\cal
    O}_{X_s,x}$.
\end{Satz}

\begin{proof}
     \Liu, Ex. 6.3.1
\end{proof}
Furthermore, we know from Proposition \ref{0028} that each point
$x \in X$ possesses an open environment $U$ which is \'etale over
some affine space $\Affin_S^n= \Spec \co_S[x_1, \ldots, x_n]$.
Therefore, the module $\Omega_{U/S}^1$ of differential forms over
$U$ is the free $\co_U$-module generated by the differentials
$dx_1, \ldots, dx_n$ (see Proposition \ref{0027}), and we may
choose the base $\{ \omega^i:=dx_i \}$ of $\Omega_{U/S}^1$
together with the corresponding dual base $\{\partial_i \}$ of
${\cal T}_{U/S}$.

The functions $f \in \co_U$ on $U$ are algebraic over the
polynomial ring ${\cal O}_S[x_1, \ldots, x_n]$, because we may
assume that $U$ is standard \'etale over $\Affin_S^n$ (see \BLR,
Prop. 2.3/3). Consequently, there is a canonical differential
calculus on $U$ with respect to the coordinates $x_1, \ldots,
x_n$. More precisely, the vector field $\partial_i$ acts on $f$ by
means of ordinary partial derivation with respect to the $i$-th
coordinate $x_i$. This may be seen as follows: On polynomials we
have clearly $\partial_i x_j^n= \delta_{ij} \cdot nx_j^{n-1}$ due
to the Leibniz rule (see Definition \ref{0012}, a)). If $f \in
\co_U$ is arbitrary, there is an algebraic equation $\sum_{j=0}^m
c_jf^j=0$ with polynomials $c_j \in {\cal O}_S[x_1, \ldots, x_n]$,
$c_m \neq 0$. It follows that $0=
\partial_i(\sum_{j=0}^m c_jf^j)=\sum_{j=0}^m (f^{j}
\partial_ic_j + c_j  j f^{j-1}\partial_if)$ which is a
linear equation in $\partial_i f$ and thus may be solved uniquely
for $\partial_if$ on the locus where $\sum_{j=0}^m c_j j f^{j-1}
\neq 0$. However, by what we already know, $\partial_ic_j$ is the
ordinary partial derivation of $c_j$ with respect to the $i$-th
coordinate $x_i$, and so we are done.

Therefore, we obtain the following local formulas on $U$ (where we
make use of Einstein's summation convention):
\begin{enumerate}
    \item[$\bullet$]
    $g = g_{ij} \omega^{i} \otimes \omega^{j}$ \quad
    with $g_{ij}=g_{ji} \in {\cal O}_{U}$

    \item[$\bullet$]
    $\nabla_{\partial_i} \partial_j =  \Gamma_{ij}^k \partial_k$
    \quad where the functions $\Gamma_{ij}^k \in {\cal O}_{U}$ are called the
    \emph{Christoffel-symbols}.

    \item[$\bullet$]
    $\Gamma_{ij}^k = \frac{1}{2} g^{kl}( \partial_i g_{jk} + \partial_j g_{ik}- \partial_k
    g_{ij})$

    \item[$\bullet$]
    $R_{\partial_i \partial_j}(\partial_k)= R^l_{kij} \partial_l$

    \item[$\bullet$]
    $R^l_{ijk} = \partial_j \Gamma_{ki}^l -  \partial_k \Gamma_{ij}^l +
    \Gamma_{ki}^r\Gamma_{jr}^l -  \Gamma_{kr}^l\Gamma_{ij}^r$

    \item[$\bullet$]
    $R_{ik}:= \text{ric}_{ik} = R^l_{ilk}$

    \item[$\bullet$]
    $R:=\text{sc}= g^{ik} R_{ik}$
\end{enumerate}
Now, the Einstein equations take their well known form
\begin{align*}
    R_{\mu \nu} - \frac{1}{2} g_{ \mu \nu} R = \kappa T_{\mu \nu}
    \quad \text{or equivalently} \quad
    R_{\mu \nu} = \kappa \left( T_{\mu \nu} - \frac{1}{2}g_{\mu \nu} T
    \right), \ \ T:=g^{\mu \nu} T_{\mu \nu},
\end{align*}
and it follows that Theorem \ref{5125a} holds.

\begin{Theorem}\label{5125a}
    Let $g$ be a metric on a smooth $S$-scheme $X \to S$.
    The Einstein equations on $X$ are universal in the following
    sense: For $x \in X$, let
    $\{ \partial_i \}$ be a base of $\Omega^1_{X/S,x}$, and let $g_{ij} \in
    \co_{X,x}$ be the components of the metric tensor at $x$.
    Assume that there exists a tensor $G \neq g$ of rank two
    such that for all $x \in X$ the following statements hold at $x$:
   \begin{enumerate}
        \item
        $G$ is a polynomial over $K$ in the variables $g_{ij}$, $\partial_k g_{ij}$
        and $\partial_k ( \partial_l g_{ij})$ which is linear in
        $\partial_k ( \partial_l g_{ij})$.

        \item
        $G$ is a symmetrical tensor.

        \item
        $\text{div}(G)=0$.
    \end{enumerate}
    Then, $G$ coincides with the Einstein tensor $\text{ric}- \frac{1}{2} \,  \text{sc}
    \cdot g$.
\end{Theorem}

\begin{Rem}\label{5125}
    Finally, let us again point to the crucial fact that the local functions $f \in {\cal
    O}_{U}$ are \emph{algebraic} functions. Therefore,
    the class of functions, which is
    available in order to solve the equations of arithmetic general
    relativity, is  much smaller than in the differential
    geometric setting. In general, it is a non-trivial task to solve
    differential equations in the algebraic setting, and it is not
    clear whether there exist solutions at all.

    But, as we will see in the next chapter, the equations of
    general relativity have the amazing property that also many
    classical, differential geometric solutions descent to the
    algebraic category. Thus, the set of \GRs $(X \to S,g)$ is not the empty set.
\end{Rem}

                                        \chapter{Solutions of the arithmetic Einstein equations}\label{5200}

                                        In chapter \ref{0010}, we deduced the algebraic geometric
                                        analogue of the differential geometric Einstein equations.
                                        However, it is not clear at all if there actually exist pairs $(X
                                        \to S, g)$ consisting of an algebraic space $X \to S$ and a metric
                                        $g$ on $X$ which solves the Einstein equations. However, we will
                                        see in this chapter that the Einstein equations already possess
                                        many physically interesting solutions in the algebraic category.
                                        This is a highly non-trivial property of the Einstein equations,
                                        because, in general, differential equations only possess solutions
                                        which are very transcendental and far from the algebraic category.
                                        A very interesting solution, which may be interpreted as the
                                        ``adelic'' Minkowski space, is deduced and studied in section
                                        \ref{6020}.

                                        As already announced earlier, we will consider two special choices
                                        for the base $S$:
                                        \begin{itemize}
                                            \item
                                            $S$ is a Zariski zero-dimensional Dedekind scheme.

                                            \item
                                            $S$ is a Zariski one-dimensional Dedekind scheme.
                                        \end{itemize}

                                        \section{The case of Zariski zero-dimensional base}\label{5201}

                                        Within this section, let $S = \Spec K$ be the spectrum of a field
                                        $K$. We are looking for \GRs $(X \to S, g)$ (see Definition \ref{5022}).
                                        However, the  condition (ii) of Definition \ref{5022} is empty in this
                                        case. Therefore, from the adelic point of view, $S = \Spec K$ is
                                        not the physically interesting case, but rather a toy model. The
                                        interesting ``adelic'' models, where all conditions of Definition \ref{5022}
                                        are non-trivial will be considered in section \ref{6000}. At
                                        least there are many \GRs for the choice $S = \Spec K$, because
                                        things are particularly easy in this case. The most easiest
                                        example is the Minkowski solution
                                        \begin{align*}
                                            (\Affin_K^n,g_0), \qquad g_0:=\text{ diag $( \pm1, \pm1, \pm1, \pm1 )$}
                                        \end{align*}
                                        However, there are further, less trivial examples which correspond
                                        to certain solutions of the classical differential geometric
                                        Einstein equations.

                                        \subsection{Kasner solution}\label{5202}

                                        Again we choose $X_K:= \Affin_K^n$ with coordinates $(t=x^0, x^1,
                                        \ldots, x^{n-1})$, but this time we choose the following
                                        non-trivial metric $g_K$
                                        \begin{itemize}
                                            \item
                                            $g_{00}=1, \qquad g_{0i}= 0, \quad i \neq 0$

                                            \item
                                            $g_{ij}:= c \delta_{ij} \cdot  t^{2 k_i}, \quad i \neq 0 \neq j$
                                        \end{itemize}
                                        where $\delta_{ij}$ denotes the Kronecker delta, and where $c \in
                                        K$ and $k_i \in \rat$ are constants. Then $g_K$ is well defined in
                                        the category of algebraic spaces, and it remains to show that we
                                        can choose the constants in such a way that the Einstein equations
                                        are fulfilled. We may do this on stalks. Recalling the remarks
                                        below Proposition \ref{0351} and the formulas stated there, we may compute
                                        the Christoffel symbols corresponding to the given metric. Due to
                                        the fact that we use orthogonal coordinates we obtain:

                                        \begin{align*}
                                            \text{$\Gamma^0_{00}=0$, $\Gamma^0_{i0}= 0$,
                                            $\Gamma^i_{ii}=\frac{1}{2}g^{ii} \partial_i g_{ii}=0$, $\Gamma^j_{ii}= -\frac{1}{2}g^{jj}\partial_j
                                            g_{ii}=0$, $\Gamma^j_{ij}= -\frac{1}{2}g^{jj}\partial_i g_{jj}=0$}.
                                        \end{align*}
                                        The only non-vanishing Christoffel symbols are
                                        \begin{enumerate}
                                            \item[$(*)$]
                                            $\Gamma^0_{ii}= -\frac{1}{2}g^{00}\partial_0 g_{ii} =- c k_i
                                            t^{2k_i-1}$

                                            \item[$(**)$]
                                            $\Gamma^i_{0i}= \frac{1}{2}g^{ii}\partial_0 g_{ii}
                                            =\frac{k_i}{t}$
                                        \end{enumerate}
                                        We consider a vacuum solution of the Einstein equations, i.e.
                                        $T_{\mu \nu}=0$. Therefore, the Ricci tensor $R_{\mu \nu}=
                                        R^{\lambda}_{\mu \lambda \nu}$ has to vanish, too. The components
                                        of the Ricci tensor are as follows:

                                        \begin{Lemma}\label{5303}
                                            The Ricci tensor is diagonal in the given coordinates, i.e. $R_{\mu \nu} =
                                            \delta_{\mu \nu} R_{\mu \nu}$. For the diagonal elements one
                                            obtains:
                                            \begin{align*}
                                                R_{00}= \frac{1}{t^2} \sum_{j \neq 0} \left( k_j - k_j^2 \right),
                                                \quad
                                                R_{ii}=-ck_i t^{2(k_i-1)} \left(  \sum_{j \neq 0} k_j - 1
                                                \right), i \neq 0.
                                            \end{align*}
                                        \end{Lemma}

                                        \begin{Cor}\label{5304}
                                            Let $X_K:=\Affin_K^n$ and $g_K$ be as stated above.
                                            Furthermore, choose the constants $k_i \in \rat$
                                            such that
                                            \begin{align*}
                                                \sum_{j \neq 0} k_j = 1 = \sum_{j \neq 0} k_j^2.
                                            \end{align*}
                                            Then $(X_K,g_K)$ is a \GRo.
                                        \end{Cor}
                                        It remains to prove Lemma \ref{5303}.
                                        \begin{proof}
                                            First of all, $R_{00} = \partial_{\mu} \Gamma_{00}^{\mu}- \partial_0
                                            \Gamma_{\mu0}^{\mu}
                                            + \Gamma_{\mu \nu}^{\mu} \Gamma_{00}^{\nu} - \Gamma_{\nu 0}^{\mu}
                                            \Gamma_{\mu0}^{\nu}$ $= - \partial_0  \Gamma_{\mu 0}^{\mu} -
                                            \Gamma_{\nu 0}^{\mu} \Gamma_{\mu 0}^{\nu}$, because $\Gamma_{00}^{\mu}=0$.
                                            Furthermore, $\Gamma_{\nu 0}^{\mu} \Gamma_{\mu 0}^{\nu}=
                                            \Gamma_{\nu 0}^0 \Gamma_{00}^{\nu} + \sum_{j \neq 0}\Gamma_{\nu 0}^j
                                            \Gamma_{j0}^{\nu}$
                                            $= \sum_{j \neq 0} \Gamma_{\nu 0}^j \Gamma_{j0}^{\nu} \delta_{\nu j}$
                                            $= \sum _{j \neq 0} ( \Gamma_{j 0}^j )^2 $ where we used $\Gamma_{00}^{\mu}=0$
                                            and formula $(**)$ in order to obtain the second equality.
                                            Thus
                                            \begin{align*}
                                                R_{00}= - \sum _{j \neq 0} \left( \partial_0 \Gamma_{j 0}^j + ( \Gamma_{j 0}^j )^2  \right)
                                            \end{align*}
                                            and the desired formula follows from $(*)$. Let now $i \neq
                                            0$. Then
                                            $R_{i0} = R^{\mu}_{i \mu 0}$
                                            $= \partial_{\mu} \Gamma_{i0}^{\mu} -\partial_{0} \Gamma_{i \mu}^{\mu}
                                            + \Gamma_{\nu \mu}^{\mu} \Gamma_{i 0}^{\nu}- \Gamma_{0 \nu}^{\mu} \Gamma_{i
                                            \mu}^{\nu}$.
                                            Recalling that only over Greek letters is summed, we obtain
                                            $\partial_{\mu} \Gamma_{i0}^{\mu} = \partial_{i} \Gamma_{i0}^{i}=0$, because
                                            the Christoffel symbols are constant with respect to $x^i$, $i \neq
                                            0$. Furthermore, it follows from $(**)$ that
                                            $\Gamma_{\nu \mu}^{\mu} \Gamma_{i 0}^{\nu} = \Gamma_{0 \mu}^{\mu} \Gamma_{i
                                            0}^{0}$ and that
                                            $\Gamma_{0 \nu}^{\mu} \Gamma_{i \mu}^{\nu} = \sum_{j \neq 0} \Gamma_{0 j}^{j} \Gamma_{i
                                            j}^{j}$. However, as well $\Gamma_{ij}^j$ as  $\Gamma_{i0}^0$
                                            are zero and we obtain
                                            \begin{align*}
                                                R_{i0}= 0.
                                            \end{align*}
                                            Similarly, $R_{ik}=0$ if $i,k \neq 0$, $i \neq k$. This may be
                                            seen as follows:
                                            $R_{ik} = R^{\mu}_{i \mu k}$
                                            $= \partial_{\mu} \Gamma_{ik}^{\mu} -\partial_{k} \Gamma_{i \mu}^{\mu}
                                            + \Gamma_{\nu \mu}^{\mu} \Gamma_{i k}^{\nu}- \Gamma_{k \nu}^{\mu} \Gamma_{i
                                            \mu}^{\nu}$. Thereby, $\partial_{k} \Gamma_{i \mu}^{\mu}=0$ because
                                            the Christoffel symbols are constant with respect to $x^k$, $k \neq
                                            0$. In addition, $\Gamma_{i k}^{\mu}=0$ for all $\mu$. Thus,
                                            it remains to show that $\Gamma_{k \nu}^{\mu} \Gamma_{i
                                            \mu}^{\nu}= \Gamma_{k k}^{0} \Gamma_{i 0}^{k} + \Gamma_{k0}^{i} \Gamma_{i
                                            i}^{0}$ is zero. But this is clear, because $\Gamma_{i
                                            0}^{k}=0$ due to  $i,k \neq 0$, $i \neq k$. Finally, let us
                                            prove that
                                            $R_{ii} = R^{\mu}_{i \mu i}$
                                            $= \partial_{\mu} \Gamma_{ii}^{\mu} -\partial_{i} \Gamma_{i \mu}^{\mu}
                                            + \Gamma_{\nu \mu}^{\mu} \Gamma_{i i}^{\nu}- \Gamma_{i \nu}^{\mu} \Gamma_{i
                                            \mu}^{\nu}$ is as claimed. For the first summand we obtain
                                            $\partial_{\mu} \Gamma_{ii}^{\mu} $ $
                                            = \partial_{0} \Gamma_{ii}^{0} = -c k_i (2k_i-1)t^{2(k_i-1)}$. The second summand vanishes, because the
                                            Christoffel symbols are constant with respect to $x^i$, $i \neq
                                            0$. The third summand reads as
                                            $\Gamma_{\nu \mu}^{\mu} \Gamma_{i i}^{\nu}= \sum_{j \neq 0} \Gamma_{0 j}^{j} \Gamma_{i i}^{0}$
                                            $ = \sum_{j \neq 0} k_j t^{-1} (-c k_i t^{2k_i-1})$. Finally,
                                            $\Gamma_{i \nu}^{\mu} \Gamma_{i \mu}^{\nu} = \Gamma_{i i}^{0} \Gamma_{i 0}^{i}+ \Gamma_{i 0}^{i} \Gamma_{i i}^{0}$
                                            $=-2c k_i^2t^{2(k_i-1)}$. Everything summed up gives the
                                            desired result.
                                        \end{proof}
                                        \subsection{Schwarzschild solution}\label{5310}

                                        The example of the Schwarzschild metric will show very clearly the
                                        general phenomenon that the Zariski topology is too coarse for
                                        physical applications and that it is necessary to work within the
                                        context of the \'etale toplogy. However, let us again start from the
                                        affine space $\Affin_K^n$ with coordinates $(t=x^0, x^1, \ldots,
                                        x^{n-1})$, but this time we consider the $K$-scheme $X_K:= \Spec
                                        K[t,x^1, \ldots, x^{n-1},r,r^{-1}]/(r^2-\sum_{i \neq 0} (x^i)^2)$,
                                        whereby $r:=\sqrt{\sum_{i \neq 0} (x^i)^2}$ should be interpreted
                                        as a spacial radius. By construction, $X_K$ is \'etale over
                                        $\Affin_K^n$. In particular, the respective differential calculi
                                        ``coincide''. We choose the following metric $g_K$ on $X_K$:
                                        \begin{itemize}
                                            \item
                                            $g_{00}=\frac{1}{1+\frac{2m}{r}}, \qquad g_{0i}= 0, \quad i \neq 0$

                                            \item
                                            $g_{ij}:= - \left(1+\frac{2m}{r}\right)^2 \delta_{ij}
                                            + \frac{x^ix^j}{r^2} \left(1+\frac{2m}{r}\right) \frac{2m}{r}$
                                        \end{itemize}
                                        where $\delta_{ij}$ denotes the Kronecker delta, and $m \in K$ is
                                        a constant. By means of a longer but standard calculation similar
                                        to the one in the Kasner case (see section \ref{5202}), one can
                                        prove that the metric above solves the vacuum Einstein equations.
                                        Thus $(X_K,g_K)$ is indeed a \GRo. But instead of deriving this
                                        result directly, we will illustrate that it corresponds to the
                                        Schwarzschild metric. This will yield the following physical
                                        interpretation: $(X_K,g_K)$ describes the exterior of a black
                                        hole. More precisely, $r$ scales the distance from the event
                                        horizon of the black hole, and the constant $m$ turns out to be
                                        the Schwarzschild diameter of the black hole.

                                        In differential geometry, one usually writes down the
                                        Schwarzschild metric in spherical coordinates. These coordinates
                                        do not make sense in algebraic geometry. But, at least in the
                                        category of Riemannian manifolds we can prove that the metric
                                        above takes the classical form in spherical coordinates. Writing
                                        \begin{align*}
                                            \Big( x^0 &= t,
                                            x^1 = r \sin(\theta) \cos(\varphi),
                                            x^2 = r \sin(\theta) \sin(\varphi),
                                            x^3 = r \cos(\theta) \Big)
                                        \end{align*}
                                        in the four dimensional case, one obtains:
                                        \begin{itemize}
                                            \item
                                            $dR^2=dr^2= \sum_{i,j \neq 0} \frac{x^ix^j}{r^2} dx^idx^j$, \quad
                                            where $R:=r+c$ for some constant $c \in K$.

                                            \item
                                            $\sum_{i\neq 0} (dx^i)^2= dr^2+ r^2 (d\theta^2 + \sin^2(\theta) d\varphi^2
                                            )$.
                                        \end{itemize}
                                        It follows that
                                        \begin{align*}
                                            &\frac{1}{1-\frac{2m}{R}} dR^2+  R^2 (d\theta^2 + \sin^2(\theta) d\varphi^2 )
                                            = \sum_{i,j \neq 0} \Big( \frac{R}{R-2m} \frac{x^ix^j}{r^2}
                                            + \frac{R^2}{r^2}\left(\delta_{ij}-\frac{x^ix^j}{r^2}\right)
                                            \Big)dx^idx^j \\
                                            &= \sum_{i,j \neq 0} \Big( \frac{R^2}{r^2} \delta_{ij} +
                                            \frac{x^ix^j}{r^2} \left(\frac{R}{R-2m} -\frac{R^2}{r^2}\right)
                                            \Big)dx^idx^j \\
                                            &= \sum_{i,j \neq 0} \Big( \left(1+\frac{c}{r}\right)^2 \delta_{ij} +
                                            \frac{x^ix^j}{r^2} \left(\frac{r+c}{r+c-2m} -\frac{(r+c)^2}{r^2}\right)
                                            \Big)dx^idx^j \\
                                            &= \sum_{i,j \neq 0} \Big( \left(1+\frac{c}{r}\right)^2 \delta_{ij}
                                            - \frac{x^ix^j}{r^2} (r+c) \frac{2r(c-m)+c(c-2m)}{(r+c-2m)r^2} \Big)dx^idx^j \\
                                            &= \sum_{i,j \neq 0} \Big( \left(1+\frac{2m}{r}\right)^2 \delta_{ij}
                                            - \frac{x^ix^j}{r^2} \left(1+\frac{2m}{r}\right) \frac{2m}{r} \Big)dx^idx^j  \qquad \text{for $c=2m$.}
                                        \end{align*}
                                        Therefore, our original metric transforms into the standard form
                                        of the Schwarzschild metric
                                        \begin{align*}
                                            ds^2=\left( 1-\frac{2m}{R} \right)dt^2- \frac{1}{1-\frac{2m}{R}} dR^2-  R^2 (d\theta^2 + \sin^2(\theta) d\varphi^2 )
                                        \end{align*}
                                        living on the manifold $M= \real \times (2m, \infty) \times S_2$
                                        with coordinates $(t,R, \theta, \varphi)$.
                                        \subsection{Electromagnetism in vacuum}\label{5320}

                                        In this subsection we will consider the electromagnetic field.
                                        This is a gauge field, and so we may apply the techniques
                                        developed in Part II of this thesis. Therefore, let us consider a
                                        smooth $S$-scheme $X$ (which represents \sto) and a smooth
                                        $S$-group scheme $G$ (see Definition \ref{1204}) representing the gauge
                                        group. The fundamental structure underlying gauge theory is an
                                        $X$-torsor $P$ under $G_X$ (see Definition \ref{1413}), where $G_X$ denotes
                                        the $X$-group scheme $G \times_S X$. Then the field strength of
                                        the gauge field is given by a differential two-form $\cal F$ on
                                        $X$ which solves the Yang-Mills equations \ref{1551}. The differential form $\cal F$
                                        takes values in the Lie-algebra $\Liealg := (\varepsilon^*
                                        \Omega_{G/S}^1)(S)$ of $G$, where $\varepsilon:S \to G$ denotes
                                        the unit section of $G$ (see section \ref{1200a}).

                                        If $S=\Spec K$ is the spectrum of a field $K$, we will illustrate
                                        in section \ref{0140} that the gauge group of electromagnetism
                                        is the algebraic torus $\mathbb{G}_{m,K}:= \Spec
                                        K[\zeta,\zeta^{-1}]$ for some variable $\zeta$. In absence of
                                        gravity we may choose $X= \ba_K^n$ as \st part, where $n \in
                                        \Natural$ denotes the dimension. Therefore, electromagnetism in
                                        vacuum may be described by means of the trivial $\ba_K^n$-torsor
                                        $P= \Affin_K^n \times_{\Affin_K^n} (\Affin_K^n \times_K
                                        \mathbb{G}_{m,K})$ under $\Affin_K^n \times_K \mathbb{G}_{m,K}$,
                                        i.e.
                                        \begin{align*}
                                            P \cong \ba_K^n \times_K \mathbb{G}_{m,K}.
                                        \end{align*}
                                        In this setting, we will first compute the Lie-algebra $\Liealg$
                                        of $G$ in order to solve the Yang-Mills equations in the
                                        electrostatic case in a second step. One obtains the following two
                                        results.

                                        \begin{Lemma}\label{5321}
                                            Let $S=\Spec R$ be the spectrum of a ring $R$, and let
                                            $\Liealg$ be the Lie algebra of the multiplicative group
                                            $\mathbb{G}_{m,S}=\Spec R[\zeta,\zeta^{-1}]$ over $R$.
                                            Then (with respect to the chosen coordinates of $\mathbb{G}_{m,S}$)
                                            there is a canonical isomorphism $\Liealg \cong R$ of
                                            $R$-modules.
                                        \end{Lemma}

                                        \begin{proof}
                                            As an $S$-scheme, $\mathbb{G}_{m,S}$ coincides with $\Spec
                                            R[\zeta,\zeta^{-1}]$. The ring $R[\zeta,\zeta^{-1}]$ is the
                                            localization $R[\zeta]_{\zeta}$ of the polynomial ring
                                            $R[\zeta]$ with respect to the variable $\zeta$. Therefore,
                                            the canonical morphism $i:\Spec R[\zeta,\zeta^{-1}]
                                            \hookrightarrow \Spec R[\zeta]= \Affin_{S}^1$ is an
                                            open immersion. As open immersions
                                            are \'etale, it follows from Proposition \ref{0027} that
                                            $\Omega_{\mathbb{G}_{m,S}/S}^1 \cong i^*
                                            \Omega_{\Affin_{S}^1/S}^1$ is an isomorphism. By definition,
                                            the Lie-algebra $\Liealg$ is the $R$-module $(\varepsilon^*
                                            \Omega_{\mathbb{G}_{m,S}/S}^1)(S)$, where $\varepsilon: S \to
                                            \mathbb{G}_{m,S}$ denotes the unit section corresponding to
                                            the morphism of $R$-algebras $\varepsilon^*: R[\zeta,\zeta^{-1}] \to
                                            R$, $\zeta \mapsto 1 \in R^*$, mapping $\zeta$ into the group of units $R^*$ of
                                            $R$. In order to compute $\Liealg$, first notice that
                                            $\varepsilon^* \Omega_{\mathbb{G}_{m,S}/S}^1 \cong \varepsilon^* i^*
                                            \Omega_{\Affin_{S}^1/S}^1$$\cong ( i \circ \varepsilon)^*
                                            \Omega_{\Affin_{S}^1/S}^1$, whereby $ i \circ \varepsilon$
                                            corresponds to the $R$-algebra homomorphism $R[\zeta] \to
                                            R$, $\zeta \mapsto 1 \in R^*$. Recalling that
                                            $\Omega_{\Affin_{S}^1/S}^1(S)$ is the free $R[\zeta]$-module
                                            $R[\zeta]\cdot    d\zeta$ generated by the differential  $d\zeta$, we obtain the
                                            following $R$-module isomorphisms:
                                            \begin{align*}
                                                \Liealg
                                                = \left(( i \circ \varepsilon)^* \Omega_{\Affin_{S}^1/S}^1 \right)(S)
                                                \cong \Omega_{\Affin_{S}^1/S}^1(S) \otimes_{R[\zeta]} R
                                                \cong {R[\zeta]} \otimes_{ \text{id}, R[\zeta], (i \circ \varepsilon)^*} R
                                                \cong {R}.
                                            \end{align*}
                                        \end{proof}

                                        \begin{Cor}\label{5322}
                                            Let $S$ be an affine scheme, let $X$ be an smooth, affine $S$-scheme, and let $G=\bg_m$. Let
                                            $P=X \times_S G$ be the trivial $X$-torsor under $G_X=G \times_S X$.
                                            Then the field strength $\cf$ may be viewed as differential
                                            two form on $X$, i.e. $\cf \in \Omega_{X/S}^2(X)$.
                                        \end{Cor}

                                        \begin{proof}(of Corollary \ref{5322}):
                                            By Definition \ref{1533}, the field strength $\cf=s^* \Omega$ is locally defined as pull back
                                            of a Lie-algebra valued global differential form $\Omega \in \Omega_{P/S}^2(P)
                                            \otimes_{\co_{S}(S)}\Liealg$ under the canonical section $s$. As $\Liealg \cong
                                            \co_{S}(S)$  due to Lemma \ref{5321}, and we may view $\Omega$ as
                                            differential form $\Omega \in \Omega_{P/S}^2(P)$ on $P$.

                                            In general, the canonical section $s$ is only defined locally in
                                            \'etale topology, but in our special  situation, $s$ coincides with the morphism
                                            obtained from the unit section $\varepsilon: S \to G$ by base
                                            change with $\pi: X \to S$. Therefore, $s$ makes the following
                                            diagram commutative.
                                            \begin{align*}
                                                \xymatrix{
                                                & X \\
                                                X \ar@{->}[r]^{s \ \ } \ar@{=}[ur]^{id} \ar[dr]_{ \varepsilon \circ \pi}
                                                & X \times_S  G \ar[d]^{p_2} \ar[u]_{p_1} \\
                                                & G
                                                }
                                            \end{align*}
                                            By Proposition \ref{0018}, we have got                  $(s^* \Omega_{P/S}^1)(X)
                                            \cong ((p_1 \circ s)^*\Omega_{X/S}^1)(X) \oplus ((p_2 \circ
                                                                        s)^*\Omega_{G/S}^1)(X)$
                                            $\cong (\text{id}^*\Omega_{X/S}^1)(X) \oplus ((\varepsilon \circ
                                                                                    \pi)^*\Omega_{G/S}^1)(X)$
                                            $\cong \Omega_{X/S}^1(X) \oplus  (\Liealg
                                            \otimes_{\co_{S}(S)}\co_X(X))$, and therefore the pull-back of
                                            differential forms under $s$ is the canonical projection
                                            \begin{align*}
                                                (s^* \Omega_{P/S}^1)(X)\cong \Omega_{X/S}^1(X) \oplus  \co_X(X)
                                                \to \Omega_{X/S}^1(X).
                                            \end{align*}
                                        \end{proof}
                                        Now, we are prepared to compute the electromagnetic field in
                                        vacuum, i.e. we will solve the corresponding Yang-Mills equations.
                                        Let $K$ be a field. As already stated above, we choose $S=\Spec
                                        K$, $G=\mathbb{G}_{m,K} \cong \Spec K[\zeta,\zeta^{-1}]$,
                                        $X=\ba_K^4$ with metric $g= \text{diag}(-1,1,1,1)$ and $P=X
                                        \times_S G$. Due to the fact that the group scheme $G$ is
                                        commutative, the Lie-bracket on $\Liealg$ vanishes (see section
                                        \ref{1230}). But then we know from Definition \ref{1613} that the
                                        covariant derivation $D_X$ on $X$ coincides with the exterior
                                        differential $d$ introduced in Theorem \ref{1156}. Furthermore,
                                        $\Omega_{X/S}^1(X) \cong \oplus_{i=0}^3 \co_X(X) dx^i$ is a free
                                        $\co_X(X)$ module, where $(x^0=t,x^1,x^2,x^3)$ denote the
                                        coordinates of $X$. Thus we may write $\cf= \sum_{\mu \nu}
                                        \cf_{\mu \nu} dx^{\mu} \wedge dx^{\nu}$ with polynomials $\cf_{\mu
                                        \nu} \in \co_X(X)=K[t,x^1,x^2,x^3]$. The components of the
                                        anti-symmetric tensor $\cf$ are usually written as matrix:
                                        \begin{align*}
                                            (\cf_{\mu \nu})=\left(%
                                        \begin{array}{cccc}
                                        0 & E_1 & E_2 & E_3 \\
                                        -E_1 & 0 & B_1 & B_2 \\
                                        -E_2 & -B_1 & 0 & B_3 \\
                                        -E_3 & -B_2 & -B_3 & 0 \\
                                        \end{array}%
                                        \right)
                                        \end{align*}
                                        The vector $\overrightarrow{E}:=(E_1,E_2,E_3)$ is called the
                                        electric field strength, $\overrightarrow{B}:=(B_1,B_2,B_3)$
                                        denotes the magnetic field strength. We are interested in an
                                        electrostatic solution, i.e. $\overrightarrow{B}=0$. In this
                                        special case the Yang-Mills equations
                                        \begin{align*}
                                            *d*\cf=0 \\
                                            d\cf=0
                                        \end{align*}
                                        take the form
                                        \begin{align*}
                                            \sum_{i \neq 0} \frac{\partial}{\partial x^i}E_i&=0 \\
                                            \frac{\partial}{\partial x^i}E_j-\frac{\partial}{\partial
                                            x^j}E_i&=0 \qquad \text{for all $i,j \neq 0$}.
                                        \end{align*}
                                        This may be seen as follows: Due to the fact that $\co_{X}(X)$ is
                                        a polynomial ring, the local version of the Yang-Mills equations
                                        \ref{1622} is not only true at physical points, but even globally.
                                        Recalling that $g= \text{diag}(-1,1,1,1)$ and that $c_{ij}^k=0$
                                        (due to the commutativity of the gauge group), the Yang-Mills
                                        equations take the form $ \sum_{\mu,\rho,\sigma=1}^n
                                        \frac{\partial}{\partial x^{\sigma}} (g^{\mu \nu} g^{\rho \sigma}
                                        \mathcal{F}_{\mu\rho})=0$. But because of $\overrightarrow{B}=0$,
                                        $\cf_{\mu \nu} \neq 0$ if and only if exactly one index is zero.
                                        Thus, the only non trivial equation is: $ 0=
                                            \sum_{\mu,\rho,\sigma=1}^n \frac{\partial}{\partial x^{\sigma}}
                                            (g^{\mu 0} g^{\rho \sigma} \mathcal{F}_{\mu\rho})
                                            = - \sum_{\sigma=1}^n \frac{\partial}{\partial x^{\sigma}}
                                            \mathcal{F}_{0 \sigma}
                                            =- \sum_{i \neq 0} \frac{\partial}{\partial x^i}E_i$.
                                        In order to derive the second equation, recall that we may write
                                        $\cf = \sum_{i\neq 0} E_i dt \wedge dx^i$. Thus, $d \cf =\sum_{i,j
                                        \neq 0} \frac{\partial E_i}{\partial x^j} dx^j \wedge dt \wedge
                                        dx^i = \sum_{i,j \neq 0} \frac{\partial E_i}{\partial x^j} dt
                                        \wedge dx^i \wedge dx^j$, and if $\partial_i$ denotes the dual of
                                        $dx^i$ it follows from Definition \ref{1151}, b) that
                                        \begin{align*}
                                            d \cf (\partial_0, \partial_k, \partial_l)
                                            &= \sum_{i,j \neq 0} \frac{\partial E_i}{\partial x^j} \cdot
                                            \frac{1}{2} dt(\partial_0) \Big( (dx^i \wedge dx^j)(\partial_k, \partial_l)
                                            -(dx^i \wedge dx^j)(\partial_l, \partial_k) \Big) \\
                                            &= \sum_{i,j \neq 0} \frac{\partial E_i}{\partial x^j} \cdot
                                            (dx^i \wedge dx^j)(\partial_k, \partial_l) \\
                                            &= \sum_{i,j \neq 0} \frac{\partial E_i}{\partial x^j} (\delta_{ik} \delta_{jl}- \delta_{il}
                                            \delta_{jk})
                                            = \frac{\partial E_k}{\partial x^l}- \frac{\partial E_l}{\partial x^k}
                                        \end{align*}
                                        It is well known that $E_i:=\frac{x^i}{r}$, $r^2:=\sum_{i \neq 0}
                                        (x^i)^2$ is a solution of these equations. Obviously, $E_i$ is
                                        neither a global section of $X$ nor of $X-\{0\}=\Affin_K^n-\{0\}$.
                                        Instead we may consider the smooth $K$-scheme, $U:= \Spec
                                        K[t,x^1,x^2,x^3,r,r^{-1}]/(r^2-\sum_{i \neq 0} (x^i)^2)$ which is
                                        \'etale over $\Affin_K^4$ and contains $r$ as global function. The
                                        \'etale-open ``sub-scheme'' $U$ of $\Affin_K^4$ describes a \st
                                        which contains a charged point particle located at $r=0$ (more
                                        precisely, $U$ is the environment of this charged particle). The
                                        electrostatic field of this point charge is described by global
                                        sections $E_i \in \co_U(U)$.

                                        \subsection{Robertson-Walker models}\label{5340}
                                        Last but not least, let us briefly mention the Robertson-Walker
                                        models. In differential geometry, these are defined in spherical
                                        coordinates by
                                        \begin{align*}
                                            g_{\mu \nu}
                                            &= \text{diag}\left(1, \frac{-S(t)^2}{1-kr^2},-S(t)^2r^2,-S(t)^2r^2  sin^2(\theta) \right)\\
                                            T_{\mu \nu}&= \text{diag}\left(\rho(t), \frac{-p(t)S(t)^2}{1-kr^2},-p(t)S(t)^2r^2,-p(t)S(t)^2r^2
                                            sin^2(\theta)\right)
                                        \end{align*}
                                        where $S(t)$ is a so called scale factor, and where $\rho$ resp.
                                        $p$ denote the density resp. pressure of energy. In the special
                                        case $p=0$, the divergence $\text{div}T$ of the energy-stress
                                        tensor vanishes if and only if the product $C:=\rho(t) \cdot
                                        S(t)^3$ is constant with respect to $t$. Then the Einstein
                                        equations $R_{\mu \nu} -\frac{1}{2} g_{\mu \nu} R = \kappa T_{\mu
                                        \nu}$ yield the differential equation
                                        \begin{align*}
                                            \left(\frac{\partial S(t)}{\partial t}\right)^2+k= \frac{\kappa
                                            C}{3 S(t)}.
                                        \end{align*}
                                        Whenever the solution $S(t)$ of this differential equation is an
                                        algebraic function, the corresponding Robertson-Walker metric
                                        would make sense in algebraic geometry. However, in general,
                                        $S(t)$ will not be algebraic. So, the classical Robertson-Walker
                                        models have no algebraic geometric analogues. But at least in the
                                        case $k=0$, we find the algebraic solution
                                        \begin{align*}
                                            S(t)= t^{\frac{2}{3}} \sqrt[3]{4/3 \cdot \kappa C}.
                                        \end{align*}
                                        Therefore, we obtain a \GR if we choose $X_K:= \Affin_K^n$ with
                                        coordinates $(t=x^0, x^1, \ldots, x^{n-1})$ as well as the
                                        following metric and energy-stress tensor:
                                        \begin{align*}
                                            g_{\mu \nu} &= \text{diag} \left( 1, - c t^{\frac{4}{3}}, - c t^{\frac{4}{3}}, - c t^{\frac{4}{3}}
                                            \right), \qquad c:=(4/3 \cdot \kappa C)^{\frac{2}{3}} \in K \\
                                            T_{\mu \nu} &= \left( \frac{4}{3t^2}, 0, 0, 0 \right).
                                        \end{align*}

                                        \section{The case of Zariski one-dimensional base}\label{6000}

                                        Within this section, let $S= \Spec {\footnotesize{\text{$\cal
                                        O$}}}$  be the spectrum of a Dedekind ring which is not a field.
                                        From the ``adelic'' point of view, this is the physically
                                        interesting case, because the condition (ii) of Definition \ref{5022} is no
                                        longer empty. Consequently, it is much harder to construct \GRos.

                                        In section \ref{6010}, we will first consider the low
                                        dimensional case, because then the Einstein equations are trivial.
                                        But as soon as the tangent spaces exceed three dimensions, this is
                                        no longer true. Then, the conditions (i) and  (ii) of Definition \ref{5022}
                                        are both non-trivial. This the physically interesting situation
                                        where we are looking for \GRs $(X \to S, g)$. An example, which
                                        may be interpreted as the ``adelic'' Minkowski space, will be
                                        studied in section \ref{6020}.

                                        \subsection{The low dimensional case}\label{6010}

                                        Let $X \to S$ be a smooth $S$-scheme of relative dimension one or
                                        two with metric $g$. We will show that the Einstein equations are
                                        trivial in this case. As it suffices to show this locally, we may
                                        choose an open sub-scheme $U$ of $X$ such that $\Omega_{U/S}^1$ is
                                        free with base $\{ \omega^i \}$. In the one dimensional case, our
                                        assertion is clear, because the curvature tensor has got only a
                                        single component, and this component vanishes due to the
                                        symmetries of the curvature tensor (see Proposition \ref{5111}). Thus, also
                                        the Einstein tensor $\text{ric}- \frac{1}{2} \, g \cdot \text{sc}$
                                        vanishes . In the two dimensional case, a small computation is
                                        necessary. Again making use of the identities of the curvature
                                        tensor several times, we obtain with respect to the given base:
                                        \begin{align*}
                                            R&:= \text{sc}
                                            = g^{\mu \nu} \text{ric}_{\mu \nu} \\
                                            &= g^{\mu \nu} R^1_{\mu 1 \nu} +  g^{\mu  \nu} R^2_{\mu 2 \nu}
                                            = g^{\mu 2} R^1_{\mu 1 2} +  g^{\mu  1} R^2_{\mu 2 1}
                                            =g^{\mu 2} g^{\nu 1} R_{\nu \mu 1 2} +  g^{\mu  1} g^{\nu 2} R_{ \nu \mu 2
                                            1}\\
                                            &=  g^{11}g^{22} R_{12 1 2}+ g^{21}g^{12} R_{21 1 2}+ g^{11}g^{22} R_{21 21}+ g^{21}g^{12}
                                            R_{1221}\\
                                            &= 2 \text{det}g \cdot R_{2121}.
                                        \end{align*}
                                        Therefore, one derives that
                                        \begin{align*}
                                            R^{\mu}_{\nu}
                                            &:= \text{Ric}^{\mu}_{\nu}
                                            = g^{\mu \lambda} \text{ric}_{\lambda \nu} \\
                                            &= g^{\mu \lambda} R^1_{\lambda 1 \nu} +  g^{\mu  \lambda} R^2_{\lambda 2 \nu}
                                            = g^{\mu \lambda} g^{1 \iota} R_{\iota \lambda 1 \nu} +  g^{\mu  \lambda} g^{2 \iota} R_{\iota \lambda 2
                                            \nu} \\
                                            &= g^{\mu 1} g^{1 2} R_{21 1 \nu}+g^{\mu 2} g^{11} R_{121 \nu}
                                            + g^{\mu 1} g^{2 2} R_{21 2 \nu} +g^{\mu 2} g^{21} R_{122
                                            \nu}\\
                                            &= R_{211 \nu} (g^{\mu 1} g^{1 2} -g^{\mu 2} g^{1 1}) + R_{212 \nu} (g^{\mu 1} g^{22} -g^{\mu 2} g^{2
                                            1})\\
                                            &= \delta_{2 \mu} \delta_{\nu2} (g^{21}g^{12}-g^{22}g^{11})
                                            R_{2112} + \delta_{\mu 1} \delta_{\nu 1}
                                            (-g^{21}g^{12}+g^{22}g^{11}) R_{2121} \\
                                            &= R_{2121} \text{det}g \cdot (\delta_{2 \mu} \delta_{\nu2}+\delta_{\mu 1} \delta_{\nu 1})
                                            = \frac{1}{2} \delta_{\mu \nu} R.
                                        \end{align*}
                                        Consequently, $R^{\mu}_{\nu}- \frac{1}{2} \delta_{\mu \nu} R =0$
                                        or equivalently $R_{\mu \nu}- \frac{1}{2} g_{\mu \nu} R =0$. In
                                        particular, there are many \GRs in the low dimensional case.

                                        \begin{Cor}\label{6011}
                                            Let $X \to S$ be a smooth $S$-scheme of relative dimension one
                                            or two such that $X$ is the N\'eron model of its generic fibre.
                                            Then $X$ gives rise to a \GRo.
                                        \end{Cor}

                                        \subsection{A higher dimensional solution}\label{6020}

                                        Again, let $X \to S$ be smooth over the Zariski one-dimensional
                                        base scheme $S$. If the relative dimension exceeds two, the
                                        Einstein equations are no longer trivial. Therefore, it is not
                                        easy to find \GRs in the ``adelic'' situation. However, in this
                                        section we will state at least one example, namely the fibred
                                        product $X:= E_1 \times_S \ldots \times_S E_n$ of smooth elliptic
                                        curves $E_i$ over $S$. In a certain way, this is the ``adelic''
                                        analogue of Minkowski space-time. Before we will finally prove in
                                        Theorem \ref{6021} that $X$ is indeed a \GRo, we need some preparations.
                                        However, the idea of the proof is as follows: Locally, every
                                        smooth curve $C$ over $S$ may be embedded in some affine space
                                        $\Affin_S^n$. Pulling back the flat metric $\text{diag}(\pm1,
                                        \ldots, \pm1)$ on $\Affin_S^n$ to $C$ and  pushing forward this
                                        metric by means of an \'etale morphism $C \to \Affin_S^1$ (which
                                        exists locally due to smoothness), we obtain the first fundamental
                                        form on $C$ in local coordinates. Analogously, we obtain the first
                                        fundamental form on a product of curves in local coordinates. This
                                        metric is diagonal, because $\Omega_{{(X_1 \times_S X_2)}/S}^1
                                        \cong \bigoplus_i p_i^*\Omega_{{X_i}/S}^1$, where $p_i$ denotes
                                        the projection onto the $i$-th factor (see Lemma \ref{6022}). It
                                        follows that the corresponding curvature tensor vanishes (see
                                        Corollary \ref{6024}). Consequently, a product of elliptic curves is a
                                        vacuum solution of the Einstein equations and it is even a \GRo,
                                        because it is the N\'eron model of its generic fibre. However, let
                                        us now make the indicated steps of the proof explicit.

                                        \begin{Lemma}\label{6022}
                                            Let $C_1, \ldots, C_n$ be $n$ smooth curves over $S$. Provide $X:= C_1 \times_S \ldots \times_S
                                            C_n$ with the first fundamental form $g$ as metric (see Proposition \ref{0201s}). Then $g$ is
                                            diagonal.
                                        \end{Lemma}

                                        \begin{proof}
                                            It suffices to prove this statement locally on $X$. Therefore,
                                            we may assume that there are $S$-immersions
                                            $j_i:C_i \hookrightarrow Y_i:=\ba_S^m$ for some $m \in \Natural$ (see Definition \ref{0021}).
                                            These $S$-immersions induce an $S$-immersion $j:X \hookrightarrow
                                            Y:=\ba_S^{nm}$. Furthermore, we know from Proposition \ref{0028} that there is an open sub-scheme
                                            $C'_i$  of $C_i$ together with an \'etale morphism $g_i:C'_i \to
                                            \ba_S^1$. Thus we obtain the following chain of maps:
                                            \begin{align*}
                                                Y_i \stackrellow{j_i}\hookleftarrow C_i \stackrellow{i}\hookleftarrow C'_i \stackrellow{g_i}\to \ba_S^1.
                                            \end{align*}
                                            Eventually shrinking $C_i$, we may assume $C_i=C'_i$ and $i= \text{id}$.
                                            Recalling that the pull back of differential forms under \'etale
                                            morphisms is an isomorphism (see Proposition \ref{0027}), these
                                            maps induce the following canonical homomorphism on the level of differential
                                            forms.
                                            \begin{equation*}
                                                \begin{array}{cclcccc}
                                                    j_i^* \Omega_{\ba_S^m/S}^1 & \stackrellow{j_i^*}{\longrightarrow}
                                                    & \Omega_{C_i/S}^1 &
                                                    \stackrel{\sim}{\stackrellow{(g_i^*)^{-1}}\longrightarrow} &  g_i^*
                                                    \Omega_{\ba_S^1/S}^1 \qquad \quad (*)
                                                \end{array}
                                            \end{equation*}
                                            This morphism maps the generators $dT_{ij}$, $j=1, \ldots, m$,
                                            of the free module $j_i^* \Omega_{\ba_S^m/S}^1$ to
                                            elements $\kappa_{ij} dT_i \in g_i^* \Omega_{\ba_S^1/S}^1$,
                                            where $dT_i$ denotes the generator of the free module $g_i^*
                                            \Omega_{\ba_S^1/S}^1$.

                                            Moreover, if $p_i:X \to C_i$ denotes the canonical projection, there is an isomorphism
                                            $\Omega_{X/S}^1 \cong \bigoplus_i p_i^*\Omega_{C_i/S}^1$ due to
                                            Proposition \ref{0018}. Analogously, if $\pi_i: Y \to Y_i$ denotes the canonical
                                            projection making the diagram
                                            \begin{align*}
                                                \xymatrix @-0.2pc {  Y_i  &
                                                C_i \ar[l]_{j_i}   \\
                                                Y \ar[u]_{\pi_i}   &  X \ar[l]_{j} \ar[u]_{p_i}   }
                                            \end{align*}
                                            commutative, there is an isomorphism
                                            $\Omega_{{\ba_S^{nm}}/S}^1 \cong \bigoplus_i
                                            \pi_i^*\Omega_{{\ba_S^m}/S}^1$. Now, the
                                            pull-back $j^*\Omega_{{\ba_S^{nm}}/S}^1 \to \Omega_{X/S}^1$
                                            decomposes into a direct sum of componentwise pull-backs, i.e.
                                            the following diagram is commutative:
                                            \begin{align*}
                                                \xymatrix{  j^*\Omega_{{\ba_S^{nm}}/S}^1 \ar[r]^{\sim \ \ \ \ }
                                                \ar[d]^{j^*}
                                                & \bigoplus_i  j^*\pi_i^*\Omega_{{\ba_S^m}/S}^1 \ar[r]^{\sim \ }
                                                & \bigoplus_i  p_i^*(j_i^*\Omega_{{\ba_S^m}/S}^1) \ar[d]^{\oplus j_i^*}  \\
                                                \Omega_{X/S}^1 \ar[rr]_{\sim}   &  & \bigoplus_i  p_i^*\Omega_{C_i/S}^1  }
                                            \end{align*}
                                            The upper line of this diagram yields a morphism
                                            \begin{align*}
                                                j^*\Omega_{{\ba_S^{nm}}/S}^1
                                                \stackrel{\sim}\longrightarrow \bigoplus_i  p_i^*(j_i^*\Omega_{{\ba_S^m}/S}^1)
                                                \stackrel{\text{pr}_i}\longrightarrow
                                                p_i^*(j_i^*\Omega_{{\ba_S^m}/S}^1) \qquad \quad (**)
                                            \end{align*}
                                            where $\text{pr}_i$ denotes the projection onto the $i$-th
                                            component.
                                            Pulling back the morphism $(*)$ with $p_i$ and composing it
                                            with the above morphism $(**)$, we finally obtain the morphism
                                            \begin{align*}
                                                \begin{array}{cclcccc}
                                                    \phi_i: j^*\Omega_{{\ba_S^{nm}}/S}^1
                                                    & \stackrel{}\longrightarrow
                                                    & (j_i \circ p_i)^*\Omega_{{\ba_S^m}/S}^1
                                                    & \stackrel{}\longrightarrow
                                                    & (g_i \circ p_i)^* \Omega_{\ba_S^1/S}^1 \\
                                                    dT_{kj}
                                                    & \mapsto
                                                    & \delta_{ik}dT_{kj}
                                                    & \mapsto
                                                    & \delta_{ik}\kappa_{kj}dT_i,
                                                \end{array}
                                            \end{align*}
                                            where $\kappa_{i j} \in \co_X$ is algebraic over $\co_S[T_i]$ and $\delta_{ik}$ denotes the Kronecker delta.
                                            Now, we have all that we need to compute the first
                                            fundamental form. On $Y={\ba_S^{nm}}$ we choose the trivial
                                            metric $g_0= \sum_{\mu \nu}\varepsilon_{\mu \nu} dT_{\mu \nu} \otimes dT_{\mu \nu}$,
                                            $\varepsilon_{\mu \nu} \in  \{ \pm 1 \}$. Then, by definition,
                                            the first fundamental form $g$ reads as follows:
                                            \begin{align*}
                                                g
                                                &= \sum_{i,j,\mu, \nu}\varepsilon_{\mu \nu} \phi_i(dT_{\mu \nu}) \otimes \phi_j(dT_{\mu
                                                \nu})
                                                = \sum_{i,j,\mu, \nu}\varepsilon_{\mu \nu} \delta_{i\mu}\kappa_{\mu \nu}
                                                \delta_{j\mu}\kappa_{\mu \nu}dT_i \otimes dT_j
                                            \end{align*}
                                            Therefore, $g= \sum_{ij}g_{ij}dT_i \otimes dT_j$ with
                                            $g_{ij}= \delta_{ij}\cdot\sum_{\nu}\varepsilon_{i \nu} \kappa_{i
                                            \nu}^2$. In particular, $g$ is diagonal.
                                        \end{proof}

                                        \begin{Ex}\label{6023}
                                            In order to illustrate the derivation of the first fundamental
                                            form, let us consider the example of a smooth elliptic
                                            curve $E$ over a field $K$ which is not of characteristic two. Let us
                                            now restrict $E$ to the affine open subset
                                            $\Spec K[X,Y,Y^{-1}] \subset \bp_K^2$, and let us furthermore
                                            assume that
                                            $E$ is described by an equation
                                            $P(Y,X):=Y^2-X^3-g_{2}X-g_{3}=0$.
                                            Then we have canonical morphisms of $K$-algebras
                                            \begin{equation*}
                                                \begin{array}{cclcccc}
                                                    K[X,Y,Y^{-1}]
                                                    & \stackrellow{j^*}{\longrightarrow}
                                                    & \co_{E}:=K[X,Y,Y^{-1}]/(P)
                                                    & {\stackrellow{g^*}\longleftarrow}
                                                    &  K[X]\\
                                                    X
                                                    & \mapsto
                                                    & \qquad \quad \qquad  \overline{X}
                                                    & \leftarrowtail
                                                    & X \\
                                                    Y
                                                    & \mapsto
                                                    & \qquad \quad \qquad \overline{Y}
                                                    &
                                                    &
                                                \end{array}
                                            \end{equation*}
                                            where $g^*$ is \'etale. Therefore, we obtain on the level of
                                            differential forms
                                            \begin{align*}
                                                \begin{array}{cclcccc}
                                                    \Omega_{K[X,Y,Y^{-1}]/K}^1 \otimes_{K[X,Y,Y^{-1}]} \co_{E}
                                                    & \stackrellow{j^*}\longrightarrow
                                                    & \Omega_{\co_{E}/K}^1
                                                    & \stackrel{\sim}{\stackrellow{(g^*)^{-1}}\longrightarrow}
                                                    & \Omega_{K[X]/K}^1 \otimes_{K[X]} \co_{E} \\
                                                    dX
                                                    & \mapsto
                                                    & d\overline{X}
                                                    & \mapsto
                                                    & dX \\
                                                    dY
                                                    & \mapsto
                                                    & d\overline{Y}
                                                    & \mapsto
                                                    & \frac{3X^2+g_2}{2Y}dX
                                                \end{array}
                                            \end{align*}
                                            where we made use of the identity $(3\overline{X}^2+g_2)d\overline{X}$
                                            $=d\overline{Y^2}=2\overline{Y}d\overline{Y}$ in
                                            $\Omega_{\co_{E}/K}^1$. This is the map $(*)$ in the proof of
                                            Lemma \ref{6022}. The dual of this map is the $\co_{E}$-linear map
                                            $\partial_X \mapsto \partial_X
                                            +\frac{3X^2+g_2}{2Y}\partial_Y$, where $\partial_X$ (resp.
                                            $\partial_Y$) denotes the dual of $dX$ (resp. $dY$). In
                                            particular, the tangent vectors of $E$ in the affine
                                            open subset $\Spec K[X,Y,Y^{-1}] \subset \bp_K^2$ may be
                                            written as vectors
                                            \begin{align*}
                                                \left(%
                                                \begin{array}{c}
                                                1 \\
                                                \frac{3X^2+g_2}{2Y} \\
                                                \end{array}%
                                                \right).
                                            \end{align*}
                                            Providing $\Spec K[X,Y,Y^{-1}]$ with the trivial metric $g_0=
                                            \text{diag}(1,1)$ and interpreting $g_0$ as bilinear form, we derive the first fundamental form $g$ on $E$:
                                            \begin{align*}
                                                g=g_0\left(
                                                \left(%
                                                \begin{array}{c}
                                                1 \\
                                                \frac{3X^2+g_2}{2Y} \\
                                                \end{array}%
                                                \right),
                                                \left(%
                                                \begin{array}{c}
                                                1 \\
                                                \frac{3X^2+g_2}{2Y} \\
                                                \end{array}%
                                                \right)
                                                \right)=1+ \frac{(3X^2+g_2)^2}{4Y^2}.
                                            \end{align*}
                                            This is manifestly the same result as in differential
                                            geometry. The procedure in the case of a product of $n$ elliptic curves is straight
                                            forward. For example, if $n=2$, we have to compose the above
                                            homomorphism of $K$-algebras with the projection map
                                            $K[X,Y,Y^{-1},Z,W] \to K[X,Y,Y^{-1}]$, $Z,W \mapsto 0$, where
                                            $Z$ and $W$ are the variables of the second elliptic curve. In
                                            this case we obtain the two tangent vectors
                                            \begin{align*}
                                                \left(%
                                                \begin{array}{c}
                                                1 \\
                                                \frac{3X^2+g_2}{2Y} \\
                                                0 \\
                                                0 \\
                                                \end{array}%
                                                \right) \quad \text{and} \quad
                                                \left(%
                                                \begin{array}{c}
                                                0 \\
                                                0 \\
                                                1 \\
                                                \frac{3Z^2+g'_2}{2W} \\
                                                \end{array}%
                                                \right).
                                            \end{align*}
                                            In particular, the metric is diagonal.
                                        \end{Ex}
                                        \begin{Cor}\label{6024}
                                            Let $X$ be as in Lemma \ref{6022}. Then $X$ is flat, i.e. the
                                            curvature tensor vanishes.
                                        \end{Cor}

                                        \begin{proof}
                                            As the curvature tensor is a global section of a sheaf, it
                                            suffices to prove the statement locally. Therefore, we may
                                            assume that the components $g_{ij}$ of the metric tensor $g$ are algebraic
                                            over some polynomial ring, say with variables $x_1, \ldots, x_n$.
                                            In particular we may use the Christoffel
                                            symbols (as well as the formulas stated at the
                                            end of section \ref{5120}) in order to compute the curvature
                                            tensor. Due to Lemma \ref{6022}, the metric is diagonal.
                                            Consequently, the Christoffel symbols read as follows:
                                            \begin{enumerate}
                                                \item[]
                                                $\Gamma^i_{ii}= \frac{1}{2}g^{ii}\partial_i g_{ii}$.

                                                \item[]
                                                $\Gamma^k_{ii}= -\frac{1}{2}g^{kk}\partial_k g_{ii} $, \quad $k\neq
                                                i$.

                                                \item[]
                                                $\Gamma^j_{ij}= \frac{1}{2}g^{jj}\partial_i g_{jj}$ , \quad \ $j\neq
                                                i$.
                                            \end{enumerate}
                                            More precisely, we saw in the proof of Lemma \ref{6022} that
                                            $g_{ij}= \delta_{ij}\cdot\sum_{\nu}\varepsilon_{i \nu} \kappa_{i
                                            \nu}^2$, where $\kappa_{i \nu}^2$ is an algebraic function
                                            which only depends on the variable $x_i$, and where $\varepsilon_{i \nu} \in \{ \pm 1 \}$
                                            is a constant. Thus, $\partial_k g_{ii}$ is
                                            zero if $k \neq i$. It follows that $\Gamma^i_{ii}$ is the
                                            only non-vanishing Christoffel symbol. In terms of the
                                            Christoffel symbols, the components of the curvature tensor may be written
                                            as $R^l_{ijk} = \partial_j \Gamma_{ki}^l -  \partial_k \Gamma_{ij}^l +
                                            \Gamma_{ki}^r\Gamma_{jr}^l -  \Gamma_{kr}^l\Gamma_{ij}^r$.
                                            In our setting, $\partial_j \Gamma_{ki}^l=
                                            \delta_{ik}\delta_{il}\partial_j \Gamma_{ii}^i=$$\delta_{ij}\delta_{ik}\delta_{il}\partial_i
                                            \Gamma_{ii}^i$, where the last equality is due to the fact
                                            that $g_{ii}$ only depends on the variable $x_i$
                                            and that $\Gamma^i_{ii}= \frac{1}{2g_{ii}}\partial_i g_{ii}$.
                                            Analogously, $\partial_k \Gamma_{ij}^l=\delta_{ij}\delta_{ik}\delta_{il}\partial_i
                                            \Gamma_{ii}^i$, i.e. $\partial_j \Gamma_{ki}^l -  \partial_k \Gamma_{ij}^l
                                            =0$. Therefore, it remains to prove that also $\Gamma_{ki}^r\Gamma_{jr}^l -
                                            \Gamma_{kr}^l\Gamma_{ij}^r=0$, where over the index $r$ is summed due to
                                            Einstein´s convention. Writing this sum explicitly, we have
                                            $\sum_r\Gamma_{ki}^r\Gamma_{jr}^l=$$\sum_r\delta_{rk}\delta_{ir}\delta_{rj}\delta_{lr}(\Gamma_{rr}^r)^2=$
                                            $\delta_{ik}\delta_{ij}\delta_{il}(\Gamma_{ii}^i)^2$.
                                            Similarly, $\sum_r\Gamma_{kr}^l\Gamma_{ij}^r =$
                                            $\sum_r\delta_{rk}\delta_{rl}\delta_{ir}\delta_{rj}\Gamma_{kr}^l\Gamma_{ij}^r=$
                                            $\delta_{ik}\delta_{ij}\delta_{il}(\Gamma_{ii}^i)^2$. This
                                            yields $R^l_{ijk}=0$.
                                        \end{proof}
                                        Let us finally prove the desired result.

                                        \begin{Theorem}\label{6021}
                                            Let $X:= E_0 \times_S \ldots \times_S E_n$ be the fibred product of
                                            smooth elliptic curves $E_i$ over $S$. Let $g$ be the
                                            first fundamental form on $X$ (see Proposition \ref{0201s}). Then $(X \to S,g)$ is a
                                            \GRo.
                                        \end{Theorem}
                                        \begin{proof}
                                            By Corollary \ref{6024}, $X$ is flat, i.e. the curvature tensor
                                            vanishes. In particular, the Ricci tensor and the scalar
                                            curvature vanish. Therefore, $(X,g)$ is a solution of the vacuum Einstein equations.
                                            Due to the fact that N\'eron models fulfill the
                                            property (ii) of \GRs
                                            (see Definition \ref{5022}), it suffices to show that $X$ is the
                                            N\'eron model of its generic fibre. Now notice that the fibred
                                            product of N\'eron models over $S$ is again the N\'eron model
                                            of its generic fibre, because the universal property of fibred
                                            products implies the universal property of N\'eron models (Definition \ref{0066}).
                                            Consequently, we are reduced to the proof that an elliptic
                                            curve over $S$
                                            is the N\'eron model of its generic fibre. Thus we are done by
                                            Proposition \ref{0124}.
                                        \end{proof}

                                        \begin{Rem}\label{6025}
                                            Let $X:= E_0 \times_S \ldots \times_S E_n$ be the fibred product of
                                            $n$ smooth elliptic curves $E_i$ over $S$ provided with first fundamental form
                                            $g$. In truth, $(X \to S,g)$ is even a \SR (see Definition \ref{5024}). In particular, all
                                            results, which are proven in chapter \ref{0080}, are true for $X$.

                                            Let us now choose $S= \Spec {\footnotesize{\text{$\cal O$}}}_K$,
                                            where ${\footnotesize{\text{$\cal O$}}}_K$ is the ring of
                                            integral numbers of an algebraic number field $K \subset
                                            \real$. Let us compare $X$ with the Minkowski space-time $\ba_S^n$.
                                            Both describe a space-time without gravity, because the
                                            curvature tensor vanishes identically. Furthermore, both
                                            carry a canonical, commutative group structure. In quantum
                                            field theory over Minkowski \sto, the inverse of the group law
                                            is interpreted as a simultaneous space and time reflection.
                                            Consequently, there is also a canonical notion of space and time
                                            reflection on $X$, too. The difference between $X$ and
                                            Minkowski \st is of topological nature. While Minkowski \st
                                            (which is induced by some affine space $\Affin_S^n$)
                                            carries a trivial topology, $X$ does not. In order to
                                            illustrate this, let us consider the time coordinates in the
                                            respective models. More precisely, let us consider the time
                                            coordinate as a complex variable (whose imaginary part represents
                                            the physical time and whose real part phenomenologically
                                            describes temperature). Then, in the Minkowski case, the time variable runs in the
                                            complex plane and therefore on a sphere if we adjoin
                                            the point at infinity. In contrast to this, the complex points of the time
                                            coordinate $E_0$ of $X$ yield a torus.

                                            This difference in the global topology, which may not be seen locally, has some
                                            interesting consequences with regard to quantum field theory:
                                                The
                                                set of archimedean points $\ba_K^n(K) \cong K^n$ of Minkowski \st is
                                                still a dense subset of the continuum $\ba_K^n(\real)\cong
                                                \real^n$. In particular, $\ba_K^n(K)$ is not a discrete subset of the continuum
                                                (it is not even a finitely
                                                generated abelian group). On the other hand, there is no discrete
                                                subgroup of $\ba_K^n(K)$ which is invariant under arbitrary
                                                $K$-isomorphisms of $\ba_K^n$. In particular, a discrete subgroup
                                                of $\ba_K^n(K)$ cannot be Lorentz-invariant.

                                                But, if we consider instead the model $X$
                                                something interesting happens: Due to a theorem of
                                                Mordell, the set $X_K(K)$ of archimedean points of $X$ is
                                                a finitely generated abelian group. Furthermore, $X_K(K)$ is
                                                invariant under arbitrary $K$-isomorphisms of $X_K$.
                                                Consequently, it makes sense to say that $X_K(K)$ is
                                                ``Lorentz-invariant''.
                                                In the case of rank zero, \st is actually a discrete
                                                set. All in all, $X_K(K)$ may indeed be interpreted as a
                                                vacuum, and the methods of lattice gauge theory yield a well defined quantum field
                                                theory on $X_K(K)$.
                                                This will be exposed in more detail in chapter
                                                \ref{2100ga}.
                                        \end{Rem}
                                        It would be interesting to consider a dynamical solution $X=E_t
                                        \times_S E_1(t) \times_S \ldots \times_S E_n(t)$ of the Einstein
                                        equations, where $E_t$ is a fixed elliptic curve and where
                                        $E_i(t)$ are elliptic curves described by equations
                                        $Y^2Z=X^3+g_{2,i}(t)XZ^2+g_{3,i}(t)Z^3$ with coefficients
                                        $g_{j,i}(t)$ depending on $E_t$. For physical reasons, one should
                                        expect that the rank of each elliptic curve $E_i(t)$ stays zero
                                        for all times $t$ if it was zero for $t=0$. Therefore, one might
                                        conjecture some connections between the rank of elliptic curves
                                        and the Einstein equations.

\chapter{Some properties of the arithmetic models}\label{0080}

Unless otherwise specified, let $(X\to S, g)$ be a \SR (see
Definition \ref{5024}), and assume that $S$ is representable by a
Dedekind scheme with field of fractions $K$.

\section{Etale-invariance}\label{0081}

In its differential geometric formulation, general relativity is
as well general covariant as diffeomorphism invariant. General
covariance means that fundamental physical laws may not depend on
the special choice of a local coordinate system and is due to the
fact that the Einstein equations are tensor equations. However,
diffeomorphism invariance means the following. If $g$ is a
solution of the vacuum Einstein equations and if $\varphi$ is any
diffeomorphisms of the \st manifold, then $\varphi^*g$ is a vacuum
solution, too. Essentially this comes down to saying that
$(\varphi^* \Omega)[g]=\Omega[\varphi^*g]$ for all diffeomorphisms
$\varphi$. Thereby, $\Omega[g]$ denotes the functional $(\text{sc}
\cdot \omega)[g]=\text{sc}[g]\omega[g]$, where $\text{sc}[g]$
resp. $\omega[g]$ denote the scalar curvature resp. the volume
form corresponding to $g$.

In our arithmetic approach, the Riemannian \st manifold is
replaced by an appropriate algebraic space. Therefore, it is
natural to study in how far general covariance and diffeomorphism
invariance are realized for \GRs. For this purpose, we first have
to find the analogue of the differential geometric diffeomorphism
in the realm of schemes. In section \ref{1301}, we will expose in
a very detailed way that the class of such coordinate
transformations is given by \'etale morphisms (see Definition
\ref{0022}). This naturally implies that the Zariski-topology has
to be replaced by the so called \'etale-topology. This naturally
leads to algebraic spaces. Central is the following
characterization of \'etale morphisms which illustrates the formal
analogy between \'etale morphisms and differential geometric
diffeomorphisms.
\begin{Satz}\label{0027}
    Let $f:X \to Y$ be an $S$-morphisms. Let $x\in X$ and set
    $y:=f(x)$. Assume that $X$ is smooth over $S$ at $x$ and that
    $Y$ is smooth over $S$ at y. Then the following conditions are
    equivalent:
    \begin{enumerate}
        \item
        $f$ is \'etale at x.
        \item
        The canonical homomorphism $(f^*\Omega_{Y/S}^1)_x \to
        (\Omega_{X/S}^1)_x$ is bijective.
    \end{enumerate}
\end{Satz}

\begin{proof}
    \BLR, Cor. 2.2/10
\end{proof}
For physical reasons, the number of points of \st should be
invariant under coordinate transformations or ``deformations'' of
\sto. Therefore, it is natural to restrict attention to the class
of finite, \'etale surjective morphisms $f :X \to Y$ which have
the property that there are bijections $Y_K(K')=X_K(K')$ and
$Y(S')=X(S')$ for all \'etale $S$-schemes $S'$ with field of
fractions $K'$. Thereby, $X_K$ and $Y_K$ denote as usually the
respective generic fibres. However, let us point to the fact that
it can happen that $f$ changes the topology.

%

Now, let us return to the setting of \GRos. The following
Proposition \ref{5500} shows that this ``adelic'', arithmetic
model for gravity is covariant and \'etale-invariant.

\begin{Satz}\label{5500}
    Let $(X \to S,g)$ be a pair consisting of a smooth
    $S$-scheme $X \to S$ and a metric $g$ on $X$,
    such that $g$ is a solution of the Einstein equations
    \ref{5121}. Let
    $\varphi:X \to X$ be an arbitrary \'etale $S$-morphism,
    let $S' \to S$ be an arbitrary \'etale $S$-scheme, and let $(X' \to
    S'$, $g')$ be the pair obtained from $(X \to S,g)$ by base
    change with $S' \to S$.
    Then the following statements hold:
    \begin{enumerate}
        \item[1)]
        If $g$ is a vacuum solution of the Einstein equations, so is $\varphi^*g$.

        \item[2)]
        If $(X \to S,g)$ is a \GRo, so is $(X' \to S'$, $g')$.
    \end{enumerate}
\end{Satz}
\begin{proof}
        It suffices to prove 1) on stalks.  Therefore, we may assume that $\varphi:X \to X$ is
        given by  functions which are algebraic over some polynomial ring
        such that the corresponding Jacobi matrix is invertible. Thus, we
        obtain an explicit expression for the components of $\varphi^*g$. Furthermore,
        we may make use of the explicit formulas stated directly above Remark \ref{5125} and
        write the Einstein tensor in the well-known form
        by means of the Christoffel symbols.
        Then, a very lengthy computation, which is the same as in differential
        geometry, finally shows that the Einstein tensor
        of $\varphi^*g$ indeed vanishes.

        Let us now consider assertion 2). Again we may prove locally
        that $g'$ satisfies the Einstein equations.
        Due to Proposition \ref{0017} there
        is a canonical isomorphism $p^*\Omega_{X/S}^1  \cong
        \Omega_{X'/S'}^1$, where $p:X' \rightarrow X$ is the
        canonical
        projection. Therefore, the components of $g':=p^* g$ are
        obtained from the components of $g$ by tensoring over $\co_S(S)$
        with $\co_S(S) \to \co_{S'}(S')$, i.e. $g'_{\mu \nu}= g_{\mu \nu} \otimes
        1$. Consequently, the Einstein tensor $G'_{\mu \nu}$ of $g'_{\mu
        \nu}$ vanishes, because $G'_{\mu \nu}=G_{\mu \nu} \otimes
        1$.

        In order to prove that $(X' \to S'$, $g')$ is a \GRo, it
        suffices to show that $X' \to S'$ is a N\'eron
        model of its generic fibre, because this implies that
        it is already a \GR (apply the universal property
        of N\'eron models). But due to \BLR, Prop. 1.2/2 N\'eron models are
        stable under \'etale base change.
\end{proof}
Proposition \ref{5500} states that \GRs are \'etale-invariant: As
well any \'etale $S$-morphism $X \to X$ (i.e. any deformation of
the universe which leaves the shape of the underlying points
invariant) as any base change by \'etale morphisms $S' \to S$
(i.e. any simultaneous deformation of the points) and consequently
any combination of these two operations transforms \GRs into
\GRos. Therefore, the described physics is invariant as it should
be.

Last but not least, let us consider the set $X(S)$ of ``adelic''
points and the set $X_K(K)$ of archimedean points of a \GR $(X\to
S,g)$. It makes sense to choose $S$ maximal in the following way:
If $\varphi: S' \to S$ is a finite, \'etale surjective morphism,
then $S'=S$ and $\varphi = \text{id}$. For example, $S=\Spec
\Ganz$ is maximal in this sense (see \Neu, Kap. III, Thm. 2.18).
However, we know from the Yoneda lemma that $X(S)$ is invariant
under arbitrary $S$-isomorphisms $X \to X$. If $S$ is furthermore
maximal in the above sense, $X(S)$ is also invariant under \'etale
base change by finite, \'etale surjective
morphisms.\footnote{Physically, it does not make sense to demand
invariance of $X(S)$ under arbitrary \'etale base-change. For
example, if $s \in S$ is a closed point, the open immersion
$i:S-\{s\} \to S$ is \'etale, but neither finite nor surjective.
Performing a base change with $i$ simply means that we throw away
the fibre of $X(S)$ over the closed point $s$. So, it is clear
that it does not make sense to demand invariance of $X(S)$ under
base change with arbitrary \'etale morphisms. The physically
``correct'' morphisms, which should leave $X(S)$ invariant, are
covering maps, i.e. finite, \'etale, surjective maps.} Therefore,
Proposition \ref{5500} tells us that the pair $(X(S),g)$ which is
obtained from $(X \to S,g)$ by evaluation at $S$-valued points is
\'etale-invariant, too. The same is true for the pair
$(X_K(K),g_K)$ which is obtained from $(X \to S,g)$ by first
taking the generic fibre and then evaluating at $K$-valued points.
Thereby, the \'etale-invariance of $X_K(K)$ follows from the
canonical bijection $X_K(K) \cong X(S)$ and from the fact that any
$K$-isomorphism $X_K \to X_K$ extends uniquely to an
$S$-isomorphism $X \to X$ (recall that $X$ is the N\'eron model of
its generic fibre $X_K$). We may summarize as follows:

\begin{Cor}\label{5500bi}
    Let $(X\to S,g)$ be a \GRo. Then the induced pairs
    \begin{align*}
        \Big( X(S),g \Big) \quad \text{and} \quad \Big( X_K(K),g_K \Big)
    \end{align*}
    are \'etale-invariant. In particular, it makes sense to
    consider $X(S)$ as ``adelic'' \st and $X_K(K)$ as archimedean
    \sto.
\end{Cor}
\section{Gravity}\label{0102}
If we do not restrict attention to the archimedean component $X_K$
of $X$, but instead consider the full ``adelic'' theory, something
interesting may be observed: The space described by $(X \to S, g)$
cannot be the flat, topologically trivial Minkowski space.

Before, we will prove this interesting fact, let us slightly
extend the notion of N\'eron models. Up to know, we demanded that
N\'eron models are of finite type. In particular, they are
quasi-compact, i.e. the inverse image of any affine open subset is
quasi-compact as a topological space. But, we may drop the
condition that they are of finite type and thus drop the
compactness-condition. But due to smoothness, they are still
locally of finite type and are called N\'eron lft-models.

\begin{Def}\label{0104}
    Let $S$ be a Dedekind scheme with ring of fractions $K$. Let
    $X_K$ be a smooth $K$-scheme. A smooth and separated $S$-model
    $X$ is called a N\'eron lft-model of $X_K$ if $X$ satisfies the
    N\'eron mapping property (see Definition \ref{0066}).
\end{Def}
In the following, we denote by $R$ a discrete valuation ring with
field of fractions $K$. We denote by $R^{\text{sh}}$ the strict
henselization of $R$ with field of fractions $K^{\text{sh}}$. Let
$\widehat{R}^{\text{sh}}$ be the strict henselization of the
completion $\widehat{R}$ of $R$, and let $\widehat{K}^{\text{sh}}$
be the field of fractions of $\widehat{R}^{\text{sh}}$.

Some parts make use of the notion of excellent rings, whose
definition I will recall below. The reader who, is not familiar to
this notion, may skip it and simply think of a Dedekind domain $A$
of characteristic zero as an example of an excellent ring (in
physical situations we will deal with Dedekind domains of
characteristic zero anyway). Let us just mention that the strict
henselization of an excellent ring is excellent again. Thus the
extensions $\widehat{K}^{\text{sh}} / \widehat{K}$ and
$\widehat{K}^{\text{sh}} / K$ are separable.

Let us first state several existing criterions. Afterwards we will
give the physical interpretation.

\begin{Satz}\label{0105}
    Let $R$ be a discrete valuation ring with field of fractions
    $K$, and let $X_K$ be a smooth commutative $K$-group scheme of
    finite type. Then the following conditions are equivalent:
    \begin{enumerate}
        \item
        $X_K$ has a N\'eron model over $R$.
        \item
        $X_K \otimes_K \widehat{K}^{\text{sh}}$ contains no
        subgroup of type $\mathbb{G}_a$ or $\mathbb{G}_m$.
        \item
        $X_K(\widehat{K}^{\text{sh}})$ is bounded in $X_K$.
        \item
        $X_K(K^{\text{sh}})$ is bounded in $X_K$.
    \end{enumerate}
    If, in addition, $R$ is excellent, the above conditions are
    equivalent to
    \begin{enumerate}
        \item[e)]
        $X_K \otimes_K K^{\text{sh}}$ contains no
        subgroup of type $\mathbb{G}_a$ or $\mathbb{G}_m$.
    \end{enumerate}
\end{Satz}

\begin{proof}
    \BLR, Thm. 10.2/1
\end{proof}
The meaning of the conditions $c)$ and $d)$ will be illustrated in
the next section. Here $b)$ or $e)$ are important. If we consider
N\'eron lft-models the existing criterion is as follows.

\begin{Satz}\label{0110}
    Let $R$ be a discrete valuation ring with field of fractions
    $K$, and let $X_K$ be a smooth commutative $K$-group scheme of
    finite type. Then the following conditions are equivalent:
    \begin{enumerate}
        \item
        $X_K$ has a N\'eron lft-model over $R$.
        \item
        $X_K \otimes_K \widehat{K}^{\text{sh}}$ contains no subgroup of type
        $\mathbb{G}_a$.
    \end{enumerate}
    If, in addition, $R$ is excellent, the above conditions are
    equivalent to
    \begin{enumerate}
        \item[c)]
        $X_K$ contains no subgroup of type $\mathbb{G}_a$.
    \end{enumerate}
\end{Satz}

\begin{proof}
    \BLR, Thm. 10.2/2
\end{proof}
In physical situation we will always consider an excellent
Dedekind scheme $S$ whose field of fractions $K$ has
characteristic zero. Then, due to Proposition \ref{0108}, the
existence of a global N\'eron lft-model (resp. of a global N\'eron
model) is equivalent to the existence of the local N\'eron
lft-models (resp. of a global N\'eron models).

\begin{Theorem}\label{0106}
    Let $S$ be an excellent Dedekind scheme with field of fractions
    $K$ with $\text{char}(K)=0$. Let $X_K$ be a smooth commutative $K$-group scheme of
    finite type. Then $X_K$ admits a N\'eron lft-model over $S$ if
    and only if $X_K$ contains no subgroup of type $\mathbb{G}_a$.
\end{Theorem}
We saw that the subgroups of type $\mathbb{G}_a$ are critical. But
what is the physics behind these groups? Well, due to the basic
principles of general relativity, we always have to consider pairs
$(X,g)$ consisting of an object $X$ representing \st and a metric
$g$ on $X$ which encodes gravity  (classically $X$ was chosen as a
differentiable manifold). More precisely, $g$ is the so called
first fundamental form on $X$ (see Proposition \ref{0201s} for the
algebraic geometric analogue of this notion). However, in our
setting, we have got $X=\mathbb{G}_a$. As a scheme, $\mathbb{G}_a$
coincides with the affine space $\Affin_{\Ganz}^1$. Therefore, the
first fundamental form on $\mathbb{G}_a^n$ is the trivial metric
$\text{diag}(\pm1, \ldots, \pm1)$. Moreover, the set of
differential geometric points represented by $\mathbb{G}_a$ is:
\begin{align*}
     \mathbb{G}_a(\mathbb{K})=\mathbb{K},  \text{\quad $\mathbb{K}=\real,
     \complex$}.
\end{align*}
Consequently, $\mathbb{G}_a^n$ has to be interpreted as algebraic
geometric analogue of the manifold $\real^n$ provided with the
trivial metric. In particular, Minkowski-space $\real^4$
corresponds to the $\rat$-group scheme $\mathbb{G}_{a, \rat }^4$
in our arithmetic approach and thus contains a subgroup
$\mathbb{G}_a$. Therefore, Minkowski space is impossible due to
Theorem \ref{0106}.

Finally let us give the announced definition of excellent rings
and schemes. We will not motivate this notion and refer the reader
to \Liu, chapter $8.2$, for more details.

\begin{Def}\label{0109}
    Let $A$ be a noetherian ring. We say that $A$ is
    \emph{excellent} if it verifies the following three
    properties:
    \begin{enumerate}
        \item
        $\Spec A$ is universally catenary (see \Liu, Definition 8.2.1)
        \item
        For each $\prim \in \Spec A$ the formal fibres of
        $A_{\prim}$ are geometrically regular, i.e. for all $x \in
        A_{\prim}$ the scheme
        $\Spec ( \widehat{A_{\prim}} \otimes_{A_{\prim}} k' )$ is regular for all
        finitely generated field extensions $k'/k(x)$. Thereby $
        \widehat{A_{\prim}}$ denotes the $\prim$-adic completion
        of $A_{\prim}$.
        \item
        For every finitely generated $A$-algebra $B$, the set of
        regular points of $\Spec B$ is open in $\Spec B$.
    \end{enumerate}
\end{Def}
\begin{Satz}\label{0108}
    Let $S$ be a Dedekind scheme whose field of fractions $K$ has
    characteristic zero. Then the existence of a global N\'eron
    lft-model (resp. of a global N\'eron model) is equivalent to the
    existence of the local N\'eron lft-models (resp. of a global N\'eron
    models).
\end{Satz}

\begin{proof}
    \BLR, page $310$
\end{proof}
\section{The structure of the archimedean component}\label{0120}

In this section we are going to analyze the structure of the
archimedean limit $X_K$ of $X$. We will find that there is an
algebraic torus $T_K$ and an abelian variety  $A_K$ which is
realized as a quotient $A_K= X_K/T_K$. In analogy to the situation
of classical, differential geometric gauge theory, this obtrudes
the interpretation that \emph{\SRs} describe gravity plus
electromagnetism.

The existence of the N\'eron model $X$ of $X_K$ is equivalent to
the existence of the local models $X \times_S \Spec {\cal
O}_{S,s}$ for each closed $s \in S$ by Proposition \ref{0108}. Due
to Proposition \ref{0110} this is equivalent to the fact that the
unipotent radical of $X_K$ is trivial. Then $X_K$ is an extension
of an Abelian variety $A_K$ by a torus $T_K$, i.e. there is an
exact sequence of $K$-group schemes.
\begin{equation*}
    \xymatrix{
    0 \ar[r] & T_K \ar[r] & X_K \ar[r] & A_K \ar[r] & 0}
\end{equation*}
Before interpreting this result physically, let us recall the
notions of Abelian varieties and tori.

\begin{Def}\label{0121}
    Let $K$ be a field. An \emph{Abelian variety} $A_K$ over $K$ is
    defined to be a $K$-group scheme which is geometrically integral and
    proper.\footnote{A scheme $X$ over $S$ is called \emph{proper}, if the canonical
    morphism $X \to S$ is universally closed, separated and of finite
    type.}
    One can show that an Abelian variety is always projective and
    commutative.
\end{Def}

\begin{Def}\label{0122}
    Let $K$ be a field. An \emph{algebraic torus} $T_K$ over $K$ is
    a commutative $K$-group scheme of finite type over $K$ which is isomorphic
    to $(\mathbb{G}_m)^r$ for some $r \in \Natural$ over an
    algebraic closure $\overline{K}$ of $K$; i.e.
    \begin{equation*}
        T_{\overline{K}}:=T_K \otimes_K \overline{K} \cong
        (\mathbb{G}_{m,\overline{K}})^r \quad \text{for some} \ r
        \in \Natural.
    \end{equation*}
    If this isomorphism can be realized over $K$, i.e.
        $
        T_K  \cong
        (\mathbb{G}_{m,K})^r \quad \text{for some} \ r
        \in \Natural$,
    the torus is said to split.
\end{Def}
For simplicity, let us assume that we may write $X_K$ as a product
$X_K=T_K \times_K A_K$. Then looking at the set of physical points
and using the fact that we can choose
$\overline{K}=K^{\text{sep}}$ due to $\text{char}(K)=0$, we obtain
an extension  of ordinary groups
\begin{equation*}
    \xymatrix{
    0 \ar[r] & T_K(K^{\text{sep}}) \ar[r] & X_K(K^{\text{sep}}) \ar[r] & A_K(K^{\text{sep}}) \ar[r] &
    0}.
\end{equation*}
Then, we obtain the decomposition
\begin{align*}
    X_K(K^{\text{sep}})
    &= A_K(K^{\text{sep}}) \times T_K(K^{\text{sep}})
    = A_K(K^{\text{sep}}) \times
    \left((\mathbb{G}_{m,K^{\text{sep}}})(K^{\text{sep}})\right)^r
\end{align*}
for some $r \in \Ganz$. Thereby, the last equality is due to the
fact that the torus splits over $K^{\text{sep}}$ by definition.
This isomorphism of groups can be interpreted as follows: The
archimedean limit $X_K$ decomposes into two parts. The first part
is given by an Abelian variety $A_K$. This represents the \st
dimensions. The second part is given by a torus $T_K$ and should
be interpreted as some kind of internal space (originating from
some gauge structure). We will see that $T_K$ is $p$-adically
unbounded.
\\ \\
On the other hand one can show that each extension of an algebraic
variety by a torus admits a N\'eron lft-model. Thus each \SR $(X
\to S, g)$ is characterized by a certain structure of its
archimedean limit.

\begin{Theorem}\label{0123}
    Let $S$ be an excellent Dedekind scheme with field of fractions
    $K$ with $\text{char}(K)=0$. Let $X_K$ be a smooth commutative $K$-group scheme of
    finite type. Then is equivalent:
    \begin{enumerate}
        \item
        $X_K$ admits a N\'eron lft-model over $S$.
        \item
        $X_K$ contains no subgroup of type $\mathbb{G}_a$.
        \item
        $X_K$ is an extension of an Abelian variety $A_K$ by a torus
        $T_K$, i.e. over an algebraic closure of $K$ there is an exact sequence
        \begin{equation*}
            \xymatrix{
            0 \ar[r] & T_K \ar[r] & X_K \ar[r] & A_K \ar[r] & 0}.
        \end{equation*}
        (Later, we will interpret $T_K$ as the gauge group part and $A_K$
        as the ``\sto'' part of $X_K$ (see also Definition \ref{0142b}).)
    \end{enumerate}
\end{Theorem}

\begin{proof}
    The equivalence of $a)$ and $b)$ is just Theorem \ref{0106}. The
    implication $a) \Rightarrow c)$ was illustrated above and $c) \Rightarrow
    a)$ follows from the following propositions.
\end{proof}

\begin{Satz}\label{0124}
    Let $S$ be a connected Dedekind scheme with field of fractions
    $K$ and let $A_K$ be an Abelian variety over $K$. Then $A_K$
    admits a global N\'eron model $A$ over $S$.
\end{Satz}

\begin{proof}
    \BLR, Thm. 1.4/3
\end{proof}

\begin{Satz}\label{0124a}
    Let $S$ be a Dedekind scheme with field of fractions $K$. Then
    any torus $T_K$ over $K$ admits a N\'eron lft-model over $S$.
\end{Satz}

\begin{proof}
    \BLR, Prop. 10.1/6
\end{proof}

\begin{Satz}\label{0124b}
    Let $S' \to S$  be a finite flat extension of Dedekind schemes
    with fields of fractions $K'$ and $K$. Let $G_K$ be a smooth
    $K$-group scheme and denote by $G_{K'}$ the $K'$-group scheme
    obtained by base change. Let $H_K$ be a closed subgroup of
    $G_K$ which is smooth. Assume that $G_{K'}$ admits a N\'eron
    lft-model $G'$ over $S'$.

    Then the N\'eron lft-model of $H_K$ over $S$ exists. More
    precisely it can be constructed as a group smoothening of the
    schematic closure of $H_K$ in the Weil restriction
    $\mathfrak{R}_{S'/S}(G')$.
\end{Satz}

\begin{proof}
    \BLR, Prop. 10.1/4
\end{proof}

\begin{Satz}\label{0124c}
    Let $S$ be a Dedekind scheme with field of fractions $K$. Let
    $G_K$ be a smooth connected algebraic $K$-group which is an
    extension of a smooth $K$-group scheme $H_K$ of finite
    type by a split torus $T_K$. Assume that $\Hom(H_K,
    \mathbb{G}_{m,K})=0$; for example, the latter is the case if
    $H_K$ is an extension of an Abelian variety by a unipotent
    group.

    Then, if $H_K$ admits a N\'eron lft-model over $S$, the same is
    true for $G_K$.
\end{Satz}

\begin{proof}
    \BLR, Prop. 10.1/7
\end{proof}
Up to now we only considered the archimedean limit $X_K$ of the
\SR $(X\to S, g)$. But we want to show that also the full
``adelic'' object $X$ has a very clear structure, too.

So let $X_K$ be the archimedean limit of $X$. Due to Theorem
\ref{0123} $X_K$ is an extension of an Abelian variety $A_K$ by a
torus $T_K$, i.e. there is  exact sequence
\begin{equation*}
    \xymatrix{
    0 \ar[r] & T_K \ar[r] & X_K \ar[r] & A_K \ar[r] & 0}.
\end{equation*}
By Definition \ref{0122} there exists a finite separable field
extension $K'/K$ such that $T_{K'}:=T_K \otimes_K K'$ splits. Thus
performing a base change with the canonical morphism $\Spec K' \to
\Spec K$, we can replace $K$ by $K'$ in the exact sequence above
and assume that $T_K$ splits, say of rank $r$. But then the
extension $X_K$ of $A_K$ by $T_K$ is given by primitive line
bundles ${\cal L}_1, \ldots {\cal L}_r$ on $A_K$ by \SerreI, Chap.
VII, $\text{n}^\text{o} 15$, Thm. $5$. A line bundle ${\cal L}$ on
a group scheme $G$ is called primitive if there is an isomorphism
\begin{equation*}
    m^*{\cal L} \cong p_1^*{\cal L} \otimes p_2^*{\cal L},
\end{equation*}
where $m$ is the group law on $G$ and where $p_i:G \times G \to G$
are the projections, $i=1,2$. Since the local rings of the N\'eron
model $A$ of $A_K$ are factorial, the line bundles ${\cal L}_i, \
i=1, \ldots, r$ extend to primitive line bundles on the identity
component $A^0$ of $A$. Thus they give rise to an exact sequence
\begin{equation*}
    \xymatrix{
    0 \ar[r] & T^0 \ar[r] & X^0 \ar[r] & A^0 \ar[r] & 0},
\end{equation*}
whose generic fibre is the exact sequence we started with.
Furthermore $X^0$ is the identity component of the N\'eron
lft-model
$X$ of $X_K$. \\ \\
So the structure of  $X^0$ is clear, if a can analyse the
structure of $A^0$. But $A_K$ is an Abelian variety, and the
structure of $A^0$ is clarified by the fundamental \emph{Theorem
on the potential semi-abelian reduction} of Abelian varieties.
Before stating this theorem let us recall the notion of reduction.

Let $G$ be a smooth group scheme of finite type over a connected
Dedekind scheme $S$. We say that $G$ has \emph{abelian reduction}
(resp. \emph{semi-abelian reduction}) at a closed point $s \in S$
if the identity component $G_s^0$ is an Abelian variety (resp. an
extension of an Abelian variety by an affine torus). In
particular, if $G$ is a N\'eron model of its generic fibre $G_K$,
where $K$ is the field of fractions of $S$, we will say that $G_K$
has abelian (resp. semi-abelian) reduction at $s \in S$ if the
corresponding fact is true for $G$. The latter amounts to the same
as saying that the local N\'eron model $G \times_S \Spec {\cal
O}_{S,s}$ of $G_K$ at $s \in S$ has abelian (resp. semi-abelian)
reduction.

If $A_K$ is an abelian variety over $K$, then $A_K$ is said to
have \emph{potential abelian reduction} (resp. \emph{potential
semi-abelian reduction}) at a closed point $s \in S$ if there is a
finite Galois extension $L$ of $K$ such that $A_L$ has abelian
(resp. semi-abelian reduction) at all points over s. More
precisely, we thereby mean that the N\'eron model $A'$ of $A_L$
over the normalization $S'$ of $S$ in $L$ has abelian (resp.
semi-abelian) reduction at all closed points $s' \in S'$ lying
over $s$.

\begin{Theorem}\label{0126}
    Each abelian variety $A_K$ over $K$ has potential semi-abelian
    reduction at all closed points of $S$.
\end{Theorem}
This theorem clarifies the structure of the \SR $(X \to S,g)$. Let
us conclude this chapter with a final remark.

\begin{Rem}\label{5510}
    In physical applications, we will consider a pair $(X \to S,g)$
    with $S= \Spec \Ganz$. Now consider the \st part $A_K$ of the archimedean component
    $X_K$ of $X$. We only know that $A_K$ has
    potentially semi-abelian reduction, but in general $A_K$
    itself will not have semi-abelian reduction.
    But if we perform the classical continuum limit $K$ (i.e. we
    consider $X_L:= X_K \otimes_K L$ and let the field $L$
    tend towards $\overline{\rat}$), then, at some place, $A_L$ happens
    to have semi-abelian reduction.

    The property to have semi-abelian reduction is a very
    strong symmetry (see Proposition \ref{0131} or section \ref{0140}).
    Thus, the ``adelic'' structure on smallest scales is much
    richer and less symmetric than the structures that we find
    in the continuum approximation. For example, we find some kind of
    entanglement of the dimensions of \st on smallest scales. The
    decomposition into distinct dimensions is in general only the
    consequence of some approximation process (see Remark \ref{0142rem} for details).
\end{Rem}

\begin{Satz}\label{0131}
    If an Abelian variety $A_K$ has semi-abelian reduction, then
    the formation of the identity component of the N\'eron model of
    $A_K$ is compatible with faithfully flat extensions of
    discrete valuation rings $R \hookrightarrow R'$.
\end{Satz}

\begin{proof}
    \BLR,  Cor. 7.4/4
\end{proof}
\section{Boundedness of the archimedean component}\label{0132}

In the previous section we saw that the archimedean limit $X_K$ of
$X$ is an extension of an Abelian variety $A_K$ by a torus $T_K$,
i.e. over an algebraic closure of $K$ there is an exact sequence
\begin{equation*}
    \xymatrix{
    0 \ar[r] & T_K \ar[r] & X_K \ar[r] & A_K \ar[r] & 0}
\end{equation*}
(by  Theorem \ref{0123}). $T_K$ represents the gauge group part
and $A_K$ represents the \st part of $X_K$.

Within this section we will prove that the archimedean limit $A_K$
is bounded. More precisely, as well $A_K(K)$ as all
non-archimedean continuum limits $A_K(\widehat{K})=A_K(K_{\prim})$
are bounded with respect to the corresponding $\prim$-adic norm
(recall that $K_{\prim}$ is a finite extension of some $\rat_p$).
In the special case $S:= \Spec \Ganz$, the set of all possible
norms (up to equivalence of norms) is given by the unique
archimedian norm $| \cdot |_{\infty}$ (also called the prime spot
at infinity) and the set of non-archimedian norms $| \cdot |_{p}$.
There is one norm $| \cdot |_{p}$ for each prime number $p \in
\Ganz$, namely the $p$-adic norm on $\complex_p$. Then the above
boundedness assertion states that the set $A_K(\rat_p)$ of points
of the $p$-adic continuum limit of $A_K$ is \emph{necessarily}
bounded (with respect to the $p$-adic norm $| \cdot |_{p}$ on
$\complex_p$) for every prime number $p$ of $\Ganz$. The objective
of this section \ref{0132} is as well the definition of the notion
of boundedness as the illustration why this finiteness occurs.

Therefore, let $S$ be  the spectrum of the ring of integers of an
algebraic number field $K$. As $S$ is a Dedekind scheme one knows
from number theory that the local rings ${\cal O}_{S,s}$ for
closed points $s\in S$ are discrete valuation rings. Let $\pi \in
{\cal O}_{S,s}$ be a generator of the maximal ideal $\prim_s$ of
${\cal O}_{S,s}$. This prime ideal is called a finite prime. $\pi$
is called a uniformizing element. Any $r\in {\cal O}_{S,s}$ can be
uniquely written in the form
\begin{equation*}
    r=\pi^n \cdot u,
\end{equation*}
with uniquely determined $n \in \Natural$ and  $u \in {\cal
O}_{S,s}^*$. Therefore each $r \in K:=\text{Frac}({\cal O}_{S,s})$
can be uniquely written in the form
\begin{equation*}
    r=\pi^n \cdot u,
\end{equation*}
where $n \in \Ganz$ and $u \in {\cal O}_{S,s}^*$. Then there is a
norm $| \cdot |_{\prim_s}$ on $\text{Frac}({\cal O}_{S,s})$
defined as follows. Let $r \in K:=\text{Frac}({\cal O}_{S,s})$ and
decompose $r$ as above: $r= \pi^n \cdot u $. Let $p=\text{char\,}
k(s)$ be the characteristic of the residue class field $k(s)$ at
$s$. Then
\begin{align*}
    |r|_{\prim_s} &:= \mathfrak{N}(\prim_s)^{-n}
\end{align*}
with $\mathfrak{N}(\prim_s) := p^{f_{\prim_s}} $ and $ f_{\prim_s}
:= [k(s): \mathbb{F}_p]$. This definition may look deterrent, if
one is not used to it. $\mathfrak{N}(\prim_s)$ occurs in this
definition in order to have certain nice number theoretic
identities. The norm $|r|_{\prim_s}$ can be written more
``transparently`` in the form
\begin{align*}
    |r|_{\prim_s} &:= q^{-n}
\end{align*}
for some $q\in \Natural$. The exact value of $q$ is important for
number theoretic identities, but does not influence the topology
induced by this norm. Thus the size of a number $r \in K$ with
respect to the norm $| \cdot |_{\prim_s}$ is only dependent on the
number of factors $\pi$ that are contained in $r$. The norm $|
\cdot
|_{\prim_s}$ is a non-archimedian norm. \\ \\
After these preparations let us introduce the notion of
\emph{boundedness}. We start with a discrete valuation ring $R$
with field of fractions $K$. Furthermore, consider a faithfully
flat extension of discrete valuation rings $R \hookrightarrow R'$
and let $K'$ be the field of fractions of $R'$. In physical
situations we will choose an \'etale morphism $R \hookrightarrow
R'$ or we will consider the strict henselization $R':=R^{sh}$ of
$R$. Then, by the above, $R$ and $R'$ give rise to absolute values
on $K$ and $K'$. Assuming that these absolute values coincide on
$K$, we denote them by $|\cdot|$. This is justified by the
following proposition.
\begin{Satz}\label{0128}
    Let $K$ be a field which is complete with respect to the
    absolute value $|\cdot|$. Let $L/K$ be an algebraic extension.
    Then there is a unique continuation of $|\cdot|$ to $L$. If $[L:K]=n <
    \infty$, then for all $\alpha \in L$
    \begin{equation*}
        |\alpha|=\sqrt[n]{|\text{N}_{L/K}(\alpha)|}.
    \end{equation*}
\end{Satz}
\begin{proof}
    \Neu, Chap. II, Thm. 4.8
\end{proof}
Now for any $K$-scheme $X_K$, for any point $x\in X_K(K')$
corresponding to the maximal ideal $\maxi_x$ and for any section
$g$ of ${\cal O}_{X_K}$ being defined at $x$, we may view the
image $g(x)$ of $g$ in ${\cal O}_{X_K,x}/ (\maxi_x) $ as an
element of $K'$. Thus the absolute value $|g(x)|$ is well defined
and it makes sense to say that $g$ is bounded on a subset of
$X_K(K')$. As an example let us consider the special case
$X_K=\Affin_K^n$. Then any section $g$ is given by a polynomial
$g=g(T_1, \ldots ,T_n)\in K[T_1, \ldots ,T_n]$ with variables
$T_1, \ldots ,T_n$. Furthermore, any point $x\in X_K(K')=(K')^n$
can be written as $x=(x_1, \ldots ,x_n)\in (K')^n$. Thus $g(x)$ is
simply the evaluation of the polynomial $g$ at $x$ and $|g(x)|$ is
the size of the number $g(x)\in K'$ with respect to the absoulte
value $|\cdot|$. Everything works like one is used to from the
real or complex case, just with the archimedian norm
$|\cdot|_{\infty}$ replaced by its non archimedian counterpart
$|\cdot|$.

Applying this procedure to the coordinate functions of the affine
$n$-space $\Affin_K^n$, we arrive at the notion of bounded subsets
of $\Affin_K^n(K')$.

\begin{Def}\label{0129}
    As before, let $R \hookrightarrow R'$ be a faithfully flat
    extension of discrete valuation rings with fields of fractions
    $K$ and $K'$. Furthermore, let $X_K$ be a $K$-scheme of finite
    type and consider a subset $E\subset X_K(K')$.
    \begin{enumerate}
    \item
    If $X_K$ is affine, $E$ is called bounded in $X_K$ if there
    exists a closed immersion $X_K \hookrightarrow \Affin_K^n$
    mapping $E$ onto a bounded subset of $\Affin_K^n(K')$.
    \item
    In the general case, $E$ is called bounded in $X_K$ if there
    exists a covering of $X_K$ by finitely many affine open
    subschemes $U_1, \ldots ,U_s \subset X_K$ as well as a
    decomposition $E=\bigcup E_i$ into subsets $E_i \subset
    U_i(K')$ such that, for each $i$, the set $E_i$ is bounded in
    $U_i$ in the sense of $a)$.
    \end{enumerate}
\end{Def}
One can show that condition $b)$ of Definition \ref{0129} is
independent of the particular affine open covering $U_i$ of $X_K$
(see \BLR, Lemma 1.1/3). Furthermore the image of a bounded set is
bounded again as one would expect intuitively.

\begin{Satz}\label{0130}
    Let $R \hookrightarrow R'$ be a faithfully flat
    extension of discrete valuation rings with fields of fractions
    $K$ and $K'$. Consider a $K$-morphism $f:X_K \to Y_K$ between
    $K$-schemes of finite type. Then, for any bounded subset $E\subset
    X_K(K')$, its image under $X_K(K') \to Y_K(K')$ is bounded in
    $Y_K$.
\end{Satz}

\begin{proof}
    \BLR, Prop. 1.1/4
\end{proof}
\begin{Satz}\label{0130}
    Let $R \hookrightarrow R'$ be a faithfully flat
    extension of discrete valuation rings with fields of fractions
    $K$ and $K'$. Consider a proper $K$-scheme $X_K$.
    Then any subset $E\subset X_K(K')$ is bounded in $X_K$.
\end{Satz}

\begin{proof}
    \BLR, Prop. 1.1/6
\end{proof}
As $A_K$ is an Abelian variety, it is proper and thus bounded due
to Proposition \ref{0130}. Due to our exposition in the previous
section \ref{0120}, $X^0$ is given by the extension
\begin{equation*}
    \xymatrix{
    0 \ar[r] & T^0 \ar[r] & X^0 \ar[r] & A^0 \ar[r] & 0},
\end{equation*}
where $A^0$ is the identity component of the N\'eron model $A$ of
$A_K$. Recalling conditions $c)$ and $d)$ of Proposition
\ref{0105} we see that the boundedness of the extended \st
dimensions is not by chance, but a fundamental property of \SRos.
The boundedness of the extended \st dimensions is necessary. In
particular, one obtains an effective $\prim$-adic infrared cutoff.

Last but not least let us consider the ``gauge part'' of $X^0$.
Due to the equality $\mathbb{G}_{m,K}(K')=(K')^*$, this is
\emph{not} a bounded set. If $\alpha \in (K')^*$ with
$|\alpha|>1$, then two points with coordinates $\alpha$ and
$\alpha^n$ are arbitrary far separated from each other with
respect to $|\cdot|$ if $n$ is big enough.
\section{Lifting of structures from the archimedean to the adelic level}\label{0140}

Fibre-bundles are an essential tool in real-valued physics (based
on differential geometry). In particular, they allow a global,
coordinate independent description of physics, e.g. of fields
mediating forces. In the framework of vector-bundles, physical
fields occur as local sections of vector-bundles with \st as base
manifold $M$. Let
\begin{equation*}
    f: N \to M
\end{equation*}
be the corresponding projection map from the vector-bundle onto
the base, whose sections are therefore physical fields.

The notion of principal bundles is furthermore a principal item of
gauge theory which we are interested in. Let $P$ be a
$G$-principal bundle over $M$, where $M$ is a Riemannian manifold
and $G$ is a Lie-group. In particular, a principal bundle gives
rise to a canonical exact sequence
\begin{equation*}
    \xymatrix{
    0 \ar[r] & G \ar[r] & P \ar[r] & M \ar[r] & 0}
\end{equation*}
and is equipped with a free and transitive $G$-action $\psi$ on
$P$
\begin{equation*}
    \psi: P \times G \to P
\end{equation*}
(see section \ref{1401}). Let us consider the analogues of these
structures in our algebraic geometric setting which is exposed in
detail in section \ref{1410}. Then the above morphism $f$
corresponds to a \emph{smooth} $K$-morphism  $f_K:X_K \to Y_K$.
The notion of a field  (e.g. the electromagnetic field) as section
of this morphism makes sense, because locally (with respect to
\'etale topology) at physical points smooth morphisms admit
sections due to Proposition \ref{0030}.

Furthermore, in algebraic geometry, the differential geometric
$G$-principal bundles of gauge theory is the so called torsor
under an algebraic $K$-group $G_K$ (see Definition \ref{1413}).
Again there is a canonical exact sequence
\begin{equation*}
    \xymatrix{
    0 \ar[r] & G_K \ar[r] & P_K \ar[r] & X_K \ar[r] & 0}
\end{equation*}
and $P_K$ is equipped with a free and transitive $G_K$-action
$\psi_K$ which is given by an isomorphism
\begin{equation*}
    \xymatrix{\psi_K: P_K \times_{X_K} G_{X_K} \ar[r]^{\quad \sim} & P_K \times_{X_K}
    P_K}
\end{equation*}
with $G_{X_K}:=G_K \times_K X_K$. But we want to do physics not
only in the archimedean  limit, but also in the ``adelic''
situation. In particular, we intend to do gauge theory in this
setting. Thus, the above structures like the morphism $f_K$ or the
$X_K$-torsor $P_K$ under $G_K$ (given by the above exact sequence)
should lift to the ``adelic'' level in certain situations.

We will see within this section that this lifting of structures
from the archimedean to the ``adelic'' level  in fact happens
under quite general assumptions. \\ \\
\textbf{Lifting of physical fields}
Let us consider an archimedean, physical field given by a section
$s_K: Y_K \hookrightarrow X_K$ of a smooth $K$-morphism $f_K: X_K
\to Y_K$ and let $X$, $Y$ the corresponding ``adelic'' \SRos. Then
due to the N\'eron mapping property there exist unique morphisms
$f:X \to Y$ and $s:Y \to X$ extending $f_K$ and $s_K$. By
functoriality, we have $f \circ s = \text{id}|_Y$, i.e. $s$ is a
section of $f$ and
thus a physical field in the ``adelic'' world. \\ \\
\textbf{Lifting of exact sequences}
There are some quite general situations where exact sequences lift
to the ``adelic'' level.

\begin{Satz}\label{0141}
    Let $R$ be a discrete valuation ring with field of fractions
    $K$, and let $0 \to {A}_K^{'} \to A_K \to {A}_K^{''} \to 0$ be an exact
    sequence of Abelian varieties. Consider the associated
    sequence of N\'eron models $0 \to A' \to A \to A'' \to 0$.
    Assume that the following condition is satisfied:

    $R$ has mixed characteristic and the ramification index
    $e=\nu(p)$ satisfies $e < p-1$, where p is the residue
    characteristic of $R$ and where $\nu$ is the valuation on $R$
    which is normalized by the condition that $\nu$ assumes the
    value $1$ at uniformizing elements of $R$.

    Then the following assertions hold:
    \begin{enumerate}
        \item
        If $A'$ has semi-abelian reduction, $A' \to A$ is a closed
        immersion.
        \item
        If $A$ has semi-abelian reduction, the sequence $0 \to A' \to A \to A'' \to
        0$ is exact.
        \item
        If $A$ has abelian reduction, the sequence $0 \to A' \to A \to A'' \to
        0$ is exact and consists of Abelian $R$-schemes.
    \end{enumerate}
\end{Satz}

\begin{proof}
    \BLR, Thm. 7.5/4
\end{proof}
\begin{Lemma}\label{0142}
    Let $0 \to {A}_K^{'} \to A_K \to {A}_K^{''} \to 0$ be an exact
    sequence of Abelian varieties over $K$. Then $A_K$ has
    semi-abelian (resp. abelian) reduction if and only if ${A}_K^{'}$
    and ${A}_K^{''}$ have semi-abelian (resp. abelian) reduction.
\end{Lemma}

\begin{proof}
    \BLR, Lemma 7.4/2
\end{proof}
\begin{Rem}\label{0142rem}
    Let $K$ be an algebraic number field, and let $S=\Spec {\footnotesize{\text{$\cal O$}}}_K$ be the
    spectrum of the ring of integral numbers of $K$.
    If the archimedean world $X_K$ has no gauge part, $X_K$ is
    already an Abelian variety. Let us  assume that there is an inductive
    decomposition of $X_K$ into lower dimensional subspaces
    $A_K^{(i)}$, $i=1, \ldots,n$ which are Abelian varieties,
    too. More precisely, assume that there are exact sequences of
    Abelian varieties
    \begin{equation*}
        \xymatrix{
        0 \ar[r] & B_K^{(i)} \ar[r] & B_K^{(i+1)} \ar[r] & A_K^{(i)} \ar[r] &
        0}, \qquad B_K^{(1)}:=A_K^{(1)},
    \end{equation*}
    such that $X_K=B_K^{(n+1)}$. For example, one may assume that
    each Abelian variety $A_K^{(i)}$ is an elliptic curve. From the physical point of view,
    the latter yields a decomposition of \st $X_K$ into the different
    dimensions. We know that each variety $A_K^{(i)}$ (resp. $B_K^{(i)}$) possesses
    a N\'eron model $A^{(i)}$ (resp. $B^{(i)}$). Physically,
    $A^{(i)}$ (resp. $B^{(i)}$) describes an ``adelic'' world whose archimedean
    component is given by $A_K^{(i)}$ (resp. $B_K^{(i)}$). However, in general,
    the Abelian varieties $A_K^{(i)}$, $B_K^{(i)}$ will not have semi-abelian reduction. Consequently,
    the above decomposition of $X_K$ will in general not lift to the
    ``adelic'' level, i.e. the induced sequences
    \begin{equation*}
        \xymatrix{
        0 \ar[r] & B^{(i)} \ar[r] & B^{(i+1)} \ar[r] & A^{(i)} \ar[r] & 0}
    \end{equation*}
    possibly fail to be exact (see Proposition \ref{0141}). Physically,
    this defect of exactness may be interpreted as some kind of entanglement of
    the dimensions on the smallest scales. This is clearly an
    ``adelic'' effect. Also with respect to the metric we meet the
    same phenomenon: At $S$-valued points, the metric takes values in
    the ring ${\footnotesize{\text{$\cal O$}}}_K$. But in general, even
    if ${\footnotesize{\text{$\cal O$}}}_K$
    is by chance a principal ideal ring, the metric
    cannot be diagonalized.
    Only upon the archimedean component $X_K$, the
    metric may be diagonalized, and we find the decomposition of $X_K$
    into distinct dimensions (see Proposition \ref{0206}).

    However, there is still another phenomenon. If we perform the
    continuum limit (i.e. we consider $X_L:= X_K \otimes_K L$ and let
    the field $L$ tend towards $\overline{\rat}$), then, as soon as $L$ becomes big enough,
    $A_L^{(i)}:=A_K^{(i)} \otimes_K L$ happens to have
    semi-abelian reduction (Theorem \ref{0126}). If ${A'}^{(i)}$ (resp. ${B'}^{(i)}$) denotes the
    N\'eron model of $A_L^{(i)}$ (resp. ${B}_L^{(i)}$) over the normalization $S'$ of $S$ in
    $L$, it follows that induced exact sequences
    \begin{equation*}
        \xymatrix{
        0 \ar[r] & B_L^{(i)} \ar[r] & B_L^{(i+1)} \ar[r] & A_L^{(i)} \ar[r] & 0}
    \end{equation*}
    actually lifts to the ``adelic'' level, because each $A_L^{(i)}$ has semi-abelian
    reduction (apply Proposition \ref{0141}). More precisely, we obtain
    exact sequences
    \begin{equation*}
        \xymatrix{
        0 \ar[r] & {B'}^{(i)} \ar[r] & {B'}^{(i+1)} \ar[r] & {A'}^{(i)} \ar[r] & 0}
    \end{equation*}
    on the ``adelic'' level, too. Therefore, the entanglement
    of the dimensions on the ``adelic'' level is lost if we
    perform the continuum limit. Furthermore, there is suddenly not
    only compatibility of $X'={A'}^{(n)}$ with unramified, but also with \emph{ramified}
    base change (see Proposition \ref{0131}).
\end{Rem}
Let us now incorporate gauge theory. Therefore, consider a
$X$-torsor $P$ under a gauge group $G_X:=G \times_S X$. We know
that the \SR $X$ is the N\'eron model of its generic fibre $X_K$.
But in general, $G$ need not be the N\'eron model of its generic
fibre. Physically the latter means that $G$ is the gauge group of
a gauge field which is limited to the quantum level. Exactly those
gauge fields, which appear as classical fields with unlimited
range (like the electromagnetic field), must posses a gauge group
which is the N\'eron model of its generic fibre. Let us now
determine all gauge groups with this property. For this purpose
assume that $G_K$ is the (commutative) gauge group of a classical
gauge field with infinite range. However, there is a canonical
gauge theory associated to the trivial $X_K$-torsor $P_K:=X_K
\times_{X_K} G_{X_K} = X_K \times_K G_K$ under $G_{X_K}:=X_K
\times_K G_K$ which fits into an exact sequence $0 \to G_K \to P_K
\to X_K \to 0$ of group schemes. This gauge theory realizes the
most simplest vacuum structure which is possible and is therefore
usually chosen in physical applications. By assumption, the
considered gauge field is classical with infinite range.
Therefore, its gauge group possesses a N\'eron model. Now there is
the following result.

\begin{Lemma}\label{6142a}
    Let $S$ be a Dedekind scheme with field of fractions $K$, and
    let
    \begin{equation*}
        \xymatrix{
        0 \ar[r] & G'_K \ar[r] & G_K \ar[r] & G''_K \ar[r] & 0}.
    \end{equation*}
    be an exact sequence of smooth $K$-group schemes of finite
    type (which are not necessarily commutative).

    \begin{enumerate}
        \item
        If $G_K$ admits a N\'eron model over $S$, the same is true for $G'_K$.

        \item
        If $G'_K$ and $G''_K$ admit a N\'eron model over $S'$, the same
        is true for $G_K$.
    \end{enumerate}
\end{Lemma}

\begin{proof}
    \BLR, Prop. 7.5/1
\end{proof}
Therefore, also $P_K$ possesses a N\'eron model $P$. It follows
from Theorem \ref{0123} that $P_K$ also fits into an exact
sequence $0 \to T_K \to P_K \to A_K \to 0$, where $T_K$ is a
torus, and where $A_K$ is an Abelian variety. Comparing the two
exact sequences it is therefore natural to interpret $T_K$ as
gauge group part of $P_K$. As higher-dimensional tori are products
of one-dimensional tori, only the one-dimensional torus is
indecomposable, i.e. describing gauge bosons which are elementary
particles. Therefore, we make the following definition.

\begin{Def}\label{0142b}
    Let $\mathbb{G}_{m,K}$ be the one-dimensional torus.
    Then the gauge field associated to the (commutative) gauge group
    \begin{align*}
        G_K:= \mathbb{G}_{m,K}
    \end{align*}
    is called the arithmetic \emph{electromagnetic field}.
\end{Def}

                                        \chapter{The dimensionality of the arithmetic models}\label{0200}

                                        Within this section let us consider a \emph{\GR} $(X \to S,g)$,
                                        and let $X_K$ be the generic fibre of $X$ (see Definition \ref{5022}).
                                        Recall that, unless otherwise specified, $K \subset \real$ is an
                                        algebraic number field (i.e. a finite algebraic extension of
                                        $\rat$), and that ${\footnotesize{\text{$\cal O$}}}_K$ is the ring
                                        of integral numbers of $K$ (i.e. the integral closure of $\Ganz$
                                        in $K$). For example, think of ${\footnotesize{\text{$\cal
                                        O$}}}_K= \Ganz$ and $K=\rat$. As already illustrated, we consider
                                        a smooth algebraic space $X_K$ over $K$ as the fundamental object
                                        underlying the archimedean \st limit (see Definition \ref{0021}). Then, in
                                        the archimedean continuum limit, we obtain a classical
                                        differentiable manifold $M$ which is realized as the set of
                                        $\real$-valued points of $X_K$, i.e. $M=X_K(\real)$ .

                                        In this chapter we will draw some conclusions from the presence of
                                        gravity concerning the dimensionality of \sto. Let us once for all
                                        denote the relative dimension of $X_{K}$ by $n$. In contrast to
                                        relativistic theories over the infinite prime spots $\real$ or
                                        $\complex$, which may be formulated for arbitrary values of $n$,
                                        it will turn out that the finite prime spots of the algebraic
                                        number field $K$ exert a wide influence on the admissible number
                                        of dimensions. Already B. Dragovich, I.V. Volovich et al.
                                        mentioned some remarkable coincidences between $p$-adics and
                                        dimensionality of \st (see \Drago, Chap. 5, concluding remarks \
                                        and \DraVol, end of section 2.1). We will meet this phenomenon in
                                        our approach, too.

                                        But first, let us  recall that there are a priori two different
                                        ways of describing the $\real$-valued Minkowski-\st $\real^n$
                                        which underlies the theory of special relativity:
                                        \begin{enumerate}
                                            \item
                                            Both, space $\vec{x}$ and time $t$, are $\real$-valued
                                            variables, i.e.
                                            $\vec{x} \in \real^{n-1}$, $t \in \real$. The
                                            metric of \st is given by the Minkowski-metric $g_{\mu \nu}= \text{diag}(1,-1,\ldots,-1)$.

                                            \item
                                            Space $\vec{x}$ is a $\real$-valued variable, but time $t$ is purely
                                            imaginary, i.e. $t \in  i \real$ (as it is proposed by the principles of quantum mechanics). The
                                            metric of \st is given by the euclidian-metric $g_{\mu \nu}= \text{diag}(1,\ldots,1)$.
                                        \end{enumerate}
                                        However, within the bounds of this chapter, only the second point
                                        of view (that time is a purely imaginary variable) will be
                                        presented. Assuming that gravity is a field of spin $2$, one may
                                        prove the following statement:
                                        \begin{enumerate}
                                            \item[$\bullet$]
                                            In the case $b)$, $n=4$ and each tangent space is equipped
                                            with the structure of a quaternion algebra with ``Pauli
                                            matrices'' as generators.
                                        \end{enumerate}
                                        Let us at least mention another issue. If one admits furthermore
                                        that next to the spin-$2$ field of gravity there is also a
                                        metrical spin-$3$ field, one can show that this implies $n\geq
                                        10$.

                                        In order to be able to talk about gravity in our algebraic
                                        geometric setting, we have to introduce the notion of a metric.
                                        For details we refer to section \ref{0200g}.
                                        \section{The privileged character of four dimensional spaces}\label{0202b}
                                        Let us consider our archimedean limit $X_K \to \Spec K$ of
                                        relative dimension $n$. We assume that gravity is a field of
                                        spin-$2$ metric. Therefore, we may choose a metric
                                        \begin{align*}
                                            g_K:{T_{X_K/K} \times_{X_K} T_{X_K/K}} & \to  \Affin_{X_K}^1
                                        \end{align*}
                                        (see Definition \ref{0201}). Thereby, $T_{X_K/K}$ denotes the tangent
                                        bundle over $X_K$, and $\Affin_{X_K}^1$ denotes the
                                        one-dimensional affine space over $X_K$. For those readers, who
                                        are not used to these notions, we may give an equivalent
                                        description of the metric in terms of matrices. For simplicity,
                                        let us assume that the module of differential forms
                                        $\Omega_{X_K/K}^1$ is globally free. If $(\omega^1, \ldots,
                                        \omega^n )$ is a global base of the differential forms
                                        $\Omega_{X_K/K}^1$, then $g_K$ corresponds to a non-degenerate
                                        symmetric bilinear form
                                        \begin{align*}
                                            g_K = \sum\limits_{1 \leq \mu,\nu \leq n} g_{\mu \nu} \, \omega^{\mu} \otimes \omega^{\nu} \in
                                            \Omega_{X_K/K}^{\otimes 2}(X_K), \quad \quad g_{\mu \nu}\in {\cal O}_{X_K}(X_K).
                                        \end{align*}
                                        Thus, the metric may be interpreted as a matrix $(g_{\mu \nu})$.
                                        At this place we will not derive the concrete structure of $g_K$,
                                        we simply work with a given structure. The reader may think that
                                        $g_{\mu \nu}$ is stipulated by the equations of general
                                        relativity. We interpret the metric as an instrument which enables
                                        us to measure \st distances.

                                        Let us study the metric in more detail. Choose a physical point
                                        $x\in X_{K}$ (i.e. a closed point of $X_K$). Then $x$ takes values
                                        in a finite separable extension $k(x) \subset \ratb$ of $K$, where
                                        $\ratb \subset \complex$ denotes an algebraic closure of $\rat$.
                                        Using the physical interpretation given in section \ref{0200g}
                                        we see that the evaluation of $g_K$ at the physical point $x\in
                                        X_{K}$ induces a bilinear form
                                        \begin{align*}
                                            g_{K}(x): T_{X_{K}/{K}}(x) \times T_{X_{K}/{K}}(x) \to k(x).
                                        \end{align*}
                                        Due to the equality
                                        \begin{align*}
                                            T_{X_{K}/{K}}(x)= \Hom_{\text{$k(x)$-lin}}\left(\Omega_{X_{K}/ K}^1 \otimes k(x), k(x) \right)
                                            = \Hom_{\text{$k(x)$-lin}}\left( k(x)^n, k(x) \right) \cong
                                            k(x)^n
                                        \end{align*}
                                        we may write $g_{K}(x)$ as a bilinear form on the $k(x)$-vector
                                        space $k(x)^n$
                                        \begin{align}
                                            g_{K}(x):k(x)^n \times k(x)^n \to k(x).
                                        \end{align}
                                        Let us by some abuse of notation denote the matrix representing
                                        the bilinear form $g_{K}(x)$ by $g_{\mu \nu}(x)$ or simply by
                                        $g_{\mu \nu}$. As $\text{char} \left(k(x)\right)=0 \neq 2$, we can
                                        equivalently write $g_{K}(x)$ as a quadratic form
                                        \begin{align*}
                                            g_{K}(x)(T_1, \ldots, T_n)= \sum\limits_{\mu, \nu} g_{\mu
                                            \nu}(x) T^{\mu}T^{\nu}, \quad g_{\mu \nu}(x) \in k(x).
                                        \end{align*}
                                        Taking into regard that the metric $g_{K}(x)$ is an instrument to
                                        perform measurements in \sto, we can make a further statement on
                                        the structure of $g_{K}(x)$: As the building stones of a classical
                                        \st are infinite small zero-dimensional point particles, arbitrary
                                        small distances in \st are realized. Thus tangent-vectors of
                                        arbitrary lengths may be constructed in the classical world. In
                                        particular, if we submit a measurable length, i.e. a number of
                                        $k(x)$, then, at least in a classical world, one can construct a
                                        tangent-vector which takes exactly this length. This amounts to
                                        saying that $g_{K}(x)$ is surjective.

                                        Consider the infinite prime spot $\real$  of the rational numbers
                                        $\rat$. Performing a base change with $K \hookrightarrow \real$,
                                        the archimedean limit $X_{K}$ with its metric $g_{K}$ induces a
                                        differentiable manifold
                                        \begin{align*}
                                            \Big( X_{\real}(\real), g_{\real} \Big),
                                        \end{align*}
                                        the \st manifold of the archimedean continuum limit.

                                        Before we proceed, let us introduce some notation which we will
                                        need below. The field $K$ possesses  valuations which enable us to
                                        measure the ``size'' of its elements. These valuations are called
                                        the prime spots of $K$ (see Definition \ref{0431}). In particular, each
                                        prime spot induces a topology on $K$. However, the topological
                                        field $K$ need not be complete with respect to a chosen prime spot
                                        $\prim$. As usual, let $K_{\prim}$ denote the completion of $K$
                                        with respect to a prime spot $\prim$. One can prove that either
                                        $K_{\prim}= \real$, $K_{\prim}= \complex$ or that there exists a
                                        prime number $p \in \Ganz$ such that $K_{\prim}$ is a finite
                                        extension of $\rat_p$. The complete fields $K_{\prim}$ are called
                                        \emph{local} fields, whereas $K$ is a \emph{global} field. The
                                        field $\real$ of real numbers is an ordered field (i.e. given two
                                        elements $x,y \in \real$, then either $x \leq y$ or $y \leq x$).
                                        In contrast to this, the $p$-adic fields are not ordered.
                                        Therefore, in the case $K_{\prim} \neq \real$, the field
                                        $K_{\prim}$ is not ordered: There does not exist a split into
                                        positive and negative numbers (i.e. the arrow of time and the
                                        split into future and past disappears. In a certain way, there
                                        exists only future). This interpretation motivates the following
                                        definition.

                                        \begin{Def}\label{0210d}
                                            Let $\alpha \in K_{\prim}$.
                                            \begin{enumerate}
                                                \item
                                                If $K_{\prim} = \real$, $\alpha$ is called \emph{strictly
                                                positive} if and only if $\alpha > 0$.

                                                \item
                                                If $K_{\prim} \neq \real$, $\alpha$ is called \emph{strictly
                                                positive} if and only if $\alpha \neq 0$.

                                            \end{enumerate}
                                        \end{Def}
                                        Finally, let us consider a quadratic form $g_K: K^n \times K^n \to
                                        K$ over $K$. Evaluation of $g_K$ at pairs $(x,y) \in K_{\prim}^n
                                        \times K_{\prim}^n$ induces a quadratic form over $K_{\prim}$
                                        which we denote by $g_{K_{\prim}}$.
                                        \subsection*{The case of euclidian gravity.}\label{0224c}

                                        Let us now illustrate the privileged character of four-dimensional
                                        relativistic models. We will assume that $( X_{\real}(\real),
                                        g_{\real} )$ describes euclidian gravity. Therefore, in inertial
                                        systems, the metric takes the form
                                        \begin{align*}
                                                    g_{\mu \nu}= \left(
                                                    \begin{array}{cccc}
                                                    1 & 0 & 0 & 0 \\
                                                    0 & 1 & 0 & 0 \\
                                                    0 & 0 & 1 & 0 \\
                                                    0 & 0 & 0 & 1 \\
                                                    \end{array} \right)
                                        \end{align*}
                                        over the prime spots at infinity. This quadratic form is
                                        anisotropic over $\real$ and represents all strictly positive real
                                        numbers (see Definition \ref{0210}). These two properties of the
                                        relativistic metric $g_{\mu \nu}$ over $\real$ motivate the
                                        following definition (where we make use of the notation which was
                                        introduced above in the paragraph containing Definition \ref{0210d}).

                                        \begin{Def}\label{0224d}
                                            Let $K$ be either a global or a local field.
                                            Let $g_K: K^n \times K^n \to K$ be a quadratic form over $K$. Then $g_K$
                                            is called \emph{\rela} if and only if there
                                            exists a finite prime spot $\prim$ of $K$, such that:

                                            \begin{enumerate}
                                                \item
                                                $g_{K_{\prim}}$ is anisotropic.

                                                \item
                                                $g_{K_{\prim}}$ represents all strictly positive numbers $\alpha \in K_{\prim}$.
                                            \end{enumerate}
                                            If $K$ does not possess a finite prime spot, $g_K$ is called \rela if
                                            a) and b) are true with $K_{\prim}:=K$.

                                        \end{Def}
                                        \begin{Rem}\label{0224e}
                                            Let $g_K$ be a quadratic form \relao.
                                            \begin{enumerate}
                                                \item
                                                More intuitively, the two conditions a) and b) of Definition \ref{0224d} may be stated as follows:
                                                \begin{enumerate}
                                                    \item[$\bullet$]
                                                    $\| x \|_{\prim}^2:=|g_{K_{\prim}}(x,x)|_{\prim}=0$ \quad $\Leftrightarrow$ \quad $x=0$.

                                                    \item[$\bullet$]
                                                    Classical scales do not have holes!
                                                \end{enumerate}

                                                \item
                                                By the Hasse-Minkowski theorem \ref{0212}, $g_K$ does not represent zero over $K$.

                                                \item
                                                If $n \geq 4$, then the second condition in Definition \ref{0224d} is actually true for \emph{all}
                                                finite prime spots of $K$ (as we will see soon).
                                                Especially, if $K \subset \real$, we deduce from
                                                the Hasse-Minkowski theorem \ref{0212} that either $g_K$ or $-g_K$ represents all strictly positive
                                                numbers $0<\alpha \in K$.

                                                This fact justifies the
                                                ``global'' statement: Classical scales do not have holes!

                                                \item
                                                If $n=4$ and $K=\real$, then $g_K =
                                                \text{diag}(1,1,1,1)$, i.e. $g_K$ describes a \sto which is relativistic in the
                                                sense of euclidian gravity.
                                                Therefore, Definition \ref{0224d} generalizes the notion of a
                                                relativistic metric (in the sense of the theory of special
                                                relativity) from $\real$ to arbitrary local and global
                                                fields.
                                            \end{enumerate}
                                        \end{Rem}
                                        Now we are prepared to incorporate gravity.

                                        \begin{Def}\label{0224f}
                                            Let $X_K \to \Spec K$ be of dimension $n$. Let $g_K$ be
                                            a spin-$2$ metric on $X_K$ describing gravity.

                                            Then $X_K$ is called \emph{\rela} if and only if
                                            $g_K(x)$ is \rela in the sense of Definition \ref{0224d}
                                            for all physical (i.e. closed) points $x \in X_K$.
                                        \end{Def}

                                        \begin{Theorem}\label{0224g}
                                            Let $X_K \to \Spec K$ be \rela  in the sense of
                                            Definition \ref{0224f} and of dimension $n$. Then the following
                                            statements are true:
                                            \begin{enumerate}
                                                \item[$\bullet$]
                                                $n=4$.

                                                \item[$\bullet$]
                                                In inertial systems, the metric may be written as
                                                \begin{align*}
                                                    g_{\mu \nu}= \left(
                                                    \begin{array}{cccc}
                                                    1 & 0 & 0 & 0 \\
                                                    0 & - \varepsilon & 0 & 0 \\
                                                    0 & 0 & - \pi & 0 \\
                                                    0 & 0 & 0 & \varepsilon \cdot \pi \\
                                                    \end{array} \right)
                                                \end{align*}
                                                with numbers $\varepsilon, \pi \in K^*$ which have the
                                                following property:

                                                There exists a finite prime spot $\prim$ of $K$ such that
                                                \begin{enumerate}
                                                    \item[(i)]
                                                    $\pi$ is a uniformizing element of $K_{\prim}$.

                                                    \item[(ii)]
                                                    The reduction of $\varepsilon$ mod $\pi$ is not a
                                                    square (i.e. $\overline{ \varepsilon } \notin
                                                    k_{\prim}^{*2}$).
                                                \end{enumerate}
                                                More precisely, $g_{\mu \nu}$ is the norm form of the
                                                non-split quaternion algebra $(\varepsilon,
                                                \pi)$ over $K_{\prim}$ (see Definition \ref{0216}).
                                            \end{enumerate}
                                        \end{Theorem}

                                        \begin{proof}
                                            In order to perform the proof in an elegant manner, let us
                                            introduce some notation. For this purpose let us consider the
                                            field $\rat_p$ of $p$-adic numbers ($p$ being a prime number).
                                            Let $a,b \in \rat_p^*$ and put \vspace{-1mm}
                                            \begin{tabbing}
                                                \qquad \= $(a,b)=1$ \quad \ \ \= if $z^2-ax^2-by^2=0$ has
                                                a solution $(z,x,y)\neq(0,0,0)$ in $\rat_p^3$. \vspace{3mm} \\
                                                \> $(a,b)=-1$ \> otherwise.
                                            \end{tabbing}
                                            \vspace{-1mm}
                                            The number $(a,b) \in \{\pm 1\}$ is called the \emph{Hilbert
                                            symbol} of $a$ and $b$ relative to $\rat_p$. The number
                                            $(a,b)$ does not change when $a$ and $b$ are multiplied by
                                            squares; thus we may consider the Hilbert symbol as a map
                                            \begin{align*}
                                                (  \cdot  ,  \cdot  ) : \rat_p^*/\rat_p^{*2} \times  \rat_p^*/\rat_p^{*2} \to \{ \pm
                                                1\}.
                                            \end{align*}
                                            Due to \SerreArit, Chap. III, §1, Thm. 1, this map has an interesting property which will be
                                            important
                                            later: It is bilinear and nondegenerate. The assertion
                                            ``nondegenerate'' has to be understood as follows: Each
                                            number $b \in \rat_p^*$ which has the property that $(a,b)=1$ for all $a \in
                                            \rat_p^*$ is a square, i.e. $b \in \rat_p^{*2}$.

                                            If $Q = \sum_{i,j=1}^n a_{ij} X_iX_j$ is a quadratic form
                                            in $n$ variables over $\rat_p$, we know from Proposition \ref{0206} that
                                            $Q$ may be diagonalized:
                                            \begin{align*}
                                                Q \sim a_1 X_1^2 + \ldots + a_n X_n^2.
                                            \end{align*}
                                            Then the two elements
                                            \begin{align*}
                                                \det(Q)&:= a_1 \cdot \ldots \cdot a_n \quad \ \, \, \ \in \rat_p^*/\rat_p^{*2} \\
                                                s(Q)&:= \prod_{1\leq i<j \leq n}(a_i,a_j) \   \in \{\pm 1\}
                                            \end{align*}
                                            are important invariants of the quadratic form $Q$.

                                            After this preparation, let us begin with the proof of Theorem \ref{0224g}.
                                            Let $x \in X_K$ be a physical point. We may assume that $K=k(x)=\rat$. By assumption, $X_K$ is
                                            \relao, i.e. there exists a prime number $p$ such that
                                            on the one hand $g_{\rat_p}(x)$ does not represent zero, but on
                                            the other hand represents all non-zero numbers $a \in \rat_p^*$. Let
                                            us prove that this already implies $n=4$. We will do so by
                                            showing that dimensions $n \neq 4$ are impossible. As we are free to perform coordinate transformations
                                            (i.e. it suffices to consider similarity classes of quadratic forms),
                                            all numbers may be viewed as elements of
                                            $\rat_p^*/\rat_p^{*2}$. Furthermore, let us use the
                                            abbreviation $Q:=g_{\rat_p}(x)$.
                                            \begin{enumerate}
                                                \item[n=1:]
                                                By Proposition \ref{0224h} b), $Q$ represents $a \in
                                                \rat_p^*$ if and only if $a = \det (Q)$. But $\rat_p^*/\rat_p^{*2} \neq \{ 1
                                                \}$ by Proposition \ref{0208}. In particular, $Q$ does not represent
                                                all $a \in \rat_p^*$.

                                                Therefore, $n=1$ is not possible.

                                                \item[n=2:]
                                                On the one hand, we know from Proposition \ref{0224h} b) that
                                                $(a, -\det(Q))=s(Q)$ for all $a \in  \rat_p^*/\rat_p^{*2}$.
                                                On the other hand, it follows from Proposition \ref{0224h}
                                                a) that $-\det(Q) \neq 1$. But, we saw above that
                                                the Hilbert symbol $(  \cdot  ,  \cdot )$ is
                                                a \emph{nondegenerate} bilinear form. Therefore, it is
                                                possible to choose a number $b \in  \rat_p^*/\rat_p^{*2}$
                                                such that $(b, -\det(Q)) \neq s(Q)$, and we arrive at a contradiction.

                                                Thus, $n=2$ is not possible.

                                                \item[n=3:]
                                                For each $a \in \rat_p^*/\rat_p^{*2}$, we must have either
                                                \begin{align*}
                                                    a \neq -\det(Q) \quad \text{or} \quad \Big( a= - \det(Q) \text{ and } s(Q)= \left(-1, - \det(Q) \right)
                                                    \Big).
                                                \end{align*}
                                                But on the other hand, we already know that $s(Q) \neq \left(-1, - \det(Q)
                                                \right)$, because
                                                otherwise $Q$ would represent zero by Proposition \ref{0224h} a).
                                                This implies the contradiction $a \neq -\det(Q)$ for all $a \in
                                                \rat_p^*/\rat_p^{*2}$.

                                                Therefore, $n=3$ is not possible.

                                                \item[$n \geq 5:$]
                                                By Proposition \ref{0224h} a), the quadratic form $Q$ is isotropic.

                                                Therefore, $n \geq 5$ is not possible either, and it remains the case
                                                $n=4$. Now the theorem follows from Proposition \ref{0211}.
                                            \end{enumerate}
                                        \end{proof}

                                        \begin{Cor}\label{0224gc}
                                            $K$-schemes $X_K \to \Spec K$ \rela exist.
                                        \end{Cor}

                                        \begin{proof}
                                            As in the proof of Theorem \ref{0224g}, we may assume that $K=k(x)=\rat$. By Theorem \ref{0224g}, we know that
                                            $X_K$ has to be four-dimensional. Therefore, Proposition \ref{0221b} shows that
                                            $K$-schemes $X_K \to \Spec K$ \rela actually exist.
                                        \end{proof}

                                        \begin{Cor}\label{0224gcg}
                                            Let $X \to S$ be a smooth $S$-scheme
                                            of relative dimension $n$ with metric $g$ such that the archimedean
                                            limit $X_K$ is \rela.
                                            Then
                                            \begin{align*}
                                                n=4.
                                            \end{align*}
                                        \end{Cor}

                                        \begin{proof}
                                            Due to  Theorem \ref{0224g}, the relative dimension $n_K$ of $X_K$ is
                                            four. But $n_K$ coincides with $n$ by Lemma \ref{0224}.
                                        \end{proof}
                                        Evoking Lemma \ref{0222} and Lemma  \ref{0223}, we finally obtain the
                                        following result.
                                        \begin{Lemma}\label{0224}
                                            Let $S$ be a Dedekind scheme with field of fractions $K$.
                                            Let $f: X \to S$ be a  smooth morphism of relative dimension $n$,
                                            and assume that $X$ is connected. Let $n_K$ be the relative dimension of the generic fibre $f_K: X_K \to
                                            \Spec K$ of $f$. Then
                                            \begin{align*}
                                                n=n_K.
                                            \end{align*}
                                        \end{Lemma}

                                        \begin{proof}
                                            As already stated, this is a direct consequence from Lemma \ref{0222} and
                                            Lemma \ref{0223}.
                                        \end{proof}

                                        \begin{Satz}\label{0224h}
                                            Let $p$ be a prime number, and let $Q$ be a quadratic form of rank $n$ on
                                            $\rat_p$. Then the following statements are true.
                                            \begin{enumerate}
                                                \item
                                                For $Q$ to represent zero it is necessary and sufficient
                                                that
                                                \begin{enumerate}
                                                    \item[(i)]
                                                    $n=2$ and $\det(Q)=-1$ (in $\rat_p^*/\rat_p^{*2}$).

                                                    \item[(ii)]
                                                    $n=3$ and $s(Q)= \left(-1, - \det (Q) \right)$.

                                                    \item[(iii)]
                                                    $n=4$ and either $\det(Q) \neq 1$ or $\Big( \det(Q)=1 \text{ and } s(Q)=(-1, -1)
                                                    \Big)$.

                                                    \item[(iv)]
                                                    $n \geq 5$.
                                                \end{enumerate}

                                                \item
                                                Let $a \in \rat_p^*$. In order that $Q$ represents $a$ it
                                                is necessary and sufficient that
                                                \begin{enumerate}
                                                    \item[(i)]
                                                    $n=1$ and $a=\det(Q)$.

                                                    \item[(ii)]
                                                    $n=2$ and $s(Q)=(a, - \det (Q))$.

                                                    \item[(iii)]
                                                    $n=3$ and either $a \neq -\det(Q) $ or $\Big( a=- \det(Q) \text{ and } s(Q)= \left(-1, - \det(Q) \right)
                                                    \Big)$.

                                                    \item[(iv)]
                                                    $n \geq 4$.
                                                \end{enumerate}
                                            \end{enumerate}
                                            Note that in this statement $a$ and $\det(Q)$ are viewed as
                                            elements of  $\rat_p^*/\rat_p^{*2}$. Also all equations have
                                            to be read in  $\rat_p^*/\rat_p^{*2}$; e.g. the inequality $a \neq
                                            -\det(Q)$ means that $a$ is not equal to the product of $-
                                            \det(Q)$ and a square.
                                        \end{Satz}

                                        \begin{proof}
                                            \SerreArit, Chap. IV, §2, Thm. 6 and the corollary to this
                                            theorem.
                                        \end{proof}

                                        \begin{Rem}\label{0225}
                                            In particular, we see that the tangent space
                                            $T_{X_K/K}(x)$ of $X_K$ at  a physical point $x \in X_K$
                                            is not only a vector space, but even a quaternion algebra if $X_K$ is \relao.
                                            We may choose a basis of $T_{X_K/K}(x)$ which is actually the basis of
                                            a non-split quaternion algebra
                                            \begin{align*}
                                                T_{X_K/K}(x) &\cong K e_0  \oplus  K e_1 \oplus K e_2 \oplus
                                                K e_3. \\
                                                e_1e_2&=e_3, \quad e_2e_1=-e_2e_1, \quad e_0=1 \\
                                                e_1^2&=\varepsilon \cdot 1 =\varepsilon,  \quad e_2^2= \pi \cdot 1
                                                =\pi
                                            \end{align*}
                                            Thus, we may write a gauge field ${\cal A}(x)$ at $x$ in the general form
                                            \begin{align*}
                                                {\cal A}(x) =  {\cal A}^0 e_0 + \sum_{i=1}^3  {\cal A}^i e_i.
                                            \end{align*}
                                            The elements $e_1$, $e_2$ and $e_3$ are analogues of the Pauli spin
                                            matrices. By means of conjugation in the
                                            quaternion algebra (see Definition \ref{0216}), the gauge field ${\cal A}(x)$
                                            gives rise to the field $\overline{{\cal A}}(x)$,
                                            \begin{align*}
                                                \overline{{\cal A}}(x) =  {\cal A}^0 e_0 - \sum_{i=1}^3  {\cal A}^i e_i.
                                            \end{align*}
                                            We call $\overline{{\cal A}}(x)$ the anti-field of ${{\cal
                                            A}}(x)$. Considering pairs $({\cal A}, \overline{{\cal A}})$,
                                            we may write
                                            \begin{align*}
                                                ({\cal A}, \overline{{\cal A}}) =
                                                {\cal A}_0
                                                \left(%
                                                \begin{array}{cc}
                                                0 & 1 \\
                                                1 & 0 \\
                                                \end{array}%
                                                \right)
                                                + {\cal A}_1
                                                \left(%
                                                \begin{array}{cc}
                                                0 & e_1 \\
                                                -e_1 & 0 \\
                                                \end{array}%
                                                \right)
                                                +{\cal A}_2
                                                \left(%
                                                \begin{array}{cc}
                                                0 & e_2 \\
                                                -e_2 & 0 \\
                                                \end{array}%
                                                \right)
                                                +{\cal A}_3
                                                \left(%
                                                \begin{array}{cc}
                                                0 & e_3 \\
                                                -e_3 & 0 \\
                                                \end{array}%
                                                \right).
                                            \end{align*}
                                            Introducing the gamma matrices
                                            \begin{align*}
                                                \gamma_0 := \left(%
                                                \begin{array}{cc}
                                                0 & 1 \\
                                                1 & 0 \\
                                                \end{array}%
                                                \right), \qquad
                                                \gamma_i:=
                                                \left(%
                                                \begin{array}{cc}
                                                0 & e_i \\
                                                -e_i & 0 \\
                                                \end{array}%
                                                \right), \quad \text{for $i=1,2,3$},
                                            \end{align*}
                                            we finally arrive at the expression
                                            \begin{align*}
                                                ({\cal A}, \overline{{\cal A}}) = \sum_{\mu=0}^3 {\cal A}^{\mu}
                                                \gamma_{\mu}.
                                            \end{align*}
                                            We claim that the gamma matrices fulfill the relations
                                            \begin{align*}
                                                \gamma_{\mu}\gamma_{\nu}+\gamma_{\nu}\gamma_{\mu} = 2
                                                g_{\mu \nu} \cdot \eins
                                            \end{align*}
                                            Thereby, $\eins$ denotes the unit matrix, and $g_{\mu \nu}$ denotes
                                            the metric at the physical point $x \in X_K$ (whose concrete form is determined by Theorem \ref{0224g}).
                                            \begin{proof}
                                                {\footnotesize{Let $i,j=1,2,3$. Then we derive that
                                                \begin{itemize}
                                                    \item
                                                    $\gamma_0 \gamma_0
                                                    =\left(%
                                                    \begin{array}{cc}
                                                    0 & 1 \\
                                                    1 & 0 \\
                                                    \end{array}%
                                                    \right)
                                                    \left(%
                                                    \begin{array}{cc}
                                                    0 & 1 \\
                                                    1 & 0 \\
                                                    \end{array}%
                                                    \right)
                                                    =\left(%
                                                    \begin{array}{cc}
                                                    1 & 0 \\
                                                    0 & 1 \\
                                                    \end{array}%
                                                    \right)
                                                    = g_{00} \cdot \eins$

                                                    \item
                                                    $\gamma_0 \gamma_i
                                                    =\left(%
                                                    \begin{array}{cc}
                                                    0 & 1 \\
                                                    1 & 0 \\
                                                    \end{array}%
                                                    \right)
                                                    \left(%
                                                    \begin{array}{cc}
                                                    0 & e_i \\
                                                    -e_i & 0 \\
                                                    \end{array}%
                                                    \right)
                                                    =\left(%
                                                    \begin{array}{cc}
                                                    -e_i & 0 \\
                                                    0 & e_i \\
                                                    \end{array}%
                                                    \right)
                                                    \quad \text{and}$ \\
                                                    $\gamma_i \gamma_0
                                                    =\left(%
                                                    \begin{array}{cc}
                                                    0 & e_i \\
                                                    -e_i & 0 \\
                                                    \end{array}%
                                                    \right)
                                                    \left(%
                                                    \begin{array}{cc}
                                                    0 & 1 \\
                                                    1 & 0 \\
                                                    \end{array}%
                                                    \right)
                                                    =\left(%
                                                    \begin{array}{cc}
                                                    e_i & 0 \\
                                                    0 & -e_i \\
                                                    \end{array}%
                                                    \right)$. \vspace{1mm} \\
                                                    Thus, $\gamma_0 \gamma_i+\gamma_i \gamma_0 = 0 = 2g_{0i} \cdot
                                                    \eins$.

                                                    \item
                                                    $\gamma_i \gamma_j
                                                    =\left(%
                                                    \begin{array}{cc}
                                                    0 & e_i \\
                                                    -e_i & 0 \\
                                                    \end{array}%
                                                    \right)
                                                    \left(%
                                                    \begin{array}{cc}
                                                    0 & e_j \\
                                                    -e_j & 0 \\
                                                    \end{array}%
                                                    \right)
                                                    =\left(%
                                                    \begin{array}{cc}
                                                    -e_i e_j & 0 \\
                                                    0 & -e_i e_j \\
                                                    \end{array}%
                                                    \right)$. \vspace{1mm} \\
                                                    Thus, $\gamma_i \gamma_j+\gamma_j \gamma_i
                                                    = - \left( e_i e_j + e_j e_i \right) \cdot \eins$. But
                                                    the generators of the quaternion algebra fulfill the
                                                    following identities:
                                                    \begin{itemize}
                                                    \item[$\bullet$]
                                                    $e_1e_2=-e_2e_1$, \quad
                                                    $e_1e_3=-e_1e_1e_2=-e_1e_2e_1=-e_3e_1$, \quad
                                                    $e_2e_3=e_2e_1e_2=-e_1e_2e_2=-e_3e_2$.

                                                    \item[$\bullet$]
                                                    $e_1^2= \varepsilon = -g_{11}$, \quad $e_2^2= \pi =-g_{22}$,
                                                    \quad $e_3^2=e_1e_2e_1e_2= - e_1^2e_2^2= - \varepsilon \pi
                                                    = - g_{33}$.
                                                    \end{itemize}
                                                    This shows that $ e_i e_j + e_j e_i =- 2 g_{ij} $. With this, everything is
                                                    proven.
                                                \end{itemize}}}
                                            \end{proof}
                                            Therefore, pairs $({\cal A}, \overline{{\cal A}})$ are
                                            elements of a Clifford algebra. In particular,
                                            we see that the archimedean limit $X_K$ of $X$ has got the property that gauge fields act
                                            canonically on spinors, if $X_K$ is \rela (in the sense of
                                            Definition \ref{0224f}). The coupling of gauge fields to
                                            spinors via gamma matrices occurs  naturally.
                                        \end{Rem}
                                        \section{Quadratic forms}
                                        Consider the standard situation of a metric $g_K$ living on a
                                        smooth $K$-scheme $X_K$ (see Definition \ref{0021}). In physical
                                        applications, the field $K$ will be an algebraic number field,
                                        i.e. a finite algebraic extension of the rational numbers $\rat$.
                                        As a consequence, the evaluation of $g_K$ at physical (i.e.
                                        closed) points of $X_K$ yields quadratic forms over algebraic
                                        number fields (see section \ref{0200g}). A good understanding of
                                        these quadratic forms is therefore a prerequisite for a successful
                                        treatment of a theory of gravity. In particular, some properties
                                        of algebraic number fields have to be collected which we are going
                                        to present now. An important tool for the analysis of algebraic
                                        number fields are valuations.

                                        \begin{Def}\label{0439k}
                                            Let $K$ be a field, and let $K^*$ be the group of units of $K$.
                                            A  \emph{valuation} of $K$ is a map
                                            $v : K \to \real \cup \{ \infty \}$
                                            which has the following properties:
                                            \begin{enumerate}
                                                \item
                                                $v(a)= \infty \Leftrightarrow a=0$.

                                                \item
                                                $v(ab)=v(a)+v(b)$ for all $a,b \in K^*$.

                                                \item
                                                $v(a+b) \geq \min \{ v(a), v(b) \}$ for all $a,b \in K^*$.
                                            \end{enumerate}
                                        \end{Def}
                                        \begin{Def}\label{0431}
                                            A prime spot $\prim$ of an algebraic number field $K$ is an
                                            equivalence class of valuations of $K$. The non-archimedean
                                            equivalence classes are called finite primes, the archimedean
                                            ones are called infinite primes. We write $\prim \nmid
                                            \infty$ (resp. $\prim \mid \infty$) if $\prim$ is finite (resp.
                                            infinite). In the case $\prim \nmid  \infty$, we write $\prim \mid
                                            p$ if the residue field $\kappa(\prim)$
                                            corresponding to $\prim$, is of characteristic $p$. Recall
                                            that
                                            \begin{align*}
                                                \kappa(\prim):=\text{\footnotesize{$\mathcal O$}}_{\prim}/
                                                \maxi_{\prim},
                                            \end{align*}
                                            where $\text{\footnotesize{$\mathcal O$}}_{\prim}:= \{ \alpha \in K
                                            \mid {\prim}(\alpha)\geq 0 \}$ and where $\maxi_{\prim} \subset
                                            \text{\footnotesize{$\mathcal O$}}_{\prim}$ is the maximal ideal.
                                        \end{Def}
                                        In analogy to the  case of finite prime spots exposed in
                                        Lemma \ref{0430}, the infinite prime spots are given by embeddings
                                        $\tau :K \hookrightarrow \complex$. An infinite prime spot $\prim$
                                        is called real or complex depending on whether the completion
                                        $K_{\prim}$ is isomorphic to $\real$ or $\complex$, and we define
                                        \begin{align*}
                                            \kappa(\prim):=K_{\prim}.
                                        \end{align*}
                                        We may associate to each prime spot $\prim$ of $K$ a canonical
                                        homomorphism
                                        \begin{align*}
                                            v_{\prim} : K^* \to \real.
                                        \end{align*}
                                        If $\prim$ is finite, we define $v_{\prim}$ to be the normed
                                        $\prim$-adic valuation on $K$ given by $v_{\prim}(K^*)=\Ganz$. If
                                        $\prim$ is infinite, we define $v_{\prim}(a):=- \log|\tau a|$,
                                        where $\tau :K \to \complex$ is the embedding defining $\prim$. \\
                                        \\
                                        Let us now return to quadratic forms. A central result is the fact
                                        that quadratic forms over fields of characteristic zero may be
                                        diagonalized.

                                        \begin{Satz}\label{0206}
                                            Every quadratic module $(V,Q)$ over a field $k$ (i.e. $V$ is a
                                            $k$-vector space with a quadratic form $Q$ on $V$) has an
                                            orthogonal basis.
                                        \end{Satz}

                                        \begin{proof}
                                            \SerreII, Chap. IV, Thm. 1
                                        \end{proof}
                                        Therefore let $(g_{\mu \nu})= \text{diag}(g_{11}, \ldots,
                                        g_{nn})=:\langle  g_{11}, \ldots, g_{nn} \rangle$ be diagonal. As
                                        we are working with  spin-$2$ gravity (i.e. with quadratic forms),
                                        and as we are only looking up to coordinate transformation (i.e.
                                        up to isometry of quadratic modules), we are free to multiply the
                                        $g_{ii}$ with elements of $K^{* 2}:= \left\{\alpha^2| \alpha \in
                                        K^* \right\}$. Therefore, it is of interest to understand fields
                                        modulo squares.

                                        \begin{Satz}\label{0208}
                                            Let $K_{\prim}$ be a $\prim$-adic field with residue class field $k_{\prim}$ and
                                            prime $\pi$.
                                            \begin{enumerate}
                                                \item
                                                If $\text{char}(k_{\prim})\neq 2$, then $K_{\prim}$ has exactly four
                                                square classes
                                                \begin{align*}
                                                    K_{\prim}^*/K^{* 2}_{\prim}= \{1, \varepsilon, \pi, \varepsilon \pi \}
                                                \end{align*}
                                                where $\varepsilon$ is a unit whose reduction mod $\pi$ is
                                                not a square, i.e. $\overline{\varepsilon} \notin k_{\prim}^{*2}$.
                                                \item
                                                If $\text{char}(k_{\prim}) = 2$ that is $K_{\prim}$ is a finite
                                                extension of $\rat_2$, then $K_{\prim}$ has exactly
                                                $2^{n+2}$ square classes for $n=[K_{\prim}:\rat_2]$.
                                                \item
                                                Up to isomorphism $K_{\prim}$ has exactly one non-split\footnote{Let $(\alpha, \beta)$ be
                                                a quaternion algebra (see Definition \ref{0216}). Then the
                                                following statements are equivalent:
                                                \begin{enumerate}
                                                    \item[$\bullet$]
                                                    $(\alpha, \beta)$ splits

                                                    \item[$\bullet$]
                                                    $(\alpha, \beta)$ is not a skew-field.

                                                    \item[$\bullet$]
                                                    The norm form $N$ of $(\alpha, \beta)$ is isotropic.
                                                \end{enumerate} } quaternion
                                                algebra. If we define the ``Hasse-Minkowski symbol'' by
                                                \begin{align*}
                                                s(\alpha, \beta):=
                                                \left\{%
                                                \begin{array}{ll}
                                                    \text{1} & \quad \text{if $(\alpha, \beta)$ splits}   \\
                                                    \text{-1} & \quad \text{if $(\alpha, \beta)$ does not splits}   \\
                                                \end{array}
                                                \right\}
                                                \end{align*}
                                                then the map $s:K_{\prim}^*/K^{* 2}_{\prim} \times K_{\prim}^*/K^{* 2}_{\prim} \to \{\pm
                                                1\}$ is nonsingular. If $\text{char}(k_{\prim}) \neq 2$, then $(\varepsilon,
                                                \pi)$ is a non-split quaternion algebra.
                                            \end{enumerate}
                                        \end{Satz}

                                        \begin{proof}
                                            \Schar, Chap. 6, Fact 4.1
                                        \end{proof}
                                        Before we state the important Hasse-Minkowski theorem, let us
                                        introduce the following notion.
                                        \begin{Def}\label{0210}
                                            Let $K$ be a field and let $Q(T_1, \ldots, T_n):=
                                            \sum\limits_{i,j}a_{ij}T^iT^j$, $a_{ij}\in K$ be a quadratic form in $n$
                                            variables over $K$.
                                            \begin{enumerate}
                                            \item
                                            We say that $Q$ represents an element $a\in
                                            K$ if there exists $t \in K^n$, $t \neq 0$, such that
                                            $Q(t)=a$.
                                            \item
                                            $Q$ is called \emph{isotropic} if it represents $0\in K$.
                                            \item
                                            $Q$ is called \emph{anisotropic} if it does not represent $0 \in K$.
                                            \end{enumerate}
                                        \end{Def}

                                        \begin{Def}\label{0216}
                                            For $a,b \in K^*$ define a $4$-dimensional $K$-algebra with
                                            basis $1, e_1, e_2, e_3$ by the following multiplication
                                            table:
                                            \begin{align*}
                                                e_1e_2=e_3, \quad e_2e_1=-e_2e_1, \quad e_1^2=a \cdot 1
                                                =a,  \quad e_2^2= b\cdot 1 =b
                                            \end{align*}
                                            where $1$ denotes the unit element which is sometimes also denoted by $e_0$. The elements of this
                                            algebra are linear combinations
                                            \begin{align*}
                                                \alpha_0 e_0 + \alpha_1 e_1 + \alpha_2 e_2 + \alpha_3 e_3,
                                                \quad \alpha_i \in K.
                                            \end{align*}
                                            This algebra is associative. It is denoted by $(a,b)=(a,b)_K$
                                            and is called a quaternion algebra over $K$.
                                            There is a canonical involution
                                            \begin{align*}
                                                ^- : (a,b) \to (a,b), \quad x= x^0 e_0 +
                                                \sum_{i=1}^3x^ie_i \mapsto \overline{x}:= x^0 e_0 -
                                                \sum_{i=1}^3x^ie_i
                                            \end{align*}
                                            which is the quaternion analogue of complex conjugation. The
                                            quadratic form
                                            \begin{align*}
                                                N : (a,b) \to K, \quad x \mapsto x\overline{x} = \overline{x}x
                                            \end{align*}
                                            is called the norm form of the quaternion algebra. There is the
                                            following isomorphism of quadratic spaces:
                                            \begin{align*}
                                                \Big( (a,b),N \Big) \cong \langle 1, -a, -b, ab \rangle.
                                            \end{align*}
                                        \end{Def}

                                        \begin{Satz}\label{0211}
                                            Let $K_{\prim}$ be a $\prim$-adic field.
                                            \begin{enumerate}
                                                \item
                                                Every form of dimension $\geq 5$ is isotropic.
                                                \item
                                                Up to isometry there exists exactly one anisotropic
                                                $4$-dimensional form, namely the norm form of the
                                                non-split quaternion algebra. Using
                                                Proposition \ref{0208}, this norm form can be written as
                                                $\langle 1, -\varepsilon, -\pi, \varepsilon \pi \rangle$.
                                            \end{enumerate}
                                            Therefore quadratic forms can be classified by dimension,
                                            determinant and Hasse-Minkowski symbol
                                            $s\langle\alpha_1, \ldots, \alpha_n\rangle:= \prod_{i<j}s(\alpha_i,
                                            \alpha_j)$.
                                        \end{Satz}

                                        \begin{proof}
                                            \Schar, Chap. 6, Thm. 4.2
                                        \end{proof}

                                        \begin{HM}\label{0212}(Local-global principle for quadratic forms.)
                                            Let $K$ be a global field of characteristic $\neq 2$ and let
                                            $Q$ be a $n$-dimensional quadratic form over $K$. Then
                                            $Q$ is isotropic if and only if $Q$ is isotropic over
                                            all completions $K_{\prim}$ with $\prim$ non-archimedean or
                                            real.
                                        \end{HM}

                                        \begin{proof}
                                            \Schar, Chap. 6, Main Theorem 4.2
                                        \end{proof}
                                        \begin{Satz}\label{0221b}
                                            Let $p$ be a prime number, let $n \geq 1$, let $d \in
                                            \rat_p^*/\rat_p^{*2}$ and let $\varepsilon \in \{ \pm 1 \} $.
                                            In order that there exists a quadratic form $Q$ of rank $n$
                                            such that:
                                            \begin{enumerate}
                                                \item[$\bullet$]
                                                $\text{det}(Q)=d$ \quad and

                                                \item[$\bullet$]
                                                $s(Q)=\varepsilon$,
                                            \end{enumerate}
                                            it is necessary and sufficient that one of the following conditions is fulfilled:
                                            \begin{enumerate}
                                                \item[(i)]
                                                $n=1$ and $\varepsilon =1$

                                                \item[(ii)]
                                                $n=2$ and $\varepsilon =1$

                                                \item[(iii)]
                                                $n=2$ and $d \neq -1$

                                                \item[(iv)]
                                                $n \geq 3$.
                                            \end{enumerate}
                                        \end{Satz}

                                        \begin{proof}
                                            \SerreArit, Chap. IV, Prop. 6
                                        \end{proof}
                                        \section{The metric tensor}\label{0200g}

                                        In Riemannian geometry the metric tensor is a smooth family of
                                        scalar products. For each point there is one scalar product, i.e.
                                        a non-degenerate, positive-definite symmetric bilinear-form living
                                        in the tangent space at the point. We would like to provide this
                                        notion in algebraic geometry, too.

                                        As usual let us  consider a smooth scheme $X$ over a Dedekind
                                        scheme $S$. Let $\omega \in \Omega_{X/S}^1(X)$ be a global
                                        differential form and let $(U_i)_{i \in I}$ an affine open
                                        covering of $X$. Consider the $\Ox(U_i)$-algebra homomorphisms
                                        \begin{align*}
                                            \Ox(U_i)[T] & \to
                                            \text{Sym}_{\Ox(U_i)}\left(\Omega_{X/S}^1(U_i)\right)\\
                                            T & \mapsto  \omega|_{U_i}
                                        \end{align*}
                                        where ``Sym'' denotes the associated symmetric algebra. This
                                        algebra homomorphism corresponds to a $X$-morphism
                                        \begin{align*}
                                            \omega_i : T_{U_i/S} \to \Affin_{U_i}^1,
                                        \end{align*}
                                        because by Definition \ref{1122e} and Definition \ref{1129} one has got
                                        \begin{align*}
                                            T_{U_i/S} & = \mathbb{V}(\Omega_{U_i/S}^1)
                                            = {\text{Spec}}\left(\text{Sym}_{\Ox(U_i)}\left(\Omega_{X/S}^1(U_i)\right)\right).\\
                                            \Affin_{U_i}^1 & = \Spec \Ox(U_i)[T]
                                        \end{align*}
                                        $T_{U_i/S}$ is called the algebraic geometric tangent bundle,
                                        whose basic properties are summarized in chapter \ref{1100} of
                                        this thesis. As $\omega$ is a global section of $\Omega_{X/S}^1$
                                        the morphisms $\omega_i$ glue to a $X$-morphism
                                        \begin{align*}
                                            \omega : T_{X/S} \to \Affin_{X}^1
                                        \end{align*}
                                        which we also denote by $\omega$ with some abuse of notation. Vice
                                        versa each such morphism $ \omega : T_{X/S} \to \Affin_{X}^1$
                                        gives rise to a global section $\omega$ of $\Omega_{X/S}^1$. This
                                        global section is the image of the variable $T$ under the
                                        associated algebra homomorphism $\Ox[T]  \to
                                        \text{Sym}_{\Ox}\left(\Omega_{X/S}^1\right)$. Thus each global
                                        differential form  $\omega \in \Omega_{X/S}^1$ can be interpreted
                                        as a $X$-morphism $ \omega : T_{X/S} \to \Affin_{X}^1$.

                                        Analogously we can perform with global sections $\omega \in
                                        \Omega_{X/S}^{\otimes n}(X)$. Writing
                                        \begin{align*}
                                            \omega =
                                            \sum\limits_{1 \leq i_1, \ldots, i_n \leq m} \omega_{i_1} \otimes
                                            \ldots \otimes \omega_{i_n}, \quad \omega_{i_j} \in \Omega_{X/S}^1(X)
                                        \end{align*}
                                        as a sum of elementary tensors, then $\omega$ corresponds to the
                                        ``multi-linear'' morphism
                                        \begin{align*}
                                            \omega: T_{X/S} \times_X \ldots \times_X T_{X/S} & \to
                                            \Affin_X^1 , \quad
                                            (t_1, \ldots, t_n)  \mapsto \sum\limits_{1 \leq i_1, \ldots, i_n \leq m} \omega_{i_1}(t_1) \cdot \ldots \cdot
                                            \omega_{i_n}(t_n).
                                        \end{align*}
                                        More precisely this morphism is obtained as follows. By what we
                                        have already seen, each $\omega_{i_j} \in \Omega_{X/S}^1(X)$ can
                                        be interpreted as a $X$-morphism $ \omega_{i_j} : T_{X/S} \to
                                        \Affin_{X}^1$. Thus $\omega$ gives rise to a canonical morphism
                                        \begin{align*}
                                            \underbrace{T_{X/S} \times_X \ldots \times_X T_{X/S}}_{\text{n-times}} & \to
                                            \left(\Affin_X^1 \times_X \ldots \times_X    \Affin_X^1\right)^{\oplus m}
                                        \end{align*}
                                        where both fibre products consist of $n$ factors. Composing first
                                        with the $n$-fold multiplication morphism
                                        \begin{equation*}
                                            \xymatrix{ \left(\Affin_X^1 \times_X \ldots \times_X    \Affin_X^1 \right)^{\oplus m}
                                            \ar[rr]^{\quad \quad \text{mult.} } & & \left(\Affin_X^1 \right)^{\oplus m}}
                                        \end{equation*}
                                        and then summing up the $m$ factors
                                        \begin{equation*}
                                            \xymatrix{ \left(\Affin_X^1 \right)^{\oplus m}
                                            \ar[r]^{\quad  \sum } & \Affin_X^1 }
                                        \end{equation*}
                                        we finally get the $X$-morphism $\omega$ as the composition of
                                        these three canonical morphisms. The multi-linear character of
                                        this morphism is illustrated in more detail in the beginning of
                                        section \ref{1150}.

                                        Let us now assume that $\Omega_{X/S}^1$ is free $\Ox$-module. For
                                        example, this is the case if $X$ is an $S$-group scheme and if $S$
                                        is in addition the spectrum of a principal ideal domain (let us
                                        remark that this will be the case in physical applications). Then
                                        choose a base $(\omega^1, \ldots, \omega^n )$ of
                                        $\Omega_{X/S}^1(X)$ where $n$ is the relative dimension of $X$
                                        over $S$. In order to construct a metric we consider an element
                                        \begin{align*}
                                            g= \sum\limits_{1 \leq i,j \leq n} g_{ij} \, \omega^i \otimes \omega^j \in \Omega_{X/S}^{\otimes
                                            2}(X), \quad \quad g_{ij}\in \Ox(X).
                                        \end{align*}
                                        By what we have seen above $g$ corresponds to a ``bilinear''
                                        $X$-morphism
                                        \begin{align*}
                                            g:{T_{X/S} \times_X T_{X/S}} & \to  \Affin_X^1.
                                        \end{align*}
                                        More precisely bilinear means the following. Let $\alpha: X' \to
                                        T_{X/S} $ be a $X$-morphism. Due to the universal property of the
                                        fibre product there is a canonical bijection
                                        \begin{align*}
                                            \Hom_X(X',T_{X/S} \times_X T_{X/S}) \cong \Hom_X(X',T_{X/S})
                                            \times \Hom_X(X',T_{X/S}).
                                        \end{align*}
                                        Thus $g$ induces a canonical map
                                        \begin{align*}
                                            g(\alpha):\Hom_X(X',T_{X/S}) \times \Hom_X(X',T_{X/S}) \to \Hom_X(X',\Affin_X^1).
                                        \end{align*}
                                        This map is bilinear. Let us make this more transparent by looking
                                        at generic fibres. \\ \\
                                        \underline{Physical interpretation:} Let $X\to S$ be smooth of
                                        relative dimension $n$. Let $S$ be a Dedekind scheme with field of
                                        fractions $K$. Let $g:{T_{X/S} \times_X T_{X/S}} \to \Affin_X^1$
                                        be as above. Performing a base change with the canonical inclusion
                                        $\Spec K \hookrightarrow S$ we arrive at the archimedean component
                                        of $g$ which is given by the $X_K$-morphism
                                        \begin{align*}
                                            g_K:{(T_{X/S} \otimes_S K) \times_{X_K} (T_{X/S} \otimes_S K)} \to
                                            \Affin_{X_K}^1.
                                        \end{align*}
                                        Recalling the properties of fibre bundles we know that
                                        \begin{align*}
                                            T_{X/S} \otimes_S K = \mathbb{V}(\Omega^1_{X/S}) \times_X X_K
                                            = \mathbb{V}(i_K^* \Omega^1_{X/S}) =
                                            \mathbb{V}(\Omega^1_{X_K/K}) = T_{X_K/K}
                                        \end{align*}
                                        where $i_K:X_K \hookrightarrow X$ is the canonical inclusion.
                                        Thereby, the first and last equality is by definition, the second
                                        one is due to Proposition \ref{1123}, and the third equality is due to
                                        Proposition \ref{0017}. Thus we can write $g_K$ in the more transparent way
                                        \begin{align*}
                                            g_K:{T_{X_K/K} \times_{X_K} T_{X_K/K}} \to
                                            \Affin_{X_K}^1.
                                        \end{align*}
                                        Let us evaluate this morphism at $L$-valued points, where the
                                        field $L$ is a separable extension of $K$. Because of
                                        $\Affin_{X_K}^1(L) = (\Affin_{\Ganz}^1 \times _{\Ganz} X_K)(L) = L
                                        \times X_K(L)$ we have got
                                        \begin{equation*}
                                            \xymatrix{ &&& L
                                            \\
                                            {T_{X_K/K}(L) \times T_{X_K/K}(L)}
                                            \ar[rr]^{\quad \quad g_K(L)} &&
                                            L \times X_K(L) \ar[ru]^{p_1} \ar[dr]_{p_2}
                                            \\
                                            &&& X_K(L)}
                                        \end{equation*}
                                        where $p_1$ and $p_2$ are the  projections onto the respective
                                        factor. Anticipating the notions and statements immediately
                                        following Remark \ref{1130}, we obtain
                                        \begin{align*}
                                            T_{X_K/K}(L)
                                            = \bigcup\limits_{x \in X_K(L)} T_{X_K/K}(x)
                                            = \bigcup\limits_{x \in X_K(L)} T_{X_K,x}
                                            \cong \bigcup\limits_{x \in X_K(L)} L^{\,
                                            n}.
                                        \end{align*}
                                        If we choose $L$ as a separable algebraic closure $\overline{K}$
                                        of $K$, this shows that $T_{X_K/K}(\overline{K})$ is actually the
                                        differential geometric tangent space (i.e. the family of tangent
                                        spaces indexed by points). There is one tangent space for each
                                        physical point. Thus $g_K(\overline{K})$ is a family of bilinear
                                        forms; for each physical point there is one bilinear form living
                                        on the tangent space of this point. More precisely $p_1 \circ
                                        g_K(\overline{K})$ is a bilinear form and $p_2 \circ
                                        g_K(\overline{K})$ gives the physical point in whose tangent
                                        space the bilinear forms lives. \\ \\
                                        Later we will consider the above bilinear forms
                                        \begin{align*}
                                            g(\alpha):\Hom_X(X',T_{X/S}) \times \Hom_X(X',T_{X/S}) \to \Hom_X(X',\Affin_X^1).
                                        \end{align*}
                                        for $S$-valued points $\alpha \in X(S)$ in order to have the
                                        notion of a metric at the ``adelic'' points of $X$.

                                        So let us continue to construct the metric. We already saw that
                                        the global section
                                        \begin{align*}
                                            g= \sum\limits_{1 \leq \mu,\nu \leq n} g_{\mu \nu} \, \omega^\mu \otimes \omega^\nu \in \Omega_{X/S}^{\otimes
                                            2}(X), \quad \quad g_{\mu \nu}\in \Ox(X).
                                        \end{align*}
                                        can be interpreted as a bilinear form. The matrix $(g_{\mu \nu})$
                                        is symmetric if and only if the corresponding $X$-morphism
                                        $g:{T_{X/S} \times_X T_{X/S}}  \to  \Affin_X^1$ is symmetric, i.e.
                                        $g(v,w) = g(w,v)$. In order to say what we mean with a
                                        non-degenerated bilinear form, we consider  the $X$-functor ${\cal
                                        T}_{X/S}:=\GHom_{\Ox}(\Omega_{X/S}^1, \Ox)$ represented by
                                        $T_{X/S}$. The $X$-morphism $g$ induces a canonical map
                                        \begin{align*}
                                            {\cal T}_{X/S} \to ({\cal T}_{X/S})^{\vee}:= \GHom_{\Ox}({\cal T}_{X/S}, \Ox) \cong \Omega_{X/S}^1 , \quad v \mapsto
                                            g(\, \cdot \, , v).
                                        \end{align*}
                                        We say the $g$ is non-degenerate if and only if this map is an
                                        isomorphism of $X$-functors. In particular the matrix $(g_{\mu
                                        \nu})$ is invertible, i.e. $\det (g_{\mu \nu}) \in \Ox(X)^*$ if
                                        $g$ is non-degenerate. We are now prepared to make the
                                        \begin{Def}\label{0201}
                                            Let $X \to S$ be smooth of relative dimension $n$ such that
                                            $\Omega_{X/S}^1$ is a free $\Ox$-module. Let
                                            \begin{align*}
                                                g= \sum\limits_{1 \leq i,j \leq n} g_{ij} \, \omega^i \otimes \omega^j \in \Omega_{X/S}^{\otimes
                                                2}(X), \quad \quad g_{ij}\in \Ox(X).
                                            \end{align*}
                                            where $(\omega^1, \ldots, \omega^n )$ is a base of
                                            $\Omega_{X/S}^1(X)$. Let
                                            \begin{align*}
                                                g:{T_{X/S} \times_X T_{X/S}} & \to  \Affin_X^1.
                                            \end{align*}
                                            the associated bilinear $X$-morphism as constructed above. Then
                                            $g$ is called a \emph{spin-$2$ metric} if and only if $g$ is
                                            symmetric and non-degenerate; i.e. if and only if:
                                            \begin{enumerate}
                                                \item
                                                The matrix $(g_{\mu \nu})$ is symmetric.
                                                \item
                                                The canonical morphism ${\cal T}_{X/S} \to ({\cal
                                                T}_{X/S})^{\vee}$ is an isomorphism of $X$-functors.
                                            \end{enumerate}
                                            In particular $\det (g_{\mu \nu}) \in \Ox(X)^*$, i.e. the matrix
                                            $(g_{\mu \nu})$ is invertible. We denote the inverse matrix
                                            of $(g_{\mu \nu})$ by $(g^{\mu \nu})$.
                                        \end{Def}

                                        \begin{Satz}\label{0201s}
                                            Let $S=\Spec R$ be an affine base scheme and let $X \to S$ be a smooth $S$-scheme of finite type.
                                            Then there exists a canonical metric $g$ on $X$, the so called \emph{first
                                            fundamental form}. Locally, this metric looks as follows:

                                            In the special case $X=\Affin_S^n=\Spec R [T_1, \ldots, T_n]$,
                                            one has got $\Omega_{X/S}^1= \bigoplus_{i=1}^n \co_X \cdot
                                            dT_i$. Then $(g_0)_{\mu \nu}:= \text{diag}(\pm 1,\ldots,\pm 1)$
                                            induces a trivial metric $g_0$ on $X$. If $j:X \to \Affin_S^n$ is a
                                            closed $S$-immersion, we may pull back the trivial metric $g_0$ via $j$ in order
                                            to obtain a metric $j^*g$ on $X$. The metric $j^*g$ is called the \emph{first
                                            fundamental form}.
                                        \end{Satz}

                                        \begin{proof}
                                            We may choose a finite open covering $(U_i)$ of $X$ by affine schemes $U_i$ together
                                            with closed $S$-immersions $\iota_i:U_i \hookrightarrow \Affin_S^n$,
                                            i.e. $U_i=V(I_i)$ is the zero set of some ideal $I_i \subset R [T_1, \ldots,
                                            T_n]$. Let $g_i:= \iota_i^* g_0$ be the first fundamental form
                                            on $U_i$ (as described above). We will show that we may glue the local metrics
                                            $g_i$ along the intersections $U_{ij}:=U_i \subset U_j$.
                                            Due to Lemma \ref{0201t}, we know that $U_{ij}= U_i \times_{\Affin_S^n} U_j$ as scheme.
                                            Let $p_k:U_{ij} \to U_k$, $k=i,j$, be the canonical
                                            projection. Then, $p_i^*g_i= p_i^* \iota_i^* g_0 = (\iota_i \circ
                                            p_i)^*g_0$$=(\iota_j \circ p_j)^*g_0 =p_j^*g_j$. Interpreting
                                            $g_i$ as section $g_i \in \Omega_{X/S}^{\otimes 2}(U_i)$, the
                                            equation above may be written as $g_i \mid_{U_{ij}}=g_j
                                            \mid_{U_{ij}}$. Thus, the local metrics $g_i$ glue to a
                                            global metric $g \in \Omega_{X/S}^{\otimes 2}(X)$. This metric
                                            is called the first fundamental form.
                                        \end{proof}

                                        \begin{Lemma}\label{0201t}
                                            Let $S=\Spec R$ be an affine base scheme, and let $X$ be a
                                            closed subset of $\ba_S^n$. Then there is a unique sheaf of
                                            ideals $\ci$ such that $V({\ci})$ realizes $X$ as closed sub-scheme of $\ba_S^n$ which is smooth over $S$.
                                        \end{Lemma}

                                        \begin{proof}
                                            We may assume that $X$ is affine. Let $I,J \subset A:=\co_{\ba_S^n}(\ba_S^n)$ be two
                                            ideals such that $V(I)$ and $V(J)$ define two closed sub-scheme structures
                                            on $X$. As $V(I)$ and $V(J)$ coincide as topological spaces, one knows from
                                            Hilbert´s Nullstellensatz that $\text{rad}(I)=\text{rad}(J)$. Due to
                                            smoothness, one has got $\text{rad}(I)=I$. This may be seen as
                                            follows: Clearly, $I \subset \text{rad}(I)$. Conversely, let
                                            $f \in \text{rad}(I)$ and $\overline{f} \in A/I$ be the
                                            residue class of $f$. There is a $m \in \Natural$ such
                                            that $f^m \in I$, i.e. $\overline{f}^m=0$. Due to smoothness,
                                            $A/I$ has no zero-divisors, and it follows that already
                                            $\overline{f}=0$, i.e. $f \in I$.
                                            All in all, we obtain $I=\text{rad}(I)=\text{rad}(J)=J$.
                                        \end{proof}

                                        \begin{Rem}\label{0201r}
                                            In exactly the same way as in differential geometry, we may
                                            furthermore define the notion of a covariant derivation. Using
                                            as well the metric as the notion of the commutator of
                                            vector fields (see Definition \ref{1158}), it makes also sense to talk
                                            about torsion free and metric preserving connections. Like in
                                            differential geometry, we obtain a unique connection which is
                                            torsion free and metric preserving, the Levi-Civita connection. It may be derived
                                            in terms of the metric by means of the Koszul formula. All in
                                            all, we may introduce the notion of curvature for relative
                                            schemes $X \to S$. This establishes a general relativity
                                            for schemes, because we may simply write down the equations of
                                            general relativity in our algebraic geometric setting. This is done in section
                                            \ref{5100}.
                                        \end{Rem}

                                        \  \\ \textbf{Ricci calculus for schemes.}\label{0202}\\
                                        Let $X \to S$ be a scheme of relative dimension $n$ with spin-$2$
                                        metric $g$. Then, locally, we can perform Ricci calculus in the
                                        same manner we are used to from differential geometry. For this
                                        purpose choose a local base $(\partial_1, \ldots ,
                                        \partial_n)$ of ${\cal T}_{X/S}$ and let $(\omega^1, \ldots, \omega^n)$ be
                                        a local base of $\Omega_{X/S}^1$ such that both bases are dual to
                                        each other. Each local section $v$ of ${\cal T}_{X/S}$, i.e. each
                                        local vector field can be written uniquely as
                                        \begin{align*}
                                            v=\sum\limits_{\mu} v^{\mu} \partial_{\mu} , \quad \text{with local sections $v^{\mu}$ of
                                            $\Ox$.}
                                        \end{align*}
                                        Let us denote the image of $v$ under the isomorphism ${\cal
                                        T}_{X/S} \to ({\cal T}_{X/S})^{\vee}$ by
                                        \begin{align*}
                                            g( \, \cdot \, , v) =\sum\limits_{\mu} v_{\mu} \omega^{\mu} , \quad \text{with local sections $v_{\mu}$ of $\Ox$. }
                                        \end{align*}
                                        On the other hand we have got the explicit description $ g( \,
                                        \cdot \, , v) = \sum\limits_{\mu,\nu} g_{\mu \nu} \omega^{\nu}(v)
                                        \cdot \omega^{\mu} = \sum\limits_{\mu,\nu} g_{\mu \nu} v^{\nu}
                                        \omega^{\mu}$. Therefore
                                        \begin{align*}
                                            g_{\mu \nu} v^{\nu} = v_{\mu}
                                        \end{align*}
                                        where over indices which occur twice is summed. Thus indices are
                                        lowered with $g_{\mu \nu}$. Vice versa indices are lifted with
                                        $g^{\mu \nu}$:
                                        \begin{align*}
                                            g^{\mu \nu} v_{\nu} = v^{\mu}.
                                        \end{align*}
                                        \ \\ \textbf{Differential forms with values in Lie-algebras} \\
                                        Let $ \pi: X \to S$ be a smooth connected $S$-scheme which admits
                                        global sections (e.g. a universe) and let $G \to S$ be a smooth
                                        $S$-group scheme. Let $\Liealg= \Lie(G/S)(S)$ denote the
                                        Lie-algebra of $G$. We know that we can write as well $\Liealg=
                                        (\varepsilon^*\Omega_{G/S}^1)(S)$. In gauge theory we are
                                        interested in differential forms which take values in the
                                        Lie-algebra $\Liealg$. This way the fields corresponding to the so
                                        called gauge-bosons are described.

                                        The morphism $ \pi: X \to S$ gives a morphism of sheaves $\pi^{-1}
                                        \Os \to \Ox$. By our premise $\pi$ has sections; in particular
                                        faithfully flat. Then, $\pi^{-1} \Os = \Os \circ \pi$ and
                                        $\pi^{-1} \Os \hookrightarrow \Ox$ is injective. Evaluating at
                                        global sections, we get a monomorphism of rings $\Os(S)
                                        \hookrightarrow \Ox(X)$. Let us consider $\Os(S)$ and $\Liealg$ as
                                        a constant sheaves on $X$, i.e. for all open subsets $U \subset X$
                                        we set:
                                        \begin{align*}
                                            \Os(S)(U) &:=\Os(S) \\
                                            \Liealg(U) &:=\Liealg
                                        \end{align*}
                                        We consider $\Ox$ as a $\Os(S)$-algebra in the following way: For
                                        all open subsets $U \subset X$ we set
                                        \begin{align*}
                                            \Os(S)(U) \to \Ox(U), \quad \alpha \mapsto \alpha|_U
                                        \end{align*} where we consider $\alpha$ as a global section of
                                        $\Ox$. Thus we can consider $\Omega_{X/S}^1$ as a $\Os(S)$-module.
                                        \begin{Def}
                                            Let $ \pi: X \to S$ be a smooth connected $S$-scheme which admits
                                            global sections (e.g. a universe) and let $G \to S$ be a smooth
                                            $S$-group scheme. Let $\Liealg$ be the Lie-algebra of $G$. A
                                            Lie-algebra valued differential form is a section of the
                                            $\Ox$-module
                                            \begin{equation*}
                                                \Omega_{X/S}^1 \otimes_{\Os(S)} \Liealg .
                                            \end{equation*}
                                        \end{Def}
                                        Regarding Proposition \ref{1135}, we have got
                                        \begin{align*}
                                            \Omega_{X/S}^1 \otimes_{\Os(S)} \Liealg
                                            &= \Omega_{X/S}^1 \otimes_{\Ox} \left(\Ox \otimes_{\Os(S)}
                                            \Liealg \right)\\
                                            &= \GHom_{\Ox}\left({\cal T}_{X/S}, \Ox\right) \otimes_{\Ox} \left(\Ox \otimes_{\Os(S)}
                                            \Liealg \right) \\
                                            &= \GHom_{\Ox}\left({\cal T}_{X/S}, \Ox \otimes_{\Os(S)}
                                            \Liealg \right)
                                        \end{align*}
                                        Thus elements of $\Omega_{X/S}^1 \otimes_{\Os(S)} \Liealg$ can be
                                        evaluated at tangent-vectors and the result is a Lie-algebra
                                        valued function. We would like to give an equivalent
                                        characterization  of global sections of $\Omega_{X/S}^1
                                        \otimes_{\Os(S)} \Liealg$ in terms of morphisms of schemes as we
                                        did above for ordinary differential forms, i.e. sections of
                                        $\Omega_{X/S}^1$. So let us consider the fibre-bundle
                                        \begin{align*}
                                            T_{X/S}^{\Liealg}:=\mathbb{V}\left(\Omega_{X/S}^1 \otimes_{\Os(S)} \Liealg\right)
                                        \end{align*}
                                        and let ${\cal T}_{X/S}^{\Liealg}$ be the associated $X$-functor.
                                        In terms of $X$-functors we can write
                                        \begin{align*}
                                            {\cal T}_{X/S}^{\Liealg}
                                            &= \GHom_{\Ox}\left(\Omega_{X/S}^1 \otimes_{\Os(S)} \Liealg, \Ox
                                            \right)\\
                                            &= \GHom_{\Ox}\left(\Omega_{X/S}^1,  \GHom_{\Ox}\left( \Ox \otimes_{\Os(S)} \Liealg,\Ox \right)
                                            \right) \\
                                            &= \GHom_{\Ox}\left(\Omega_{X/S}^1,  \left( \Ox \otimes_{\Os(S)} \Liealg \right)^{\vee} \right)
                                        \end{align*}
                                        where we have used again Proposition \ref{1135}. Simply by replacing
                                        $\Omega_{X/S}^1$ through $\Omega_{X/S}^1 \otimes_{\Os(S)} \Liealg$
                                        in our former calculations, we see that we can identify global
                                        sections of $\Omega_{X/S}^1 \otimes_{\Os(S)} \Liealg$, i.e.
                                        elements of $\Omega_{X/S}^1(X) \otimes_{\Os(S)} \Liealg$, with
                                        $X$-morphisms
                                        \begin{align*}
                                            T_{X/S}^{\Liealg} \to \Affin_{X/S}^1.
                                        \end{align*}
                                        Analogously we can perform with more general tensors with values
                                        in the Lie-algebra, i.e. with global sections $\omega \in
                                        \Omega_{X/S}^{\otimes n}(X) \otimes_{\Os(S)} \Liealg$. Writing
                                        \begin{align*}
                                            \omega =
                                            \sum\limits_{1 \leq i_1, \ldots, i_n \leq m} \omega_{i_1} \otimes
                                            \ldots \otimes \omega_{i_{n-1}} \otimes (\omega_{i_n} \otimes r_{i_n}), \quad \omega_{i_j} \in
                                            \Omega_{X/S}^1(X), \ r_{i_n} \in \Liealg
                                        \end{align*}
                                        as a sum of elementary tensors, then $\omega$ corresponds to the
                                        following ``multi-linear'' morphism.
                                        \begin{align*}
                                            \omega:\underbrace{ T_{X/S} \times_X \ldots \times_X T_{X/S}}_{\text{$(n-1)$-times}} \times_X T_{X/S}^{\Liealg}
                                            &\to \Affin_X^1 \\
                                            (t_1, \ldots, t_n)  &\mapsto \sum\limits_{1 \leq i_1, \ldots, i_n \leq m} \omega_{i_1}(t_1) \cdot \ldots \cdot
                                            \omega_{i_{n-1}}(t_{n-1}) \cdot (\omega_{i_n} \otimes
                                            r_{i_n})(t_n)
                                        \end{align*}

                                        \begin{Rem}\label{6100}
                                            Let $K$ be local field of characteristic zero. Let $X_K$ be a
                                            scheme which is locally of finite type over $K$. If
                                            $\overline{K}$ denotes an algebraic closure of $K$, the set
                                            $X_K(\overline{K})$ may be endowed with an analytic structure. In
                                            the case $\overline{K}=\complex$, we may consider
                                            $X_K(\overline{K})$ as complex analytic space, and in the $p$-adic
                                            case $\overline{K}=\complex_p$ we obtain a rigid analytic space.

                                            In some analytic theories, the metric is introduced as a powerful
                                            notion which contains further geometric information. This is the
                                            motivation to introduce this notion on the algebraic level, too
                                            (as presented in section \ref{0200g}). If $X_K$ is in addition
                                            smooth over $K$ (or more generally reduced), we know from
                                            Proposition \ref{0201s} that there exists a metric on $X_K$, the first
                                            fundamental form. So, let us assume that there is a metric
                                            $g_K:T_{X_K} \times_{X_K} T_{X_K} \to \ba_{X_K}^1$ on $X_K$. Then
                                            we consider the following category: The objects are pairs
                                            $(X_K,g_K)$ and a morphism $(Y_K,h_K) \to (X_K,g_K)$ is defined as
                                            a morphism $f_K:Y_K \to X_K$ of schemes such that $h_K=f_K^*g_K$.
                                            This category might be called the category of \emph{Riemannian
                                            schemes} over $K$ (of course, the same definition makes sense if
                                            we replace $K$ by an arbitrary base scheme). We may consider the
                                            category of smooth schemes (or reduced schemes which are locally
                                            of finite type) as a full subcategory of Riemannian schemes by
                                            endowing the respective scheme with the first fundamental form as
                                            metric.

                                            Let us now evaluate Riemannian schemes $(X_K,g_K)$ at
                                            $\overline{K}$-valued points and study the analytic objects that
                                            we obtain this way. In order to do this, we assume that $X_K$ is
                                            smooth. As mentioned above, the evaluation at
                                            $\overline{K}$-valued points of a scheme $X_K$ yields a set which
                                            is the zero set of some polynomials, and which may be endowed with
                                            the structure of an analytic space. But if we start from
                                            Riemannian schemes instead, it is possible that the resulting
                                            analytic object describes the zero set of functions which are
                                            \emph{not} necessarily algebraic but may be transcendent.

                                            \begin{Ex}\label{6101}
                                                Let $K=\real$ and $X_K:=\ba_K^2$ with coordinates $(t,r)$. With
                                                respect to the global base $dt, dr$ of $\Omega_{X_K/K}^1$ choose
                                                the metric
                                                $g_K=\left(%
                                                \begin{array}{cc}
                                                p(r)^2 & 0 \\
                                                0 & p'(r)^2 \\
                                                \end{array}%
                                                \right)$
                                                on $X_K$, where $p(r)$ is a polynomial in
                                                $r$, and $p'(r):=\frac{\partial p(r)}{\partial r}$ denotes the first
                                                partial derivative of $p(r)$ with respect to $r$.
                                                Evaluation of the Riemannian scheme $(X_K,g_K)$ at
                                                $\overline{K}$-valued points yields a complex analytic space,
                                                whose $K$-valued points constitute the Riemannian manifold
                                                $M=(\real^2, g_{\mu \nu})$ with $g_{\mu
                                                \nu}=\text{diag}(p(r)^2,q(r)^2)$. We claim that $M$
                                                is isomorphic (in the category of Riemannian manifolds)
                                                to a zero set of non-algebraic, transcendent functions. In
                                                order to prove this consider an embedding
                                                \begin{align*}
                                                    \phi: \real^2 \to \real^4, \quad (t,r) \mapsto (t,r, \varphi_r \varphi_t, \varphi_r \psi_t)
                                                \end{align*}
                                                where $\varphi_r, \varphi_t,\psi_t  \in C^{\infty}(\real)$
                                                will be chosen later. Thereby, $\varphi_r$ only depends on $r$,
                                                and $\varphi_t, \psi_t \in C^{\infty}(\real)$ only depend on
                                                $t$. We want to choose $\varphi_r, \varphi_t$ and $\psi_t$ in
                                                such a way that $\phi(\real^2)$ endowed with the first
                                                fundamental form is isomorphic to $M$. If we introduce the abbreviations
                                                $\dot{\varphi_t}=\frac{\partial \varphi_t}{\partial t}$ and
                                                ${\varphi'_r}=\frac{\partial \varphi_r}{\partial r}$, we obtain the following
                                                basis-vectors $\partial_t$, $\partial_r$ of the tangent space of
                                                $\phi(\real^2)$:
                                                \begin{align*}
                                                    \partial_t=\left(%
                                                    \begin{array}{c}
                                                    1 \\
                                                    0 \\
                                                    \varphi_r \dot{\varphi}_t\\
                                                    \varphi_r \dot{\psi}_t\\
                                                    \end{array}%
                                                    \right) \quad \text{and} \quad
                                                    \partial_r=\left(%
                                                    \begin{array}{c}
                                                    0 \\
                                                    1 \\
                                                    \varphi'_r \varphi_t\\
                                                    \varphi'_r \psi_t\\
                                                    \end{array}%
                                                    \right).
                                                \end{align*}
                                                Choosing the trivial metric diag$(1,1,1,1)$ in $\real^4$ and providing
                                                $\phi(\real^2)$ with the first fundamental form, we
                                                obtain the following differential equations.
                                                \begin{align*}
                                                    p^2&=g_{tt}= \varphi_r^2
                                                    (\dot{\varphi}_t^2+\dot{\psi}_t^2)\\
                                                    0&=g_{tr}= \varphi_r \varphi'_r
                                                    (\varphi_t \dot{\varphi}_t +\psi_t \dot{\psi}_t)\\
                                                    {p'}^2&=g_{rr}= {\varphi'}_r^2
                                                    ({\varphi}_t^2+{\psi}_t^2)
                                                \end{align*}
                                                One easily checks that $\varphi_r:=p(r)$,
                                                $\varphi_t:=\sin(t)$ and $\psi_t:=\cos(t)$ is a solution of
                                                these equations. Thus, if $(t,r,y,z)$ are the variables of
                                                $\real^4$, we see that we may realize $M \cong
                                                \varphi(\real^2)$ as zero set in $\real^4$ of the $C^{\infty}$-functions $y-p(r)
                                                \sin(t)$ and $z-p(r)\cos(t)$. However, these functions are
                                                manifestly not algebraic.
                                            \end{Ex}
                                        \end{Rem}

                                        \ \\
                                        \textbf{Some background material from algebraic number theory}\\
                                        We already introduced the notion of a valuation of an algebraic
                                        number field in Definition \ref{0439k}. Important is the following result.

                                        \begin{Lemma}\label{0430}
                                            Let $K \hookrightarrow L$ be an algebraic extension of fields.
                                            Let $v$ be a valuation on $K$, and let $w$ be a valuation on
                                            $L$ which extends $v$. Then there exists a $K$-embedding $\tau : L \to
                                            \overline{K_v}$ such that
                                            \begin{align*}
                                                w = \overline{v} \circ \tau,
                                            \end{align*}
                                            where $ \overline{v}$ is the canonical continuation of $v$ to
                                            an algebraic closure $\overline{K_v}$ of the completion $K_v$ of
                                            $K$ with respect to $v$. Furthermore, if $L_w$ denotes the
                                            completion of $L$ with respect to $w$, one has got
                                            \begin{align*}
                                                L_w = L K_v.
                                            \end{align*}
                                        \end{Lemma}

                                        \begin{proof}
                                            The first statement is \Neu, Chap. II, Fortsetzungssatz 8.1. Therefore, it
                                            remains to prove the equality $L_w = L K_v$. As a finite
                                            extension of a complete field is complete again (see \Neu, Chap. II, Thm. 4.8),
                                            the finite field extension $L K_v \subset L_w $ of $K_v$ is complete with respect
                                            to $w$. On the other hand $L \subset L K_v$, and thus $L K_v$ must
                                            already coincide with the completion $L_w$ of $L$ with respect to $w$.
                                        \end{proof}
                                        Let us recall that a prime spot $\prim$ of an algebraic number
                                        field $K$ is an
                                            equivalence class of valuations of $K$. The non-archimedean
                                            equivalence classes are called finite primes, the archimedean
                                            ones are called infinite primes. We write $\prim \nmid
                                            \infty$ (resp. $\prim \mid \infty$) if $\prim$ is finite (resp.
                                            infinite). In the case $\prim \nmid  \infty$, we write $\prim \mid
                                            p$ if the residue field $\kappa(\prim)$
                                            corresponding to $\prim$, is of characteristic $p$. Recall
                                            that
                                            \begin{align*}
                                                \kappa(\prim):=\text{\footnotesize{$\mathcal O$}}_{\prim}/
                                                \maxi_{\prim},
                                            \end{align*}
                                            where $\text{\footnotesize{$\mathcal O$}}_{\prim}:= \{ \alpha \in K
                                            \mid {\prim}(\alpha)\geq 0 \}$ and where $\maxi_{\prim} \subset
                                            \text{\footnotesize{$\mathcal O$}}_{\prim}$ is the maximal ideal.
                                        In analogy to the  case of finite prime spots exposed in
                                        Lemma \ref{0430}, the infinite prime spots are given by embeddings
                                        $\tau :K \hookrightarrow \complex$. An infinite prime spot $\prim$
                                        is called real or complex depending on whether the completion
                                        $K_{\prim}$ is isomorphic to $\real$ or $\complex$, and we define
                                        \begin{align*}
                                            \kappa(\prim):=K_{\prim}.
                                        \end{align*}
                                        We may associate to each prime spot $\prim$ of $K$ a canonical
                                        homomorphism
                                        \begin{align*}
                                            v_{\prim} : K^* \to \real.
                                        \end{align*}
                                        If $\prim$ is finite, we define $v_{\prim}$ to be the normed
                                        $\prim$-adic valuation on $K$ given by $v_{\prim}(K^*)=\Ganz$. If
                                        $\prim$ is infinite, we define $v_{\prim}(a):=- \log|\tau a|$,
                                        where $\tau :K \to \complex$ is the embedding defining $\prim$.
                                        \begin{Def}\label{0432}
                                            If $L/K$ is a finite  extension of $K$, let us denote
                                            the prime spots of $L$ with $\Prim$. We will write $\Prim \mid
                                            \prim$ if the restriction of $\Prim$ to $K$ gives $\prim$. In
                                            this situation we define:
                                            \begin{enumerate}
                                                \item
                                                the ramification index $e_{\Prim \mid \prim}$ and the
                                                inertia index $f_{\Prim \mid \prim}$ by:
                                                \begin{align*}
                                                    e_{\Prim \mid \prim}&:=
                                                    \left\{%
                                                    \begin{array}{llll}
                                                        \left(\Prim(L^*):\prim(K^*)\right) & \quad \text{if $\prim \mid p$}   \\
                                                        \quad \quad \quad \ 1 & \quad \text{if $\prim \mid \infty$} \\
                                                    \end{array}
                                                    \right\} \\
                                                    f_{\Prim \mid \prim}&:=
                                                        \left[\kappa(\Prim):\kappa(\prim)\right] \\
                                                    f_{\prim}&:=f_{\prim \mid p}
                                                \end{align*}
                                                \item
                                                the absolute norm $\mathfrak{N}$ by:
                                                \begin{align*}
                                                    \mathfrak{N}(\prim)&:=
                                                    \left\{%
                                                    \begin{array}{llll}
                                                        p^{f_{\prim}} & \quad \text{if $\prim \mid p$}   \\
                                                        e^{f_{\prim}} & \quad \text{if $\prim \mid \infty$} \\
                                                    \end{array}
                                                    \right\}
                                                \end{align*}
                                                where $e=\sum_n 1/{n!}$ is Euler's number.
                                                \item
                                                the $\prim$-adic norm $|\cdot|_{\prim}:K \to \real$ by:
                                                \begin{align*}
                                                |a|_{\prim}&:=
                                                \left\{%
                                                \begin{array}{llll}
                                                    \mathfrak{N}(\prim)^{-v_{\prim}(a)} & \quad \text{if $\prim \mid p$  and $a \neq 0$}   \\
                                                    \ \, \ 0 & \quad \text{if $\prim \mid p$  and $a = 0$}   \\
                                                    \ |\tau a|& \quad \text{if $\prim \mid \infty$ is real} \\
                                                    \ |\tau a|^2& \quad \text{if $\prim \mid \infty$ is complex} \\
                                                \end{array}
                                                \right\}
                                                \end{align*}
                                            \end{enumerate}
                                        \end{Def}
                                        \begin{Satz}\label{0433}
                                            For arbitrary prime spots $\Prim \mid \prim$ the following
                                            identities hold:
                                            \begin{align*}
                                                \sum_{\Prim \mid \prim} e_{\Prim \mid \prim} f_{\Prim \mid \prim}
                                                &= \sum _{\Prim \mid \prim}  \left[ L_{\Prim}:K_{\prim} \right]
                                                = \left[ L :K \right] \\
                                                \mathfrak{N}(\Prim)
                                                &=\mathfrak{N}(\prim)^{f_{\Prim \mid \prim}} \\
                                                v_{\Prim}(a)&= e_{\Prim \mid \prim} v_{\prim}(a)
                                                &&\text{for  } a \in K^*, \\
                                                v_{\prim}\left( N_{L_{\Prim} \mid K_{\prim}}(a) \right)
                                                &= f_{\Prim \mid \prim} v_{\Prim}(a)  &&\text{for  } a \in L^*, \\
                                                |a|_{\Prim}&= |N_{L_{\Prim} \mid K_{\prim}}(a)|_{\prim}
                                                &&\text{for  } a \in  L^*.
                                            \end{align*}
                                        \end{Satz}

                                        \begin{proof}
                                            \Neu, Chap. III, Satz 1.2
                                        \end{proof}

                                        \begin{Satz}\label{0434}
                                            Let $K$ be an algebraic number field.
                                            For all $a \in K^*$ there are only finitely many
                                            prime spots $\prim$ such that $|a|_{\prim} \neq 1$.
                                            Furthermore,
                                            \begin{align*}
                                                \prod_{\prim} |a|_{\prim} =1.
                                            \end{align*}
                                        \end{Satz}

                                        \begin{proof}
                                            \Neu, Chap. III, Satz 1.3
                                        \end{proof}
                                        \section{Metrical fields of higher spin}\label{0239}
                                        In general relativity gravity is described by a field of spin $2$.
                                        In this section we will argue that in principle it is
                                        mathematically possible that there also exists a metrical field of
                                        spin-$3$. The resulting model is then at least ten dimensional.
                                        The global results of this thesis are formulated in a language
                                        which leaves open the possibility to work with pure spin-$2$
                                        gravity or with a combined spin-$2$ and spin-$3$ metrical field.
                                        Nevertheless, those parts of this thesis which contain
                                        computations in local coordinates will work with pure spin-$2$
                                        gravity only. Therefore, this section \ref{0239} is thought as a
                                        brief prospect for metrical fields with higher spin and will not
                                        be that detailed as the
                                        exposition concerning spin-$2$ gravity in section \ref{0202b}. \\ \\
                                        Let us consider a \GR $(X\to S, g)$ of relative dimension $n$. In
                                        the preceding section \ref{0202b} we assumed that gravity is
                                        describable by a spin-$2$ metric
                                        \begin{align*}
                                            g:{T_{X/S} \times_X T_{X/S}} & \to  \Affin_X^1
                                        \end{align*}
                                        (see Definition \ref{0201}). If $(\omega^1, \ldots, \omega^n )$ is a local
                                        base of $\Omega_{X/S}^1$, then $g$ corresponds to a non-degenerate
                                        symmetric bilinear form
                                        \begin{align*}
                                            g= \sum\limits_{1 \leq \mu,\nu \leq n} g_{\mu \nu} \, \omega^{\mu} \otimes \omega^{\nu} \in \Omega_{X/S}^{\otimes
                                            2}(X), \quad \quad g_{\mu \nu}\in \Ox(X).
                                        \end{align*}
                                        and is thus given by a matrix $(g_{\mu \nu})$. This definition is
                                        motivated by our intuitive experience that e.g. the length of
                                        vector $x=(x_1, \ldots, x_n) \in \real^n$ should be measured using
                                        the theorem of Pythagoras in order to get a ``physically
                                        sensible'' result:
                                        \begin{align*}
                                            \|x\|_2&:=\sqrt{{\sum_{i=1}^n x_i^2}}.
                                            \end{align*}
                                        But in principle one could also interpret
                                        \begin{align*}
                                            \|x\|_r:=\left({\sum_{i=1}^n x_i^r}\right)^{\frac{1}{r}} \quad \quad \text{ with } r\in \real, \ 0< r <
                                            \infty,
                                        \end{align*}
                                        as a length of the vector $x$ as this metric yields the same
                                        topology on $\real^n$. As we would interpret the index $r$ as the
                                        spin of an corresponding gauge boson, we are reduced to values $r
                                        \in \{2,3,4, \ldots \} \subset \Natural$. We have to exclude $r=1$
                                        as we want to relate gravity with curvature. The general form of a
                                        the metric in $\real^n$ would then be written as a symmetric form
                                        of homogeneous degree $r$.
                                        \begin{align*}
                                            \|x\|_r^r:={\sum_{\mu_1, \ldots, \mu_r} g_{\mu_1 \ldots \mu_r} \cdot x_{\mu_1} \ldots x_{\mu_r}}
                                            \quad \quad \text{ with } r\in  \{2,3,4, \ldots \}.
                                        \end{align*}
                                        In our setting the spin-$r$ field would be given by a symmetric
                                        rank $r$ tensor $g_{\mu_1 \ldots \mu_r}$ or more precisely by a
                                        morphism
                                        \begin{align*}
                                            g:{\underbrace{T_{X/S} \times_X \ldots \times_X T_{X/S}}_{\text{$r$-times}}} & \to
                                            \Affin_X^1.
                                        \end{align*}
                                        Locally this could then be written as
                                        \begin{align*}
                                            g= \sum\limits_{1 \leq \mu_1, \ldots, \mu_r \leq n} g_{\mu_1 \ldots \mu_r}
                                            \, \omega^{\mu_1} \otimes \ldots \otimes \omega^{\mu_r} \in \Omega_{X/S}^{\otimes
                                            r}(X), \quad \quad g_{\mu_1, \ldots, \mu_r}\in \Ox(X).
                                        \end{align*}
                                        As the tensor product is associative and commutative for symmetric
                                        tensors, we can decompose $g_{\mu_1 \ldots \mu_r}$ into a
                                        tensor-product of symmetric tensors of smaller rank, e.g.:
                                        \begin{align*}
                                            g_{\mu_1 \ldots \mu_i \ldots \mu_r}= g_{\mu_1 \ldots \mu_{i}} \otimes g_{\mu_{i+1} \ldots
                                            \mu_{r}}.
                                        \end{align*}
                                        According to this decomposition we would interpret the ``spin-$r$
                                        field'' as some kind of overlap of a ``spin-$i$ field'' $g_{\mu_1
                                        \ldots \mu_{i}}$ and a ``spin-$(r-i)$ field'' $g_{\mu_{i+1} \ldots
                                        \mu_{r}}$. As $r$ is restricted to values $r \in \{2,3,4, \ldots
                                        \} \subset \Natural$ we see that the ``spin-$2$ field'' $g_{\mu
                                        \nu}$ and the ``spin-$3$ field'' $g_{\mu \nu \lambda}$ are the
                                        only fields which are indecomposable in this sense. They are
                                        irreducible in this regard. Therefore only  $g_{\mu \nu}$ and
                                        $g_{\mu \nu \lambda}$ can be interpreted as elementary particles.

                                        This means that next to the massless spin-$2$ graviton $g_{\mu
                                        \nu}$ of general relativity there could  also exist a spin-$3$
                                        field $g_{\mu \nu \lambda}$. Due to the principles of quantum
                                        field theory, this should be an repulsive force which is limited
                                        to small distances if we provide it with a big mass. In the same
                                        way as in the case of the pure spin-$2$ gravity of the previous
                                        section \ref{0202b}, let us now analyze the effects of a
                                        metrical spin-$3$ field on the dimensionality of \sto. As  we will
                                        only work with the spin-$2$ graviton in the section on gauge
                                        theory, we will do this very shortly and not that detailed as in
                                        the spin-$2$ case.

                                        The number-theoretic background for the analysis of a spin-$3$
                                        field is the theory of cubic forms. Of course the cubic form
                                        $g_{\mu \nu \lambda}$ describing the potential of the spin-$3$
                                        field should be non-singular. But then there is a central result
                                        due to D.R. Heath-Brown:

                                        \begin{Satz}\label{0240}
                                            \begin{enumerate}
                                            \item
                                            Every non-singular cubic form over the rational numbers in at
                                            least $10$ variables represents zero.

                                            \item
                                            There exist non-singular cubic forms over the rational numbers
                                            in $9$ variables that do not represent zero.
                                            \end{enumerate}
                                        \end{Satz}

                                        \begin{proof}
                                            For item $a)$ see \Heath, and for item $b)$ see \Mordell.
                                        \end{proof}
                                        By similar arguments as given in the previous section
                                        \ref{0202b} on pure spin-$2$ gravity, this implies:

                                        \begin{Cor}\label{0241}
                                            Let $X \to S$ be smooth of relative dimension $n$. If
                                            there is a spin-$3$ field then $n \geq 10$.
                                        \end{Cor}

\chapter{Discreteness of geometry and quantization of gauge fields}\label{0300}

Let $(X \to S,g)$ be a \SR (see Definition \ref{5024}). The
classical choice for the base is $S= \Spec \mathbb{C}$. In this
case, $X(S)$ may be considered as a manifold, and the points of
$X(S)$ form a continuum. However, as we represent the adelic point
of view, we make a different choice which also takes into
consideration the finite prime spots. Therefore, from now on, let
$K \subset \real$ be an algebraic number field with ring of
integral numbers ${\footnotesize{\text{$\cal O$}}}_K$, and let
$S=\Spec {\footnotesize{\text{$\cal O$}}}_K$. Recall that $X_K(K)$
denotes the archimedean component of $X(S)$ (with $X_K$ being the
generic fibre of $X \to S$).

Within this chapter, we will finally illustrate that $X(S)$ indeed
carries a  discrete geometry in the ``adelic'' case  $S=\Spec
{\footnotesize{\text{$\cal O$}}}_K$ (in contrast to the situation
over the real or complex numbers). Because of the canonical
bijection $X_K(K) \cong X(S)$, the same is true for the
archimedean component $X_K(K)$. More precisely, both $X_K(K)$ and
$X(S)$ are finitely generated abelian groups. If the rank of these
groups is zero, they are actually finite groups.

In particular, gauge fiels on $X_K(K)$ may be treated by means of
the methods of lattice gauge theory. This is particularly
interesting, because lattice gauge theory offers the possibility
to perform a non-perturbative, well-defined quantization of gauge
fields.
\section{The discrete geometry of the arithmetic models}\label{2000}

From now on let us consider a smooth, separated $K$-scheme $X_K$
of relative dimension $n$ over an algebraic number field $K$. Let
us assume that $X_K$ gives rise to a \SR $(X \to S,g)$ (see
Definition \ref{5024}). In this section, we will study the
archimedean component $X_K(K)$ of $X$. At first, we will
illustrate that $X_K(K)$ indeed carries a discrete structure. The
argument is as follows. Due to Theorem \ref{0123}, $X_K$ is an
extension of an Abelian variety $A_K$ by a torus $T_K$. Following
the ideas of chapter \ref{0080}, we interpret the torus as the
gauge group part and $A_K$ as the \st part of $X_K$. As we are
interested in the structure of \sto, we will therefore assume that
$X_K$ is an Abelian variety. Therefore, the Mordell-Weil theorem
\ref{2001a} tells us that the archimedean limit $X_K(K)$ of $X(S)$
is indeed finitely generated abelian group. Because of the
bijection $X_K(K) \cong X(S)$, the same is true for the ``adelic''
world $X(S)$.

\begin{Theorem}\label{2001a}
    Let $K$ be an algebraic number field and let $A_K$ be an
    algebraic variety over $K$. Then the set $A_K(K)$ of $K$-valued points
     of $A_K$ is a finitely generated abelian group.
\end{Theorem}
As usually we will let $\Ganz$ denote the additive group of
integers, and we will let $\Ganz_m$ denote the cyclic group
$\Ganz/m\Ganz$ of integers mod $m$. Then the fundamental theorem
of finitely generated abelian groups tells us that $X_K(K)$ looks
like
\begin{align*}
    X_K(K) \cong \Ganz^d \oplus \Ganz_{p_1^{\nu_1}} \oplus \cdots
    \oplus \Ganz_{p_s^{\nu_s}}
\end{align*}
for some prime numbers $p_i \in \Natural$ and integers $d,s,\nu_i
\in \Natural$. As we are interested in a physical theory which is
completely free of infinities, it is appealing to demand that
$d=0$, because in this special case, $X_K(K) \cong X(S)$ is
actually a finite group. Consequently, there are only finitely
many \st points.

In order to exploit some more physical properties of $X_K(K)$, let
us assume that $X:= E_0 \times_S \ldots \times_S E_n$ is the
fibred product of $n$ smooth elliptic curves $E_i$ over $S$
endowed with the first fundamental form $g$. We know from Theorem
\ref{6021} that $(X \to S,g)$ is a \SRo. In the case $d > 0$, the
set $X_K(K)$ is already dense in the continuum $X_K(\real)$. If
$d=0$, $X_K(K)$ is a group of finite order. In the special case
$K=\rat$, the Nagell-Lutz theorem tells us that the coordinates of
the points of $X_{\rat}(\rat)$ are actually integers. Therefore
the generator $\ell:=1 \in {\footnotesize{\text{$\cal O$}}}_K$ of
the group ${\footnotesize{\text{$\cal O$}}}_K=\Ganz$ appears as
smallest possible length (Planck length) as one should expect.

Next to the existence of a discrete geometry, $(X \to S,g)$
sometimes has the interesting property to consists of only
finitely many points. We already met this phenomenon in the case
that $X_K$ is the product of elliptic curves. Therefore, let us
next assume that $X_K$ decomposes into a fibred product
\begin{align*}
    X_K = C_K^{(1)} \times_K \ldots \times_K C_K^{(n)}
\end{align*}
of one-dimensional projective, smooth curves $C_K^{(i)}$, $i=1,
\ldots, n$. For example, in the case $n=4$, think of $C_K^{(1)}$
as time coordinate and think of $C_K^{(2)}, C_K^{(3)}$ and
$C_K^{(4)}$ as space coordinates. First notice that the genus of
the curves cannot be smaller than 1. This may be seen as follows:
Assume that the genus of a curve $C_K^{(i)}$ is smaller than one
for some $i$. However, $C_K^{(i)}$ is present on the level of
archimedean \st if and only if it contains a non-virtual point,
i.e. if and only if $C_K^{(i)}(K) \neq \emptyset$. Then
Proposition \ref{2003} implies that $C_K^{(i)}$ is already the
one-dimensional projective space $\Proj_K^1$ over $K$.
Thereby, we also made use of the fact that the arithmetic and the
geometric genus of the curve $C_K^{(i)}$ agree:

\begin{Rem}\label{2004}
    Let $C$ be a smooth, geometrically connected, projective
    curve over a field $k$. Let us remind the reader that
    \begin{align*}
        p_a(C):= \dim_k H^1 \left( C, {\cal O}_C \right)
    \end{align*}
    is called the \emph{arithmetic} genus, and that
    \begin{align*}
        p_g(C):= \dim_k H^0 \left( C, {\Omega}_{C/k}^1 \right)
    \end{align*}
    is called the \emph{geometric} genus of $C$. Then one can
    prove that $p_a(C) = p_g(C)$.
\end{Rem}

\begin{proof}
    \Liu, Rem. 7.3.28
\end{proof}
\begin{Satz}\label{2003}
            Let $C$ be a geometrically integral projective curve
            over a field $k$ of arithmetic genus $p_a\leq 0$. Then
            we have the following properties:
            \begin{enumerate}
                \item
                The curve $C$ is a smooth conic over $k$, i.e. $C$
                is smooth over $k$ and there exists a homogeneous
                polynomial of degree $2$ such that $C=V_+(F) \subset
                \Proj_k^2$.
                \item
                We have $C \cong \Proj_k^1$ if and only if $C(k) \neq
                \emptyset$.
            \end{enumerate}
\end{Satz}

\begin{proof}
    \Liu, Prop. 7.4.1
\end{proof}
But the projective spaces $\Proj_K^r$, $r \in \Natural$, cannot be
the archimedean component of a \GRo, because not every
$K$-automorphism of $\Proj_K^r$ extends to an $S$-automorphism of
$\Proj_S^r$. For instance consider the following counterexample:

\begin{Ex}\label{2005}
    Let $R$ be a discrete valuation ring with field of fractions
    $K$. Consider a $K$-automorphism $u_K: \mathbb{P}_K^r \stackrel{\sim}
    \rightarrow \mathbb{P}_K^r$. Using a set of homogeneous coordinates $x_0,
    \ldots, x_r$ of  $\mathbb{P}_K^r$, we can describe $u_K$ by
    \begin{align*}
        x_i \mapsto  \sum_{j=0}^r a_{ij} x_j, \qquad i=0, \ldots, r,
    \end{align*}
    where $A:=(a_{ij})$ is a matrix in Gl$_{r+1}(K)$. We may assume
    that all coefficients $a_{ij}$ belong to $R$. Then, by the theory
    of elementary divisors, there are matrices $S,T \in
    \text{Gl}_{r+1}(K)$ and integers $0 \leq n_0 \leq \ldots \leq n_r$
    such that $S A T$ is the diagonal matrix $\text{diag}( \pi^{n_0},
    \ldots, \pi^{n_r} )$. Hence there exist sets of homogenous
    coordinates $x_0, \ldots, x_r$ and $x'_0, \ldots, x'_r$ of
    $\mathbb{P}_R^r$, such that $u_K$ is described by
    \begin{align*}
        x_i \mapsto \pi^{n_i} x'_i,
    \end{align*}
    where we may assume $n_0=0$.

    If $n_0= \ldots = n_r =0$, it is clear that  $u_K: \mathbb{P}_K^r
    \stackrel{\sim} \to \mathbb{P}_K^r$ extends to an automorphism $u:
    \mathbb{P}_R^r \stackrel{\sim}\rightarrow \mathbb{P}_R^r$. However, if
    $n_0= \ldots = n_s =0$ and $n_{s+1}, \ldots, n_r > 0$ for some
    $s<r$, then $u_K$ extends only to an $R$-rational map  $u:
    \mathbb{P}_R^r \stackrel{\sim} \dashrightarrow \mathbb{P}_R^r$.
    Namely, $u$ is defined on the $R$-dense open subscheme $V  \subset
    \mathbb{P}_R^r$ which consists of the generic fibre and of the
    open part $V_k \subset \mathbb{P}_k^r$ complementary to the linear
    subspace $Q_k$ where $x_0, \ldots, x_s$ vanish. In fact, if $Q'_k$
    is the linear subspace in $\mathbb{P}_k^r$ where $x'_{s+1},
    \ldots, x'_r$ vanish, we can view $u_k$ as projection of
    $\mathbb{P}_k^r$ to $Q'_k$ with center $Q_k$.
\end{Ex}
Therefore the projective spaces $\Proj_K^r$, $r \in \Natural$,
cannot occur. This reflects the fact that \GRs cannot be flat
spaces with trivial topology, and may be regarded as projective
version of the fact that e.g. N\'eron models cannot contain affine
spaces (Proposition \ref{0110}).

Therefore, the genus of the curves $C_K^{(i)}$ is bigger than
zero. This is the case of a non-trivial vacuum topology. In the
case of genus bigger than one, we may finally evoke Falting's
theorem:

\begin{Faltings}\label{2004}
    Any non-singular projective curve of genus $g>1$ defined over
    an algebraic number field $K$ contains only finitely many $K$-valued
    points.
\end{Faltings}
All in all, in the case of curves $C_K^{(i)}$ of genus bigger than
one, we see that
\begin{align*}
    X_K(K) = C_K^{(1)}(K) \times \ldots \times C_K^{(n)}(K)
\end{align*}
is a finite set. Only in the case of genus one, it is still
possible that $X_K(K)$ contains infinitely many points. However,
we have furthermore proved the following result.

\begin{Vac}\label{5600}
    Let us assume that the archimedean limit $X_K$ decomposes into a product
    \begin{align*}
        X_K = C_K^{(1)} \times_K \ldots \times_K C_K^{(n)}
    \end{align*}
    of one-dimensional projective, smooth curves $C_K^{(i)}$, $i=1,
    \ldots, n$. Then the genus of the curves is strictly bigger than zero,
    i.e. $X_K$ carries a non-trivial topology. Due to
    Corollary \ref{6024}, the curvature tensor of $X_K$ vanishes.
    Therefore, $X_K$ is a vacuum which carries a non-trivial topology.
\end{Vac}
The next step is the incorporation of gauge fields.
\section{Lattice gauge theory in the arithmetic setting}\label{2100ga}

In the regime of high energy physics, gravity may be neglected.
Therefore, one usually chooses Minkowski space $\real^n$ as \st
manifold and considers gauge fields upon $\real^n$. In a next step
one has to quantize these gauge field. Essentially, there are two
different approaches. First, there is a perturbative approach
based on Feynman diagrams. Second, there is the non-perturbative
approach of lattice gauge theory. The first one is manifestly
Lorentz-invariant, but in order to quantize the gauge fields (e.g.
by means of the Fadeev-Popov method) one has to make a gauge
fixing. This breaks the gauge invariance and one has to make sure
that physical entities are independent from the chosen gauge. In
contrast to this, lattice gauge theory conserves gauge invariance.
But one replaces the continuum of Minkowski space by a lattice,
and this way Lorentz invariance is broken.

However, Minkowski \st is not the only solution of Einstein´s
field equations of general relativity which describes an empty,
static universe. Every direct product of one dimensional manifolds
is as well a vacuum solution, because the curvature tensor
vanishes globally. Nevertheless the different vacua will in
general differ with respect to their topological structure. While
Minkowski \st is the product of curves of genus zero (and
therefore of trivial vacuum structure), it is as well admissible
to consider a product of curves of higher genus. In the case of
genus one, we obtain a product of elliptic curves.

In this section we will illustrate that a \st manifold consisting
of a product of elliptic curves (or more generally an Abelian
variety) possesses a canonical ``lattice'' which is even invariant
under algebraic diffeomorphisms (see Definition \ref{2100gb}, b)).
Therefore, in this case, lattice gauge theory offers a possibility
to quantize gauge field in such a way that neither
gauge invariance nor \st symmetries are broken. \\ \\
From now on, we will provide as well the flat space $\real^n$ as
the affine space $\Affin_K^n$ with the euclidean metric diag$(1,
\ldots, 1)$, i.e. instead of working with the hyperbolic Minkowski
\st we will consider an euclidean \st. This is usually done in
lattice gauge theory. Therefore, from now on, all manifolds will
be considered as Riemannian manifolds.
\begin{Def}\label{2100gb}
    \begin{enumerate}
        \item
        Let $M$ be a differentiable manifold and $i: \Gamma
        \hookrightarrow M$ a subset ($i$ denotes the canonical inclusion).
        Let $D$ be a set of diffeomorphisms of $M$. Then
        $\Gamma$ is called \emph{invariant under D} if the following
        holds: For every $\varphi \in D$ there is a bijection
        ${\varphi}|_{\Gamma}: \Gamma \stackrel{\sim} \to \Gamma$ of sets such that
        the  following diagram is commutative.
        \begin{align*}
            \xymatrix{ \Gamma \ar@{^(->}[d]^i   \ar[r]^{{\varphi}|_{\Gamma}} & \Gamma \ar@{^(->}[d]^i    \\
            M \ar[r]^{\varphi} & M}
        \end{align*}

        \item
        Let $K \subset \real$ be an algebraic number field and let
        $X_K$ be a smooth $K$-scheme. Let  $i: \Gamma  \hookrightarrow
        X_K(\real)$ be a subset ($i$ denotes the canonical inclusion).
        Then  $\Gamma$ is called \emph{$K$-isomorphism invariant}
        if the following holds: For every $K$-isomorphism $\varphi: X_K \to
        X_K$ there is a bijection ${\varphi}(\real)|_{\Gamma}: \Gamma \stackrel{\sim} \to
        \Gamma$ of sets such that the following  diagram is
        commutative.
        \begin{align*}
            \xymatrix{ \Gamma \ar@{^(->}[d]^i   \ar[r]^{{\varphi}(\real)|_{\Gamma}} & \Gamma \ar@{^(->}[d]^i    \\
            X_K(\real) \ar[r]^{\varphi(\real)} & X_K(\real)}
        \end{align*}
        Thereby,  $\varphi(\real): X_K(\real) \to X_K(\real)$ denotes the
        morphism which is obtained by evaluation of $\varphi$ at
        $\real$-valued points.
    \end{enumerate}
\end{Def}
Due to the smoothness of $X_K$ over $K$, we may endow $X_K(\real)$
with the structure of a differentiable manifold. Thus, if $\Gamma$
is a  $K$-isomorphism invariant subset, then it is invariant under
the set of diffeomorphisms  $\varphi(\real): X_K(\real) \to
X_K(\real)$ of $X_K(\real)$ which are induced by isomorphisms
$\varphi:X_K \to X_K$ of schemes.

In the special case of  the flat \st  $\real^n$ we may consider
the set of Lorentz-transformations. A Lorentz-transformation is
linear diffeomorphism which describes  a rotation in $\real^n$.
Each Lorentz-tranformation may be written as a finite product of
matrices of the following type.
\begin{align*}
    L_{\theta}:=\left(%
    \begin{array}{cccccccc}
    \ddots &  &  &  &  &  &  & \\
     & 1 & 0 & 0 & 0 & 0 & 0 & \\
     & 0 & 1 & 0 & 0 & 0 & 0 & \\
     & 0 & 0 & \cos(\theta) & \sin(\theta) & 0 & 0 & \\
     & 0 & 0 & - \sin(\theta) & \cos(\theta) & 0 & 0 & \\
     & 0 & 0 & 0 & 0 & 1 & 0 &  \\
     & 0 & 0 & 0 & 0 & 0 & 1 &  \\
     &  &  &  &  &  &  & \ddots \\
    \end{array}%
    \right)
\end{align*}
Let $D_{Lor}$ denote the set of all Lorentz-transformations. A
subset $\Gamma \subset \real^n$ is called \emph{Lorentz invariant}
if it is invariant under $D_{Lor}$ in the sense of Definition
\ref{2100gb}, a). However, whereas the notion of Lorentz
invariance is especially adapted to $\real^n$, the notion of
$K$-isomorphism invariance applies to every scheme $X_K$ over $K$.
But in the special case of $\real^n$ one can show that
$K$-isomorphism invariance already implies Lorentz-invariance for
a quite big class of sets $\Gamma$ (see Proposition \ref{2100gc}).
Therefore, in the setting of non-trivial manifolds $X_K(\real)
\neq \real^n$, where the notion of Lorentz-invariance does in
general not make sense, it is natural to look for $K$-isomorphism
invariant structures.

\begin{Satz}\label{2100gc}
    Let $X_K=\Affin_K^n$ and consider $X_K(\real) \cong \real^n$ as
    flat manifold in a canonical way. Let $\Gamma \subset X_K(\real) $
    be a closed or an open subset. If $\Gamma$ is $K$-isomorphism invariant, it is
    Lorentz-invariant, too.
\end{Satz}

\begin{proof}
    Let us first assume that $\Gamma \subset X_K(\real)$ is a closed
    subset. Choose $x \in \Gamma$. As every Lorentz transformation may be written
    as a finite product of matrices $L_{\theta}$, it suffices to show,
    that $L_{\theta}(x) \in \Gamma$ for all the Lorentz transformations
    of type $L_{\theta}$. Thus it remains to show that
    $\inf_{\gamma \in \Gamma}|L_{\theta}(x) - \gamma|=0$,
    because $\Gamma \subset X_K(\real) $ is a
    closed subset. Thereby, $| \cdot |$ denotes the canonical norm
    on $\real^n$.

    In order to prove this consider the  sphere of radius one
    $S^1 \subset \real^2$. For
    every $z \in S^1$ there is a $\theta \in [0,2\pi]$ such that
    $z=(\cos(\theta),\sin(\theta)) \in \real^2$.
    By Lemma \ref{2100gd}, we know that the set of $K$-valued points of
    $S^1$ is dense in $S^1$ (with respect to the canonical topology on
    $S^1$). Consequently,  we may choose a point
    $z_{\varepsilon}=(\cos(\theta_{\varepsilon}),\sin(\theta_{\varepsilon})) \in K^2$
    such that $|\cos(\theta)-\cos(\theta_{\varepsilon})|< \varepsilon$ and
    $|\sin(\theta)-\sin(\theta_{\varepsilon})|< \varepsilon$ for every
    ${\varepsilon}>0$. In particular,
    \begin{align*}
        |L_{\theta}(x) - L_{\theta_{\varepsilon}}(x)| \leq \sqrt{2}
        \varepsilon |x|
        \qquad \text{for all ${\varepsilon}>0$.}
    \end{align*}
    But each Lorentz transformations
    $L_{\theta_{\varepsilon}}$ is induced by a $K$-isomorphisms of
    $X_K$, because the matrix $L_{\theta_{\varepsilon}}$ has only
    entries in $K$. Thus $L_{\theta_{\varepsilon}}(x) \in \Gamma$
    due to our assumption, and it follows that
    $\inf_{\gamma \in \Gamma}|L_{\theta}(x) - \gamma|=0$.

    Let us now assume that $\Gamma \subset X_K(K)$ is open. Then the complement $\Gamma^c:= X_K(K) \backslash
    \Gamma$ of $\Gamma$ in $X_K(K)$ is a closed subset. Let us first show that $\Gamma^c$ is
    $K$-isomorphism invariant. In order to do this, assume that
    there is a $K$-isomorphism $\varphi:X_K \to X_K$ and a $y \in
    \Gamma^c$ such that $x:=\varphi(y) \in \Gamma$. But then $\varphi^{-1}(x)=y \notin \Gamma$
    which contradicts the $K$-isomorphism invariance of $\Gamma$.
    Consequently, $\Gamma^c$ is even Lorentz invariant by what we have
    already shown.
    Let us now prove the Lorentz-invariance of $\Gamma$. For this purpose choose a point $x \in
    \Gamma$ and assume that $y:=L_{\theta}(x) \notin \Gamma$ for some
    Lorentz transformation $L_{\theta}$. Therefore, the Lorentz
    transformation $L_{-\theta}=L_{\theta}^{-1}$ maps $y \in
    \Gamma^c$ to $x \in \Gamma$. This contradicts the
    Lorentz-invariance of $\Gamma^c$. Consequently, $\Gamma$ is
    Lorentz invariant.
\end{proof}

\begin{Lemma}\label{2100gd}
    Let $K \subset \real$ be an algebraic number field. Let $S^1:=\{(x,y) \in \real^2|x^2+y^2=1\} \subset
    \real^2$ be the sphere of radius one endowed with its canonical topology.
    Let $S^1(K) := \{(x,y) \in K^2|x^2+y^2=1\} \subset S^1$ be
    the set of $K$-valued points of $S^1$. Then $S^1(K)$ is a dense
    subset of $S^1$.
\end{Lemma}

\begin{proof}
    First notice that for any $ t \in K$
    \begin{align*}
        z_t:=\left( \frac{1-t^2}{1+t^2}, \frac{2t}{1+t^2}
        \right) \in S^1(K).
    \end{align*}
    Identifying $\real^2$ with the complex numbers $\complex$ in
    a canonical way, we may write $z_t= e^{i \varphi_t}$
    for some angle $\varphi_t$. If we let $t \to 0$, we see that
    there are points $z_t \in S^1(K)$ with arbitrary small, but non-vanishing angle
    $\varphi_t \neq 0$.

    Let now $z= e^{i \varphi} \in S^1$ and $\varepsilon > 0$.
    By what we have seen, there is an $m \in \Natural$ and a point
    $z_t= e^{i \varphi_t} \in S^1(K)$ such that $|\varphi- m
    \varphi_t| \leq \varepsilon$ (simply choose $\varphi_t$ sufficiently small). But $e^{i m \varphi_t} \in
    S^1(K)$, because $e^{i  m \varphi_t} = \left(e^{i
    \varphi_t}\right)^m$. Therefore we are done because $\varepsilon >
    0$ is arbitrary.
\end{proof}
However, in our ``adelic'' approach to physics, \st is given by
the set $X_K(K)$ of $K$-valued points of a $K$-scheme $X_K$. As
motivated above, one should expect for physical reasons that
$X_K(K)$ is $K$-isomorphism invariant. This is indeed always true:

\begin{Satz}\label{2100ge}
    Let $K \subset \real$ be an algebraic number field, and
    let $X_K$ be a $K$-scheme. Then $X_K(K) \subset X_K(\real)$ is
    $K$-isomorphism invariant (see Definition \ref{2100gb}, b)).
\end{Satz}

\begin{proof}
    Let $\Ccat$ be the category of $K$-schemes. Each object
    $X\in \Ccat$ gives rise to its functor of points
    $
        h_X:\Ccat \to \left( Sets \right)
    $
    which associates to any $T\in \Ccat$ the set
    $
        h_X(T):=X(T):={\Hom} (T,X)
    $
    of $T$-valued points of $X$. Each morphism $X\to X'$ in $\Ccat$
    induces a morphism $h_X\to h_{X'}$ of functors by the composition
    of morphisms in $\Ccat$. In this way one gets a covariant functor
    \begin{align*}
        f: \Ccat \to {\Hom} \left( \Ccat^0, {\left( Sets \right)} \right)
    \end{align*} of $\Ccat$ to the category of covariant functors from
    $\Ccat^0$ (the dual of $\Ccat$) to the category of sets which
    (by Proposition \ref{1201}) defines $\Ccat$ as full subcategory of
    $\Hom(\Ccat^0$,(Sets)).

    In particular, each $K$-isomorphism $\varphi:X_K \to X_K$
    gives rise to a natural transformation $f(\varphi):h_{X_K} \to
    h_{X_K}$. The latter means that for all $K$-morphisms $T \to T'$ there is a commutative diagram
    \begin{align*}
        \xymatrix{ X_K(T') \ar[d]  \ar[r]^{\sim} & X_K(T') \ar[d]    \\
        X_K(T) \ar[r]^{\sim} & X_K(T)}
    \end{align*}
    whereby the horizontal maps are isomorphisms. If we
    consider the special case of the canonical $K$-morphism
    $\Spec \real \to \Spec K$ corresponding to the inclusion $K \subset
    \real$, we obtain the commutative diagram
    \begin{align*}
        \xymatrix{ X_K(K) \ar@{^(->}[d]^i   \ar[r]^{\sim} & X_K(K) \ar@{^(->}[d]^i  .  \\
        X_K(\real) \ar[r]^{\sim} & X_K(\real)}
    \end{align*}
    Thereby $i: X_K(K) \hookrightarrow X_K(\real)$ denotes the
    canonical inclusion. This means exactly that $X_K(K)$ is $K$-isomorphism
    invariant.
\end{proof}

\begin{Rem}\label{2100gf}
    \begin{enumerate}
        \item
        Consider the affine space $X_K=\Affin_K^n$. In quantum field theory,
        one usually chooses $\Affin_K^n(\real)^n \cong \real^n$ as underlying
        \st manifold. As already mentioned the lattices $\Gamma
        \cong \Ganz^n \subset \real^n$ are obviously not
        Lorentz-invariant. In particular, we know from Proposition \ref{2100gc} that they are not
        $K$-isomorphism invariant. Conversely, the canonical $K$-isomorphism invariant
        set $\Affin_K^n(K)$ is neither a finitely generated abelian group nor a lattice, because
        $\Affin_K^n(K)=K^n$ is a dense subset of $\real^n$.

        \item
        Let us now consider a product $X_K= E_K^{(1)} \times_K \ldots \times_K E_K^{(n)}$ of elliptic curves
        over $K$. Like Minkowski \st, also
        $X_K(\real)= E_K^{(1)}(\real) \times \ldots \times
        E_K^{(n)}(\real)$ is a solution of Einstein's equations
        which describes a vacuum without gravity. But this
        time we know from Theorem \ref{2001a} that $X_K(K)$ is a
        finitely generated abelian group. If the rank of this
        group is zero, $X_K(K)$ is a discrete set. Furthermore,
        the set $X_K(K)$ is $K$-isomorphism invariant.

        All in all, it makes sense to interpret the set $X_K(K)$ as
        vacuum (as it is suggested from the adelic point of view anyway).
        Gauge fields may be quantized in a non-perturbative way by
        means of lattice gauge theory.

        \item
        If there is an Abelian variety $A_K$ over $K$, which is
        solution of the Einstein equations, the statements of b)
        remain essentially true. The only difference is that
        gravity need not be trivial, i.e., in general $A_K(K)$ is not a vacuum.
    \end{enumerate}
\end{Rem}
Let us from now on consider the Abelian variety $X_K:= E_K^{(1)}
\times_K \ldots \times_K E_K^{(n)}$ defined as  a product of
smooth elliptic curves over $K$. Let us consider the
$K$-isomorphism invariant set $X_K(K) \subset X_K(\real)$ as
vacuum. If $X_K(K)$ is a group of rank zero, it is a discrete set.
In the continuum, gauge fields are solutions of the continuum
Yang-Mills equations. In the lattice formulation, the respective
expressions are obtained from the continuum Yang-Mills theory by
replacing the infinitesimal differential operators by finite
difference operators. There is a canonical way to introduce
difference operators: Each elliptic curve $E_K^{(i)}$ induces a
one dimensional Riemannian manifold $M^i \cong E_K^{(i)}(\real)$
(if we forget the point at infinity). For each manifold $M^i$ we
may choose an embedding $\gamma^i:\real \to \real^2$. The
corresponding first fundamental form on $M^i$ induces a
Levi-Civita connection, and we may perform parallel transports of
tangent vectors along $\gamma^i$. If $M^i_{\gamma^i(t)}$ denotes
the tangent space of $M^i$ at $\gamma^i(t)$, the parallel
transport from $M^i_{\gamma^i(t)}$ to $M^i_{\gamma^i(s)}$ is
usually denoted by $\tau_{s,t}^i$. Then the covariant derivation
in the continuum may be written as
\begin{align*}
    \nabla_{(\gamma^i)'(t)}\vecv= \stackrellow{ h \to 0}\lim \frac{1}{h} \Big(
    \tau_{t,t+h}^i\vecv(t+h)-\vecv(t) \Big).
\end{align*}
The set $E_K^{(i)}(K)$ is a finite, discrete subgroup of the
continuum $E_K^{(i)}(\real)$. Consequently, there is a finite set
$\{t_1, \ldots, t_m\}$ of real numbers $t_j < t_{j+1} \in \real$
whose image under $\gamma^i$ is $E_K^{(i)}(K)$. At the $K$-valued
point $\gamma^i(t_j) \in E_K^{(i)}(K)$, we may therefore introduce
the following discrete notions.

\begin{enumerate}
    \item
    $(\nabla_i f)(t_j):= \frac{1}{t_{j+1}-t_{j}} \Big( f(t_{j+1})-f(t_{j})
    \Big)$ \qquad \quad \quad for all functions $f$.

    \item
    $(\overline{\nabla}_i f)(t_j):= \frac{1}{t_{j}-t_{j-1}} \Big( f(t_{j})-f(t_{j-1})
    \Big)$ \qquad \quad \quad for all functions $f$.

    \item
    $(\nabla_i \vecv)(t_j):= \frac{1}{t_{j+1}-t_{j}} \Big( \tau_{t_j,t_{j+1}}^i
    \vecv(t_{j+1})-\vecv(t_{j}) \Big)$ $\quad \! \: $ for all vector-fields
    $\vecv$.

    \item
    $(\overline{\nabla}_i \vecv)(t_j):= \frac{1}{t_{j}-t_{j-1}} \Big(
    \vecv(t_{j})-\tau_{t_j,t_{j-1}}^i\vecv(t_{j-1}) \Big)$ $\quad \! \: $ for all vector-fields
    $\vecv$.

    \item
    $\int f := \sum_j (t_{j+1}-t_{j}) f(t_j)$
\end{enumerate}
On the set $X_K(K)=E_K^{(1)}(K) \times \ldots \times
E_K^{(n)}(K)$, $\nabla_i$ resp. $\nabla_i$ is the discrete
analogue of the partial derivative in the $i$-th direction.

These definitions naturally extend the classical notions of
discrete calculus in Minkowski space, and are the starting point
of a lattice gauge theory for Abelian varieties. However, it is
beyond the scope of this thesis to work out this lattice gauge
theory explicitly. Instead, in a first step, we will treat
Yang-Mills theory over arbitrary base schemes. This in done in
Part II of this thesis.

                                        \chapter{Appendix I}\label{5300}
                                        \section{N\'eron models}\label{0052}

                                        We already mentioned in section \ref{baaa} that a \GR $(X\to S,
                                        g)$ is almost a N\'eron model. A \SR (see Definition \ref{5024}) actually is a
                                        N\'eron model. These N\'eron models are universal objects which we
                                        will introduce in this section.
                                        \begin{Def}\label{0066}
                                            Let $X_K$ be a smooth and separated $K$-scheme of finite type.
                                            A \emph{N\'eron model} of $X_K$ is an
                                            $S$-model $X$ which is smooth, separated and of finite type,
                                            and which satisfies the following universal property, the so called N\'eron mapping
                                            property:

                                            For each smooth $S$-scheme $Y$ and each $K$-morphism $u_{K}:Y_K \to
                                            X_K$ there is an unique $S$-morphism $u:Y \to X$ extending
                                            $u_K$.
                                        \end{Def}
                                        It is highest non-trivial to prove the existence of the N\'eron
                                        model, and in general it will not exist without further
                                        assumptions. At least, due to its universal property, it is clear
                                        that the N\'eron model is unique up to canonical isomorphism.  Due
                                        to its uniqueness and its universal property, the formation of
                                        N\'eron models is a functor.
                                        \begin{Rem}\label{0068}
                                            If the archimedean limit $X_K$ admits a N\'eron model $X$,
                                            and if $X_K$ in addition carries a group structure (e.g. if $X \to S$
                                            is a \SRo), this group structure extends to $X$ due to the N\'eron mapping property.
                                            Interpreting the formation of the inverse with respect to the group structure
                                            as a simultaneous parity and time-reversal operation, we see
                                            that there is a canonical notion of parity and time-reversal on $X$, too.
                                            Thus, parity and time-reversal lift to the ``adelic'' level.
                                        \end{Rem}
                                        We finish this section with the statement of a further fundamental
                                        property of N\'eron models whose significance will become clear in
                                        the next chapter, too. It is the so called \emph{extension
                                        property for \'etale points}. Physically, this property guarantees
                                        that the archimedean zero-dimensional \st points appear as
                                        projections of the ``adelic'' \st points to their archimedean
                                        component.

                                        \begin{Def}\label{0069}
                                            Let $X$ be a scheme over a Dedekind scheme $S$. Then we say
                                            that $X$ satisfies the extension property for \'etale points at
                                            a closed point $s \in S$ if, for all \'etale local ${\cal
                                            O}_{S,s}$-algebra $R'$ with field of fractions $K'$, the
                                            canonical map $X(R') \to X_K(K')$ is surjective.
                                        \end{Def}

                                        \begin{Satz}\label{0070}
                                            Let $S$ be a Dedekind scheme with field of fractions $K$, and
                                            let $X_K$ be a smooth and separated $K$-scheme of finite type
                                            which admits a N\'eron model $X$. Furthermore, let $S'$ be a
                                            second Dedekind scheme with field of fractions $K'$ such that
                                            $S' \to S$ is \'etale.  Then there is a bijection
                                            \begin{align*}
                                                X(S') \cong X_K(K').
                                            \end{align*}
                                        \end{Satz}

                                        \begin{proof}
                                            Due to the N\'eron mapping property, each morphism $\Spec K' \to X_K$ extends to a unique morphism
                                            $S' \to X$.
                                        \end{proof}
                                        Important is the fact that N\'eron models are local on the base
                                        which is stated by means of  the following proposition.

                                        \begin{Satz}\label{0098}
                                            Let $S$ be a Dedekind scheme and let $X$ be an $S$-scheme of
                                            finite type. Then the following assertions are equivalent:
                                            \begin{enumerate}
                                                \item
                                                $X$ is a N\'eron model of its generic fibre.
                                                \item
                                                For each closed point $s \in S$, $X \times_S \Spec {\cal
                                                O}_{S,s}$ is a N\'eron model of its generic fibre.
                                            \end{enumerate}
                                        \end{Satz}

                                        \begin{proof}
                                            \BLR, Prop. 1.2/4
                                        \end{proof}

                                        \begin{Satz}\label{0099}
                                            Let $X$ be a smooth $R$-group scheme of finite type or a
                                            torsor under a smooth $R$-group scheme of finite type. Then
                                            the following conditions are equivalent:
                                            \begin{enumerate}
                                                \item
                                                $X$ is a N\'eron model of its generic fibre $X_K$.
                                                \item
                                                $X$ is separated and the canonical map $X(R^{\text{sh}}) \to X_K(K^{\text{sh}})$ is
                                                surjective.
                                                \item
                                                The canonical map $X(R^{\text{sh}}) \to X_K(K^{\text{sh}})$ is bijective.
                                            \end{enumerate}
                                        \end{Satz}

                                        \begin{proof}
                                            \BLR, Thm. 7.1/1
                                        \end{proof}
                                        The surjectivity of the canonical map $X(R^{\text{sh}}) \to
                                        X_K(K^{\text{sh}})$ is due to the definition of the strict
                                        henselisation equivalent to the extension property for \'etale
                                        points in the case of a local base scheme $\Spec R$. Furthermore,
                                        due to the valuative criterion of separatedness, the surjection is
                                        even a bijection as stated in Proposition \ref{0099} $c)$.

                                        \section{Further statements on the archimedean component}\label{0051}
                                        Let us consider a \GR $(X \to S, g)$ in the sense of Definition \ref{5022}.
                                        Furthermore assume that $K \subset \real$ is an algebraic number
                                        field. The purpose of this subsection is to illustrate in more
                                        detail how the archimedean continuum limit $\mathfrak{X}_{\infty}$
                                        (see section \ref{5030}) emerges. Let us for simplicity assume
                                        that $X \to S$ is representable by a smooth $S$-scheme of relative
                                        dimension $n$. Consider the archimedean component of $X$, i.e. the
                                        generic fibre $X_K$ of $X$. Due to smoothness, there are local
                                        $\Spec K$-immersion $X_{K} \hookrightarrow \Affin_{K}^n$ (see
                                        Definition \ref{0021}). Let us for simplicity assume that this local immersion is
                                        already a global, closed immersion. Consequently, $X_{K}= V(I)=
                                        \Spec K[T_1, \ldots, T_n]/I$ is the zero set of an ideal $I$.
                                        Recall that for any field $L \supset K$, $X_K(L):=
                                        $$\text{Hom}_{\Spec K}(\Spec L, X_K) \cong $ $\text{Hom}_{K}(K[T_1,
                                        \ldots, T_n]/I,L) $. Thus, due to the universal property of
                                        quotients, we see that
                                        \begin{equation*}
                                            X_{K}(L)= \left\{ (x_1, \ldots ,x_n)\in L^n
                                            \mid p(x_1, \ldots ,x_n)=0 \ \text{for all} \ p \in I
                                            \right\}.
                                        \end{equation*}
                                        Choosing $L=\real$, we obtain a set which may be endowed with the
                                        structure of a differentiable manifold. This is the archimedean
                                        continuum limit $\mathfrak{X}_{\infty}:=X_K(\real)$ of $X$.
                                        However, a point $x_{\infty} \in  \mathfrak{X}_{\infty}$ is in
                                        general not induced by a point of the underlying scheme $X_K$. As
                                        we will illustrate below, only those points $x_{\infty} \in
                                        \mathfrak{X}_{\infty}$ origin from prime ideals of $X_K$, whose
                                        coordinates take values in an algebraic closure $\overline{K}$ of
                                        $K$. These point are elements of the set $X_{K}(K^{\text{sep}})=
                                        X_{K}(\overline{K}) $, where $K^{\text{sep}}$ is a separable
                                        algebraic closure of $K$. Thereby, we may write $K^{\text{sep}}=
                                        \overline{K}$ because $ \text{char}K =0$.

                                        Let $x=(x_1, \ldots ,x_n) \in X_{K}(K^{\text{sep}})$. Then we can
                                        associate to it the maximal ideal $\maxi_x:=(T_1 - x_1, \ldots ,
                                        T_n -x_n) \subset K^{\text{sep}}[T_1, \ldots, T_n]$. Using Taylor
                                        expansion around $x$ we see that $I \cdot K^{\text{sep}}[T_1,
                                        \ldots, T_n]  \subset \maxi_x$, and therefore $\maxi_x$
                                        corresponds to a closed point of the scheme $X_{K}$. Let us denote
                                        this point by $x$, too. More precisely any element of
                                        $X_{K}(K^{\text{sep}})$ is uniquely determined by the data
                                        consisting of a closed point $x \in X_{K}$ and a $K$-algebra
                                        homomorphism $K \hookrightarrow K^{\text{sep}}$. Alternatively,
                                        one can prove the following result.

                                        \begin{Satz}\label{0092}
                                            Let $X_K$ be a scheme over a field $K$. Then there is a
                                            bijection
                                            \begin{align*}
                                                X_K(K) & \cong \{x \in X_K \mid k(x)=K\}
                                            \end{align*}
                                        \end{Satz}

                                        \begin{proof}
                                            Let $\sigma \in X_K(K)$ and let $x$ be the image of the point
                                            of $\Spec K$. The homomorphism $\sigma_x^*$ induces a field
                                            homomorphism $k(x)\to K$. As $k(x)$ is $K$-algebra, this
                                            implies that $k(x)=K$.
                                            Let conversely be $x\in X_K$ with $k(x)=K$. Composing the
                                            closed immersion $\Spec k(x) \to \Spec {\cal O}_{X_K,x}$ with
                                            the canonical morphisms $\Spec {\cal O}_{X_K,x} \to X_K$, one
                                            obtains the desired section $\Spec K \to X_K$ whose image is
                                            $x$.
                                        \end{proof}
                                        However, due to the algebraic nature of the underlying scheme, the
                                        set $X_{K}(K^{\text{sep}} \cap \real)$ is already a dense subset
                                        of the manifold $\mathfrak{X}_{\infty}$ with respect to the
                                        canonical topology on $\mathfrak{X}_{\infty}$. In particular,
                                        $\mathfrak{X}_{\infty}$ may be reconstructed from
                                        $X_{K}(K^{\text{sep}})$. On the other hand, we saw that every
                                        point of $X_{K}(K^{\text{sep}})$ $ \subset M$ gives rise to a
                                        maximal ideal, i.e. \emph{closed} point of $X_{K}$. But these
                                        points also lie dense in $X_K$ with respect to Zariski topology:
                                        \begin{Satz}\label{0058}
                                            If $X$ is a smooth scheme over a field $k$, the set of closed
                                            points $x$ of $X$ such that $k(x)$ is a separable extension of
                                            $k$ is dense in $X$.
                                        \end{Satz}

                                        \begin{proof}
                                            \BLR, Cor. 2.2/13
                                        \end{proof}
                                        This motivates the following notion.
                                        \begin{Def}\label{0057}
                                            Let $f:X \to S$ be a smooth scheme. Let $x \in X$ and $s:=f(x)$. Then $x$ is called
                                            a \emph{physical point}, if:
                                            \begin{enumerate}
                                                \item
                                                $x$ is closed in the fibre $X_s=X \times_S \Spec k(s)$.
                                                \item
                                                $k(s) \hookrightarrow k(x)$ is a separable.
                                            \end{enumerate}
                                        \end{Def}

                                        \begin{Rem}\label{0094}
                                            \begin{enumerate}
                                                \item
                                                Let $X_K$ be a variety over $K$ (e.g. our smooth \st
                                                continuum)and let $K'/K$ a Galois extension (e.g. $K'= K^{\text{sep}}$)
                                                with Galois group $G$ acting on $X_K$ in the canonical
                                                way: Any $\sigma \in G$ induces an automorphism of
                                                $K$-schemes $\Spec K \to \Spec K$. Thus there is a
                                                canonical action of $G$ on $X_K':=X_K \times_{\Spec K} \Spec
                                                K'$ which is the identity on the first component.
                                                If $X_K$ is as in $a)$, then for any $\sigma \in G$ and
                                                for any $x=(x_1, \ldots , x_n)$ we have $\sigma(x) = \sigma(x_1), \ldots
                                                \sigma(x_n)$. The set of classes $G \setminus X_K(K')$ injects into $X_K$.

                                                More precisely the physical points of $X_K$ are Galois orbits of
                                                physical points of $X_K(K^{\text{sep}})$ (i.e. of the points of
                                                the archimedean continuum limit of \st). Thus $X_K$ occurs in
                                                a canonical way as a base space for some kind of ``gauge theory'' (of \sto)
                                                whose ``gauge group'' is a  Galois group.

                                                \item
                                                Intuitively, we may think of the bijections
                                                \begin{equation*}
                                                X_K(K)  \cong \{x \in X_K \mid k(x)=K\} \quad
                                                \text{or} \quad
                                                X_K(K^{\text{sep}}) \cong \{ \text{physical points of} \
                                                    X_{K^{\text{sep}}}\}
                                                \end{equation*}
                                                as some kind of ``wave-particle duality'' on \st level.
                                                For example, the elements of $X_K(K^{\text{sep}})$ are sections of the
                                                structure morphism $X_{K^{\text{sep}}} \to
                                                K^{\text{sep}}$. But morphisms of schemes have similar
                                                symmetries like the differential geometric morphisms of
                                                vector bundles (e.g. one needs a one-cocycle condition
                                                in order to glue). Therefore, $X_K(K^{\text{sep}})$
                                                represents the wave interpretation of \st points, because sections of
                                                vector bundles are classically interpreted as fields.

                                                But the physical points of $X_{K^{\text{sep}}}$ are
                                                points of a topological space. They represent the particle
                                                interpretation of \st points.
                                            \end{enumerate}
                                        \end{Rem}

\part{Arithmetic Yang-Mills theory}

\chapter{Some essential techniques}\label{1300}

In ordinary (i.e. $\real$-valued) Yang-Mills theory, gauge fields
are described by co-vector-fields. If we want to generalize the
$\real$-valued, differential geometric Yang-Mills theory to
arbitrary commutative rings $R$ (or even base schemes $S$), we
therefore have to make use of the notion of the algebraic
geometric tangent bundle. In particular, we need a suitable notion
of pull-back and push-forward of tensor fields in the realm of
algebraic geometry. This chapter is devoted to the development of
such a notion (see section \ref{1320}). We start with some
introductory remarks on general coordinate transformations in the
setting of algebraic geometry.

\section{General coordinate transformations}\label{1301}

Consider a classical physical situation: $S= \Spec \br$ and $\psi:
X'\to X$ is a morphism of smooth algebraic varieties $X,X'$ over
$\real$ (Definition \ref{0021}). Then the sets of $\real$-valued
points $X(\real), X'(\real)$ may be considered as differential
manifolds, and $\psi$ gives rise to a morphism $\psi(\real):
X(\real) \to X'(\real)$ of differential manifolds (which was
illustrated in the physical interpretation after Definition
\ref{0021}). We saw within the bounds of the physical
interpretation following Definition \ref{0014} that on
$\real$-valued points $x \in X(\real)$ the pull-back map
$\psi^\ast \Omega_{X/S}^1\to \Omega_{X/S}^1$ of differential forms
coincides with the ordinary, differential geometric notion of
pull-back of 1-forms: It is described by the Jacobian matrix
$D\psi$. If we choose especially $X=X'$, then $X(\real)=X'(\real)$
and $\psi(\real):X(\real) \to X(\real)$ is called a diffeomorphism
(of $X(\real)$ in the differential geometric sense), if $D \psi$
is an invertible matrix. However, physics should be general
covariant, i.e. if we pass from one observer to another one by
means of a coordinate transformation, the physical laws should be
invariant. General relativity is even only determined up to
diffeomorphisms.

It is our goal to evoke a Yang-Mills theory over arbitrary numbers
(given by a commutative ring $R$) which is therefore no longer
limited to the real numbers $\real$. In particular, we have to
find the algebraic geometric analogue of diffeomorphisms. Starting
from the differential geometric characterization of
diffeomorphisms as morphisms with invertible Jacobi matrix and
recalling our analysis at the beginning of this section, we will
see in Corollary \ref{1303} that the algebraic geometric analogues
of diffeomorphisms are exactly the \'etale morphisms.

\begin{Def}\label{1302}
    A morphism of schemes $f:X \to S$ is called \emph{\'etale} (at a
    point) if it is smooth (at the point) of relative dimension
    $0$ (see Definition \ref{0021} for the notion of smoothness).
\end{Def}

\begin{Cor}\label{1303}
    Let $f: X\to Y$ be an $S$-morphism. Let $x$ be a point and let
    $y=f(x)$. Assume that $X$ is smooth over $S$ at $x$ and that $Y$
    is smooth over $S$ at $y$. Then the following conditions are
    equivalent:
    \begin{enumerate}
        \item[a)] $f$ is \'etale at $x$.
        \item[b)] The canonical homomorphism $(f^\ast \Omega_{Y/S}^1)_x
                  \to (\Omega_{X/S}^1 )_x$ is bijective.
    \end{enumerate}
\end{Cor}

\begin{proof}
    \BLR, Cor. 2.2/10
\end{proof}

\begin{Cor}\label{1304}
    Let $f:X\to Y$ be an $S$-morphism of smooth $S$-schemes $X$ and $Y$. Then is equivalent:
    \begin{enumerate}
        \item[a)] $f$ is \'etale.
        \item[b)] The canonical homomorphism $f^\ast \Omega_{Y/S}^1\to \Omega_{X/S}^1$ is bijective.
        \item[c)] The canonical homomorphism $(f^\ast \Omega_{Y/S}^1)_x\to (\Omega_{X/S}^1 )_x$ is bijective
                  for all $x\in X$.
    \end{enumerate}
\end{Cor}
This feature of \'etale morphism has a far reaching consequence
and leads to a phenomenon which does not occur in differential
geometry: We already mentioned that we should only look at ``\st''
and ``physical objects'' up to coordinate transformation. Thus
from a physical point of view one should not distinguish between
$X'$ and $X$ if $X'\to X$ is surjective and \'etale. But in
contrast to the differential geometric setting, an \'etale,
surjective morphism  $X'\to X$ will in general not be an
isomorphism if we work over arbitrary rings $R$ (instead of e.g.
the complex numbers $\complex$). Thus, $X$ and $X'$ are different
with respect to Zariski topology. But from the physical point of
view, we should not distinguish between $X$ and $X'$. Therefore,
Zariski topology is physically not appropriate. At least, one
knows that \'etale morphisms are open due to flatness. But, we
have to extend the Zariski topology in such a way that $X'$ is
``open'' in $X$ with respect to the bigger ``topology'' if $X' \to
X$ is \'etale. If $X' \to X$ is \'etale, surjective we want to
consider $X'$ as an ``open'' covering of $X$ consisting of only
one element. In such a case we will consider them as equal in a
certain way: ``$X'=X$''. We may say this mathematically more
correctly as follows:
\begin{align*}
    \textit{We have to work with respect to \'etale topology instead of Zariski topology.}
\end{align*}
Etale topology is an example of a so called Grothendieck topology
whose definition we are going to recall next.

\begin{Def}\label{1305}
    A \emph{Grothendieck topology} $\mathfrak{T}$ consists of a category
    $\text{Cat} \, \mathfrak{T}$ and a set  $\text{Cov} \, \mathfrak{T}$
    of families $(U_i \to U)_{i \in I}$ of morphisms in
    $\text{Cat} \, \mathfrak{T}$, called \emph{covering}, such that the following holds:
    \begin{enumerate}
        \item
        If $\Phi: U \to V$ is an isomorphism in $\text{Cat} \, \mathfrak{T}$,
        then $(\Phi) \in \text{Cov} \, \mathfrak{T}$.

        \item
        If $(U_i \to U)_{i \in I}$ and $(V_{ij} \to U_i)_{j \in J_i}$ for $i \in I$
        belong to  $\text{Cov} \, \mathfrak{T}$, then the same is
        true for the composition $(V_{ij} \to U_i \to U)_{i \in I, \, j \in
        J_i}$.

        \item
        If $(U_i \to U)_{i \in I}$ is in $\text{Cov} \, \mathfrak{T}$
        and if $V \to U$ is a morphism in $\text{Cat} \, \mathfrak{T}$,
        then the fibred products $U_i \times_U V$ exist in $\text{Cat} \, \mathfrak{T}$,
        and  $(U_i \times_U V \to V)_{i \in I}$ belongs to
        $\text{Cov} \,  \mathfrak{T}$.
    \end{enumerate}
\end{Def}
We may think of the objects of $\text{Cat} \, \mathfrak{T}$ as of
the open sets of our topology and of the morphisms of $\text{Cat}
\, \mathfrak{T}$ as of the inclusions of open sets. Furthermore, a
family $(U_i \to U)_{i \in I}$ of $\text{Cov} \, \mathfrak{T}$ has
to be interpreted as a covering of $U$ by the $U_i$ abd a fibred
product of $U_i \times_U V$ as the intersection of $U_i$ and $V$.
Thinking along this lines an ordinary topological space $X$ is
canonically equipped with a Grothendieck topology: $\text{Cat} \,
\mathfrak{T}$ is the category of open subsets of $X$ with
inclusions as morphisms, and $\text{Cov} \, \mathfrak{T}$ consists
of all open covers of open subsets of $X$.

\begin{Def}\label{1306}
    For a scheme $X$, let $\text{Cat}  \, \mathfrak{T}$ be the category
    of \'etale $X$-schemes with \'etale morphisms as morphisms, and let $\text{Cov} \, \mathfrak{T}$ be the
    set of families $(U_i \to U)_{i \in I}$ of \'etale morphisms
    such that $U=\bigcup_{i \in I} U_i$. Then $\mathfrak{T}$ is
    called the \emph{\'etale topology} on $X$.
\end{Def}
It should be pointed out that the ``intersection'' of ``open''
sets is dealt with in condition $c)$ of Definition \ref{1305},
whereas we have refrained from giving any sense to the union of
``open'' sets. In fact, even in examples, where the union of
``open'' sets does make sense, such a union will not necessarily
yield an ``open'' set again.

The notion of a Grothendieck topology has been designed in such a
way that the notion of presheaf or sheaf can easily be adapted to
such a situation:

\begin{Def}\label{1307}
    Let $\mathfrak{T}$ be a Grothendieck topology and
    $\mathfrak{C}$ be a category which admits cartesian products.
    A \emph{presheaf} on $\mathfrak{T}$ with values in
    $\mathfrak{C}$ is a contravariant functor $\cf : \text{Cat} \, \mathfrak{T} \to
    \mathfrak{C}$. We call $\cf$ a \emph{sheaf} if the diagram
    \begin{align*}
        \cf(U) \rightarrow  \prod\limits_{i\in I}^{ \ } \cf(U_i)
        \rightrightarrows  \prod\limits_{i,j\in I}^{ \ } \cf(U_i \times_U U_j)
    \end{align*}
    is exact for any covering $(U_i \to U)_{i \in I}$ in
    $\text{Cov} \, \mathfrak{T}$.
\end{Def}
Having defined the ``open subsets'' of a scheme $X$ with respect
to \'etale topology, we may now introduce the notion of
neighborhoods of points.

\begin{Def}\label{1308}
    A morphism $X'\to X$ of schemes is called an
    {\it \'etale neighborhood} of a point $x\in X$, if $X'\to X$
    is \'etale and if $x$ is contained in the image of $X'$.
\end{Def}
The following Proposition \ref{1309} finally gives the
justification, why it is physically admissible to consider \'etale
morphisms $X \to Y$ as inclusions.

\begin{Satz}\label{1309}
    Let $f: X\to Y$ be locally of finite type. Let $x\in X$ and $y:=f(x)$.
    If $f$ is \'etale at $x$ (resp. unramified at $x$), there exists
    an \'etale neighborhood $Y'\to Y$ of $y$ such that, locally at
    each point of $X:=X \times_Y Y'$ above $x$, the morphism $f':X'\to
    Y'$ obtained by base change $Y'\to Y$ is an open immersion (resp.
    an immersion).
\end{Satz}

\begin{proof}
    \BLR, Prop. 2.3/8
\end{proof}
Therefore, the algebraic geometric analogue of the differential
geometric diffeomorphism should be the \'etale morphism. However,
if we perform a coordinate transformation or any ``deformation''
of \sto, the number of points should be invariant under coordinate
transformations. In this situation, we should therefore restrict
attention to the class of finite, \'etale surjective morphisms $f
:X \to Y$ which have the property that there are bijections
$Y_K(K')=X_K(K')$ and $Y(S')=X(S')$ for all \'etale $S$-schemes
$S'$ with field of fractions $K'$. Thereby, $X_K$ and $Y_K$ denote
as usually the respective generic fibres.
%
%
However, let us point to the fact that it can happen that $f$
changes the topology.

Working with respect to \'etale topology, there is the
\emph{theorem of the inverse function} for smooth schemes. This is
crucial if we want to perform analysis, and we have to make use of
it later when we develope Yang-Mills theory over rings. It takes
the following form.

\begin{Satz}\label{1312}
    Let $f:X \to S$ be a smooth morphism. Let $s$ be a point of
    $S$, and let $x$ be a closed point of the fibre $X_s=X \times_S \Spec
    k(s)$ such that $k(x)$ is a separable extension of $k(s)$.
    Then there exists an \'etale morphism $g:S' \to S$ and a point
    $s' \in S'$ above $s$ such that the morphism $f': X \times_S S' \to
    S'$ obtained from $f$ by the base change $S' \to S$ admits a
    section $h: S' \to X \times_S S'$, where $h(s')$ lies above
    $x$, and where $k(s')=k(x)$.
\end{Satz}

\begin{proof}
    \BLR, Prop. 2.2/14
\end{proof}

\begin{Satz}\label{1351}
    Let $X$ be an $S$-scheme. Then the notion of physical
    points is stable under \'etale base change.
\end{Satz}

\begin{proof}
    Consider a physical point $x\in X$ over $s \in S$, and let $S' \to S$ be an \'etale morphism.
    There is a closed immersion $ \{ x \} \hookrightarrow X_s=X\otimes_S k(s)$. Therefore
    the morphism $S'\times_S \Spec k(x)\hookrightarrow (X\times_S S')\times_S k(s)$
    obtained by base change with $S' \to S$ is a closed immersion, too.
    Furthermore, $S'\times_S$ Spec~$k(x)\to$ Spec~$k(x)$ is \'etale.
    In particular, $ S'\times_S \Spec k(x)$ is a finite set. More precisely
    it is a finite disjoint union of separable field extensions of $k(x)$,
    and thus $S'\times_S \Spec k(x)$ is equipped with discrete topology.
    It follows that each point of $S' \times_S \Spec k(x)$ is a closed
    point. Let $h(s')\in S'\times_S \Spec k(x)$ be a point over $s' \in
    S'$, where $s'$ is over $s$.  Due to the closed immersion
    $S'\times_S \Spec k(x)\hookrightarrow (X\times_S S')\times_S
    k(s)$, $h(s')$ is a point of $(X\times_S S') \times_{S'} k(s')$ which is closed in $(X\times_S S')\times_S
    k(s)$. But then, the continuity of the canonical inclusion
    \begin{align*}
        \xymatrix{ (X\times_S S') \times_S k(s) \ar@{=}[r]  &  (X\times_S S') \times_{S'} (S'\times_S k(s)) \\
        (X\times_S S') \times_{S'} k(s')\ar@{=}[r] & (X\times_S S')\times_{S'}  (k(s')\times_S
        k(s)) \ar@{^{(}->}[u]
        }
    \end{align*}
    shows that $h(s')$ is already closed in $(X\times_S
    S')\times_{S'} k(s')$.
    \item It remains to show that $k(h(s'))$ is separable over
    $k(s')$.
    By assumption $ S'\to S$ is \'etale and it follows that $k(s) \hookrightarrow k(s')$ is separable.
    Furthermore $X\times_S S'  \stackrel{pr_1}{\longrightarrow}  X$ is
    \'etale, too. But $h(s')$ is mapped to $x$ under this morphism
    which implies that $k(x)\hookrightarrow k(h(s'))$ is
    separable. Finally $k(s) \hookrightarrow k(x)$ is separable,
    because $x \in X_s$ is a physical point. Due to the canonical
    commutative diagram
    \begin{align*}
        \xymatrix{ k(h(s'))  &  k(x) \ar@{_{(}->}[l] \ , \\
        k(s')\ar@{^{(}->}[u] & k(s) \ar@{_{(}->}[l]  \ar@{^{(}->}[u]
        }
    \end{align*}
    $k(h(s'))$ has to be separable over  $k(s')$, because separable field
    extensions are transitive. All in all we see that all points,
    which lie over the physical point  $x$ after \'etale base
    change, are physical points, too.
\end{proof}
\section{Pull-back and push-forward of tensor fields}\label{1320}

During this section, we assume $f: X\to Y$ to be a smooth morphism
of smooth $S$-schemes. We already introduced the pull-back of
differential forms in Definition \ref{0014}. Within this section
we will prove at first that smooth schemes allow a local (with
respect to \'etale topology) push-forward of vector fields.
Finally we will use these notions as a basis for an inductive
definition of pull-back and push-forward of tensor fields. We
begin with the statement of some preparing results.

\begin{Lemma}\label{1321}
    If $X$ is a smooth scheme over a field $k$, the set of closed points
    $x\in X$ such that $k(x)$ is a separable extension of $k$ is dense in
    $X$. Recall that we call these points \emph{physical points}.
\end{Lemma}

\begin{proof}
     \BLR, Cor. 2.2/13
\end{proof}
For all  $S$-schemes $Z \to S$, let us denote the fibre over a
point $s \in S$ by $Z_s$. For each morphism $f: X\to Y$ of
$S$-schemes and for all $s \in S$, denote by $f_s: X_s \to Y_s$
the induced morphism on the fibre over $s\in S$. With this
notation we may state the following result.

\begin{Satz}\label{1322}
    Let $f: X\to Y$ be a morphism of $S$-schemes such that $X_s$ and $Y_s$ are
    algebraic varieties over $k(s)$ for all $s \in S$.
    Let $x\in X$ be a physical point. Then $y:=f(x)$ is a physical point, too.
\end{Satz}

\begin{proof}
    By assumption, the morphism $X \to Y$ induces
    a family of morphisms of algebraic varieties $X_s \to Y_s$ over
    $k(s)$, $s \in S $.
    Let $x$ be a physical point, $y=f(x)$.
    Looking at the local rings, we have got commutative diagrams
    \begin{align*}
        \xymatrix{ k(x) &  k(y) \ar@{^{(}->}[l] \\
                   k(s) \ar@{^{(}->}[ru] \ar@{^{(}->}[u]  }
    \end{align*}
    As $k(s)\hookrightarrow k(x)$ is a separable algebraic extension, the same is
    true for $k(s)\hookrightarrow k(y)$. In particular $y\in Y_s$ is a
    closed point in $Y_s$, because for algebraic varieties $Z$ over a
    field $k$ one knows the following fact (e.g. due to \Liu, Ex. 2.5.9): $z\in Z$ is closed
    if and only if $ k\hookrightarrow k(z)$ is algebraic.
\end{proof}

\begin{Satz}\label{1323}
    Let $f: X\to Y$ be a smooth morphism of $S$-schemes. Let
    $y\in f(X)$ be a physical point. Then there exists a
    physical point $x\in X$ which is mapped to $y$.
\end{Satz}

\begin{proof}
Let $s$ be the image of $y$ under the structure morphism $Y\to S$.
We first remark that we write $x_y$ (resp. $x_s$) if we consider
$x\in X$ as a point of $X_y$ (resp. of $X_s$). As $X_y\to k(y)$ is
smooth and $X_y \not= \varnothing$, we know by Lemma \ref{1321}
that there exists  a $x_y\in X_y$ such that $x_y$ is closed in
$X_y$. and such that $ k(y)\hookrightarrow k(x_y)$ is separable.
By \Liu, Ex. 2.5.9 this is equivalent to the statement that
$k(y)\hookrightarrow k(x_y)$ is a separable algebraic extension.
If $s$ denotes the image of $y$ under the structure morphism $Y\to
S$ we know furthermore by assumption that: $k(s) \hookrightarrow
k(y_s)$ is separable algebraic. So we are done by the following
lemma if we apply \Liu, Ex. 2.5.9 once again.
\end{proof}

\begin{Lemma}\label{1324}
    Let $f:X\to Y$ be a morphism of schemes. Let $x\in X$ and
    $y:=f(x)$. If we denote by $x_y$ the point $x$ considered as a
    point of the fibre $X_y$, then:
    \begin{align*}
        k(x_y)=k(x).
    \end{align*}
\end{Lemma}

\begin{proof}
    Considered as a scheme, the fibre $X_y$ is isomorphic to  $X \times_Y \Spec
    k(y)$. Therefore we get for the local ring at $x_y \in X_y$
    \begin{align*}
        \co_{X_y,x_y} = \co_{X,x} \otimes_{\co_{Y,y}} k(y)= \co_{X,x}/\m_y
        \co_{X,x}.
    \end{align*}
    In particular the maximal ideal $\m_{x_y}$ of $\co_{X_y,x_y}$
    is given by $\m_{x_y} = \m_x/\m_y \co_{X,x}$. Therefore
    \begin{align*}
        k(x_y)=\co_{X_y,x_y}/\m_{x_y} = (\co_{X,x}/\m_y
        \co_{X,x})/(\m_x/\m_y \co_{X,x}) = \co_{X,x}/\m_x=k(x).
    \end{align*}
\end{proof}
Consider now the smooth morphism $f:X\to Y$ of $S$-schemes. Let us
consider a (local) vector field $\vecv: U {\longrightarrow}
T_{X/S}$ on $X$. We would like to define a local push-forward of
$\vecv$ under $f$ to a local vector field $f_* \vecv$ on $Y$.

This should be done in a canonical way using the pull-back of
co-vector fields. Due to the duality between vector fields and
co-vector fields (see Remark \ref{1130}) we may identify
$\vecv:U\to T_{X/S}$ with its corresponding
$\co_U$-module-homomorphism $\vecv:\Omega_{X/S}^1\to \co_U$ which
we will also denote by $v$. Also using the contravariant functor
$\bv$ introduced in Proposition \ref{1123}, we will finally arrive
at the following diagram as explained below:
\begin{align}
    \begin{array}{cccclcl}
        f(U) & \stackrel{f}{\longleftarrow} & U & \stackrel{\vecv}{\longrightarrow}
        & T_{X/S} & \stackrel{T_{X/S}(f)}{\longrightarrow} & T_{Y/S} \times_Y X \\[1ex]
        f^{-1}\co_{f(U)} & \stackrel{f^\ast}\longrightarrow & \co_U &
        \stackrel{\vecv}{\longleftarrow} & \Omega_{X/S}^1 &
        \stackrel{f^\ast}{\longleftarrow} & f^\ast \Omega_{Y/S}^1.
    \end{array}
\end{align}
Here $f^*$ denotes as well the canonical morphism
$f^{-1}\co_{f(U)} \longrightarrow  \co_U$ as the pull-back of
co-vector fields $\omega\in\Omega_{Y/S}^1$
\begin{align*}
    \begin{array}{lcl}
        f^\ast \Omega_{Y/S}^1 & \stackrel{f^\ast}{\longrightarrow} & \Omega_{X/S}^1 \\[1ex]
        \omega\otimes 1 & \longmapsto & f^\ast \omega
    \end{array},
\end{align*}
which we sometimes simply abbreviate by $\omega  \mapsto  f^\ast
\omega$.  The pull-back is defined globally, i.e. if we consider a
co-vector field $\omega$ living on all of $Y$, we can pull it back
to a co-vector field $f^\ast \omega$ living on {\it all} of $X$.

In contrast to this, (even globally defined) vector fields $\vecv$
may in general only be pushed forward to vector fields living on
(\'etale-)open subsets of $f(X)$. This is analogously to the
situation upon real-valued manifolds. As an example consider the
projection $p: S^2 \subset \real^3 \to\br^2$, $(x,y,z)\mapsto
(x,y)$ and a vector field on the two-dimensional sphere $S^2$. Let
us denote the corresponding flow by $\phi_t$. But in general, $p
\circ \phi_t$ will not be injective. Then, the push-forward does
not exist globally. More precisely, we have to face the following
problem, whenever we are trying to define of a push-forward
$f_\ast \vecv$ of a vector field $\vecv$: In general there is only
the following canonical diagram
\begin{align*}
    \xymatrix{ f(U) & U  \ar[r]^{ \vecv \ \ } \ar[l]_{ \ \ \ \ f } &
    T_{X/S}  \ar[rr]^{pr \, \circ \, T_{X/S}(f)} && T_{Y/S} },
\end{align*}
but in general there is even no chance to get a local section of
$f$. Thus we will not find a $Y$-morphism $V \to T_{Y/S}$ with
$V\subset f(U)$ Zariski-open which is induced by $\vecv$. Thus
there is no chance to define the push-forward $f_\ast v$ of the
given vector field $\vecv$ within Zariski-topology.

The way out of trouble is performing a local coordinate
transformation, i.e. performing an \'etale base change $Y'\to Y$.
Intuitively one may think that a badly chosen coordinate system
may lead to singularities which do not exist in ``reality'', but
which are due to the bad choice of the coordinates. Such
``imaginary'' singularities may be resolved by performing a
suitable coordinate transformation. As an example consider a flat
\st $\br^2$ with metric $ds^2=dx^2+dy^2$ and introduce the new
coordinate system $(x',y')=({x^3}/{27},y)$. Then
$dx'=d({x^3}/{27})={1}/{9} \cdot x^2 dx=x^{'2/3} dx$ and therefore
\begin{align*}
    ds^2=x'^{-4/{3}} dx^{'2}+dy^2
\end{align*}
which is singular at $x'=0$. But in truth there is ``no''
physically relevant difference between the points $(0,y)$ and
$(1,y)$ of empty \st $\real^2$.

So let us perform an \'etale base change $Y'\to Y$ as follows: Let
$y\in f(U)$. Thus $f^{-1}(y) = X_y\not= \emptyset$, and by Lemma
\ref{1321} we may choose a closed point $x\in X_y$ such that
$k(y)\hookrightarrow k(x)$ is separable, i.e. we may choose a
physical point $x \in X_y$. Then we are in the situation of
Proposition \ref{1312}, and we infer that there exists an \'etale
morphism $g: Y'\to Y$ such that $f':{X\times_Y Y'} \to Y'$ has a
section $h$. Thereby $f'$ is the morphism induced by $f$ by base
change with $g: Y'\to Y$.

We want to use this section in order to define a push-forward of
$\vecv$. For this purpose let us first prove that the canonical
diagram $(9.1)$

\begin{equation*}
    \begin{array}{cclcccc}
        Y & \stackrel{f}{\longleftarrow} & X & \stackrel{\vecv}{\longrightarrow} & T_{X/S} &
        \stackrel{T_{X/S}(f)}\longrightarrow &  T_{Y/S} \times_Y X \\[1ex]
        f^{-1} \co_Y & \stackrel{f^\ast}\longrightarrow & \co_X & \stackrel{\vecv}{\longleftarrow} &
        \Omega_{X/S}^1 & \stackrel{f^\ast}\longleftarrow & f^\ast \Omega_{Y/S}^1
    \end{array}
\end{equation*}
is stable under local coordinate transformations, i.e. compatible
with \'etale base change.

So let us apply the above base change $g: Y'\to Y$ to the diagram
(9.1), and let $f': X'\to Y'$ be as above ($X':=X\times_Y Y'$). We
claim that we arrive at the diagram
\begin{equation}
    \begin{array}{ccccccc}
        Y' & \stackrel{f'}{\longleftarrow} & X' & \stackrel{ \, \vecv'}{\longrightarrow} &
        T_{X'/S} & \stackrel{T_{X'/S}(f')}\longrightarrow & T_{Y'/S} \times_{Y'} X' \\[1ex]
        f^{'-1} \co_Y & \stackrel{{f'}^\ast}\longrightarrow & \co_{X'} &
        \stackrel{ \, \vecv'}{\longleftarrow} & \Omega_{X'/S}^1 & \stackrel{ \ f'^\ast}\longleftarrow &
        f^{'\ast} \Omega_{Y'/S}^1
    \end{array}
\end{equation}

\begin{proof}
    Consider the cartesian diagram
    \begin{align*}
        \xymatrix @-0.2pc { X' \ar[d]_j \ar[r]^{f'} &  Y' \ar[d]^g \\
        X  \ar[r]^f  &  Y }
    \end{align*}
    As $g$ is \'etale, $j$ is \'etale, too (because $j= \text{id}_X \times
    g$). Thus we know from Corollary \ref{1304} that
    \begin{eqnarray*}
        g^\ast \Omega_{Y/S}^1 & = & \Omega_{Y'/S}^1. \\
        j^\ast \Omega_{X/S}^1 & = & \Omega_{X'/S}^1.
    \end{eqnarray*}
    These relations and the equation $f\circ j=g\circ f'$, which expresses
    the commutativity of the above cartesian diagram, yield
    \begin{eqnarray*}
        (f\circ j)^\ast \Omega_{Y/S}^1 & = & (g\circ f')^\ast \Omega_{Y/S}^1 \\
        & = & f^{'\ast} (g^\ast \Omega_{Y/S}^1) \\
        & = & f^{'\ast} \Omega_{Y'/S}^1.
    \end{eqnarray*}
    Also using  Proposition \ref{1123}, we deduce:
    \begin{itemize}
        \item
        \small{$T_{X/S} \times_Y Y' = T_{X/S} \times_X X'  =  \bv (j^\ast \Omega_{X/S}^1) =
        \bv(\Omega_{X'/S}^1)=T_{X'/S}$}.

        \item
        \small{$(T_{Y/S}\times_Y X)\times_Y Y' = T_{Y/S} \times_Y X'  =  \bv ((f\circ j)^\ast
        \Omega_{Y/S}^1)
         = \bv (f^{'\ast} \Omega_{Y'/S}^1)
        =  T_{Y'/S}\times_{Y'} X'$}.
    \end{itemize}
    So we see that we arrive at the diagram (9.2) if we apply
    the base change $g: Y'\to Y$ to the diagram (9.1).
\end{proof}

Let us emphasize that the base-scheme $S$ is  {\it not} effected
by this process, i.e. $f'$ is actually a restriction of $f$ (with
respect to \'etale topology). The ``adelic'' \st points (which are
isomorphic to $S$) remain unaltered; we are only looking closer at
a region of the universe. Thus, with some abuse of notation, we
may suppress the ``dashes'' in (9.2), and we may assume (after
eventually shrinking $Y$ and performing a coordinate
transformation) that $f$ has a section $h$. Then we derive the
following canonical morphism $\widetilde{f_\ast \vecv}: Y\to
T_{Y/S}\times_Y X$ induced by $f$ and $\vecv$:
\begin{align*}
    \xymatrix{ \widetilde{f_\ast \vecv}: \quad Y \ \,
    \ar@<2pt>@{^{(}->}[rr]^{ \ \ \ \ h} & & \ X \ar@<2pt>[ll]^{ \ \
    \ \ f} \ar[r]^{\vecv \ \,} & T_{X/S} \ar[rr]^{T_{X/S}(f) \quad \ }
    && T_{Y/S} \times_Y X}
\end{align*}
This induces a canonical morphism on the sheaves
\begin{align*}
    \begin{array}{ccccccc}
        \widetilde{(f_\ast \vecv)}: \ \ f^\ast \Omega_{Y/S}^1 &
        \stackrel{f^*}\longrightarrow & \Omega_{X/S}^1 &
        \stackrel{\vecv}{\longrightarrow}
        & \co_X & \stackrel{h^*}\longrightarrow & h_* \co_Y, \\[1ex]
        \qquad \quad \ \, \omega\otimes 1 & \mapsto & f^\ast \omega &
        \mapsto & \vecv(f^\ast\omega) & \mapsto & \widetilde{(f_\ast \vecv)}
        (\omega\otimes 1)
    \end{array}
\end{align*}
where $\omega \in \Omega_{Y/S}^1$. So far we found a canonical
morphism $\widetilde{f_\ast \vecv}: f^\ast \Omega_{Y/S}^1\to h_*
\co_Y$ such that
\begin{align*}
    \widetilde{(f_\ast \vecv)} (\omega\otimes 1)
    =( h^* \circ \vecv \circ f^\ast ) ( \omega)
    = h^*\left(\vecv(f^\ast \omega)\right)
    \stackrellow{\mbox{\scriptsize by duality}}{=} h^*\left((f^\ast \omega)(\vecv)\right).
\end{align*}
We would like to interpret $\widetilde{f_\ast \vecv}$ as our
push-forward of $\vecv$, but actually $\widetilde{f_\ast \vecv}$
is not even a vector field. Luckily, Proposition \ref{1325} shows
that the $\co_X$-module homomorphism $\widetilde{f_\ast \vecv}:
f^\ast \Omega_{Y/S}^1\to h_*\co_Y$ corresponds to a uniquely
determined $\co_Y$-module homomorphism ${f_\ast \vecv}:
\Omega_{Y/S}^1 \to \co_Y$, i.e. a vector field on $Y$. This will
be a push-forward of $\vecv$ under $f$.

\begin{Satz}\label{1325}
    Let $f: X\to Y$ be a morphism of schemes and let $\cg$ be a
    $\co_Y$-module on $Y$. Then the pull-back $\co_X$-module $f^\ast\cg$
    on $X$ fulfills the universal property that there is a functorial isomorphism
    \begin{align*}
        \Hom_{\co_Y} (\cg,f_\ast\cf) \stackrel{\sim}{\longrightarrow} \Hom_{\co_X} (f^\ast\cg,\cf)
    \end{align*}
    functorial in $\co_X$-modules $\cf$ on $X$.
\end{Satz}

\begin{proof}
    \Liu, Ex. 5.1.1
\end{proof}
Let us describe this functorial isomorphism shortly. For this
purpose, we assume that $\cg$ is quasi-coherent. Let $\varphi \in
\Hom_{\co_Y}(\cg,f_\ast \cf)$, i.e. $\varphi$ consists of a
collection of morphisms
\begin{align*}
    \varphi(V): \cg(V) \longrightarrow f_*\cf(V):=\cf(f^{-1}(V)),
    \qquad \text{$V$ in $Y$ open,}
\end{align*}
which is compatible with restrictions. For open subsets $U \subset
X$ with $U \subset f^{-1}(V)$, this yields morphisms
\begin{align*}
    \varphi(V): \cg(V) \longrightarrow \cf(f^{-1}(V)) \stackrel{\text{res}}\longrightarrow \cf(U),
    \qquad \text{$V$ in $Y$ open,}
\end{align*}
where ``res'' denotes the canonical restriction. Let now $U$ be an
affine open subset of $X$ such that $f(U)$ is contained in an
affine open subset $V$ of $Y$. Then one knows (e.g. by \Liu, Prop.
5.1.14) that
\begin{align*}
    f^*\cg|_U \cong \left(\cg(V) \otimes_{\co_Y(V)}
    \co_X(U) \right)^{\sim}.
\end{align*}
We consider $\cf(U)$ as an $\co_Y(V)$-module by means of the
canonical morphism $\co_Y(V) \longrightarrow \co_X(f^{-1}(V))
\stackrel{\text{res}}\longrightarrow \co_X(U)$. The
$\co_Y(V)$-linear map $\varphi(V): \cg(V) \to \cf(U)$ gives rise
to a $\co_Y(V)$-bilinear map
\begin{align*}
    \cg(V) \times \co_X(U) \to \cf(U), \qquad (g,a) \mapsto
    \varphi(V)(g) \cdot a.
\end{align*}
Using the universal property of tensor products, we finally arrive
at a $\co_X(U)$-linear map
\begin{align*}
    \widetilde{\varphi} (U):\cg(V) \otimes_{\co_Y(V)} \co_X(U) \to \cf(U), \qquad g \otimes a \mapsto
    \varphi(V)(g) \cdot a.
\end{align*}
As $(f^*\cg)(U) = \cg(V) \otimes_{\co_Y(V)} \co_X(U)$, the
$\co_X(U)$-linear maps $\widetilde{\varphi} (U)$ give rise to the
searched $\co_X$-module homomorphism $\widetilde{\varphi} \in
\Hom_{\co_X} (f^\ast\cg,\cf)$. Thus the functorial isomorphism of
Proposition \ref{1325} may be described as follows:
\begin{align*}
        \Hom_{\co_Y}(\cg,f_\ast\cf)    \stackrel{\sim}{\longrightarrow}   \Hom_{\co_X}
        (f^\ast\cg,\cf), \qquad \varphi \mapsto
        \widetilde{\varphi}, \text{ where \ }
        \widetilde{\varphi}(g \otimes a)  := \varphi(g) \cdot a.
\end{align*}
Now we may apply this result to our morphism $\widetilde{f_\ast
\vecv}: f^\ast \Omega_{Y/S}^1\to h_*\co_Y$. Let us denote its
image under the functorial isomorphism of Proposition \ref{1325}
by ${f_\ast \vecv}: \Omega_{Y/S}^1\to f_*h_*\co_Y$. One knows that
the direct image of sheaves is a covariant functor (see e.g. \Liu,
Ex. 2.2.6). In particular $f_*h_*\co_Y = (f \circ h)_*\co_Y
=\co_Y$ because $h$ is a section of $f$, i.e. $f \circ h =
\text{id}$. All in all we arrive at a $\co_Y$-module homomorphism
${f_\ast v}: \Omega_{Y/S}^1\to \co_Y$, i.e. at a vector field on
$Y$, which by construction fulfills the equation ${f_\ast
\vecv}(\omega)=\widetilde{(f_\ast \vecv)} (\omega\otimes 1) =( h^*
\circ \vecv \circ f^\ast ) ( \omega)$, i.e.:
\begin{align*}
    {f_\ast \vecv} = h^* \circ \vecv \circ f^\ast
\end{align*}

\begin{Rem}\label{1326}
    \begin{enumerate}
        \item
        The smoothness of the morphism $f:X \to Y$ was only needed
        in order to guarantee the existence of local sections of $f$ (where local
        is meant with respect to \'etale topology). If we assume
        that there is already a section $h$ of $f$, we may drop
        the smoothness assumption of $f$. Then the construction of the
        push-forward works as described above, and every global vector field
        $\vecv \in \Gamma(T_{X/S}/X)$ on $X$ yields a \emph{global} vector field $f_*
        \vecv = h^* \circ \vecv \circ f^\ast$ on $Y$.

        \item
        If $f:X \to Y$ is an isomorphism, then $h:=f^{-1}$ is a
        global section of $f$, and the push-forward of every
        vector field $\vecv \in \Gamma(T_{X/S}/X)$ on $X$ fulfills
        the equation:
        \begin{align*}
            {f_\ast \vecv} = {(f^{-1})}^* \circ \vecv \circ
            f^\ast.
        \end{align*}
        This is exactly the definition of the push-forward of
        vector fields which is considered in differential geometry.
        Thus the differential geometric notion is contained as
        a special case.

        \item
        Let $f:X \to Y$ and $g:Y \to Z$ be morphisms with sections
        $i_f$ and $i_g$, and let $\vecv \in \Gamma(T_{X/S}/X)$.
        Then
        \begin{align*}
            (g \circ f)_*\vecv = g_* (f_* \vecv).
        \end{align*}
        This may be seen as follows:
        \begin{align*}
            ((g \circ f)_*\vecv ) (\omega) &= (i_f \circ i_g)^* \circ \vecv \circ  (g\circ
            f)^*(\omega) = i_g^* \circ i_f^* \circ \vecv \circ f^* \circ g^*
            (\omega). \\
            (g_*(f_* \vecv))(\omega) &= i_g^* ((f_*
            \vecv)(g^*\omega))= i_g^* (i_f^* (\vecv(f^*
            g^*\omega))).
        \end{align*}
    \end{enumerate}
\end{Rem}

\begin{Def}\label{1327}
    Let $X$ and $Y$ be smooth $S$-schemes, and let $f: X\to Y$ be a
    $S$-morphism which locally (with respect to \'etale topology on
    $Y$) admits a section $h: Y \hookrightarrow X$ (e.g. $f$ is
    smooth). Let $\omega\in\Omega_{Y/S}^1$ be a co-vector field on $Y$
    and $v: X\to T_{X/S}$ be an vector field on $X$. Then, locally (in
    \'etale topology), there exists a vector field $f_\ast \vecv$ on $Y$
    fulfilling the following identity
    \begin{align*}
        (f_\ast \vecv)(\omega) =h^\ast
        \left(\vecv\left(f^\ast\omega\right)\right).
    \end{align*}
    This vector field is called the \emph{push-forward} of $\vecv$ under $f$ (with respect to $h$).
\end{Def}
If $y \in Y$ and if we take the inductive limit over all \'etale
neighborhoods of $y$ in $Y$ (see Definition \ref{1308}), we get a
push-forward  $(f_\ast \vecv)(y) \in \Hom_{\co_{Y,y}}(\Omega_{Y/S,
\, y}^1, \co_{Y,y})$ on stalks
\begin{align*}
    (f_\ast \vecv)_y(\omega_y)
    :=(f_\ast \vecv)(\omega)(y)
    :=(f_\ast \vecv)(y)(\omega)
    =h^*_y(\vecv(f^*\omega)(h(y)))
    =h^*_y(\vecv_{h(y)}(f^*\omega)_{h(y)}),
\end{align*}
where $h^*_y: \co_{X,h(y)} \to \co_{Y,y}$ is the canonical map.

\begin{Rem}\label{1327r}
    On stalks there is also a well-defined notion of push-forward
    for sections of smooth and separated morphism. In order to see
    this, let $\pi:X \to Y$ be a smooth and separated morphism of
    $S$-schemes, and let $s:Y \hookrightarrow X$ be a section of
    $\pi$. In particular, $s$ is a closed immersion. As already
    described above, every vector field $\vecv \in T_{Y/S}$ gives
    rise to a $\co_Y$-linear morphism
    \begin{align*}
        \begin{array}{ccccccc}
            \widetilde{(s_\ast \vecv)}: \ \ s^\ast \Omega_{X/S}^1 &
            \stackrel{s^*}\longrightarrow & \Omega_{Y/S}^1 &
            \stackrel{\vecv}{\longrightarrow}
            & \co_Y , \\[1ex]
            \qquad \quad \ \, \omega\otimes 1 & \mapsto & s^\ast \omega &
            \mapsto & \vecv(s^\ast\omega)
        \end{array}
    \end{align*}
    which corresponds to a $\co_X$-linear morphism $ \Omega_{X/S}^1 \longrightarrow s_*
    \co_Y$. If $x=s(y)$ is in the image of $s$, then we get on
    stalks at $x$ an $\co_{X,x}$-linear morphism

    \begin{align*}
        \Omega_{X/S,x}^1 \longrightarrow (s_* \co_Y)_x =
        \co_{Y,y}.
    \end{align*}
    The last equality is due to the fact that
    \begin{align*}
        (s_* \co_Y)_x=
        \left\{%
        \begin{array}{llll}
            0 & \quad \text{if $x \notin s(Y)$}   \\
            \co_{Y,y} & \quad \text{if $x=s(y)$} \\
        \end{array}
        \right\},
    \end{align*}
    because $s$ is a closed immersion (see e.g. \Liu, Prop. 2.2.24).
    Due to the flatness of $\pi$, the canonical morphism $\pi^*_x: \co_{Y,y} \to
    \co_{X,x}$ is injective, because flat morphisms of local rings
    are injective. Thus we may embed $\co_{Y,y}$ into $\co_{X,x}$
    and finally get a $\co_{X,x}$-linear homomorphism
    \begin{align*}
        (s_* \vecv)_x:\Omega_{X/S,x}^1 \longrightarrow \co_{X,x}.
    \end{align*}
    This gives a well-defined stalkwise push-forward of vector fields for sections
    $s$ of smooth and separated morphisms. Again we obtain the formula
    \begin{align*}
        (s_* \vecv)_x = \pi^*_x \circ \vecv_y \circ
        s^*_y,
    \end{align*}
    where $x =s(y)$.\footnote{Viewing an element $\alpha \in \co_X$
    as a function $\alpha:X \to \coprod\limits_{x \in X}
    \co_{X,x}$, $x \mapsto \alpha_x$, we may even write
    \begin{align*}
        (s_* \vecv)(\omega)=\Big( \vecv(s^*\omega) \Big) \circ \pi
        \qquad \quad \text{for all $\vecv \in \ct_{Y/S}$, $\omega \in \Omega_{X/S}^1$.}
    \end{align*}
    This may be seen as follows. As we are working on stalks, we
    may assume that $X$ and $Y$ are affine, say $X=\Spec B$ and
    $Y=\Spec A$. For $x \in X$ and $y=\pi(x) \in Y$ let $B_x$ and $A_y$ be the localization
    of $B$ and $A$ at the corresponding prime ideals. Then
    there is a canonical commutative diagram
    \begin{align*}
        \xymatrix{ A \ar[r]^{\pi^*} \ar[d]  & B \ar[d] &
        \alpha  \ar@{{|}->}[r] \ar@{{|}->}[d] & \pi^*(\alpha) \ar@{{|}->}[d]. \\
        A_y \ar@{^{(}->}[r]  & B_x & \alpha(\pi(x)) \ar@{{|}->}[r]  &  \pi^*(\alpha)(x)
        }
    \end{align*}
    As $A_y \to B_x$ is injective, we  may identify $A_y$ with a
    subring of $B_x$, whence yielding the desired equation $\pi^*(\alpha)=\alpha \circ
    \pi$ of functions from $X$ to  $\coprod\limits_{x \in X}
    \co_{X,x}$.
    } In particular we may push-forward vector fields along constant
    morphism. Thereby an $S$-morphism $f:X \to Y$ of $S$-schemes is called
    \emph{constant}, if there is a section $s: S \hookrightarrow Y$ such
    that $f= s \circ \pi$, where $\pi:X \to S$ is the canonical
    morphism.
\end{Rem}

\begin{Def}\label{1328}
    Let $f: X\to Y$ be an isomorphism of smooth $S$-schemes. Let $\vecv: Y\to T_{Y/S}$ be a vectorfield on $Y$
    and $\omega \in \Omega_{X/S}^1$ a covectorfield on $X$. Then we define:
    \begin{enumerate}
        \item[a)]
        The pull-back $f^\ast \vecv$ of $\vecv$ to $X$ to be the push-forward $(f^{-1})_\ast \vecv$.
        \item[b)]
        The push-forward $f_\ast\omega$ of $\omega$ to $Y$ to be the pull-back $(f^{-1})^\ast
        \omega$.
    \end{enumerate}
\end{Def}

\begin{Def}\label{1329}
    Let $X$ and $Y$ be smooth $S$-schemes. A \emph{$r$-times
    covariant, $s$-times contravariant (relative) tensor field} on $Y$
    over $S$ is a global section of
    \begin{align*}
        (\ct_{Y/S})_s:=
        \underbrace{(\Omega_{Y/S}^1)^\vee\otimes\ldots\otimes(\Omega_{Y/S}^1)^\vee}_{r\mbox{-times}}
        \otimes \underbrace{\Omega_{Y/S}^1\otimes\ldots\otimes
        \Omega_{Y/S}^1}_{s\mbox{-times}}.
    \end{align*}
    Let $f:X\to Y$ be an $S$-morphism and $h:Y \hookrightarrow X$ a
    local section of $f$ (where local is meant with respect to \'etale
    topology on $Y$). Then, whenever the following makes sense, we
    define:
    \begin{enumerate}
        \item[a)]
        Let $\vect \in (\ct_{X/S})_s^r, \; \vect=\vecv_1\otimes\ldots\otimes
        \vecv_r\otimes\omega^1\otimes\ldots\otimes\omega^s$, be a tensor field,
        and let $\nu^1, \ldots, \nu^r \in \Omega_{Y/S}^1$
        and let $ \vecw_1, \ldots, \vecw_s \in (\Omega_{Y/S}^1)^{\vee}$. Then
        \begin{eqnarray*}
        (f_\ast \vect)(\nu^1,\ldots,\nu^r, \vecw_1,\ldots,\vecw_s)
        & := & \prod_{i=1}^r h^* ( \vecv_i(f^\ast \nu^i)) \cdot \prod_{j=1}^s \vecw_j(h^\ast \omega^j)
        \end{eqnarray*}
        is called the \emph{push-forward} of the tensor field $\vect$.
        \item[b)]
        Let $\vect \in (\ct_{Y/S})_s^0, \; \vect=\omega^1\otimes\ldots\otimes\omega^s$,
        be a tensor field consisting of tensor powers of differential forms,
        and let $ \vecw_1, \ldots, \vecw_s \in (\Omega_{X/S}^1)^{\vee}$. Then
        \begin{eqnarray*}
        (f^* \vect)(\vecw_1,\ldots,\vecw_s)
        & := & \prod_{j=1}^s \vecw_j(f^\ast \omega^j)
        \end{eqnarray*}
        is called the \emph{pull-back} of the tensor field $\vect$.
        \item[c)]
        Let $\vect\in (\ct_{Y/S})_s^r$ and let $f$ be an isomorphism.
        Then $f^\ast \vect:=(f^{-1})_\ast \vect$ is called the
        \emph{pull-back} of the tensor field $\vect$.
    \end{enumerate}
\end{Def}

\subsection{Some formulas}\label{1350}

Within this subsection let us collect some formulas concerning the
pull-back or push-forward with respect to some types of morphisms.
More precisely let us consider
\begin{enumerate}
    \item
    constant morphism.

    \item
    morphisms of the type $(f,g)$ or $f \times g$ involving fibre products.
\end{enumerate}

\begin{Satz}\label{1352}
    Let $f:X\to Y$ be a morphism of smooth $S$-schemes which
    allows a section $h:Y \hookrightarrow X$. Let $S' \to S$ be
    an unramified morphism. Assume that there is a factorization
    \begin{align*}
        \xymatrix{ X \ar[r]^f \ar[d]_{p} &  Y  \\
        S' \ar[ur]_s }
    \end{align*}
    Then $f^* \omega = 0$ for all $\omega \in \Omega_{Y/S}^1$. In
    particular, $f_\ast \vecv = 0$ for all $\vecv \in
    \Gamma(X,T_{X/S})$, and this equation holds for any
    choice of the section $h$.
    (Recall that $f_* \vecv$ denotes the push-forward of $\vecv$ under
    $f$ with respect to $h$ in the sense of Definition \ref{1327}.)
\end{Satz}

\begin{proof}
    $(f_\ast \vecv)(\omega)=h^*(\vecv(f^\ast \omega))$ for all
    $\omega\in\Omega_{Y/S}^1$. Thus it suffices to show that $f^\ast
    \omega=0$ for all $\omega\in\Omega_{Y/S}^1$. The above diagram
    induces a commutative diagram
    \begin{align*}
        \xymatrix{ f^\ast\Omega_{Y/S}^1 \ar[r]  &  \Omega_{X/S}^1 . \\
        p^\ast (s^\ast\Omega_{Y/S}^1) \ar@{=}[u] \ar[r] & \pi^\ast
        \Omega_{S'/S}^1 \ar[u]  }
    \end{align*}
    By Proposition \ref{0020}, $\Omega_{S'/S}^1=0$ and we are done.
\end{proof}
The ``adelic'' \st points are elements of $X(S)$. Thus we see that
constant maps $X(S) \to Y(S)$ have zero differential (as we are
used to from differential geometry). Let us illustrate this in
more detail

\begin{Def}\label{1353}
    Let $f:X \to Y$ be a  morphism of smooth $S$-schemes.
    \begin{enumerate}
        \item
        $f$ is called \emph{constant}, if there exists a factorization
        \begin{align*}
            \xymatrix{ X \ar[r]^f \ar[d]_{p} &  Y  \\
            S \ar@{^{(}->}[ur]_s }
        \end{align*}
        where $s:Y \hookrightarrow Y$ is a section of the
        structure morphism $Y \to S$.

        \item
        $f$ is called \emph{locally constant}, if any point $y \in Y$
        possesses an \'etale neighborhood $V$, such that $f \times \text{id} : X \times_Y V \to
        V$ is constant.
    \end{enumerate}
\end{Def}

\begin{Satz}\label{1355}
    Let $f:X \to Y$ a smooth morphism of smooth $S$-schemes. Let
    $S$ be a Dedekind scheme and let $X$ and $Y$ be a N\'eron models of
    their generic fibre. Let $y$ be a physical point of $Y$ over
    $s \in S$ which lies in the image of $f$. Let $x \in X$
    be a physical point which is mapped to $y$ under $f$ (such a
    point exists by Proposition \ref{1323}).

    Then $x$ extends to an $S$-valued point of $X$,
    whose image under $f$ is an $S$-valued point $\alpha:S
    \hookrightarrow Y$ of $Y$ containing $y$.
    In particular, there exists a constant morphism $g: X \to Y$ which maps $X$ to $\alpha$.
    Furthermore, there exists an \'etale environment $V \to Y$ of
    $y$, such that the restriction
    $g|_V=g \times \text{id}_V: X \times_Y V \to V$ of $g$ induces a vanishing push-forward:
    \begin{align*}
        (g|_V)_\ast \vecv = 0 \quad \text{ for all }  \vecv \in
        \Gamma(X \times_Y V,T_{X \times_Y V/S}).
    \end{align*}
    With some abuse of notation, we denote the constant morphism $g$ by $y:X \to
    Y$.
\end{Satz}

\begin{proof}
    As we are working with respect to \'etale topology, the local
    rings are strictly henselian. Let us first prove that $x$ lifts to an $S$-valued point of $X$.
    If $s$ is the generic point of $S$, this is true due to the N\'eron mapping property (Definition \ref{0066}).
    Therefore let $s$ be a closed point of $S$. Then $x$ lifts to a
    $\co_{S,s}$-valued point $x'$ of $X$ by Proposition \ref{0030a} which itself may be lifted
    to an $S$-valued point $\beta$ of $X$ (again apply N\'eron mapping
    property in order to extend the point of $x'$ which lies over the generic point of
    $S$). Due to the uniqueness assertion of the N\'eron mapping
    property, the point of the image of $\beta$ which lies over $s$ is $x$. If
    $p: X \to S$ is the structure morphism, $\alpha \circ p$ is
    the searched constant morphism, where $\alpha:= f \circ \beta$.

    By Proposition \ref{1312}, there exists an \'etale open environment $V \to
    Y$ of $y$, such that $f|_V=f \times \text{id}: X \times_Y V \to
    V$ admits a section $h|_V$. By construction, $h|_V$ is also a
    section of $g|_V$. Thus the push-forward of vector fields under $g|_V$
    exists. In order to prove that it is zero, it suffices to show that
    the corresponding pull-back of differential forms is zero, because
    $(g|_V)_*\vecv= (h|_V)^* \circ \vecv \circ (g|_V)^*$. Like in the proof of
    Proposition \ref{1352}, the diagram
    \begin{align*}
        \xymatrix{ X \times_Y V \ar[d] \ar[r] & V \ar[d]^{\varphi_V} \\
            X  \ar[d] \ar[r] & Y \\
            S \ar[ur]
        }
    \end{align*}
    shows that the pull-back of differential forms $(g|_V)^*$ is
    zero. Thereby one has to use that $(\varphi_V)^*$ is an
    isomorphism, because $\varphi_V$ is \'etale.
\end{proof}

\begin{Satz}\label{1356}
    Let $f:X\to Y, \; g:X\to Z$ be smooth morphisms of $S$-schemes.
    Let $x\in X$ be a physical point and also denote by $f(x)$ (resp.
    $g(x)$) the constant map $X\to Y, \; q\mapsto f(x)$ (resp. $X\to
    Z, \; q\mapsto g(x)$) (see Proposition \ref{1355}). Then in the stalk at
    $(f(x),g(x))=(f,g)(x)$ there holds the equation
    \begin{align*}
        (f,g)_\ast = (f(x),g)_\ast + (f,g(x))_\ast.
    \end{align*}
\end{Satz}

\begin{proof}
    We have a commutative diagram
    \begin{align*}
        \xymatrix{ & Y \\
        X \ar[r]^{(f,g) \ \, } \ar[ur]^f \ar[dr]_g  & Y \times_S Z \ar[u]_{p_1} \ar[d]^{p_2} \\
        & Z
        }
    \end{align*}
    Therefore, we also get a commutative diagram
    \begin{align*}
        \xymatrix{ (f,g)^\ast \Omega_{Y\times_S Z/S}^1 \ar[r] &
        \Omega_{X/S}^1 \\
        (f,g)^\ast (p_1^\ast \Omega_{Y/S}^1 \oplus p_2^\ast
        \Omega_{Z/S}^1)  \ar@{=}[r] \ar[u]^{\wr} &  f^\ast \Omega_{Y/S}^1\oplus g^*
        \Omega_{Z/S}^1 \ar[u] \\
        p_1^\ast \omega_1+p_2^\ast\omega_2 \ar@{|->}[r] &
        f^\ast \omega_1 + g^\ast \omega_2 \\
        (\omega_1,\omega_2)  \ar@{=}[r] \ar@{|->}[u] &  (\omega_1,\omega_2) \ar@{|->}[u]
        }
    \end{align*}
    Thus we can write:
    \begin{align*}
        (f,g)^\ast (\omega_1,\omega_2)=f^\ast\omega_1+g^\ast \omega_2.
    \end{align*}
    In particular, this equation holds in the stalk at $(f(x),g(x))$.
    But in the stalk at $(f(x),g(x))$, we obtain analogously equations
    \begin{align*}
        (f(x),g)^\ast (\omega_1,\omega_2)   &=f(x)^\ast\omega_1+g^\ast \omega_2
        =g^* \omega_2 \quad {and} \\
        (f,g(x))^\ast (\omega_1,\omega_2)   &=f^\ast\omega_1+g(x)^\ast
        \omega_2 =f^\ast\omega_1.
    \end{align*}
    Thereby, we made use of the fact that the pull-back of
    differential forms under constant morphisms is zero
    (Proposition \ref{1352}). This finishes the proof.
\end{proof}

\begin{Satz}\label{1357}
    Let $f:X\to Y$ and $g:Z\to T$ be smooth morphisms of $S$-schemes.
    Let $x\in X$ and $z\in Z$ be physical points and also denote by
    $f(x)$ (resp. $g(z)$) the constant maps $X\to Y, \; q\mapsto f(x)$
    (resp. $Z\to T, \; q\mapsto g(z)$). Then in the stalk at $(f(x),g(z))=(f\times
    g)(x,z)$ there holds the equation:
    \begin{align*}
        (f\times g)_\ast = (f(x)\times g)_\ast + (f\times g(z))_\ast.
    \end{align*}
\end{Satz}

\begin{proof}
    We have a commutative diagram
    \begin{align*}
        \xymatrix{ X \ar[r]^f & Y \\
        X\times_S Z \ar[r]^{ f \times g \ \, } \ar[u]^{\pi_1} \ar[d]_{\pi_2}  & Y \times_S T \ar[u]_{p_1} \ar[d]^{p_2} \\
        Z \ar[r]^g & T
        }
    \end{align*}
    Therefore, we also get a commutative diagram
    \begin{align*}
        \xymatrix{ (f\times g)^\ast \Omega_{Y\times_S T/S}^1 \ar[r] & \Omega_{X\times_S Z/S}^1 \\
        (f\times g)^\ast (p_1^\ast \Omega_{Y/S}^1 \oplus p_2^\ast \Omega_{T/S}^1)
        \ar[r] \ar[u]^{\wr} \ar@{=}[d]
        &  \pi_1^\ast \Omega_{X/S}^1 \oplus \pi_2^\ast \Omega_{Z/S} \ar[u]^{\wr}  \\
        \pi_1^\ast f^\ast \Omega_{Y/S}^1 \oplus \pi_2^\ast g^\ast
        \Omega_{T/S}^2 \\
        p_1^\ast\omega_1+p_2^\ast\omega_2 \ar@{|->}[r] & (f\times g)^\ast p_1^\ast\omega_1 + (f\times g)^\ast p_2^\ast\omega_2 \\
        (\omega_1,\omega_2)
        \ar@{|->}[r] \ar@{|->}[u] \ar@{=}[d]
        &  (f^\ast \omega_1, g^\ast \omega_2) \ar@{|->}[u]  \\
        (\omega_1,\omega_2)
        }
    \end{align*}
    Thus we can write:
    \begin{align*}
        (f\times g)^\ast
        (\omega_1,\omega_2)=(f^\ast\omega_1,g^\ast\omega_2)=\pi_1^\ast
        f^\ast\omega_1+\pi_2^\ast g^\ast\omega_2
    \end{align*}
    In particular, this equation holds in the stalk at $(f \times
    g)(x,z)=(f(x),g(z))$. But in the stalk at $(f(x),g(z))$, we obtain
    analogously equations
    \begin{align*}
        (f(x) \times g)^\ast (\omega_1,\omega_2)
        &=\pi_1^\ast f(x)^\ast\omega_1+\pi_2^\ast g^\ast\omega_2
        = \pi_2^\ast g^\ast\omega_2 \quad \text{and} \\
        (f \times g(z))^\ast (\omega_1,\omega_2)
        &=\pi_1^\ast f^\ast\omega_1+\pi_2^\ast g(z)^\ast\omega_2
        = \pi_1^\ast f^\ast\omega_1.
    \end{align*}
    Thereby, we made use of the fact that the pull-back of
    differential forms under constant morphisms is zero
    (Proposition \ref{1352}). This finishes the proof.
\end{proof}

                                        \chapter{Gauge theory}\label{1400}

                                        \section{The classical, differential geometric theory}\label{1401}

                                        The fundamental object underlying classical, differential
                                        geometric Yang-Mills theory are principal bundles. Before we state
                                        the definition of the algebraic geometric analogue of principle
                                        bundles in the next section \ref{1410}, let us briefly recall
                                        some differential geometric notions which are essential for gauge
                                        theory. Throughout this section let $G$ be a Liegroup, and let $P$
                                        be a differentiable manifold.

                                        \begin{Def}\label{1402}
                                            Let $\psi: P\times G\to P$ be a differentiable map and denote by $\psi_g: P \to P$ resp.
                                            $\psi_p : G \to P$ the induced maps $\psi_g(\cdot):=\psi( \cdot
                                            ,g)$ resp. $\psi_p(\cdot):=\psi(p,\cdot)$ where $g \in G$ and $p \in P$. Let us also introduce
                                            the abbreviation $p \cdot g := \psi(p,g)$.
                                            Then $\psi$ is called a {\it right group action} if and only if:
                                            \begin{itemize}
                                                \item[a)]
                                                $\psi_g:P\to P$ is a diffeomorphism for all $g\in G$.

                                                \item[b)]
                                                $\psi_a\circ\psi_b(p)=\psi_{ba}(p)$; i.e.: $(p\cdot b)\cdot a=p\cdot (b\cdot a)$.
                                            \end{itemize}
                                        \end{Def}

                                        \begin{Def}\label{1403}
                                            A group action $\psi: P\times G\to P$ is called {\it free}, if and
                                            only if the following holds: If there exists a $p \in P$ and a $g
                                            \in G$ such that $p \cdot g = p$, then $g=e$ is the unit element
                                            of $G$.
                                        \end{Def}

                                        \begin{Satz}\label{1404}
                                            Let $G$ act on $P$ on the right by means of $\psi: P\times G\to P$. Let $A_e\in \Liealg$, where
                                            $\Liealg$ denotes the Lie-algebra of $G$, and let $\exp: \Liealg
                                            \to G$ be the exponential map. Then $A_e$ induces a vectorfield
                                            $(A_e)_\ast$ on $P$, the so called \emph{{ killing-vectorfield}}
                                            or \emph{{fundamental vectorfield}}. More precisely:
                                            \begin{align*}
                                                \left((A_e)_\ast f\right)(p)
                                                =\frac{d}{dt} \Big|_{t=0} \left(f\circ \psi_{\exp (t A_e)} (p)\right)
                                            \end{align*}
                                            for all $f\in C^\infty (P)$ and $p \in P$.
                                        \end{Satz}
                                        \begin{Rem}\label{1405}
                                            Let $G$ act on $P$ on the right by means of $\psi: P\times G\to P$. Denote
                                            by $0 \in\Gamma(P,TP)$ the zero section of tangent bundle $TP$ of
                                            $P$. Let $A_e\in\Liealg$ and denote by $A$ the unique left
                                            invariant vectorfield on $G$ corresponding to $A_e$. Then we may consider
                                            the vectorfield $0 \times A$ living on $P \times G$, and we
                                            get for all $p \in P$:
                                            \begin{eqnarray*}
                                                \left((\psi_p)_\ast A_e\right)(p)
                                                =  {\left((A_e)_\ast\right)}(p)
                                                = \left(\psi_\ast (0\times A)\right)(p),
                                            \end{eqnarray*}
                                            where $\Big(\psi_\ast (0 \times A)\Big)(p)=\left((\psi_\ast)_{(p \cdot g^{-1},g)} (0_{p \cdot g^{-1}}\times
                                            A_{g})\right)(p)$ for all $g \in G$.
                                        \end{Rem}

                                        \begin{proof}
                                            Denote by $\psi_p:G \to P$, $g \mapsto \psi_p(g):=p \cdot g$ the translation by
                                            $p$. For all $g \in G$ let $A_g:=A(g) \in T_gG$ be the tangent vector of $A$ at $g$, and consider
                                            the integral curve $g(t):= g \cdot \exp(tA_e)= L_g \circ \exp
                                            (tA_e)$ representing $A_g$ (where $L_g$ is the left translation by
                                            $g$). Then we obtain
                                            \begin{align*}
                                                \Big(((\psi_p)_*A_g)(f)\Big)(p\cdot g) =\frac{d}{dt}
                                                \Big|_{t=0}f \circ \psi_p \circ g(t)
                                            \end{align*}
                                            for all $f\in C^\infty (P)$. In particular
                                            \begin{align*}
                                                \Big(((\psi_p)_*A_e)(f)\Big)(p)
                                                &=  \frac{d}{dt}\Big|_{t=0}f \circ \psi_p \left(
                                                    \exp(tA_e)\right)
                                                =  \frac{d}{dt}\Big|_{t=0}f \circ \psi_{\exp(tA_e)}(p)
                                                =  {\left((A_e)_\ast f \right)}(p).
                                            \end{align*}
                                            This shows the first equality. On the other hand, $(0 \times A)(p,g)=0_p \times
                                            A_g$ for all $(p,g) \in P \times G$. If $p(t)$ is an integral
                                            curve of $0_p$, i.e. $p(t)=p$ for all $t \in \real$, then we
                                            get for all $f\in C^\infty (P)$:
                                            \begin{align*}
                                                \Big(\psi_*(0_p \times A_g)(f) \Big)(p\cdot g)
                                                &=  \frac{d}{dt}\Big|_{t=0}f \circ \psi(g(t),p(t))
                                                =  \frac{d}{dt}\Big|_{t=0}f \circ \psi(g(t),p)\\
                                                &=  \frac{d}{dt}\Big|_{t=0}f \circ \psi_p(g(t))
                                                =  ((\psi_p)_*A_g)(f)(p\cdot g).
                                            \end{align*}
                                            As $\psi$ is a group action, we know that $(p\cdot g) \cdot h = p \cdot
                                            (gh)$ for all $p\in P$ and $g,h \in G$. It follows that
                                            \begin{align*}
                                                \psi_{p \cdot h} \circ (h^{-1}g)(t)
                                                &=  \psi_{p \cdot h}(h^{-1}g \cdot \exp(tA_e))
                                                =   (p \cdot h) \cdot (h^{-1}g \cdot \exp(tA_e))\\
                                                &=   p \cdot (g \cdot \exp(tA_e))
                                                =   \psi_p \circ g(t),
                                            \end{align*}
                                            and thus we obtain
                                            \begin{align*}
                                                ((\psi_p)_*A_g)(f)(p\cdot g)
                                                &=  \frac{d}{dt}\Big|_{t=0}f \circ \psi_p(g(t))
                                                &=  \frac{d}{dt}\Big|_{t=0}f \circ \psi_{p\cdot g}(\exp(tA_e))
                                                &=  {\left((A_e)_\ast f \right)}(p\cdot g).
                                            \end{align*}
                                            All in all we see that
                                            \begin{align*}
                                                \Big(\psi_*(0_p \times A_g) \Big)_{p \cdot g}(f)
                                                &=  {\left((A_e)_\ast \right)}_{p\cdot g}(f)
                                                =  \Big(\psi_*(0_{p\cdot h} \times A_{h^{-1}g}) \Big)_{p \cdot g}(f)
                                            \end{align*}
                                            for all $p\in P$ and $g,h \in G$. Thus it indeed makes sense
                                            to write
                                            \begin{align*}
                                                \Big(\psi_*(0 \times A) \Big)(p \cdot g)=\psi_*(0_p \times
                                                A_g).
                                            \end{align*}
                                        \end{proof}
                                        \section{Torsors}\label{1410}

                                        Now we are going to transfer these structures to algebraic
                                        geometry, in order to provide the basis for a Yang-Mills theory
                                        over arbitrary commutative rings. Let $G$ be a group scheme over
                                        $X$ and $P$ and $X$-scheme. Then an action of $G$ on $P$ is a
                                        $X$-morphism
                                        \begin{align*}
                                            \psi: P\times_X G \to P
                                        \end{align*}
                                        that induces an action of the group $G(T)$ on the set $P(T)$ for
                                        all $S$-scheme $T$ (especially $(p\cdot g)\cdot h=p\cdot (g \cdot
                                        h)$ for all $p\in P(T)$, $g,h\in G(T)$).

                                        \begin{Satz}\label{1411}
                                            Let $G$ act on a scheme $P$. Let $\pi:P\to X$ be the structure
                                            morphism. Then it is equivalent:
                                            \begin{enumerate}
                                                \item[a)]
                                                The scheme $P$ is faithfully flat and locally of finite-type over $X$ and there is an isomorphism
                                                \begin{align*}
                                                    P\times_X G \stackrel{\sim}{\longrightarrow} P \times_X P,
                                                    \quad (p,g)\mapsto (p,p\cdot g).
                                                \end{align*}

                                                \item[b)]
                                                There is a covering $(U_i\to X)$ for the flat topology
                                                \footnote{See Definition \ref{1305} for the notion of a Grothendieck topology.}
                                                on $X$, such that, for each $i$, there is an isomorphism
                                                \begin{align*}
                                                    \chi_i: \pi^{-1} (U_i):=P_{U_i}:=U_i\times_X P\to U_i \times_X G=: G_{U_i}
                                                \end{align*}
                                                respecting the $G_{U_i}$-action.
                                            \end{enumerate}
                                        \end{Satz}

                                        \begin{proof}
                                            \Milne, Chap. III, Prop. 4.1
                                        \end{proof}

                                        \begin{Rem}\label{1412} \
                                            \begin{itemize}
                                            \item
                                            Let $G$ be a smooth group scheme. Then any torsor under $G$ is smooth
                                            (see Proposition \ref{1414}). Using the fact that any $X$-torsor $P$ under $G$, which allows a $X$-section,
                                            is trivial, Proposition \ref{0030} shows that we may choose the
                                            covering $(U_i\to X)$ in $b)$ for the \'etale topology.
                                            This is important in regard to physical applications as we already
                                            saw in section \ref{1301} that we have to work with respect to \'etale topology in
                                            order to incorporate the principle of general covariance.
                                            In physical situations we will furthermore assume that there exists a covering
                                            $(U_i\to X)$ consisting of only  finitely many maps $U_i \to
                                            X$. This will allow us to glue sheaves. Then the
                                            global  and the family of local versions of a physical theory are
                                            equivalent (as one would expect).
                                            \item
                                            Condition $b)$ says that there are commutative diagrams \\
                                            \begin{align*}
                                                \xymatrix @-0.2pc { \pi^{-1}(U_i) \ar[dr]_{\pi|_{U_i}:=\text{id}_{U_i} \times \pi}
                                                \ar[rr]^{\chi_i} & &  U_i \times_X G \ar[dl]^{pr_1} \\
                                                & U_i  }
                                            \end{align*}

                                            \begin{align*}
                                                \xymatrix  {  P_{U_i}\times_{U_i} G_{U_i} \ar@{=}[r]  &
                                                U_i\times_X P\times_X G \ar[d]_{\text{id} \times \psi} \ar[r]^{\chi_i \times \text{id}} &
                                                U_i \times_X G\times_X G \ar@{=}[r]  \ar[d]^{\text{id} \times m}
                                                & G_{U_i} \times_{U_i} G_{U_i}  \\
                                                & \quad U_i \times_X P \quad \ar[r]^{\chi_i}  & \quad U_i \times_X G \quad  }
                                            \end{align*}
                                            where $m:G\times_X G\to G$ denotes the group multiplication.
                                            Evaluated at $T$-valued points $p$ of $\pi^{-1}(U_i)$ and $g$ of $G$ (where $T$ is an $S$-scheme), the
                                            latter commutative diagram may be stated as follows:
                                            \begin{align*}
                                                (pr_2 \circ \chi_i) (p\cdot g)=(pr_2\circ\chi_i(p))\cdot g.
                                            \end{align*}
                                            Thereby $pr_2$ denotes the canonical projection $U_i \times_X G
                                            \to G$.

                                            \item Condition $a)$ implies that the group action $\psi:
                                            P\times_X G\to P$ is free (see Definition \ref{1403}) and transitive on fibres over $X$. This
                                            may be seen as follows: The morphism $ P\times_X G
                                            \stackrel{\sim}{\longrightarrow} P\times_X P$, $(p,g)\mapsto
                                            (p,p\cdot g)$ is an isomorphism by definition. Due to injectivity,
                                            the preimage of a pair $(p, p \cdot g)$ with $p \cdot g =p$
                                            consists of the pair $(p,e)$, where $e$ denotes the unit element
                                            of $G$. This shows that the action is free.
                                            The surjectivity implies that for all $p,p' \in P$ with
                                            $\pi(p)=\pi(p')$ (i.e. for all $(p,p')\in P\times_X P)$ there
                                            exists a $g\in G$ such that  $p'=pg$.
                                            \end{itemize}
                                        \end{Rem}

                                        \begin{Def}\label{1413}
                                            A scheme $P$, on which acts $G$, and that satisfies the
                                            equivalent statements of Proposition \ref{1411} is called a {\it principal homogeneous space}
                                            or an $X${\it-torsor under} $G$.
                                        \end{Def}
                                        In physical applications we are going to consider smooth schemes
                                        provided with \'etale topology. Thus we see that the notion of
                                        torsors is exactly corresponds to the $G$-principal bundles in
                                        ordinary real-valued differential geometry.

                                        \begin{Satz}\label{1414}
                                            If $G$ is smooth respectively \'etale over $X$, then so also any
                                            $X$-torsor under $G$.
                                        \end{Satz}

                                        \begin{proof}
                                            \Milne, Chap. III, Prop. 4.2
                                        \end{proof}

                                        \begin{Satz}\label{1415}
                                            Let us denote the set of all isomorphism classes of $X$-torsors
                                            under $G$ by $PHS(G/X)$. Then there is a canonical injection
                                            $PHS(G/X) \hookrightarrow \check{H}^1(X_{fl},G)$. This map is
                                            an isomorphism if $G$ is affine over $X$ (i.e. the preimage
                                            of every affine subscheme of $X$ is an affine subscheme of $G$).
                                        \end{Satz}

                                        \begin{proof}
                                            Let us shortly summarize how a $X$-torsor $P$ under $G$ gives
                                            rise to a 1-cocycle. For the details of the proof see
                                            \Milne, Chap. III, Cor. 4.7. As we are interested in physical
                                            applications let us directly work with \'etale topology. For
                                            this purpose we assume that $G$ is smooth. Then $P$ is smooth,
                                            too, by Proposition \ref{1414}, and due to Proposition \ref{0030} we may assume
                                            that there exists an \'etale covering $\mathfrak{U}:=(U_i \to X)$ that
                                            trivializes $P$, so that $P(U_i)$ is non-empty for all $i$.
                                            Then choose a section $s_i \in P(U_i)$ for all $i$. As the
                                            chosen covering trivializes $P$, there exist unique $g_{ij} \in
                                            G(U_{ij})$ (where $U_{ij}:=U_i \times_X U_j$), such that
                                            $s_i|_{U_{ij}} \cdot g_{ij} = s_j|_{U_{ij}}$. Omitting
                                            restriction signs we therefore obtain $s_ig_{ij}g_{jk}=s_k=s_i
                                            g_{ik}$, and so $g_{ij}g_{jk}=g_{ik}$, because the
                                            $G$-action on $P$ is free. One says that the family $(g_{ij})$ is a
                                            1-cocycle. If  $s'_i \in P(U_i)$ is another family of
                                            sections, then there is a family $(h_i)$, $h_i \in G(U_i)$,
                                            such that $s_i=s'_i \cdot h_i$ for all $i$, because the chosen
                                            covering trivializes $P$. Again omitting restriction signs, we get
                                            $s'_i  h_i g_{ij}=s_i  g_{ij} = s_j = s'_j  h_j = s'_i  g'_{ij}
                                            h_j$, and so $g'_{ij}=h_i g_{ij} h_j^{-1}$. One says that $(g_{ij})$
                                            and $(g'_{ij})$ are cohomologous 1-cocycles. The property of 1-cocycles to be
                                            cohomologous is an equivalence relation, and the set of
                                            cohomology classes is denoted by $\check{H}^1(\mathfrak{U},G)$.
                                            Of course the cohomology class is unaltered if $P$ is replaced by an isomorphic torsor.
                                            Thus $P$ defines an element of $\check{H}^1(X_{\acute{e}t},G)$ which is defined to
                                            be $\varinjlim\limits \check{H}^1(\mathfrak{U},G)$ where the
                                            limit runs over all coverings $\mathfrak{U}$ obtained by refinement.
                                        \end{proof}

                                        \begin{S-valued-point}\label{1370} \ \\
                                        Let $X$ be a smooth $S$-scheme, let $G$ be a smooth $S$-group
                                        scheme and let $G_X:= G \times_S X$. Let $\psi: P {\times}_X G_X
                                        \to P$ be the group action of $G_X$ on a $X$-torsor $P$ under
                                        $G_X$. Consider an $S$-valued point $\alpha: S \hookrightarrow P$
                                        of $P$.
                                        Let $f:P\to P$ and $g:P\to G$ be $S$-morphisms. Then $f_{\alpha}
                                        :=f\circ\alpha$ and $g_{\alpha}:=g\circ\alpha$ are $S$-valued
                                        points, too.

                                        The group action induces  two kinds of translation by $S$-valued
                                        points. First, we consider the translation $ \psi_{f(\alpha)}$ by
                                        the  $S$-valued point  $f_{\alpha}$ of $P$:
                                        \begin{eqnarray*}
                                            \psi_{f(\alpha)}: & & G \stackrel{(\pi,id)}{\longrightarrow} S \times_S G
                                            \stackrel{f_{\alpha} \times id}{\longrightarrow}
                                            P {\times}_S G \stackrel{\psi}{\longrightarrow} P.
                                        \end{eqnarray*}
                                        In formulas we may write $\psi_{f(\alpha)} :=  \psi \circ
                                        (f_{\alpha} \times \mbox{id})\circ (\pi,\mbox{id}) = \psi \circ
                                        (f_{\alpha} \circ \pi, \mbox{id})$, where $\pi : G \to S$ is the
                                        canonical morphism. If  $f(\alpha)$ denotes the constant map
                                        making the  diagram
                                        \begin{align*}
                                            \xymatrix{ G \ar[r]^{f(\alpha)} \ar[d]^{\pi} & P \\
                                            S \ar[ur]_ {f\circ\alpha}
                                            }
                                        \end{align*}
                                        commutative, then we may state the translation $
                                        \psi_{f(\alpha)}:$ in the following way:
                                        \begin{align*}
                                            \psi_{f(\alpha)}=\psi \circ (f(\alpha),\mbox{id}): G \to P.
                                        \end{align*}
                                        Analogously we define the translation $ \psi_{g(\alpha)} $ by the
                                        $S$-valued point $g_{\alpha}$ of $G$:
                                        \begin{align*}
                                            \psi_{g(\alpha)}: P \stackrel{(\mbox{\scriptsize
                                            id},\pi)}{\longrightarrow} P \times_S S
                                            \stackrel{\mbox{\scriptsize id}\times g_{\alpha}}{\longrightarrow}
                                            P \times_S G \stackrel{\psi}{\longrightarrow} P,
                                        \end{align*}
                                        where $\pi$ is this time the canonical morphism $P \to S$. So we
                                        get $\psi_{g(\alpha)}: =  \psi\circ (\mbox{id}\times
                                        g_{\alpha})\circ (\mbox{id},\pi)  =  \psi\circ
                                        (\mbox{id},g_{\alpha}\circ \pi)$. Using the constant map
                                        $g(\alpha)$ making the diagram
                                        \begin{align*}
                                            \xymatrix{ P \ar[r]^{g(\alpha)} \ar[d]^{\pi} & G \\
                                                S \ar[ur]_ {g\circ\alpha}
                                        }
                                        \end{align*}
                                        commutative, we arrive at the identity
                                        \begin{align*}
                                        \psi_{g(\alpha)}=\psi \circ (\mbox{id},g(\alpha)): P \to P.
                                        \end{align*}
                                        Furthermore, if $h_P: P\to P$ and $h_G:P\to G$ are morphisms of
                                        $S$-schemes, we get identities:
                                        \begin{eqnarray*}
                                        \psi_{f(\alpha)} \circ h_G & = & \psi\circ (f(p),h_G). \\
                                        \psi_{g(\alpha)} \circ h_P & = & \psi \circ (h_P,g(p)).
                                        \end{eqnarray*}

                                        \end{S-valued-point}

                                        \section{Gauge transformations}\label{1420}

                                        Throughout this section let $X$ be a smooth $S$-scheme, and let
                                        $G$ be a smooth $X$-group scheme.

                                        \begin{Def}\label{1421}
                                        Let $P$ be a $X$-torsor $P$ under $G$. A $X$-automorphism
                                        $\hat{\theta}: P \to P$ is called a {\it gauge transformation} or
                                        a {\it vertical automorphism}, if $\hat{\theta}$ is compatible
                                        with the group action $\psi$ of $G$ on $P$; i.e., there are the
                                        following commutative diagrams.
                                        \begin{align*}
                                            \xymatrix @-0.2pc {   P\times_X G \ar[d]_{\hat{\theta}\times id} \ar[r]^{\quad \psi} &
                                            P \ar[d]^{\hat{\theta}}  & &
                                            \hat{\theta}:P \ar[dr] \ar[rr]^{\sim} & &  P \ar[dl] \\
                                            P\times_X G \ar[r]^{ \quad \psi}  &  P  & & & X  }
                                        \end{align*}
                                        We denote be the group of gauge transformations by $\cg$.
                                        \end{Def}
                                        Furthermore, let us call a morphism $u:P \to G$
                                        \emph{equivariant}, if $u(p\cdot g)=g^{-1} u(p)g$ for all
                                        $T$-valued points $p \in P(T)$ and $g \in G(T)$ and for all
                                        $S$-schemes $T$.

                                        \begin{Satz}\label{1422}
                                            There is a one-to-one correspondence between gauge
                                            transformations and equivariant $X$-morphisms $u: P\to G$.
                                        \end{Satz}

                                        \begin{proof}
                                            As usual let us denote the formation of the inverse respectively
                                            the multiplication on the group scheme $G$ by $\iota$ respectively $m$.
                                            Let assume that an equivariant $X$-morphisms $u$ is given.
                                            Then we define the following $X$-morphism $\hat{\theta}_u$.
                                            \begin{align*}
                                                \begin{array}{ccccccc}
                                                    \hat{\theta}_u: \ \ P &
                                                    \stackrel{(\text{id},u)}\longrightarrow & P \times_X G &
                                                    \stackrel{\psi}{\longrightarrow}
                                                    & P \\[1ex]
                                                    \qquad  p & \mapsto & (p,u(p)) &
                                                    \mapsto & p \cdot u(p)
                                                \end{array}
                                            \end{align*}
                                            Of course, $\hat{\theta}_u$ is a $X$-isomorphism with inverse
                                            $\hat{\theta}_{u^{-1}}$, where $u^{-1}:= \iota \circ u: G \to
                                            G$. Furthermore $\hat{\theta}_u$ is compatible with the
                                            group action of $G$ on $P$. This may be seen as follows:
                                            By assumption, $u(p \cdot g)=g^{-1}u(p)g$ and therefore
                                            \begin{align*}
                                                \hat{\theta}_u(p \cdot g)= (p \cdot g) \cdot u(p \cdot g)
                                                = (p \cdot g) \cdot (g^{-1} u(p)g) = (p \cdot u(p)) \cdot
                                                g = \hat{\theta}_u(p) \cdot g.
                                            \end{align*}
                                            Thus $\hat{\theta}_u$ is indeed a gauge transformation.

                                            Conversely, let now a gauge transformation $\hat{\theta}$ be
                                            given. Then we define the following $X$-morphism $u_{\hat{\theta}}$.
                                            \begin{align*}
                                                \begin{array}{ccccccc}
                                                    u_{\hat{\theta}}: \ \ P &
                                                    \stackrel{id \times \hat{\theta}}\longrightarrow & P \times_X P &
                                                    \stackrel{\sim}{\longrightarrow}
                                                    & P \times_S G & \stackrel{pr_2}{\longrightarrow} & G \\[1ex]
                                                \end{array}
                                            \end{align*}
                                            Thereby $P \times_X P  \stackrel{\sim}{\longrightarrow} P
                                            \times_S G$ denotes the inverse of the canonical isomorphism induced
                                            by the group action $\psi$ (see Proposition \ref{1411} $a)$). Let us show that this morphism is
                                            equivariant. By definition of $u_{\hat{\theta}}$, we have got
                                            \begin{align*}
                                                \text{\small{$(p \cdot g, \hat{\theta}(p \cdot g))= (p \cdot g, \hat{\theta}(p)\cdot g)
                                                = \left(p \cdot g, (p \cdot u_{\hat{\theta}}(p))\cdot g \right)
                                                = \left(p \cdot g, (p \cdot g) \cdot (g^{-1} u_{\hat{\theta}}(p)g) \right) \in P \times_X P$}}.
                                            \end{align*}
                                            Thus the image of $p \cdot g$ under $u_{\hat{\theta}}$  is indeed $g^{-1}
                                            u_{\hat{\theta}}(p)g$. Furthermore, the given associations are inverse to each other:
                                            \begin{align*}
                                                p \cdot u_{\hat{\theta}_u}(p)
                                                = {\hat{\theta}_u}(p)
                                                = p \cdot u(p), \quad \text{ i.e. \ \ }   u_{\hat{\theta}_u}(p)
                                                = u(p) \text{ for all $p$},
                                            \end{align*}
                                            because the group action is free. On the other hand
                                            \begin{align*}
                                                \hat{\theta}_{u_{\hat{\theta}}}(p) = p \cdot u_{\hat{\theta}}(p) =\hat{\theta}(p)
                                                \quad  \text{ for all $p$}.
                                            \end{align*}
                                        \end{proof}

                                        \section{Horizontality}\label{1430}

                                        Throughout this section let $X$ be a smooth $S$-scheme, and let
                                        $G$ be a smooth $S$-group scheme. Furthermore let $X$ and $G$ be
                                        separated, let $G_X:=G \times_S X$ and consider an $X$-torsor $P$
                                        under $G_X$ (see Proposition \ref{1411}). Let $T$ be an arbitrary
                                        $S$-scheme. We already know from Proposition \ref{1414} that $P$ is smooth
                                        and separated over $X$ and therefore also over $S$. Denote by
                                        $\Liealg:=(\varepsilon^* \Omega_{G/S}^1)(S)$ the Lie-algebra of
                                        $G$, where $\varepsilon:S \hookrightarrow G$ denotes the unit
                                        section.

                                        By definition of torsors there is a commutative diagram
                                        \begin{align*}
                                            \xymatrix{
                                            & & P \\
                                            P \times_S G \ar@{=}[r]
                                            &   P \times_X G_X \ar[r]^{\sim}_{\pi_1 \times \psi} \ar[ur]^{\psi} \ar[dr]_{\pi_1}
                                            &   P \times_X  P \ar[d]^{p_1} \ar[u]_{p_2} \\
                                            & & P
                                            }
                                        \end{align*}
                                        where $\pi_i$ and $p_i$ are the canonical projections of the
                                        respective schemes onto the $i$-th factor, and where $\pi_1 \times
                                        \psi$ is an isomorphism. As usual, $\psi$ denotes the group
                                        action. Then the diagram shows that the diagonal $\Delta: P \to P
                                        \times_X P$ induces a canonical global section $s:= (\pi_1 \times
                                        \psi)^{-1} \circ \Delta$ of $\psi$ which is given by $p \mapsto
                                        (p,1)$ on $T$-valued points. Therefore, by Remark \ref{1326}, every
                                        global vector field $\vecv$ on $P \times_X G$ may be pushed
                                        forward to a global vector field $\psi_* \vecv$ on $P$ which is
                                        given by
                                        \begin{align*}
                                            \psi_* \vecv = s^* \circ \vecv \circ \psi^\ast.
                                        \end{align*}
                                        Drawing inspiration from Remark \ref{1405}, we use this push-forward,
                                        in order to define the killing-vector field. For this purpose
                                        consider an element $\vecv_e \in \Liealg$ and denote by $\vecv$
                                        the uniquely determined left-invariant vector field induced by
                                        $\vecv_e$ (see Proposition \ref{1241}). Furthermore, let $0 \in
                                        \Gamma(P,T_{P/S})$ be the vector field which is constant zero
                                        (i.e. the zero-section). Recalling the identity $T_{P \times_S
                                        G/S} = T_{P/S} \times_S T_{G/S}$ from Proposition \ref{1123} or as well the
                                        isomorphism $\pi_1^*\Omega_{P/S}^1 \oplus \pi_2^*\Omega_{G/S}^1
                                        \stackrel{\sim}\longrightarrow \Omega_{P \times_S G/S}^1$
                                        from Proposition \ref{0018}, we may interpret the cartesian product $0 \times
                                        \vecv$ as a vector field on $P \times_X G$. Then we define a map
                                        \begin{eqnarray*}
                                            \widetilde{\sigma}: \Liealg \ &  \longrightarrow & \Gamma(P, T_{P/S})=\Hom_{\co_{P}}(\Omega_{P/S}^1,\co_{P}), \\
                                            \vecv_e & \mapsto &  \psi_\ast (0 \times \vecv)
                                        \end{eqnarray*}
                                        which turns out to be $\co_S(S)$-linear. By tensoring over
                                        $\co_S(S)$ with $\co_P$, $\widetilde{\sigma}$ induces a
                                        $\co_P$-linear map
                                        \begin{eqnarray*}
                                            \sigma: \co_P \otimes_{\co_S(S)}  \Liealg & \longrightarrow & \Gamma(P,T_{P/S})=\Hom_{\co_P}(\Omega_{P/S}^1,\co_P) \\
                                            r \otimes \vecv_e  \quad \  & \mapsto & r\cdot \psi_\ast (0 \times
                                            \vecv) = r \cdot \widetilde{\sigma}(\vecv_e).
                                        \end{eqnarray*}

                                        \begin{Rem}\label{1430r}
                                            By Proposition \ref{1235}, $\Liealg$ carries the structure of a
                                            $\co_S(S)$-Lie-algebra by means of the $\co_S(S)$-linear
                                            Lie-bracket $[\cdot,\cdot]$. Therefore we may endow $\co_P \otimes_{\co_S(S)}
                                            \Liealg$ with the following structure of a
                                            $\co_P$-Lie-algebra: For $r \in \co_P$ and $\vecv_e \in
                                            \Liealg$ let
                                            \begin{align*}
                                                [r \otimes \vecv_e \, , \ \cdot \ ]: \co_P \otimes_{\co_S(S)}
                                                \Liealg \to \co_P \otimes_{\co_S(S)} \Liealg, \quad s
                                                \otimes \vecw_e \mapsto rs \otimes [\vecv_e,\vecw_e].
                                            \end{align*}
                                            This map is $\co_P$-linear and induces a Lie-bracket on $\co_P \otimes_{\co_S(S)}
                                            \Liealg$.
                                        \end{Rem}

                                        \begin{Satz}\label{1431}
                                            $\widetilde{\sigma}: \Liealg \to \Gamma(P,T_{P/S})$ is
                                            $\co_S(S)$-linear. In particular $\sigma$ is $\co_P$-linear.
                                        \end{Satz}

                                        \begin{proof}
                                        Let $\vecv_e, \vecw_e\in\Liealg$. The canonical map $\ve^\ast
                                        \Omega_{G/S}^1 \longrightarrow p_\ast \Omega_{G/S}^1$,
                                        $\vecv_e\mapsto \vecv$, extending sections in $\ve^\ast
                                        \Omega_{G/S}^1$ to left invariant sections of $\Omega_{G/S}^1$, is
                                        a $\co_S(S)$-linear map (as illustrated in the proof of
                                        Proposition \ref{1241} or simply by Proposition \ref{1241a}). In particular, we know
                                        that
                                        \begin{eqnarray*}
                                            \vecv_e+\vecw_e & \mapsto &  \vecv+\vecw     \qquad    \text{and} \\
                                            a \cdot \vecv_e & \mapsto &  a \cdot \vecv   \qquad \ \ \, \text{for all $a \in \co_S(S)$ }
                                        \end{eqnarray*}
                                        under this map. Due to the decomposition $\pi_1^*\Omega_{P/S}^1
                                        \oplus \pi_2^*\Omega_{G/S}^1 \stackrel{\sim}\longrightarrow
                                        \Omega_{P \times_S G/S}^1$ of Proposition \ref{0018}, we know that $(0
                                        \times (\vecv + \vecw))=(0 \times \vecv) + (0 \times \vecw)$. Thus
                                        we conclude that

                                        \begin{eqnarray*}
                                            \widetilde{\sigma}(a \vecv_e+ a' \vecw_e)(\omega)
                                            & = & \psi_\ast({0 \times (a \vecv+ a' \vecw)}) (\omega) \\
                                            & = & s^* \left(\left(0 \times (a \vecv+ a' \vecw)\right)(\psi^\ast\omega)\right) \\
                                            & = & s^*\left((0 \times a \vecv)(\psi^\ast \omega)\right)+ s^*\left(( 0 \times a' \vecw)(\psi^\ast \omega)\right) \\
                                            & = & s^*\left(a(0 \times \vecv)(\psi^\ast \omega)\right)+ s^*\left(a'( 0 \times \vecw)(\psi^\ast \omega)\right) \\
                                            & = & a \, s^*\left((0 \times \vecv)(\psi^\ast \omega)\right)+ a's^*\left(( 0 \times \vecw)(\psi^\ast \omega)\right) \\
                                            & = & a \, \widetilde{\sigma}(\vecv_e)(\omega) + a' \widetilde{\sigma}(\vecw_e)(\omega).
                                        \end{eqnarray*}
                                        for all $\omega \in \Omega_{P/S}^1$. This shows that
                                        $\widetilde{\sigma}$ is $\co_S(S)$-linear. As $\sigma(r \otimes
                                        \vecv_e)=r \cdot \widetilde\sigma(\vecv_e)$ by definition, we are
                                        done.
                                        \end{proof}

                                        \begin{Def}\label{1432}
                                            Let $P$ be a $X$-torsor under $G_X$. Let $\vecv_e \in
                                            \Liealg$. Then the vector field $\widetilde{\sigma}(\vecv_e)
                                            ={\sigma}(1  \otimes \vecv_e ) \in \Gamma(T_{P/S}/P)$ is called the
                                            \emph{killing-vector field} of the group action.
                                        \end{Def}
                                        Up to now, we considered the killing-vector field globally on all
                                        of $P$. Let us now, analyze the local properties of
                                        $\widetilde{\sigma}(\vecv_e)$.

                                        By means of the the canonical isomorphism stated in Proposition \ref{1135}
                                        a), let us interpret $\widetilde{\sigma}(\vecv_e)$ as a
                                        differential form. Locally (with respect to \'etale topology), $P$
                                        allows a split $P=X \times_S G$. So, as we are now interested in a
                                        local consideration, let us assume that $P$ splits and let
                                        \begin{align*}
                                            \rho_1&: X \times_S G \to X \\
                                            \rho_2&: X \times_S G \to G
                                        \end{align*}
                                        be the canonical projections. Due to the isomorphism
                                        $\rho_1^*\Omega_{X/S}^1 \oplus \rho_2^*\Omega_{G/S}^1
                                        \stackrel{\sim}\longrightarrow \Omega_{P/S}^1$ (Proposition \ref{0018}), we
                                        may decompose $\widetilde{\sigma}(\vecv_e)$ into a direct sum
                                        $(\omega_1,\omega_2) \in \rho_1^*\Omega_{X/S}^1 \oplus
                                        \rho_2^*\Omega_{G/S}^1$. We claim that the component $\omega_1$ is
                                        equal to zero. Therefore $\widetilde{\sigma}(\vecv_e)$ lives only
                                        in the group part $G$ of $P$ and is killed if it is pushed forward
                                        to the base $X$ using the structure morphism $\pi:P \to X$.

                                        Let us call the $X$-component in the local split $P=X \times_S G$
                                        the \emph{horizontal} component, and let us call the group part
                                        $G$ the \emph{vertical} component.
                                        In particular, we see that (locally) the killing vector field
                                        lives in the vertical component of $P$. But before defining the
                                        notion of horizontality and verticality for vector fields, let us
                                        first introduce the notion of a connection form.

                                        \begin{Def}\label{1437}
                                            Let $P$ be an $X$-torsor under $G_X$, let $\kil: \Liealg \to
                                            \Gamma(T_{P/S}/P)$ be the killing-form, and let $d$ be the exterior differential (see Theorem \ref{1156}).
                                            Furthermore, let $\omega$ be a global section $\con$ of $\Omega_{P/S}^1 \otimes_{\co_S(S)} \Liealg$. Then
                                            we introduce the following two $\co_p$-linear morphisms $\text{hor}, \text{ver}
                                            \in \Hom_{\co_p}(\Omega_{P/S}^1,\Omega_{P/S}^1)$:
                                            \begin{align*}
                                                \text{ver}&:= \sigma \circ \omega. \\
                                                \text{hor}&:= \text{id} - \sigma \circ \omega.
                                            \end{align*}
                                            $\omega$ is called a \emph{connection form}
                                            (or more shortly a \emph{connection}) if and only if:
                                            \begin{enumerate}
                                                \item
                                                $\con \circ \kil = \text{id}: \co_P \otimes_{\co_S(S)}  \Liealg \to \co_P \otimes_{\co_S(S)}  \Liealg$.

                                                \item
                                                $\psi_a^* \omega= \mathrm{Ad}(a^{-1}) \omega$ \quad for all $S$-valued points $a \in G(S)$.\footnote{
                                                This equation has to be understood in the sense of  $(\psi_a^* \omega)(\vecv)=\mathrm{Ad}(a^{-1})
                                                (\omega(\vecv))$ for all $\vecv \in \Gamma(T_{P/S}/P)$,
                                                where $\psi_a$ denotes the translation by the $S$-valued point $a$ (see \ref{1370} above). The map $\text{Ad}$
                                                was introduced in subsection \ref{1230}.}

                                                \item
                                                $\omega(\text{hor} \, \vecv) = 0$ \quad for all $\vecv \in \Gamma(T_{P/S}/P)$.

                                                \item
                                                $\omega : \ct_{P/S} \to \co_P \otimes_{\co_S(S)} \Liealg$
                                                is a homomorphism of Lie-algebras.\footnote{This means that
                                                $\omega([\vecv,\vecw])= [\omega(\vecv),\omega(\vecw)]$ for all vector fields $\vecv, \vecw \in
                                                \ct_{P/S}$. Thereby $\ct_{P/S}$ is considered as a Lie-algebra
                                                with the commutator as Lie-bracket (see Definition \ref{1158}). The Lie-algebra
                                                structure of $\co_P \otimes_{\co_S(S)} \Liealg$ was introduced in
                                                Remark \ref{1430r}. }
                                            \end{enumerate}
                                        \end{Def}

                                        \begin{Def}\label{1438}
                                            A vector field $\vecv \in
                                            \Gamma(T_{P/S}/P)$ is called \emph{horizontal} (resp.
                                            \emph{vertical}), if $ \vecv = \text{hor} \, \vecv$ (resp. $ \vecv =
                                            \text{ver}\, \vecv$).
                                        \end{Def}

                                        \begin{Satz}\label{1439}
                                            Let $\omega \in \Omega_{P/S}^1 \otimes_{\co_S(S)} \Liealg$ be a connection
                                            form. Let $\psi_a$ be the translation by an $S$-valued point $a$ (induced by the group action $\psi$).
                                            Then:
                                            \begin{enumerate}
                                                \item
                                                $ \text{hor} = \text{hor} \circ \text{hor} $ \qquad \qquad $\text{hor} \circ \text{ver} = 0$.  \\
                                                $ \text{ver} \, = \text{ver} \circ \text{ver} \, $ \qquad  \qquad $\text{ver} \circ \text{hor} = 0$.

                                                \item
                                                $(\psi_a)_* \vecv$ is horizontal for all horizontal
                                                vector fields $\vecv$ on $P$. \\
                                                $(\psi_a)_* \vecv$ is vertical for all vertical
                                                vector fields $\vecv$ on $P$.

                                                \item
                                                $\text{hor} \, \circ (\psi_a)_* =(\psi_a)_* \circ \text{hor} \,$ \\
                                                $\text{ver} \, \circ (\psi_a)_* =(\psi_a)_* \circ \text{ver} \,$

                                            \end{enumerate}
                                        \end{Satz}

                                        \begin{proof}
                                            If $\omega$ is a connection form, then $\omega \circ \sigma =
                                            \text{id}$. Therefore $\text{ver} \circ \text{ver} = \sigma
                                            \circ \omega \circ \sigma \circ \omega = \sigma \circ \omega
                                            =\text{ver}$ and $\text{hor} \circ \text{hor} = (\text{id}- \text{ver}) \circ (\text{id}-
                                            \text{ver})= \text{id} - 2 \cdot \text{ver} + \text{ver} \circ
                                            \text{ver}= \text{id} - \text{ver} = \text{hor}$. In
                                            particular, it follows that $\text{hor} \circ \text{ver} = (\text{id} - \text{ver}) \circ \text{ver}=
                                            0= \text{ver} \circ (\text{id} - \text{ver}) = \text{ver} \circ
                                            \text{hor}$.
                                            In order to prove $b)$ consider first a horizontal vectorfield
                                            $\vecv$. Then
                                            \begin{align*}
                                            \text{ver} \, ((\psi_a)_* \vecv)
                                            &= \sigma \circ \omega ((\psi_a)_* \vecv)
                                            =  \sigma \left( (\psi_a^{-1})^* ((\psi_a^*\omega)(\vecv))    \right)
                                            =  \sigma \left( (\psi_a^{-1})^* ((\text{Ad}(a^{-1})\omega)(\vecv)) \right)\\
                                            &= \sigma \left( \text{Ad}(a^{-1}) (\psi_a^{-1})^* (\omega(\vecv))
                                            \right).
                                            \end{align*}
                                            Thus $\text{hor} \, ((\psi_a)_* \vecv)= (\psi_a)_* \vecv$, because
                                            $\omega(\vecv)=0$ for horizontal vector fields $\vecv$.

                                            Let
                                            now be $\vecv$ be a vertical vector field, i.e.
                                            $\vecv = \text{ver} \, (\vecv) = \sigma (\omega( \vecv))$ .
                                            Writing $\omega( \vecv)= r \otimes \vecw_e$ with $r \in \co_P$
                                            and $\vecw_e \in \Liealg$ and denoting the unique left-invariant vector field associated to $\vecw_e$ by
                                            $\vecw$, we obtain
                                            \begin{align*}
                                                (\psi_a)_* \vecv
                                                &= (\psi_a)_* (\sigma (\omega( \vecv)))
                                                = (\psi_a)_* (r \cdot \psi_* (0 \times \vecw))
                                                \stackrellow{(*)}= ((\psi_a^{-1})^* r) \cdot (\psi_a)_*\psi_* (0 \times \vecw),
                                            \end{align*}
                                            where $(*)$ will be proven below. Recalling
                                            the notion of the
                                            right-translation $\tau'_a$ by $a$ on $G$ from subsection
                                            \ref{1240}, we may write $\psi_a \circ \psi = \psi \circ (\text{id} \times \tau'_a): P \times_S G \to
                                            P$, $(p,g) \mapsto p \cdot (ga)$, and we get
                                            \begin{align*}
                                                (\psi_a)_* \vecv
                                                &= ((\psi_a^{-1})^* r) \cdot \psi_* (\text{id} \times \tau'_a)_* (0 \times \vecw)
                                                = ((\psi_a^{-1})^* r) \cdot \psi_* (0 \times (\tau'_a)_* \vecw)
                                            \end{align*}
                                            As left- and right-translations on $G$ commute with each
                                            other, $(\tau'_a)_* \vecw$ is left invariant again. Therefore
                                            $(\tau'_a)_* \vecw$ is the unique left invariant vector field on $G$ corresponding to an element
                                            $((\tau'_a)_* \vecw)_e \in \Liealg$, and, using Definition \ref{1437} c), this yields
                                            \begin{align*}
                                                (\psi_a)_* \vecv
                                                &= \sigma\left( (\psi_a^{-1})^* r \otimes ((\tau'_a)_* \vecw)_e \right)
                                                =  (\sigma \circ \omega \circ \sigma) \left( (\psi_a^{-1})^* r \otimes ((\tau'_a)_* \vecw)_e \right)
                                                =  \text{ver} \, ((\psi_a)_* \vecv).
                                            \end{align*}
                                            In order to finish the proof of b), it remains to explain the equality $(*)$. Let $\vect$ be a
                                            vector field on $P$ and let $\alpha \in \Omega_{X/S}^1$. Then
                                            \begin{align*}
                                                \left((\psi_a)_* (r \cdot \vect)\right)(\alpha)
                                                &= (\psi_a^{-1})^* \left( (r \cdot \vect) (\psi_a^*\alpha) \right)
                                                =  (\psi_a^{-1})^* \left(  \vect (r \cdot \psi_a^*\alpha)  \right) \\
                                                &= (\psi_a^{-1})^* \left(  \vect (\psi_a^*((\psi_a^{-1})^* r \cdot \alpha)  \right)
                                                =  \left( (\psi_a)_*  \vect \right) ((\psi_a^{-1})^* r \cdot  \alpha)
                                                && (*)\\
                                                &= (\psi_a^{-1})^* r \cdot \left( (\psi_a)_* (\vect) \right) ( \alpha).
                                            \end{align*}
                                            Finally let us prove c). But this is an immediate consequence of
                                            b):
                                            \begin{align*}
                                                \text{hor} \, (\psi_a)_*\vecv
                                                &=\text{hor} \, (\psi_a)_* (\text{hor} \, \vecv + \text{ver} \, \vecv)
                                                =\text{hor} \, ((\psi_a)_* \text{hor} \, \vecv )+  \text{hor} \,((\psi_a)_* \text{ver} \, \vecv )
                                                =(\psi_a)_* (\text{hor} \, \vecv). \\
                                                \text{ver} \, (\psi_a)_*\vecv
                                                &=\text{ver} \, (\psi_a)_* (\text{hor} \, \vecv + \text{ver} \, \vecv)
                                                =\text{ver} \, ((\psi_a)_* \text{hor} \, \vecv )+  \text{ver} \,((\psi_a)_* \text{ver} \, \vecv )
                                                =(\psi_a)_* (\text{ver} \, \vecv).
                                            \end{align*}
                                            \
                                        \end{proof}
                                        \section{Covariant derivation}\label{1450}

                                        \begin{Def}\label{1451}
                                            Let $\alpha$ be a differential $k$-form on $P$ with values in
                                            $\Liealg$, i.e. a global section of $\Omega_{P/S}^k \otimes_{\co_S(S)}\Liealg$.
                                            Furthermore, let $d$ be the exterior differential (see Theorem \ref{1156}).
                                            Then each connection form $\omega$ on $P$ defines a covariant
                                            derivative $D\alpha$
                                            \begin{align*}
                                                (D\alpha)(\vecv_1, \ldots, \vecv_{k+1}):=(d\alpha)
                                                (\text{hor} \,\vecv_1, \ldots, \text{hor} \, \vecv_{k+1})
                                            \end{align*}
                                            with vector fields $\vecv_1, \ldots, \vecv_{k+1} \in
                                            \Gamma(T_{P/S}/P)$.
                                        \end{Def}
                                        $D\alpha$ is a differential $2$-form on $P$ with values in
                                        $\Liealg$, because $\text{hor} \in
                                        \Hom_{\co_P}(\Omega_{P/S}^1,\Omega_{P/S}^1)$.
                                        \begin{Def}\label{1457}
                                            Let $\omega$ be a curvature form. Then we write $\Omega:=D \omega$.
                                        \end{Def}
                                        $D \omega$ has some physically important properties which we are
                                        going to summarize next. For this purpose we need the following
                                        lemma.
                                        \begin{Lemma}\label{1453}
                                            Let $\phi:X \to X'$ be an isomorphism of $S$-schemes. Then the
                                            exterior differential $d$ commutes with $\phi$:
                                            \begin{align*}
                                                \phi^* d = d \phi^*.
                                            \end{align*}
                                        \end{Lemma}

                                        \begin{proof}
                                            First, we prove that $[\phi_* \vecv,\phi_* \vecw]= \phi_*[\vecv,
                                            \vecw]$ (see Definition \ref{1158} for the notion of the commutator). This may be seen as follows:
                                            \begin{align*}
                                                (\phi_* \vecv)_d(f)= (\phi_* \vecv)(df)
                                                = (\phi^{-1})^{*}\left(\vecv (\phi^*(df))\right)
                                                = (\phi^{-1})^{*}\left(\vecv (d(\phi^*f))\right),
                                            \end{align*}
                                            where the last equality is due to the fact that the exterior differential $d$
                                            coincides with $d_{X/S}$ on
                                            functions (Theorem \ref{1156}), and due to the fact that the
                                            differential $d_{X/S}$ commutes (by definition) with pull-back  on functions.
                                            Therefore
                                            \begin{align*}
                                                (\phi_* \vecv)_d= (\phi^{-1})^{*} \circ \vecv_d \circ
                                                \phi^*,
                                            \end{align*}
                                            where $(\phi^{-1})^{*}$ and $\phi^*$ both denote the
                                            pull-back of functions. Suppressing the index $d$, we obtain
                                            \begin{align*}
                                                [\phi_* \vecv, \phi_*\VECW](f) &
                                                = \phi_* \vecv  (\phi_*\VECW (f)) -\phi_* \VECW  (\phi_*\vecv (f))
                                                \\
                                                &= \phi_* \vecv \big( (\phi^{-1})^* (\VECW (\phi^* f)) \big) - \phi_* \VECW
                                                \big((\phi^{-1})^* (\vecv (\phi^* f)) \big)
                                                \\
                                                & =(\phi^{-1})^* \vecv \big(\phi^* \big( (\phi^{-1})^* (\VECW (\phi^* f))
                                                \big)\big) -(\phi^{-1})^* \VECW \big(\phi^* \big( (\phi^{-1})^* (\vecv
                                                (\phi^* f)) \big)\big)
                                                \\
                                                & =(\phi^{-1})^* \big((\VECV \VECW-\VECW \VECV)(\phi^* f) \big)=
                                                (\phi_*[\VECV,\VECW])(f).
                                            \end{align*}
                                            in terms of derivations. Therefore $[\phi_* \vecv, \phi_*\VECW] =
                                            \phi_*[\VECV,\VECW]$ is also true for vector fields. Now we
                                            conclude as follows:
                                            \begin{align*}
                                                &(\phi^{-1})^*\left((\phi^*(d \alpha))(\vecv_0, \ldots,
                                                \vecv_k)\right) \\
                                                &=  (d \alpha) (\phi_*\vecv_0, \ldots, \phi_*\vecv_k) \\
                                                &=  \sum_{i=0}^k (-1)^{i} \underbrace{(\phi_* \vecv_i)}\limits_{ =(\phi^{-1})^* \vecv_i \phi^*}
                                                    \left( \alpha(\phi_*\vecv_0, \ldots,
                                                    \check{\phi_*\vecv_i}, \ldots, \phi_*\vecv_k) \right)
                                                    \\
                                                    &+ \sum_{0 \leq i < j \leq k} (-1)^{i+j}
                                                    \alpha( \underbrace{[\phi_* \vecv_i, \phi_* \vecv_j]}\limits_{=\phi_*[\vecv_i,  \vecv_j]}, \phi_*\vecv_0, \ldots,
                                                    \check{\phi_*\vecv_i}, \ldots, \check{\phi_*\vecv_j}, \ldots,
                                                    \phi_*\vecv_k) \\
                                                &=  \sum_{i=0}^k (-1)^{i} (\phi^{-1})^* \vecv_i
                                                    \left( (\phi^* \alpha)(\vecv_0, \ldots,
                                                    \check{\vecv_i}, \ldots, \vecv_k) \right)
                                                    \\
                                                    &+ \sum_{0 \leq i < j \leq k} (-1)^{i+j}
                                                    (\phi^{-1})^* ( \phi^* \alpha) ( [\vecv_i,  \vecv_j], \vecv_0, \ldots,
                                                    \check{\vecv_i}, \ldots, \check{\vecv_j}, \ldots,
                                                    \vecv_k) \\
                                                &=  (\phi^{-1})^*\left((d(\phi^* \alpha))(\vecv_0, \ldots,
                                                    \vecv_k)\right).
                                            \end{align*}
                                        \end{proof}

                                        \begin{Satz}\label{1454}
                                            $\psi_a^*  \Omega = \mathrm{Ad}(a^{-1}) \Omega$ for all $S$-valued points $a \in G(S)$.
                                        \end{Satz}

                                        \begin{proof}
                                            The translation by $a$ is an isomorphism. Recalling the notion of
                                            pull-back of differential forms of higher degree from Definition \ref{1329},
                                            and using the fact that $(\psi_a)_*$ commutes with $\text{hor}$ (see
                                            Proposition \ref{1439} c)), we obtain:
                                            \begin{align*}
                                                (\psi_a^{-1})^* ((\psi_a^* \Omega)(\vecv,\vecw))
                                                &= \Omega((\psi_a)_*\vecv,(\psi_a)_*\vecw)
                                                =  (d\omega)(\text{hor} \, (\psi_a)_*\vecv, \text{hor} \,
                                                (\psi_a)_*\vecw)\\
                                                &= (d\omega)((\psi_a)_* \text{hor} \, \vecv, (\psi_a)_* \text{hor} \, \vecw)
                                            \end{align*}
                                            Therefore
                                            \begin{align*}
                                                (\psi_a^* \Omega)(\vecv,\vecw)
                                                &= ((\psi_a)^*d\omega)( \text{hor} \, \vecv, \text{hor} \,
                                                \vecw)\\
                                                &= (d(\psi_a)^*\omega)( \text{hor} \, \vecv, \text{hor} \,
                                                \vecw) \qquad \quad \text{by Lemma \ref{1453}} \\
                                                &=\mathrm{Ad}(a^{-1}) (d\omega)( \text{hor} \, \vecv, \text{hor} \,
                                                \vecw)\\
                                                &= \mathrm{Ad}(a^{-1}) \Omega(\vecv,\vecw).
                                            \end{align*}
                                        \end{proof}

                                        \begin{structure}\label{1455}
                                            $\Omega(\vecv,\vecw) = d\omega(\vecv,\vecw) +
                                            [\omega(\vecv),\omega(\vecw)] $, where $[\cdot,\cdot]$ denotes the
                                            Lie-bracket of $\co_P \otimes_{\co_S(S)} \Liealg$ (see Remark \ref{1430r}).
                                        \end{structure}

                                        \begin{proof}
                                            First, let us consider the Lie-algebra valued vector field
                                            $[\omega,\omega]$ which is defined by
                                            $([\omega,\omega])(\VECV,\VECW):=
                                            [\omega(\VECV),\omega(\VECW)]-
                                            [\omega(\VECW),\omega(\VECV)]=2[\omega(\VECV),\omega(\VECW)]$. Then we
                                            proposition may be stated as the following equation of differential forms:
                                            \begin{align*}
                                                \Omega=d\omega +\frac{1}{2} [\omega,\omega].
                                            \end{align*}
                                            Due to the $\co_P$-multi-linearity of differential forms, it therefore suffices to
                                            consider the following three cases:
                                            \begin{itemize}
                                                \item[(i)]
                                                \underline{$\VECV$ and $\VECW$ are vertical:} If we set $\omega(\vecv)=:r \otimes \vect_e$,
                                                then we may write $\vecv = \sigma (\omega(\vecv))= \sigma(r \otimes \vect_e) = r \cdot \sigma(1 \otimes
                                                \vect_e)$. Then, again due to $\co_P$-multi-linearity, it
                                                suffices to make the proof in the case $r=1$; i.e. we may
                                                assume that $\vecv = \sigma (1 \otimes \vect_e)$. For the
                                                same reason let $\vecw = \sigma (1 \otimes \vecu_e)$. Thus,
                                                by definition of the exterior differential $d$
                                                (see Corollary
                                                \ref{1157}), we conclude:
                                                \begin{align*}
                                                    (d\omega)(\vecv,\vecw)
                                                    = \vecv(
                                                    \omega(\vecw))-\vecw(\omega(\vecv))-\omega([\vecv,\vecw])
                                                    = \vecv(1 \otimes \vecu_e)-\vecw(1 \otimes
                                                    \vect_e)-\omega([\vecv,\vecw]).
                                                \end{align*}
                                                Now, $\vecv(1 \otimes \vecu_e)=0=\vecv(1 \otimes \vect_e)$, because
                                                the derivations $\vecv$ and $\vecw$ act on constant Lie-algebra
                                                valued functions. Thus we are done, because $\omega$ is a
                                                homomorphism of Lie-algebras.

                                                \item[(ii)]
                                                \underline{$\VECV$ and $\VECW$ are horizontal:} {In this
                                                special case, the statement of the proposition reduces to the definition of
                                                $\Omega$.}

                                                \item[(iii)]
                                                \underline{$\VECV$ is vertical and $\VECW$ horizontal:} We get
                                                \begin{align*}
                                                    (d\omega)(\vecv,\VECW)=\vecv( \omega(\VECW))- \VECW
                                                    (\omega(\vecv))-\omega([\vecv,\VECW]).
                                                \end{align*}
                                                Let us explain, why each of the summands is zero. The first one is zero,
                                                because $ \omega(\VECW)=0$ due to horizontality of
                                                $\vecw$. Therefore, $\omega([\vecv,\VECW])=[\omega(\vecv),\omega(\VECW)]$ vanishes, too.
                                                The second summand is zero due to the arguments already given
                                                in $(i)$.
                                            \end{itemize}
                                        \end{proof}

                                        \begin{Bianchi}\label{1456}
                                            $D\Omega =0$.
                                        \end{Bianchi}

                                        \begin{proof}
                                            Notice that
                                            \begin{align*}
                                                (d[\omega,\omega])(\vecu,\vecv,\vecw)
                                                &= \vecu([\omega,\omega](\vecv,\vecw))
                                                -  \vecv([\omega,\omega](\vecw,\vecu))
                                                +  \vecw([\omega,\omega](\vecu,\vecv)) \\
                                                & \ \ \ \, - [\omega,\omega]([\vecu,\vecv],\vecw) + [\omega,\omega]([\vecu,\vecw],\vecv)
                                                -  [\omega,\omega]([\vecv,\vecw],\vecu) \\
                                                &= 0
                                            \end{align*}
                                            if $\vecu,\vecv$ and $\vecw$ are horizontal vector fields.
                                            Using the structure-equation \ref{1455} and $d \circ d=0$ (see Theorem \ref{1156}), we finally obtain
                                            \begin{align*}
                                                (D\Omega)(\vecu,\vecv,\vecw)
                                                &=  (d\Omega)(\text{hor} \, \vecu, \text{hor} \, \vecv, \text{hor} \, \vecw)\\
                                                &=   (dd\omega)(\text{hor} \, \vecu, \text{hor} \, \vecv, \text{hor} \, \vecw)
                                                +   (d[\omega,\omega])(\text{hor} \, \vecu, \text{hor} \, \vecv, \text{hor} \, \vecw)
                                                =   0.
                                            \end{align*}
                                        \end{proof}

                                        \begin{Def}\label{1452}
                                            $\Omega=D \omega$ is called the \emph{curvature form} of the
                                            connection $\omega$, if $\psi^* \Omega$ is horizontal in the following
                                            sense: On stalks we can write
                                            \begin{align*}
                                                \psi^* \Omega = \sum_i (\nu_i, 0) \wedge (\nu'_i,0)
                                            \end{align*}
                                            where $(\nu_i, 0),(\nu'_i, 0) \in \pi_1^*\Omega_{P/S}^1 \oplus \pi_2^*\Omega_{G/S}^1 \cong \Omega_{P \times_S
                                            G/S}^1$ for all $i$.
                                        \end{Def}
                                        %
                                        %
                                        \textbf{Remark.} In the situation of Definition \ref{1452}, the condition
                                        $\psi^* \Omega = \sum_i (\nu_i, 0) \wedge (\nu'_i,0)$ will
                                        guarantee that the local field strengths $\cf_i$ corresponding to
                                        $\Omega$ (see Definition \ref{1533}), actually glue to a global field
                                        strength $\cf$ which lives on the whole universe.

                                        In particular, the condition $\psi^* \Omega = \sum_i (\nu_i, 0)
                                        \wedge (\nu'_i,0)$ may be omitted, if one considers a  gauge
                                        theory which is induced by a globally trivial $X$-torsor $P$ under
                                        $G$. In this case, every connection $\omega$ gives rise to a
                                        curvature form $\Omega:=D \omega$.

\chapter{Yang-Mills equation}\label{1500}

In this chapter we will finally establish the Yang-Mills theory
over commutative rings $R$. Pulling-back the connection form
$\omega$ and the curvature form $\Omega$ (introduced in Definition
\ref{1437} and Definition \ref{1452}) to \sto, we are led to the
gauge potential $\ca$ and the field strength $\cf$. The main
result of this chapter is that  the gauge potential and the field
strength, which are a priori only defined locally (with respect to
\'etale topology) on \sto, actually glue to global objects of
\sto. This proof will be done by means of Grothendiecks theory of
faithfully flat descent. Then the Yang-Mills action and the
Yang-Mills equation may be established over the base scheme $S:=
\Spec R$, generalizing the known classical theory over $R:=
\real$.

More precisely, in our physical applications, $R$ will be a
one-dimensional Dedekind ring which is also a principal ideal
domain. So, we  assume from now on that $R$ is of this type. As a
consequence, the sheaves of relative differential forms over
$S=\Spec R$ will be free sheaves of modules (Theorem \ref{1244}).
In particular, the Lie-algebra $\Liealg$ of a group scheme over
$S$ will be a free $\co_S(S)$-module and therefore flat. As usual,
let $X \to S$ be a smooth $S$-scheme, let $p:G \to S$ be a smooth
$S$-group scheme with unit section $\varepsilon :S \hookrightarrow
G$, and let $G_X:= G \times_S X$. Let $P$ be a $X$-torsor under
$G_X$ with group action $\psi$. Furthermore, we assume for
physical reasons that $X$ and $G$ are connected.
\section{The Maurer-Cartan form}\label{1510}

Due to the connectedness of $G$, the constant presheaves
\begin{align*}
    U &\rightsquigarrow \co_S(S) \\
    U &\rightsquigarrow \Liealg:=(\varepsilon^* \Omega_{G/S}^1)(S)
\end{align*}
are sheaves, and due to flatness,
\begin{align*}
    U \rightsquigarrow \co_G(U) \otimes_{\co_S(S)} \Liealg
\end{align*}
is a sheaf, too, which we denote by $\co_G \otimes_{\co_S(S)}
\Liealg $. In particular, the direct image of this sheaf under $p$
is given by $p_* \left( \co_G \otimes_{\co_S(S)} \Liealg \right)=
p_*\co_G  \otimes_{\co_S(S)} \Liealg $. We want to find a
left-invariant differential form $\Theta$ which assigns to every
vector field $\vecv \in \ct_{G/S}(G)$ the constant Lie-algebra
valued function $1 \otimes \vecv_e \in \co_G(G) \otimes_{\co_S(S)}
\Liealg $ where $\vecv_e$ denotes the value of $\vecv$ at the unit
element $e$ of $G$. \footnote{More precisely, $\vecv_e$ denotes
the pull-back of $\vecv$ with respect to the unit section
$\varepsilon$ and is then considered as an element of the
Lie-algebra $\Liealg$ (recall Proposition \ref{1137}).} Let us
prove that such a differential form exists, and that it is
uniquely determined. This is due to the following sequence of
canonical isomorphism which will be explained below.
\begin{align*}
    \Hom_{\co_S(S)} \left(   \Liealg, \co_G(G) \otimes_{\co_S(S)} \Liealg   \right)
&=  \Hom_{\co_S} \left(   \varepsilon^* \Omega_{G/S}^1,  p_*\co_G \otimes_{\co_S(S)} \Liealg   \right) \\
&=  \Hom_{\co_S} \left(   \varepsilon^* \Omega_{G/S}^1,  p_* \left( \co_G \otimes_{\co_S(S)} \Liealg  \right)  \right) \\
&=  \Hom_{\co_G} \left( p^*  \varepsilon^* \Omega_{G/S}^1,   \co_G \otimes_{\co_S(S)} \Liealg  \right)  \\
&=  \Hom_{\co_G} \left(\Omega_{G/S}^1,   \co_G \otimes_{\co_S(S)}
\Liealg  \right)
\end{align*}
The first isomorphism is (\Liu, Ex. 5.1.5), the second  one is due
to the remark in the introduction above. The third states the
adjointness of the functors $p^*$ and $p_*$ (Proposition
\ref{1325}), and the last one originates from extending elements
of the Lie-algebra to left-invariant sections (Proposition
\ref{1241a}). Also recalling Proposition \ref{1137}, this
bijection induces a bijection
\begin{align*}
    \Hom_{\co_S(S)} \left(   (\varepsilon^* \ct_{G/S})(S), \co_G(G) \otimes_{\co_S(S)} \Liealg   \right)
=  \Hom_{\co_G} \left(\ct_{G/S},   \co_G \otimes_{\co_S(S)}
\Liealg  \right)
\end{align*}

\begin{Def}\label{1511}
The \ uniquely \ determined \ Lie-algebra \ valued \ differential
\ form \ $\Theta \in \Hom_{\co_G} \left(\ct_{G/S},   \co_G
\otimes_{\co_S(S)} \Liealg \right) $ corresponding to the
$\co_S(S)$-linear homomorphism
\begin{align*}
    (\varepsilon^* \ct_{G/S})(S) &\longrightarrow  \co_G(G) \otimes_{\co_S(S)} \Liealg \\
    \vecv_e &\mapsto 1 \otimes \vecv_e
\end{align*}
is called the \emph{Maurer-Cartan form}. Thereby $1 \in \co_G(G)$
denotes the unit element.

In particular, one obtains
\begin{align*}
    \Theta(\vecv) = 1 \otimes \varepsilon^*\vecv
\end{align*}
for all left-invariant vector fields $\vecv \in
\Gamma(T_{G/S}/G)$.
\end{Def}
\section{The field strength of a gauge potential}\label{1520}

In classical gauge theory, gauge fields are described by a gauge
potential. The physical information is encoded in the associated
antisymmetric field strength tensor: For example in the case of
electromagnetism, the components of the field strength tensor are
exactly the components of the electric and the magnetic field.
Therefore the field strength has to be a global object living upon
the whole \st manifold, if the theory claims to be physically
sensible.

Let us prove that the field strength is also a global object in
our algebraic geometric setting of a \st $X(S)$ consisting of
``adelic'' \st points (isomorphic to the base
scheme $S$). \\ \\
Let $P,X,G$ and $S$ be as stated at the beginning of this chapter.
Let $\hat{\theta}$ be a gauge transformation (Definition
\ref{1421}), and let $u:=u_{\hat{\theta}}$ be the corresponding
equivariant morphism (Proposition \ref{1422}). Let $\omega$ be a
connection form (Definition \ref{1437}). Also recall the notion of
translation by $S$-valued points by means of the group action from
\ref{1370} above.
\subsection{Gauge transformations of the curvature
form}\label{1521}

\begin{Satz}\label{1522}
Let $p\in P$ be a physical point. Then the stalk of the curvature
form $\Omega$ at $p$ transforms as follows under gauge
transformations.
    \begin{align*}
        ( \hat{\theta}^* \Omega)_p &= \mathrm{Ad}(u(p)^{-1}) \Omega_p
    \end{align*}
\end{Satz}

\begin{proof}
    By definition of $u$ (see proof of Proposition \ref{1422}) we have got $\hat{\theta} = \psi \circ
    (\text{id},u)$. Let $p:P \to P$ resp. $u(p):P \to G$ be the
    constant maps in the sense of Proposition \ref{1355} mapping $P$ to
    the $S$-valued point containing $p$ resp. $u(p)$. Writing $\hat{p}:=\hat{\theta}(p)$, we
    know from Proposition \ref{1356} that at the stalk at $p$
    \begin{align*}
        (\hat{\theta}^* \Omega)_p = (\text{id},u)^* \psi^* \Omega_{\hat{p}} = (p,u)^*  \psi^* \Omega_{\hat{p}} +
        (\text{id},u(p))^* \psi^* \Omega_{\hat{p}}.
    \end{align*}
    Thus we conclude
    \begin{align*}
        (\hat{\theta}^*\Omega)_p
        &= \psi_{u(p)}^* \Omega_{\hat{p}} +  (p, u)^* \psi^*
        \Omega_{\hat{p}}.
    \end{align*}
    By Proposition \ref{1454}, $\psi_{u(p)}^* \Omega_{\hat{p}}= \text{Ad}(u(p)^{-1})
    \Omega_p$. Thus it suffices to show that the second summand
    vanishes. As $(p,u)=(p \times \text{id}) \circ (\text{id},u)$, it
    suffices to show that $(p \times \text{id})^* \psi^* \Omega_{\hat{p}}
    =0$. But $\Omega$ is a curvature
    form (see Definition \ref{1452}) and therefore we may write on stalks
    \begin{align*}
        \psi^* \Omega = \sum_i (\nu_i, 0) \wedge (\nu'_i,0)
    \end{align*}
    where $(\nu_i, 0),(\nu'_i, 0) \in \pi_1^*\Omega_{P/S}^1 \oplus \pi_2^*\Omega_{G/S}^1 \cong \Omega_{P \times_S
    G/S}^1$ for all $i$. Therefore
    \begin{align*}
        (p \times \text{id})^* \psi^* \Omega_{\hat{p}}
        &= \sum_i (p \times \text{id})^*(\nu_i, 0) \wedge (p \times \text{id})^*(\nu'_i,0)
        &= \sum_i (p^* \nu_i, 0) \wedge (p^* \nu'_i,0) = 0,
    \end{align*}
    because the pull-back of differential forms under constant morphisms
    is zero by Proposition \ref{1352}.
\end{proof}
\subsection{The curvature form in a local trivialisation}\label{1530}

Let $\pi:P \to X$ be the canonical projection. Locally in \'etale
topology on $X$, there exists a section $\bar{s}: U
\hookrightarrow P \times_X U =: \pi^{-1}(U)$ of $\pi$ over each
point $x \in X$ (Proposition \ref{0030} and Proposition
\ref{0029}). We may assume that the torsor $P$ is trivial over
$U$, i.e. $\pi^{-1}(U)= U \times_S G=:G_U$. If $p:G \to S$ is the
canonical morphism and if $\varepsilon: S \hookrightarrow G$ is
the unit section, then $\bar{s}$ induces a section
\begin{align*}
    s: U
    \stackrel{\bar{s}}\hookrightarrow \pi^{-1}(U)= U \times_S G
    \stackrel{\text{id} \times p}\longrightarrow U \times_S S
    \stackrel{\text{id} \times \varepsilon}\longrightarrow U \times_S
    G= \pi^{-1}(U).
\end{align*}
This section $s$ is called the \emph{canonical section}. \\ \\
Furthermore we will make use of the following morphisms:
\begin{enumerate}\label{1531}
    \item
    $\chi: \pi^{-1} \to U \times_S G $ denotes the trivialisation.

    \item
    $p_1: U \times_S G \to U$ and $p_2: U \times_S G \to G$ denote the canonical
    projections.

    \item
    Let $\kappa:= p_2 \circ \chi$. As $\pi= p_1 \circ \chi$ and as
    $\chi$ is an isomorphism, we get in particular the cartesian
    diagram
    \begin{align*}
        \xymatrix{\pi^{-1}(U) \ar[r]^{ \ \ \, \pi} \ar[d]_{\kappa} & U \ar[d] \\
        G \ar[r] & S}.
    \end{align*}

    \item
    Let $m: G \times_S G \to G$ be the group multiplication.
    Recall from the definition of torsors that the trivialisation respects the
    group actions; i.e.
    \begin{align*}
        \kappa \circ \chi = m \circ (\kappa \times \text{id})
    \end{align*}
\end{enumerate}

\begin{Lemma}\label{1532}
    Let $s$ be the canonical section, and let $f:= \psi \circ (s\circ \pi, \kappa): P \to P$. Then $f =
    \text{id}$.
\end{Lemma}

\begin{proof}
    There is a commutative diagram
    \begin{align*}
        \xymatrix{P \times_X G \ar[r]^{ \ \ \, \psi} \ar[d]_{\pi_1} & P \ar[d]^{\pi} \\
        P \ar[r]^{\pi} & X}.
    \end{align*}
    Therefore we conclude
    \begin{align*}
        \pi \circ f
        &= \pi \circ \psi \circ (s\circ \pi, \kappa)
        = \pi \circ \pi_1 \circ (s\circ \pi, \kappa)
        = \pi \circ s\circ \pi
        = \pi.
    \end{align*}
    Furthermore,
    \begin{align*}
        \kappa \circ f
        &= m \circ (\kappa \times \text{id}) \circ (s\circ \pi, \kappa)
        =  m (\kappa \circ s \circ \pi, \kappa) \\
        &=  m (p_2 \circ \chi \circ \chi^{-1} \circ (\text{id} \times \varepsilon)
           \circ (\text{id} \times p ) \circ \chi \circ \bar{s} \circ \pi,
           \kappa) \\
        &=  m (\varepsilon \circ p \circ p_2 \circ \chi \circ \bar{s} \circ \pi, \kappa)
        =  m \circ ( ( \varepsilon \circ p)  \times \text{id} ) \circ
            (\kappa \circ  \bar{s} \circ \pi, \kappa) \\
        &=  p_2(\kappa \circ  \bar{s} \circ \pi, \kappa)
        =  \kappa.
    \end{align*}
    Thus (by $c)$) we know that $f$  is the uniquely determined
    morphism making the diagram
    \begin{align*}
        \xymatrix{
        & U \\
        \pi^{-1}(U) \ar@{-->}[r]^{f} \ar[ur]^{\pi} \ar[dr]_{\kappa} & \pi^{-1}(U) \ar[u]_{\pi} \ar[d]^{\kappa} \\
        & G}
    \end{align*}
    commutative; i.e. $f= \text{id}$.
\end{proof}

\begin{Def}\label{1533}
    Let $s:U \hookrightarrow \pi^{-1}(U)$ be the canonical
    section, let $\omega$ be a connection form, and let $\Omega$
    be a curvature form. Then
    \begin{align*}
        \ca &:= s^* \omega \qquad \text{is called the local \emph{gauge potential}, and}\\
        \cf &:= s^* \Omega \qquad \text{is called the local \emph{field strength}.}
    \end{align*}
    These are differential forms on $U$ with values in the Lie-algebra $\Liealg$.
\end{Def}

\begin{Satz}\label{1534}
    Let $p\in P$ be a physical point. Then on stalks at $p$ we get
    the identity
    \begin{align*}
        \Omega_p &= \mathrm{Ad} (\kappa(p)^{-1}) (\pi^* \mathcal{F})_p
    \end{align*}
\end{Satz}

\begin{proof}
    Let $(s \circ \pi)(p):P \to P$ resp. $\kappa(p):P \to G$ be the
    constant maps in the sense of Proposition \ref{1355} mapping $P$ to
    the $S$-valued point containing $(s \circ \pi)(p)$ resp. $\kappa(p)$.
    By Lemma \ref{1532}, we know that $\psi \circ (s\circ \pi, \kappa) = \text{id}$.
    Due to Proposition \ref{1356} we therefore get at the stalk at $p$
    \begin{align*}
        \Omega_p
        &=  (s\circ \pi, \kappa)^* \psi^* \Omega_{p}
        =   \left((s \circ \pi)(p),\kappa \right)^*  \psi^* \Omega_{p} +
            \left(s\circ \pi, \kappa(p) \right)^* \psi^* \Omega_{p}.
    \end{align*}
    Using the fact that $\left(s\circ \pi, \kappa(p) \right)= \left(\text{id}, \kappa(p) \right) \circ s\circ
    \pi$, we conclude
    \begin{align*}
        \Omega_p
        &= \pi^* s^* \psi_{\kappa(p)}^* \Omega_{p} +   \left((s \circ \pi)(p),\kappa \right)^*  \psi^*
        \Omega_{p}.
    \end{align*}
    Exactly the same argument as in the proof of Proposition \ref{1522} shows that
    the second summand $\left((s \circ \pi)(p),\kappa \right)^*  \psi^*
    \Omega_{p}$ vanishes. For the first summand we obtain with Proposition \ref{1454}
    \begin{align*}
        \pi^* s^* \psi_{\kappa(p)}^* \Omega_{p}
        &=  \pi^* s^* \text{Ad}(\kappa(p)^{-1}) \Omega_{p \cdot \kappa(p)^{-1}}
        =   \text{Ad}(\kappa(p)^{-1}) \pi^* (s^*  \Omega)_{\pi(p)}
        =   \text{Ad}(\kappa(p)^{-1}) (\pi^* \cf )_{p},
    \end{align*}
    and we are done.
\end{proof}
Let us now choose an \'etale covering $(U_i \to X)_i$ which
trivializes $P$. For each \'etale open subset $U_i \to X$ let
$\chi_i$ be the respective trivialisation, and let $s_i:U_i
\hookrightarrow \pi^{-1}(U_i):= P \times_X U_i$ be the canonical
section of $\kappa_i:= p_2 \circ \chi_i : \pi^{-1}(U_i) \to U_i$.
Let $\pi_i: \pi^{-1}(U_i) \to U_i$ be the canonical projection.

Then we get local gauge potentials and local field strengths
\begin{align*}
    \ca_i&:=s_i^* \omega, \\
    \cf_i&:=s_i^* \Omega
\end{align*}
and by Proposition \ref{1534} we have got
\begin{align*}
    \Omega_p &= \mathrm{Ad} (\kappa_i(p)^{-1}) (\pi_i^*
    \mathcal{F}_i)_p \qquad \text{for all $i$.}
\end{align*}
Pulling back this identity with the canonical projection
$U_{ij}:=U_i \times_X U_j \to U_i$ (which is an \'etale morphism),
and using the fact that the notion of physical points in stable
under \'etale base change (Proposition \ref{1351}), it follows
that
\begin{align*}
    \mathrm{Ad} (\kappa_j(p)^{-1}) (\pi_j^*
    \mathcal{F}_j)_p
    =\Omega_p
    = \mathrm{Ad} (\kappa_i(p)^{-1}) (\pi_i^*
    \mathcal{F}_i)_p
\end{align*}
for all physical points $p \in \pi^{-1}(U_{ij}):= P \times_X
U_{ij}$, if we omit restriction signs. As $\pi_i$ and $\pi_j$ are
obtained from the global morphism $\pi$ by base change with $U_i
\to X$ resp. $U_j \to X$,  $\pi_i$ and $\pi_j$ coincide on
$U_{ij}$, and their restrictions to $U_{ij}$ may both be denoted
by $\pi_{ij}$. Choosing a section $\hat{s}$ of $\pi_{ij}$ (which
exists, as there are already sections over $U_i$ or $U_j$), and
applying this section to the equation above, we obtain for all
physical points $p \in \pi^{-1}(U_{ij})$ over $x \in U_{ij}$
\begin{align*}
    \mathrm{Ad} (\kappa_j(p)^{-1}) (\mathcal{F}_j)_x
    = \mathrm{Ad} (\kappa_i(p)^{-1}) ( \mathcal{F}_i)_x.
\end{align*}
Recalling from Proposition \ref{1223} that $\text{Ad}$ is
compatible with the group law on $G$, i.e. $\mathrm{Ad}
(\kappa_i(p)^{}) \circ \mathrm{Ad} (\kappa_j(p)^{-1}) =
\mathrm{Ad} (\kappa_i(p) \cdot \kappa_j(p)^{-1}) $, we get

\begin{Satz}\label{1535}
    Let $x \in U_{ij}$ be a physical point. Then (omitting restriction
    signs), one has got
    \begin{align*}
        (\cf_j)_x = \mathrm{Ad}(\rho_{ij}^{-1}(x)) (\cf_i)_x,
    \end{align*}
    where the family of all $\rho_{ij}:U_{ij} \to G$ is a $1$-cocycle in
    the sense of Proposition \ref{1415}.
\end{Satz}

\begin{proof}
    It remains to prove the statement on the morphism
    $\rho_{ij}$. For this purpose it suffices to show that there
    exists a morphism $\rho_{ij}:U_{ij} \to G$ such that
    \begin{align*}
        \rho_{ij} \circ \pi = m \circ (\kappa_i, \iota \circ \kappa_j)
    \end{align*}
    where $\pi: \pi^{-1}(U_{ij})=U_{ij} \times_S G \to U_{ij}$ is the
    canonical projection, where $m$ is the group law on $G$, and where $\iota$ is the
    formation of the inverse in $G$. In order to define
    $\rho_{ij}$ choose an arbitrary section of $G \to S$. By base change, this
    induces a section $\epsilon$ of $\pi: \pi^{-1}(U_{ij})=U_{ij} \times_S G \to
    U_{ij}$. Then we set
    \begin{align*}
        \rho_{ij} := m \circ (\kappa_i, \iota \circ \kappa_j) \circ
        \epsilon.
    \end{align*}
    This is a $1$-cocycle and it remains to prove that $\rho_{ij} \circ
    \pi = m \circ (\kappa_i, \iota \circ \kappa_j)$.
    By the Yoneda lemma (Proposition \ref{1201}), we may check
    this on $T$-valued points, where $T \to S$ is an arbitrary
    $S$-scheme. It suffices to show that
    $\left( m \circ (\kappa_i, \iota \circ \kappa_j) \right)(p \cdot g)=
    \left( m \circ (\kappa_i, \iota \circ \kappa_j) \right)(p)$
    for all $T$-valued points $p$ of $P$ and $g$ of $G$. Using the
    fact that the trivialization isomorphisms of torsors respect the group
    action, the following computation indeed shows that
    \begin{align*}
        \left( m \circ (\kappa_i, \iota \circ \kappa_j) \right)(p \cdot g)
        &=  (\kappa_i)(p \cdot g) \cdot \left((\kappa_j)(p \cdot
            g)\right)^{-1} \\
        &=  (\kappa_i)(p) \cdot g \cdot g^{-1} \cdot (\kappa_j)(p)^{-1}
        =   (\kappa_i)(p)\cdot (\kappa_j)(p)^{-1} \\
        &=  \left( m \circ (\kappa_i, \iota \circ \kappa_j)
            \right)(p).
    \end{align*}
\end{proof}
\subsection{The global field strength}\label{1540}

Let us finally prove that the collection of local field strengths
$\cf_i$ (considered in the previous subsection \ref{1530}) glue to
a global field strength $\cf$ living on all of $X(S)$.

The starting point is the family of \'etale-local sections
$(\cf_i)_i$, where $\cf_i=s_i^* \Omega \in \Omega_{U_i/S}^2({U_i})
\otimes _{\co_S(S)} \Liealg$. We already know from Proposition
\ref{1535} that
\begin{align*}
    (\cf_j)_x = \mathrm{Ad}(\rho_{ij}^{-1}(x)) (\cf_i)_x
\end{align*}
upon $U_{ij}$ for all physical points $x \in U_{ij}$. Using the
identification of sections of a locally free sheaf and sections of
an associated vector bundle, we may interpret $\cf_i$ as a
morphism $f_i:U_i \to \mathbb{V}(\ct_{U_i/S}^{\otimes 2}
\otimes_{\co_S(S)} \Liealg)$. Therefore
$\mathrm{Ad}(\rho_{ij}^{-1}(x))$ induces an automorphism of
$\mathbb{V}(\ct_{U_i/S}^{\otimes 2} \otimes_{\co_S(S)} \Liealg)$,
because $\mathrm{Ad} \circ \rho_{ij}^{-1} : U_{ij} \to
\Aut_{\co_S-lin}(\Lie(G/S))$ by the expositions in subsection
\ref{1230}. Due to the above equation, the restriction of the
morphisms $f_j$ and $ (\mathrm{Ad} \circ \rho_{ij}^{-1}) f_i$ to
$U_{ij}$ coincide on physical points. Let us denote by $X'$ the
$X$-scheme given by the disjoint union
\begin{align*}
    \coprod_{\text{ physical points } x \in X } V(\cj_x)
\end{align*}
where $\cj_x$ is the sheaf of ideal corresponding to  the closed
point $x$. The \'etale covering $(U_i \to X)$ induces an \'etale
covering  $(U'_i \to X')$ of $X'$. By construction, the morphisms
$f_j$ and $(\mathrm{Ad} \circ \rho_{ij}^{-1}) f_i$ coincide on
$U'_{ij}$. As $\rho_{ij}$ is a $1$-cocycle, the morphisms
$(f'_i:=f_i|_{U'_i})$ are morphisms of schemes with descent datum.
Therefore, by Theorem \ref{1383}, the family of morphism $(f'_i)$
descends to a morphism $f:X' \to \mathbb{V}(\iota ^*
\ct_{X/S}^{\otimes 2} \otimes_{\co_S(S)} \Liealg)$, where
$\iota:X' \hookrightarrow X$ is the canonical injection. $f$
corresponds to a global section of a locally free sheaf $\cf$ over
$X'$ and is already the searched global field strength, because
set theoretically $X'=X(S)$.

\begin{Rem}\label{1541}
    As the schemes $X, G$ and $P$ are N\'eron (lft)-models, the
    above calculations may be performed as well directly on
    $S$-valued point instead of physical points. All results of
    section \ref{1520} remain true if the word ``physical point'' is replaced by
    ``$S$-valued point''.

    This way it is possible to construct the global field strength
    on $X(S)$, where $X(S)$ is this time endowed with the following structure of a
    $X$-scheme:
    \begin{align*}
    X(S)=\coprod_{\text{ $S$-valued points } \alpha \in X(S) }
    V(\cj_{\alpha}).
    \end{align*}
    Thereby, $\cj_{\alpha}$ is the sheaf of ideals
    realizing the image of $\alpha$ as a closed subscheme of $X$.
\end{Rem}

\section{Yang-Mills equation}\label{1550}

Within this section, let us assume that the torsor $\pi:P \to X$
is trivial. Then the field strength $\cf$ exists as a global
section of $\Omega_{X/S}^2 \otimes_{\co_S(S)} \Liealg$; i.e. $\cf
\in \Omega_{X/S}^2(X) \otimes_{\co_S(S)} \Liealg$ due to our
assumption on $S$.  Let  $s$ be the canonical section of $\pi:P
\to X$ (introduced in the beginning of subsection \ref{1530}).
Then we may pull back the covariant derivation $D$ (see Definition
\ref{1451}) with respect to $s$ to $X$:
\begin{align*}
    D_X(s^* \alpha):=s^*(D\alpha)
\end{align*}
for all $\alpha \in \Omega_{P/S}^k(P) \otimes_{\co_S(S)} \Liealg$.
Let us remark that (due to Proposition \ref{0018}) there is a
decomposition $s^* \Omega_{P/S}^1 \cong s^* p_1^* \Omega_{X/S}^1
\oplus s^* p_2^* \Omega_{G/S}^1 \cong \Omega_{X/S}^1 \oplus p_X^*
\left(\varepsilon^* \Omega_{G/S}^1\right)$, where $\varepsilon:S
\hookrightarrow G$ is the unit section and where $p_X:X \to S$ is
the canonical morphism. In particular $\Omega_{X/S}^1(X)
\hookrightarrow (s^* \Omega_{P/S}^1)(X)$, i.e. each differential
form $\nu \in \Omega_{X/S}^1(X)$ may be written as pull-back $s^*
\alpha$ in our setting. Thus it makes sense to define

\begin{YME}\label{1551}
    $* \, D_X * \cf =0$,
\end{YME}
where $*$ is the Hodge-star operator (Definition \ref{1159}).
Furthermore we deduce from the Bianchi-identity \ref{1456} the
following proposition.

\begin{Satz}\label{1552}
    $D_X \cf =0$.
\end{Satz}

\begin{proof}
    $D_X \cf = D_X (s^* \Omega)= s^* (D \Omega) =0$.
\end{proof}

\begin{Satz}\label{1553}
    $\cf_x(\vecv,\vecw) = (d \ca)_x(\vecv,\vecw) +
    [\ca_x(\vecv),\ca_x(\vecw)]$ for all $x \in X$.
\end{Satz}

\begin{proof}
    Recalling the definition of the stalkwise push-forward of vector fields
    by means of closed immersions from Remark \ref{1327r}, we know that
    the identity  $s_* \vecv = \pi^* \circ \vecv \circ s^*$ holds on
    stalks. Let us first prove that stalkwise $s^*$ commutes with the exterior
    differential (recall that this is true globally in the case of diffeomorphisms). One finds
    on stalks
    \begin{align*}
        [s_*\VECV,s_*\VECW ]
        &=  \pi^* \circ  \VECV \circ s^* \circ \pi^* \circ \VECW
        \circ s^* - \pi^* \circ  \VECW \circ s^* \circ \pi^* \circ \VECV \circ
        s^*
        \\
        &=  \pi^* \circ  \VECV \circ \VECW \circ s^* - \pi^* \circ  \VECW \circ \VECV
        \circ s^*
        \\
        &= s_*[\VECV,\VECW]\;.
    \end{align*}
    Now we conclude as follows:
    \begin{align*}
        &\pi^*\left((s^*(d \alpha))(\vecv_0, \ldots,
        \vecv_k)\right) \\
        &=  (d \alpha) (s_*\vecv_0, \ldots, s_*\vecv_k) \\
        &=  \sum_{i=0}^k (-1)^{i} \underbrace{(s_* \vecv_i)}\limits_{ =\pi^* \vecv_i s^*}
            \left( \alpha(s_*\vecv_0, \ldots,
            \check{s_*\vecv_i}, \ldots, s_*\vecv_k) \right)
            \\
            &+ \sum_{0 \leq i < j \leq k} (-1)^{i+j}
            \alpha( \underbrace{[s_* \vecv_i, s_* \vecv_j]}\limits_{=s_*[\vecv_i,  \vecv_j]}, s_*\vecv_0, \ldots,
            \check{s_*\vecv_i}, \ldots, \check{s_*\vecv_j}, \ldots,
            s_*\vecv_k) \\
        &=  \sum_{i=0}^k (-1)^{i} \pi^* \vecv_i
            \left( (s^* \alpha)(\vecv_0, \ldots,
            \check{\vecv_i}, \ldots, \vecv_k) \right)
            \\
            &+ \sum_{0 \leq i < j \leq k} (-1)^{i+j}
            \pi^* \left(( s^* \alpha) ( [\vecv_i,  \vecv_j], \vecv_0, \ldots,
            \check{\vecv_i}, \ldots, \check{\vecv_j}, \ldots,
            \vecv_k) \right) \\
        &=  \pi^*\left((d(s^* \alpha))(\vecv_0, \ldots,
            \vecv_k)\right).
    \end{align*}
    As the ring-homomorphism $\pi^*$ is injective, we
    obtain the desired commutativity of $s^*$ with $d$. Also recalling the structure-equation \ref{1455}, it follows
    \begin{align*}
        \pi^* \left( \cf(\vecv,\vecw) \right)
        &=  \pi^* \left( (s^* \Omega) (\vecv,\vecw) \right)
        =   \Omega (s_*\vecv,s_*\vecw)
        =   d\omega (s_*\vecv,s_*\vecw) +
            [\omega(s_*\vecv),\omega(s_*\vecw)]\\
        &=  \pi^* \left(\underbrace{(s^*d\omega)}\limits_{=ds^*\omega} (\vecv,\vecw)\right) +
            [\pi^* \left((s^*\omega)(\vecv)\right),\pi^* \left((s^*\omega)(\vecv)\right)]\\
        &=  \pi^* \left((d\ca) (\vecv,\vecw)\right) +
            [\pi^* \left(\ca(\vecv)\right),\pi^* \left(\ca(\vecv)\right)]\\
        &=  \pi^* \left((d\ca) (\vecv,\vecw)\right) +
            \pi^*([\ca(\vecv),\ca(\vecw)]).\\
    \end{align*}
    Once again making use of the injectivity of the ring-homomorphism $\pi^*$, we are done.
\end{proof}

                                        \chapter{Yang-Mills theory in local coordinates}\label{1600}

                                        In the following, let us use the same notations as in chapter
                                        \ref{1500}. In particular, we consider a datum of an $X$-torsor
                                        $P$ under $G_X:=G \times_S X$ underlying the universe. In order to
                                        determine the field strength $\cf \in \Omega_{X/S}^2(X)
                                        \otimes_{\co_S(S)} \Liealg$ of a gauge field, it suffices to
                                        determine all stalks $\cf_x$ for all $x \in X$. Thus we may assume
                                        that the torsor $P$ is trivial.

                                        Within this chapter, we will express the global equations
                                        governing the gauge field of section \ref{1550} on stalks. Thus
                                        we obtain
                                        the Yang-Mills theory in local coordinates. \\ \\
                                        For clarity, let us fix some notations: Let the smooth $S$-scheme
                                        $X \to S$ underlying \st be of relative dimension $n$, and let us
                                        denote the relative dimension of the gauge group $G \to S$ by $N$.
                                        Then let

                                        \begin{align*}
                                            \{ dx^{\mu} \}_{\mu=1}^n  \, &  \text{ be a base of }
                                            \Omega_{X/S}^1(X),\\
                                            \{ \frac{\partial}{\partial x^{\mu}} \}_{\mu=1}^n&  \text{ be a base of }
                                            \ct_{X/S}(X), \text{ which is dual to }  \{ dx^{\mu} \}_{\mu=1}^n,\\
                                            \{ b_i \}_{b=1}^N \ \ &  \text{ be a base of }
                                            \quad \Liealg.
                                        \end{align*}
                                        Then we may write the stalk of the gauge potential and the field
                                        strength at $x \in X$ in the form
                                        \begin{align*}
                                            \ca_x&= \sum\limits_{\mu=1}^{n} \sum\limits_{i=1}^{N}
                                            \ca_{\mu,x}^i dx^{\mu} \otimes b_i \\
                                            \cf_x&= \sum\limits_{\mu,\nu=1}^{n} \sum\limits_{i=1}^{N}
                                            \frac{1}{2}\cf_{\mu \nu,x}^i dx^{\mu} \wedge dx^{\nu} \otimes b_i
                                        \end{align*}
                                        with $\ca_{\mu,x}^i, \cf_{\mu \nu,x}^i \in \co_{X,x}$. Recall that
                                        we embed $\co_{X,x}$ into a ring of formal power series if $x$
                                        is a physical point (see Proposition \ref{0351}).
                                        Therefore,
                                        $\frac{\partial}{\partial x^{\mu}} \ca_{\mu,x}= \frac{\partial
                                        \ca_{\mu,x}}{\partial x^{\mu}}$ for all physical points $x=(x_1,
                                        \ldots, x_n)$ of $X$, where $\frac{\partial \ca_{\mu,x}}{\partial
                                        x^{\mu}}$ denotes the ordinary partial derivative of the power
                                        series $\ca_{\mu,x}$ with respect to the variable $x^{\mu}$. The
                                        analogous statement is true for the components $\cf_{\mu \nu,x}^i$
                                        of the field strength. In order to simplify the notation let us
                                        write $\ca_{\mu}^i$ (resp. $\cf_{\mu \nu}^i$) instead of
                                        $\ca_{\mu,x}^i$ (resp. $\cf_{\mu \nu,x}^i$) whenever no confusion
                                        is possible.

                                        Furthermore let us introduce the structure coefficients $c_{ij}^k$
                                        for the chosen base of the Lie-algebra $\Liealg$ which are defined
                                        as follows: \ \, $[b_i,b_j]=:\sum\limits_{i=1}^N c_{ij}^k b_k$.
                                        \section{Relations between gauge potential and field
                                        strength}\label{1610}

                                        \begin{Satz}\label{1611}
                                            Let $x \in X$ be a physical point. Then the following equalities hold at the
                                            stalk at $x$.
                                            \begin{enumerate}
                                                \item
                                                \qquad $
                                                    \mathcal{F}^k_{\nu_1\nu_2}=
                                                    \frac{\partial\mathcal{A}^k_{\nu_2} }{\partial x^{\nu_1}} -
                                                    \frac{\partial  \mathcal{A}^k_{\nu_1}}{\partial x^{\nu_2}} +
                                                    \sum_{i,j=1}^N c_{ij}^k  \mathcal{A}^i_{\nu_1}
                                                    \mathcal{A}^j_{\nu_2} =-\mathcal{F}^k_{\nu_2\nu_1} \;.
                                                $

                                                \item
                                                \qquad $
                                                    \frac{\partial \mathcal{F}_{\mu\nu}}{\partial x^\rho}
                                                    +   \frac{\partial \mathcal{F}_{\rho\mu}}{\partial x^\nu}
                                                    +   \frac{\partial \mathcal{F}_{\nu\rho}}{\partial x^\mu}
                                                    +   [\mathcal{A}_\rho,\mathcal{F}_{\mu\nu}]
                                                    +   [\mathcal{A}_\nu,\mathcal{F}_{\rho\mu}]
                                                    +   [\mathcal{A}_\mu,\mathcal{F}_{\nu\rho}] =0 \;.
                                                $

                                            \end{enumerate}
                                        \end{Satz}

                                        \begin{proof}
                                            Item $a)$ may be derived as follows:
                                            \begin{align*}
                                                &\sum_{i=1}^{N} \frac{1}{2} \big(\mathcal{F}^i_{\nu_1\nu_2}-
                                                \mathcal{F}^i_{\nu_2\nu_1}\big) b_i
                                                \\
                                                &=  \sum_{\mu_1,\mu_2=1}^n \sum_{i=1}^{N}
                                                \frac{1}{2}\mathcal{F}^i_{\mu_1 \mu_2} \Big\{dx^{\mu_1}
                                                \Big(\frac{\partial}{\partial x^{\nu_1}}\Big) dx^{\mu_2}
                                                \Big(\frac{\partial}{\partial x^{\nu_2}}\Big) -dx^{\mu_1}
                                                \Big(\frac{\partial}{\partial x^{\nu_2}}\Big) dx^{\mu_2}
                                                \Big(\frac{\partial}{\partial x^{\nu_1}}\Big) \Big\} b_i
                                                \\
                                                &=  \sum_{\mu_1,\mu_2=1}^n \sum_{i=1}^{N} \frac{1}{2}
                                                \mathcal{F}^i_{\mu_1 \mu_2} (dx^{\mu_1} \wedge dx^{\mu_2})\Big(
                                                \frac{\partial}{\partial x^{\nu_1}}, \frac{\partial}{\partial
                                                x^{\nu_2}}\Big) b_i
                                                \\
                                                &=\mathcal{F}_x\Big(\frac{\partial}{\partial x^{\nu_1}},
                                                \frac{\partial}{\partial x^{\nu_2}}\Big)
                                                \\
                                                &= (d\mathcal{A})_x\Big(\frac{\partial}{\partial x^{\nu_1}},
                                                \frac{\partial}{\partial x^{\nu_2}}\Big) +
                                                \Big[\mathcal{A}_x\Big(\frac{\partial}{\partial x^{\nu_1}}\Big),
                                                \mathcal{A}_x\Big(\frac{\partial}{\partial x^{\nu_2}}\Big)\Big]
                                                \qquad \text{ by Proposition \ref{1553}}
                                                \\
                                                &= \frac{\partial}{\partial x^{\nu_1}} \mathcal{A}_x\Big(
                                                \frac{\partial}{\partial x^{\nu_2}}\Big) -
                                                \frac{\partial}{\partial x^{\nu_2}} \mathcal{A}_x\Big(
                                                \frac{\partial}{\partial x^{\nu_1}}\Big)
                                                -\mathcal{A}_x\Big(\Big[\frac{\partial}{\partial x^{\nu_1}} ,
                                                \frac{\partial}{\partial x^{\nu_2}}\Big]\Big) +
                                                \Big[\mathcal{A}_x\Big(\frac{\partial}{\partial x^{\nu_1}}\Big),
                                                \mathcal{A}_x\Big(\frac{\partial}{\partial x^{\nu_2}}\Big)\Big]
                                                \\
                                                &= \sum_{i=1}^N \frac{\partial\mathcal{A}^i_{\nu_2} }{\partial
                                                x^{\nu_1}} b_i - \frac{\partial  \mathcal{A}^i_{\nu_1}}{\partial
                                                x^{\nu_2}} b_i + \sum_{i,j=1}^N \big[\mathcal{A}^i_{\nu_1} b_i,
                                                \mathcal{A}^j_{\nu_2} b_j\big]\;.
                                            \end{align*}
                                            This is already the desired equation if we make use of the
                                            structure coefficients $c_{ij}^k$.
                                            Item $b)$ may be seen as follows. By  Proposition \ref{1614} we get the following equation on stalks:
                                            \begin{align*}
                                                0
                                                &=(D_X \cf)(\vecv_1,\vecv_2,\vecv_3) \\
                                                &= d \mathcal{F}(\vecv_1,\vecv_2,\vecv_3) \\
                                                & + [\mathcal{A}(\vecv_1),\mathcal{F}(\vecv_2,\vecv_3)]+
                                                [\mathcal{A}(\vecv_2),\mathcal{F}(\vecv_3,\vecv_1)]+
                                                [\mathcal{A}(\vecv_3),\mathcal{F}(\vecv_1,\vecv_2)] \;.
                                            \end{align*}
                                            Choosing $\vecv_{\mu}=\frac{\partial}{\partial x^{\mu}}$ we
                                            are done.
                                        \end{proof}
                                        In section \ref{1550} we introduced a canonical notion of
                                        covariant derivation on $X$ by pulling back a covariant derivation
                                        on the $X$-torsor $P$ under $G_X$ by means of the canonical
                                        section $s$. Alternatively we could have used the following less
                                        transparent but more explicit definition of covariant derivation
                                        on $X$.

                                        \begin{Def}\label{1611d}
                                            Let $Y$ be a smooth $S$-scheme. Let $\eta \in \Omega_{Y/S}^1(Y) \otimes_{\co_S(S)} \Liealg$ be
                                            a $\Liealg$-valued differential $1$-form, and let
                                            $\vartheta \in \Omega_{Y/S}^{k}(Y) \otimes_{\co_S(S)} \Liealg$
                                            be a $\Liealg$-valued differential $k$-form. Then we denote by  $[\eta,\vartheta]$ the
                                            following $\Liealg$-valued differential $(k+1)$-form:
                                            \begin{align*}
                                                [\eta,\vartheta](\vecv_1,\ldots,\vecv_{k+1})
                                                &:=\sum_{i=1}^{k+1}(-1)^{i+1} [\eta(\vecv_i),
                                                {\vartheta}(\vecv_1,\dots,\check{\vecv_i},\dots,\vecv_{k+1})].
                                            \end{align*}
                                        \end{Def}

                                        \begin{Def}\label{1613}
                                            Let $\alpha$ be a differential $k$-form  on $X$ with values in $\Liealg$.
                                            Then we define the covariant derivation $D_X \alpha$ of $\alpha$ at $x \in
                                            X$ in the following way:
                                            \begin{align*}
                                                D_X {\alpha} &:= d {\alpha} + [\mathcal{A}, \alpha].
                                            \end{align*}
                                        \end{Def}
                                        Let us use this notion of covariant derivation on $X$ in this
                                        section about Yang-Mills theory in local coordinates. Then
                                        Yang-Mills equation may be written as well with respect to the
                                        covariant derivation $D_X$ of Definition \ref{1613}, and furthermore one
                                        proves:

                                        \begin{Satz}\label{1614}
                                            Let $D_X$ be as in Definition  \ref{1613}. Then $D_X \cf =0$.
                                        \end{Satz}

                                        \begin{proof}
                                            It suffices to show the statement of the proposition on
                                            stalks, i.e.: $(D_X \cf)_x =0$ for all $x \in X$. Therefore let us
                                            perform the following computations in the stalk at $x$, but let us
                                            suppress the index $x$ in order to simplify the notation.

                                            Let $\omega$ be a connection form, and let $\Omega$ be
                                            the corresponding curvature form such that $\ca=s^*\omega$ and
                                            $\cf=s^*\Omega$. Then
                                            \begin{align*}
                                                D_X \cf &=  ds^*\Omega + [s^*\omega,s^*\Omega].
                                            \end{align*}
                                            By Lemma \ref{1615}, we know that $[s^*\omega,s^*
                                            \Omega]=s^*[\omega,\Omega]$, and furthermore $s^*$ commutes
                                            with the exterior differential $d$ (the latter was shown in the proof of
                                            Proposition \ref{1553}). It follows that
                                            \begin{align*}
                                                D_X \cf
                                                &=   s^* \Big( d\Omega+[\omega,\Omega] \Big).
                                            \end{align*}
                                            Then the statement of the proposition follows from
                                            Proposition \ref{1616}.
                                        \end{proof}

                                        \begin{Lemma}\label{1615}
                                            Let $s:X \hookrightarrow P$ be a section of the smooth and
                                            separated $X$-torsor $\pi:P \to X$ under $G_X:=G \times_S X$.
                                            Let $\eta \in \Omega_{P/S}^1(P) \otimes_{\co_S(S)} \Liealg$ be
                                            a $\Liealg$-valued differential $1$-form, and let
                                            $\vartheta \in \Omega_{P/S}^{k}(P) \otimes_{\co_S(S)} \Liealg$
                                            be a $\Liealg$-valued differential $k$-form. Consider the
                                            $\Liealg$-valued differential $(k+1)$-form
                                            \begin{align*}
                                                [\eta,\vartheta](\vecv_1,\ldots,\vecv_{k+1})
                                                &:=\sum_{i=1}^{k+1}(-1)^{i+1} [\eta(\vecv_i),
                                                {\vartheta}(\vecv_1,\dots,\check{\vecv_i},\dots,\vecv_{k+1})].
                                            \end{align*}
                                            Then on stalks the following identity holds:
                                            \begin{align*}
                                                [s^*\eta,s^*\vartheta]=s^*[\eta,\vartheta].
                                            \end{align*}
                                        \end{Lemma}

                                        \begin{proof}
                                            \begin{align*}
                                                &\pi^* \Big( (s^*[\eta, \vartheta])(\vecv_1,\ldots,\vecv_{k+1}) \Big)
                                                =  [\eta, \vartheta](s_*\vecv_1,\ldots,s_*\vecv_{k+1}) \\
                                                &=  \sum_{i=1}^{k+1}(-1)^{i+1}
                                                    [\eta(s_*\vecv_i), {\vartheta}(s_*\vecv_1,\dots,\check{s_*\vecv_i},\dots,s_*\vecv_{k+1})] \\
                                                &=  \sum_{i=1}^{k+1}(-1)^{i+1}
                                                    [\pi^*(\underbrace{(s^*\eta)(\vecv_i)}\limits_{=: \, \sum_j r_{ij} \otimes g_{ij}}),
                                                    \pi^*(\underbrace{({s^*\vartheta})(\vecv_1,\dots,\check{\vecv_i},\dots,\vecv_{k+1})}
                                                    \limits_{=: \, \sum_l r'_{il} \otimes g'_{il} })] \\
                                                &=  \sum_{i=1}^{k+1}(-1)^{i+1}
                                                    \sum_{j,l}[(\pi^* \otimes \text{id})(r_{ij} \otimes g_{ij}),
                                                    (\pi^* \otimes \text{id})(r'_{il} \otimes g'_{il})]     \\
                                                &=  \sum_{i=1}^{k+1}(-1)^{i+1}
                                                    \sum_{j,l}[(\pi^* r_{ij} \otimes g_{ij}),(\pi^* r'_{il} \otimes g'_{il})]
                                                =   \sum_{i=1}^{k+1}(-1)^{i+1}
                                                    \sum_{j,l}\Big( \underbrace{\pi^* r_{ij} \cdot \pi^* r'_{il}}\limits_{=\pi^*(r_{ij} \cdot r'_{il})}
                                                    \otimes [g_{ij},g'_{il}] \Big) \\
                                                &   \quad \, \text{(where we used the Lie-algebra structure on
                                                    $\co_P \otimes \Liealg$ introduced in Remark \ref{1430r})} \\
                                                &=  \sum_{i=1}^{k+1}(-1)^{i+1}
                                                    \sum_{j,l}(\pi^* \otimes \text{id}) ( r_{ij} \cdot r'_{il} \otimes [g_{ij},g'_{il}])\\
                                                &=  \sum_{i=1}^{k+1}(-1)^{i+1}
                                                    (\pi^* \otimes \text{id})
                                                    \Big( \Big[\sum_{j} r_{ij} \otimes g_{ij},\sum_{l} r'_{il} \otimes g'_{il}\Big] \Big) \\
                                                &=  \sum_{i=1}^{k+1}(-1)^{i+1}
                                                    \pi^*\Big([(s^*\eta)(\vecv_i),({s^*\vartheta})(\vecv_1,\dots,\check{\vecv_i},\dots,\vecv_{k+1})] \Big) \\
                                                &=  \pi^* \Big( [(s^*\eta),({s^*\vartheta})](\vecv_1,\ldots,\vecv_{k+1}) \Big)
                                            \end{align*}
                                            As the ring-homomorphism $\pi^*$ is injective, we are done.
                                        \end{proof}

                                        \begin{Satz}\label{1616}
                                            $d \Omega+[\omega,\Omega]=0$.
                                        \end{Satz}

                                        \begin{proof}
                                            The structure-equation \ref{1455} states that $\Omega = d
                                            \omega+\frac{1}{2}[\omega,\omega]$. Therefore
                                            \begin{align*}
                                                d \Omega+[\omega,\Omega]
                                                &=  dd\omega + \frac{1}{2}d[\omega,\omega] +
                                                    [\omega,d\omega]+
                                                    \frac{1}{2}[\omega,[\omega,\omega]] \;.
                                            \end{align*}
                                            The first summand is zero, because $d\circ d=0$. Let us show that the second and the third summand add to zero.
                                            Due to the definition of the exterior differential (Corollary \ref{1157}), the second summand
                                            reads as follows:
                                            \begin{align*}
                                                \left(\frac{1}{2}d[\omega,\omega]\right)(\vecu,\vecv,\vecw)
                                                &=  \frac{1}{2} \vecu([\omega,\omega](\vecv,\vecw))
                                                    -  \frac{1}{2} \vecv([\omega,\omega](\vecw,\vecu))
                                                    +  \frac{1}{2} \vecw([\omega,\omega](\vecu,\vecv)) \\
                                                    & \ \ \ \, - \frac{1}{2} [\omega,\omega]([\vecu,\vecv],\vecw)
                                                    + \frac{1}{2} [\omega,\omega]([\vecu,\vecw],\vecv)
                                                    -  \frac{1}{2}[\omega,\omega]([\vecv,\vecw],\vecu) \\
                                                &=  \vecu([\omega(\vecv),\omega(\vecw)]) -
                                                    \vecv([\omega(\vecw),\omega(\vecu)])+
                                                    \vecw([\omega(\vecu),\omega(\vecv)])\\
                                                &   \ \ \ \, - [\omega([\vecu,\vecv]),\omega(\vecw)] +
                                                    [\omega([\vecu,\vecw]),\omega(\vecv)]-  [\omega([\vecv,\vecw]),\omega(\vecu)]
                                            \end{align*}
                                            For the third summand we obtain
                                            \begin{align*}
                                                [\omega,d\omega](\vecu,\vecv,\vecw)
                                                &=  [\omega(\vecu),(d\omega)(\vecv,\vecw)]
                                                    -[\omega(\vecv),(d\omega)(\vecu,\vecw)]
                                                    +[\omega(\vecw),(d\omega)(\vecu,\vecv)] \\
                                                &=  [\omega(\vecu),(d\omega)(\vecv,\vecw)]
                                                    +[\omega(\vecv),(d\omega)(\vecw,\vecu)]
                                                    +[\omega(\vecw),(d\omega)(\vecu,\vecv)] \\
                                                &=  [\omega(\vecu),
                                                    \vecv(\omega(\vecw))-\vecw(\omega(\vecv))-\omega([\vecv,\vecw])]\\
                                                &   \quad \,  + [\omega(\vecv),
                                                    \vecw(\omega(\vecu))-\vecu(\omega(\vecw))-\omega([\vecw,\vecu])]\\
                                                &   \quad \, + [\omega(\vecw),
                                                    \vecu(\omega(\vecv))-\vecv(\omega(\vecu))-\omega([\vecu,\vecv])]\\
                                                &=  [\omega(\vecu),
                                                    \vecv(\omega(\vecw))] - [\omega(\vecu),\vecw(\omega(\vecv))] - [\omega(\vecu),\omega([\vecv,\vecw])]\\
                                                &   \quad \,  + [\omega(\vecv),
                                                    \vecw(\omega(\vecu))]-[\omega(\vecv),\vecu(\omega(\vecw))]-[\omega(\vecv),\omega([\vecw,\vecu])]\\
                                                &   \quad \, + [\omega(\vecw),
                                                    \vecu(\omega(\vecv))]-[\omega(\vecw),\vecv(\omega(\vecu))]-
                                                    [\omega(\vecw),\omega([\vecu,\vecv])] \\
                                                &=  \Big([\omega(\vecu), \vecv(\omega(\vecw))] + [\vecv(\omega(\vecu)),\omega(\vecw)]
                                                    \Big) - [\omega(\vecu),\omega([\vecv,\vecw])] \\
                                                &   \quad \,  + \Big([\omega(\vecv), \vecw(\omega(\vecu))] + [\vecw(\omega(\vecv)),\omega(\vecu)]
                                                    \Big) +[\omega(\vecv),\omega([\vecu,\vecw])]\\
                                                &   \quad \,  + \Big([\omega(\vecw), \vecu(\omega(\vecv))] + [\vecu(\omega(\vecw)),\omega(\vecv)]
                                                    \Big) - [\omega(\vecw),\omega([\vecu,\vecv])]
                                            \end{align*}
                                            The derivation $\vecu$  on $\co_P \otimes_{\co_S(S)} \Liealg$ is by definition of the
                                            form $\vecu = \vect_{\vecu} \otimes \text{id}$, where
                                            $\vect_{\vecu}$ is a derivation on $\co_P$. An analogous statement is of course true  for $\vecv$ and
                                            $\vecw$. Recalling the Lie-algebra structure of $\co_P \otimes_{\co_S(S)}
                                            \Liealg$ from Remark \ref{1430r}, we see that the equation
                                            \begin{align*}
                                                \vecu([a, b])= [\vecu (a), b]+ [a,\vecu (b)] && (*)
                                            \end{align*}
                                            holds for all $a,b \in \co_P \otimes_{\co_S(S)}
                                            \Liealg$. Thus we find indeed
                                            $\frac{1}{2}d[\omega,\omega]+[\omega,d\omega]=0$, and it only
                                            remains to prove the relation $(*)$. Writing $a=\sum_i f_i \otimes
                                            r_i$ and $b=\sum_j g_j \otimes s_j$ this may be seen as
                                            follows:
                                            \begin{align*}
                                                \vecu([a, b])
                                                &=  \vecu \Big( \Big[ \sum_i f_i \otimes r_i, \sum_j g_j \otimes s_j \Big] \Big)
                                                =   \sum_{i,j}(\vect_{\vecu} \otimes \text{id})(\underbrace{[ f_i \otimes r_i, g_j \otimes s_j]}
                                                    \limits_{= f_i \cdot g_j \otimes [r_i,s_j]}) \\
                                                &=  \sum_{i,j} \vect_{\vecu}(f_i \cdot g_j) \otimes [r_i,s_j]
                                                =   \sum_{i,j} \Big( \vect_{\vecu}(f_i) \cdot g_j + f_i \cdot \vect_{\vecu}(g_j) \Big) \otimes
                                                    [r_i,s_j]\\
                                                &=  \sum_{i,j}  \vect_{\vecu}(f_i) \cdot g_j \otimes
                                                    [r_i,s_j] +   \sum_{i,j} f_i \cdot \vect_{\vecu}(g_j)  \otimes
                                                    [r_i,s_j] \\
                                                &=  \sum_{i,j}  \Big[ \vect_{\vecu}(f_i) \otimes r_i,  g_j \otimes
                                                    s_j \Big] +   \sum_{i,j} \Big[ f_i \otimes r_i,  \vect_{\vecu}(g_j)  \otimes
                                                    s_j \Big] \\
                                                &=  \Big[ \sum_i {\vecu}(f_i \otimes r_i),  \sum_j g_j \otimes
                                                    s_j \Big] +   \Big[ \sum_i f_i \otimes r_i, \sum_j {\vecu}(g_j  \otimes
                                                    s_j) \Big] \\
                                                &=  [\vecu (a), b]+ [a,\vecu (b)].
                                            \end{align*}
                                            Let us finally prove that also $[\omega,[\omega,\omega]]=0$.
                                            This follows from the Jacobi-identity of Lie-algebras:
                                            \begin{align*}
                                                [\omega,[\omega,\omega]](\vecu,\vecv,\vecw)
                                                &=  [\omega(\vecu),[\omega,\omega](\vecv,\vecw)]
                                                    -[\omega(\vecv),[\omega,\omega](\vecu,\vecw)]
                                                    +[\omega(\vecw),[\omega,\omega](\vecu,\vecv)]\\
                                                &=  2 [\omega(\vecu),[\omega(\vecv),\omega(\vecw)]]
                                                    - 2 [\omega(\vecv),[\omega(\vecu),\omega(\vecw)]]
                                                    + 2 [\omega(\vecw),[\omega(\vecu),\omega(\vecv)]]\\
                                                &=  2 \Big( [\omega(\vecu),[\omega(\vecv),\omega(\vecw)]]
                                                    + [\omega(\vecv),[\omega(\vecw),\omega(\vecu)]]
                                                    + [\omega(\vecw),[\omega(\vecu),\omega(\vecv)]]
                                                    \Big)\\
                                                &=  0.
                                            \end{align*}
                                        \end{proof}

                                        \begin{Rem}\label{1617}
                                            Let $\alpha$ be a differential $k$-form on the $X$-torsor $P$
                                            under $G_X$, and let $\omega$ be a connection form. Then the results of this section motivate the
                                            following alternative definition of covariant derivation on
                                            $P$. We define the covariant derivation $D\alpha$ of $\alpha$ with respect to $\omega$ by the
                                            formula
                                            \begin{align*}
                                                D \alpha := d \alpha + [\omega,\alpha].
                                            \end{align*}
                                            In particular, Lemma \ref{1615} and the commutativity of the exterior derivation $d$ with pull-backs
                                            under closed immersions (see proof of Proposition \ref{1553})
                                            show that this definition of the covariant derivation $D$ on
                                            $P$ yields the identity
                                            \begin{align*}
                                                D_X(s^* \alpha)=s^*(D\alpha),
                                            \end{align*}
                                            where $s:X \hookrightarrow P$ is a section of $\pi:P \to X$,
                                            and where $D_X$ is the covariant derivation on $X$ in the
                                            sense of Definition \ref{1613}.
                                            Thus the covariant derivation  $D_X$ on $X$ may again be interpreted
                                            as pull-back of the covariant derivation $D$ on $P$.
                                        \end{Rem}
                                        \section{Yang-Mills equation in local coordinates}\label{1620}

                                        Let $g: T_{X/S} \times_X T_{X/S} \to \Affin_X^1$ be the metric
                                        which may be considered as well as a $\co_X(X)$-bilinear
                                        homomorphism $g:{\cal T}_{X/S}(X) \times {\cal T}_{X/S}(X)
                                        \longrightarrow \co_X(X)$ due to our expositions at the beginning
                                        of section \ref{1150}. In particular, we obtain a family of
                                        $\co_{X,x}$-bilinear morphisms
                                        \begin{align*}
                                            g_x:{\cal T}_{X/S,x} \times {\cal T}_{X/S,x} \longrightarrow
                                            \co_{X,x}
                                        \end{align*}
                                        on stalks for all $x\in X$ which may also be considered as a
                                        family of $\co_{X,x}$-linear isomorphisms $g_x:{\cal T}_{X/S,x}
                                        \stackrel{\sim}\longrightarrow {\Omega}_{X/S,x}^1$. One has got
                                        $g_{\mu\nu,x} :=  g_x\Big(\frac{\partial}{\partial  x^\mu}\Big)
                                        \Big(\frac{\partial}{\partial x^\nu}\Big) =   \sum_{\rho=1}^n
                                        g_{\mu\rho,x} d x^\rho \Big(\frac{\partial}{\partial x^\nu}\Big)$,
                                        i.e.
                                        \begin{align*}
                                            g_x(\frac{\partial}{\partial x^\mu}) = \sum_{\rho=1}^n
                                            g_{\mu\rho,x} dx^\rho.
                                        \end{align*}
                                        Then the inverse map $g_x^{-1}: {\Omega}_{X/S,x}^1 \to {\cal
                                        T}_{X/S,x}$ is given by $\frac{\partial}{\partial x^\mu}=
                                        \sum_{\rho=1}^n g_{\mu\rho,x} g^{-1}_x(dx^\rho)$. Let
                                        $g^{\mu\nu}_x$ be the components of the inverse matrix of
                                        $(g_{\mu\nu,x})_{\mu \nu}$, i.e.\ $\sum_{\rho=1}^n g_{\mu\rho}
                                        g^{\rho\nu}=\delta^\nu_\mu$. Then
                                        \begin{align*}
                                            g^{-1}_x(dx^\mu)=
                                            \sum_{\rho=1}^n g^{\mu\rho}_x \frac{\partial}{\partial
                                            x^\rho}.
                                        \end{align*}
                                        Again, let us write $g_{\mu \nu}$ (resp. $g^{\mu \nu}$) instead of
                                        $g_{\mu \nu,x}$ (resp. $g^{\mu \nu}_x$) whenever no confusion is
                                        possible. Now we are prepared to derive the Hodge-star operator in
                                        local coordinates. Let $\alpha \in \Omega_{X/S}^k(X)$ be a
                                        differential $k$-form on $X$. Then by Definition \ref{1159}  we have got
                                        \begin{align*}
                                            * \, \alpha:= (\Lambda^k g^{-1})(\alpha) \rfloor v_g.
                                        \end{align*}
                                        Writing
                                        \begin{align*}
                                            \alpha_x=\frac{1}{k!} \sum_{\mu_1,\dots,\mu_k=1}^n
                                            \alpha_{\mu_1\mu_2\dots\mu_k} dx^{\mu_1}\wedge dx^{\mu_2}\wedge
                                            \dots \wedge  dx^{\mu_k}\;,
                                        \end{align*}
                                        where the coefficients $\alpha_{\mu_1\mu_2\dots\mu_k}$ are totally
                                        antisymmetric, we obtain the first part of the Hodge-star operator
                                        \begin{align*}
                                            (\Lambda^k g^{-1} \alpha)_x = \frac{1}{k!}
                                            \sum_{\mu_1,\dots,\mu_k,\nu_1,\dots,\mu_k=1}^n g^{\mu_1\nu_1}
                                            \cdots g^{\mu_k\nu_k} \alpha_{\mu_1\dots\mu_k}
                                            \frac{\partial}{\partial x^{\nu_1}} \wedge \dots \wedge
                                            \frac{\partial}{\partial x^{\nu_k}} \;.
                                        \end{align*}
                                        Let us now choose a volume form (which exists due to
                                        Theorem \ref{1244}). As we are working with respect to \'etale topology,
                                        the local rings $\co_{X,x}$ are strictly henselian. Therefore
                                        $\sqrt{\text{det} \, g_x} \in \co_{X,x}$, where $\text{det} \,
                                        g_x$ denotes the determinant of the matrix $(g_{\mu\nu,x})_{\mu
                                        \nu}$. Then
                                        \begin{align*}
                                            v_{g,x}:= \sqrt{\text{det} \, g_x} \ dx^1 \wedge \ldots \wedge
                                            dx^n
                                        \end{align*}
                                        is a volume form which is independent of the choice of the base
                                        $\{ dx^{\mu} \}_{\mu=1}^n$ of $\Omega^1_{X/S,x}$. We obtain
                                        \begin{align*}
                                            &(*\alpha)_x \Big(\frac{\partial}{\partial x^{\nu_{k+1}}},
                                            \dots,\frac{\partial}{\partial x^{\nu_{n}}}\Big)
                                            = \Big( \Lambda^k g^{-1} (\alpha)
                                            \rfloor v_g \Big)_x \Big(\frac{\partial}{\partial x^{\nu_{k+1}}},
                                            \dots,\frac{\partial}{\partial x^{\nu_{n}}}\Big)
                                            \\
                                            &= \frac{1}{k!} \sum_{\mu_1,\dots,\mu_k,\nu_1,\dots,\nu_k=1}^n
                                            \sqrt{\det g_x}\, g^{\mu_1\nu_1} \cdots g^{\mu_k\nu_k}
                                            \alpha_{\mu_1\dots\mu_k} (dx^1 \wedge \ldots  \wedge dx^n)\Big(
                                            \frac{\partial}{\partial x^{\nu_1}},\dots,
                                            \frac{\partial}{\partial x^{\nu_{n}}}\Big)
                                            \\
                                            &= \frac{1}{k!} \sum_{\mu_1,\dots,\mu_k,\nu_1,\dots,\nu_k=1}^n
                                            \sqrt{\det g_x}\, g^{\mu_1\nu_1} \cdots g^{\mu_k\nu_k}
                                            \alpha_{\mu_1\dots\mu_k} \underbrace{\sum_{\pi \in S_n}
                                            \sigma_\pi \delta_{\nu_1}^{\pi(1)}\cdots
                                            \delta_{\nu_n}^{\pi(n)}}_{ =: \, \epsilon_{\nu_1\dots\nu_n}}
                                            \\
                                            &= \frac{1}{k!(n-k)!}
                                            \sum_{\mu_1,\dots,\mu_k,\rho_1,\dots,\rho_n=1}^n \sqrt{\det g_x}\,
                                            g^{\mu_1\rho_1} \cdots g^{\mu_k\rho_k}
                                            \epsilon_{\rho_1\dots\rho_n} \alpha_{\mu_1\dots\mu_k} \cdot
                                            \\*
                                            & \quad \ \cdot (dx^{\rho_{k+1}}\wedge \ldots \wedge
                                            dx^{\rho_n})\Big(\frac{\partial}{\partial x^{\nu_{k+1}}},
                                            \dots,\frac{\partial}{\partial x^{\nu_{n}}}\Big)
                                        \end{align*}
                                        and we may summarize as follows.

                                        \begin{Satz}\label{1621}
                                            Let $x\in X$ be a physical point, let $\alpha \in \Omega_{X/S}^k(X)$ be a differential $k$-form, and
                                            let $*$ be the Hodge-star operator. Then  the following equation holds at $x$:
                                            \begin{align*}
                                                (*\alpha)_x
                                                &= \frac{1}{k!(n-k)!}
                                                \sum_{\mu_1,\dots,\mu_k,\rho_1,\dots,\rho_n=1}^n \sqrt{\det g_x}\,
                                                g^{\mu_1\rho_1} \cdots g^{\mu_k\rho_k}
                                                \epsilon_{\rho_1\dots\rho_n} \alpha_{\mu_1\dots\mu_k}
                                                dx^{\rho_{k+1}}\wedge \ldots \wedge dx^{\rho_n} \\
                                                &= \frac{1}{k!(n-k)!}
                                                \sum_{\rho_1,\dots,\rho_n=1}^n \sqrt{\det g_x}\,
                                                \epsilon_{\rho_1\dots\rho_n} \alpha^{\rho_1\dots\rho_k}
                                                dx^{\rho_{k+1}}\wedge \ldots \wedge dx^{\rho_n}\;.
                                            \end{align*}
                                        \end{Satz}

                                        \begin{YME}\label{1622}
                                            Let $x\in X$ be a physical point. Then the Yang-Mills equation
                                            at $x$ reads as follows:
                                            \begin{align*}
                                                0=\sum\limits_{\mu,\rho,\sigma=1}^n\Big(\frac{1}{\sqrt{\det g_x}} \, \frac{\partial}{\partial x^{\sigma}} \Big(
                                                \sqrt{\det g_x}\, g^{\mu \nu} g^{\rho \sigma}
                                                \mathcal{F}^k_{\mu\rho}\Big) +  \sum_{i,j=1}^N c_{ij}^k
                                                g^{\mu \nu} g^{\rho\sigma}
                                                \mathcal{A}^i_{\sigma} \mathcal{F}^j_{\mu\rho}\Big)
                                            \end{align*}
                                            (where $\nu$ and $k$ are non-contracted indices).
                                        \end{YME}

                                        \begin{proof}
                                            Due to Proposition \ref{1621}, the Hodge-star of the field strength at $x$ is
                                            \begin{align*}
                                                (*\mathcal{F})_x
                                                = \frac{1}{2!(n-2)!} \sum_{\mu_1,\mu_2,\rho_1,\dots,\rho_n=1}^n
                                                \sqrt{\det g_x}\, g^{\mu_1\rho_1} g^{\mu_2\rho_2}
                                                \epsilon_{\rho_1\dots\rho_n} \mathcal{F}_{\mu_1\mu_2}
                                                dx^{\rho_3}\wedge \ldots \wedge dx^{\rho_n}.
                                            \end{align*}
                                            By Definition \ref{1613} one
                                            has got
                                            \begin{align*}
                                                (D_X (*\mathcal{F}))_x \big(\tfrac{\partial}{\partial x^{\nu_2}},
                                                \dots,\tfrac{\partial}{\partial x^{\nu_n}}\big) &= (d
                                                (*\mathcal{F}))_x \big(\tfrac{\partial}{\partial x^{\nu_2}},
                                                \dots,\tfrac{\partial}{\partial x^{\nu_n}}\big)
                                                \\
                                                &+ \sum_{i=2}^n (-1)^i \big[\mathcal{A}_x(\tfrac{\partial}{\partial
                                                x^{\nu_i}}), (*\mathcal{F})_x(\tfrac{\partial}{\partial x^{\nu_2}},
                                                \dots,\check{\tfrac{\partial}{\partial
                                                x^{\nu_i}}},\dots, \tfrac{\partial}{\partial
                                                x^{\nu_n}})\big] \;.
                                            \end{align*}
                                            The first summand on the right hand side is
                                            \begin{align*}
                                                &(d (*\mathcal{F}))_x
                                                \\
                                                &= \frac{1}{2!(n-2)!} \sum_{\mu_1,\mu_2,\mu_3,
                                                \rho_1,\dots,\rho_n=1}^n \frac{\partial}{\partial x^{\mu_3}} \Big(
                                                \sqrt{\det g_x}\, g^{\mu_1\rho_1} g^{\mu_2\rho_2}
                                                \epsilon_{\rho_1\dots\rho_n} \mathcal{F}_{\mu_1\mu_2}\Big)
                                                dx^{\mu_3} \wedge dx^{\rho_3}\wedge \dots \wedge
                                                dx^{\rho_n},
                                            \end{align*}
                                            and the second one may be written as follows:
                                            \begin{align*}
                                                &\sum_{\iota=2}^n (-1)^{\iota} \big[\mathcal{A}_x(\tfrac{\partial}{\partial
                                                x^{\nu_{\iota}}}), (*\mathcal{F})_x(\tfrac{\partial}{\partial x^{\nu_2}},
                                                \dots,{\check{\tfrac{\partial}{\partial x^{\nu_{\iota}}}}},\dots, \tfrac{\partial}{\partial
                                                x^{\nu_n}})\big]
                                                \\
                                                &= \sum_{i,j,k=1}^N c_{ij}^k b_k \sum_{{\iota}=2}^n (-1)^{\iota}
                                                \mathcal{A}^i_x(\tfrac{\partial}{\partial
                                                x^{\nu_{\iota}}}) (*\mathcal{F}^j)_x(\tfrac{\partial}{\partial x^{\nu_2}},
                                                \dots,{\check{\tfrac{\partial}{\partial x^{\nu_{\iota}}}}},\dots, \tfrac{\partial}{\partial
                                                x^{\nu_n}})
                                                \\
                                                &= \sum_{i,j,k=1}^N c_{ij}^k b_k (\mathcal{A}^i \wedge
                                                (*\mathcal{F}^j))_x(\tfrac{\partial}{\partial x^{\nu_2}},
                                                \dots,\tfrac{\partial}{\partial x^{\nu_n}}) \; .
                                            \end{align*}
                                            Thereby
                                            \begin{align*}
                                                &(\mathcal{A}^i \wedge (*\mathcal{F}^j))_x
                                                \\
                                                &= \frac{1}{2!(n-2)!}
                                                \sum_{\mu_1,\mu_2,\mu_3,\rho_1,\dots,\rho_n=1}^n \sqrt{\det g_x}\,
                                                g^{\mu_1\rho_1} g^{\mu_2\rho_2} \epsilon_{\rho_1\dots\rho_n}
                                                \mathcal{A}^i_{\mu_3} \mathcal{F}^j_{\mu_1\mu_2} dx^{\mu_3} \wedge
                                                dx^{\rho_3}\wedge \dots \wedge dx^{\rho_n}
                                            \end{align*}
                                            due to the above expression for $(*\cf)_x$. After substitution of these relations
                                            into the above formula for $(D_X (*\mathcal{F}))_x$, we
                                            finally have to apply the Hodge-star operator, in order to
                                            obtain the differential form occurring in  Yang-Mills equation.
                                            $(D_X (*\mathcal{F}))_x$ is a differential $(n-1)$-form,
                                            and thus the Hodge-star of this form is the following differential
                                            $1$-form:
                                            \begin{align*}
                                                &(* \, D_X \, *\mathcal{F})_x
                                                = \sum_{k=1}^N b_k  * \Big( d(*\mathcal{F}_k)+ \sum_{i,j=1}^N
                                                c_{ij}^k \mathcal{A}^i \wedge (*\mathcal{F}^j)\Big)
                                                \\
                                                &= \frac{1}{2!(n-2)!}  \sum_{k=1}^N b_k
                                                \sum_{\mu_1,\mu_2,\sigma_2,\rho_1,\dots,\rho_n,\nu_1,\dots,\nu_n}
                                                \sqrt{\det g_x} \ g^{\sigma_2\nu_2}g^{\rho_3\nu_3}\cdots
                                                g^{\rho_n\nu_n} \epsilon_{\nu_1\dots\nu_n}
                                                \epsilon_{\rho_1\dots\rho_n}
                                                \\
                                                & \quad \, \cdot \Big(\frac{\partial}{\partial x^{\sigma_2}} \Big(
                                                \sqrt{\det g_x}\, g^{\mu_1\rho_1} g^{\mu_2\rho_2}
                                                \mathcal{F}^k_{\mu_1\mu_2}\Big) +  \sum_{i,j=1}^N c_{ij}^k
                                                \sqrt{\det g_x}\, g^{\mu_1\rho_1} g^{\mu_2\rho_2}
                                                \mathcal{A}^i_{\sigma_2} \mathcal{F}^j_{\mu_1\mu_2}\Big)
                                                dx^{\nu_1}
                                                \\
                                                &= \frac{1}{2!(n-2)!}  \sum_{k=1}^N b_k
                                                \sum_{\mu_1,\mu_2,\sigma_1,\sigma_2,\sigma_3, \rho_1,\dots,\rho_n}
                                                \sqrt{\det g_x} \ \underbrace{\sum_{\nu_1,\dots,\nu_n} g^{\sigma_1\nu_1}
                                                g^{\sigma_2\nu_2}g^{\rho_3\nu_3}\cdots g^{\rho_n\nu_n}
                                                \epsilon_{\nu_1\dots\nu_n}}\limits_{= (\det g_x^{-1}) \cdot \epsilon_{\sigma_1\sigma_2\rho_3\dots\rho_n} }
                                                \epsilon_{\rho_1\dots\rho_n}
                                                \\
                                                & \quad \, \cdot \Big(\frac{\partial}{\partial x^{\sigma_2}} \Big(
                                                \sqrt{\det g_x}\, g^{\mu_1\rho_1} g^{\mu_2\rho_2}
                                                \mathcal{F}^k_{\mu_1\mu_2}\Big) +  \sum_{i,j=1}^N c_{ij}^k
                                                \sqrt{\det g_x}\, g^{\mu_1\rho_1} g^{\mu_2\rho_2}
                                                \mathcal{A}^i_{\sigma_2} \mathcal{F}^j_{\mu_1\mu_2}\Big)
                                                g_{\sigma_1\sigma_3} dx^{\sigma_3}
                                                \\
                                                &\text{ \quad \, (where we also made use of the relation
                                                \ $g^{\sigma_1 \nu_1} g_{\sigma_1\sigma_3} dx^{\sigma_3}
                                                = \delta^{\nu_1}_{\sigma_3} dx^{\sigma_3}
                                                = dx^{\nu_1}$)}
                                                \\
                                                &= \frac{1}{2!(n-2)!}  \sum_{k=1}^N b_k
                                                \sum_{\mu_1,\mu_2,\sigma_1,\sigma_2,\sigma_3,\rho_1,\rho_2}
                                                \frac{1}{\sqrt{\det g_x}} \, g_{\sigma_1\sigma_3}
                                                \underbrace{\sum_{\rho_3,\dots,\rho_n} \epsilon_{\sigma_1\sigma_2\rho_3\dots\rho_n}
                                                \epsilon_{\rho_1 \rho_2 \rho_3 \dots\rho_n}}\limits_{= \delta^{\sigma_1}_{\rho_1} \delta^{\sigma_2}_{\rho_2} -
                                                \delta^{\sigma_1}_{\rho_2} \delta^{\sigma_2}_{\rho_1} }
                                                \\
                                                &  \quad \, \cdot \Big(\frac{\partial}{\partial x^{\sigma_2}} \Big(
                                                \sqrt{\det g_x}\, g^{\mu_1\rho_1} g^{\mu_2\rho_2}
                                                \mathcal{F}^k_{\mu_1\mu_2}\Big) +  \sum_{i,j=1}^N c_{ij}^k
                                                \sqrt{\det g_x}\, g^{\mu_1\rho_1} g^{\mu_2\rho_2}
                                                \mathcal{A}^i_{\sigma_2} \mathcal{F}^j_{\mu_1\mu_2}\Big)
                                                dx^{\sigma_3}
                                                \\
                                                &= \frac{1}{2!(n-2)!}  \sum_{k=1}^N b_k
                                                \sum_{\mu_1,\mu_2,\sigma_1,\sigma_2,\sigma_3,\rho_1,\rho_2}
                                                \frac{1}{\sqrt{\det g_x}} \, g_{\sigma_1\sigma_3}
                                                (\delta^{\sigma_1}_{\rho_1} \delta^{\sigma_2}_{\rho_2} -
                                                \delta^{\sigma_1}_{\rho_2} \delta^{\sigma_2}_{\rho_1})
                                                \\
                                                & \quad \, \cdot \Big(\frac{\partial}{\partial x^{\sigma_2}} \Big(
                                                \sqrt{\det g_x}\, g^{\mu_1\rho_1} g^{\mu_2\rho_2}
                                                \mathcal{F}^k_{\mu_1\mu_2}\Big) +  \sum_{i,j=1}^N c_{ij}^k
                                                \sqrt{\det g_x}\, g^{\mu_1\rho_1} g^{\mu_2\rho_2}
                                                \mathcal{A}^i_{\sigma_2} \mathcal{F}^j_{\mu_1\mu_2}\Big)
                                                dx^{\sigma_3}
                                                \\
                                                &= \frac{1}{(n-2)!}  \sum_{k=1}^N b_k
                                                \sum_{\mu_1,\mu_2,\sigma_1,\sigma_2,\sigma_3} \frac{1}{\sqrt{\det
                                                g_x}} \, g_{\sigma_1\sigma_3}
                                                \\
                                                &  \quad \, \cdot \Big(\frac{\partial}{\partial x^{\sigma_2}} \Big(
                                                \sqrt{\det g_x}\, g^{\mu_1\sigma_1} g^{\mu_2\sigma_2}
                                                \mathcal{F}^k_{\mu_1\mu_2}\Big) +  \sum_{i,j=1}^N c_{ij}^k
                                                \sqrt{\det g_x}\, g^{\mu_1\sigma_1} g^{\mu_2\sigma_2}
                                                \mathcal{A}^i_{\sigma_2} \mathcal{F}^j_{\mu_1\mu_2}\Big)
                                                dx^{\sigma_3},
                                            \end{align*}
                                            because $\delta^{\sigma_1}_{\rho_1} \delta^{\sigma_2}_{\rho_2} -
                                            \delta^{\sigma_1}_{\rho_2} \delta^{\sigma_2}_{\rho_1}$
                                            yields a factor $2$ due to the antisymmetry $\mathcal{F}^i_{\mu\nu}
                                            =-\mathcal{F}^i_{\nu\mu}$ of the field strength. As $\{
                                            dx^{\mu}\}_{\mu=1}^n$ forms a base of $\Omega_{X/S}^1(X)$, as $\{b_i\}_{i=1}^N$ forms a base of $\Liealg$,
                                            and due to the fact that the metric $(g_{\mu\nu})_{\mu \nu}$ is an
                                            isomorphism, we see that the Yang-Mills equation reads as
                                            claimed.
                                        \end{proof}

\chapter{Appendix II}\label{1100}

In ordinary (i.e. $\real$-valued) Yang-Mills theory, gauge fields
are described by co-vector-fields. If we want to generalize the
$\real$-valued, differential geometric Yang-Mills theory to
arbitrary commutative rings $R$ (or even base schemes $S$), we
therefore have to supply the notion of the tangent bundle in the
realm of algebraic geometry. In this chapter \ref{1100} we are
going to recall some fundamental results which we will make use of
later. First we remember of the more special notion of Zariski
tangent-vectors, and then we will recall the general notion as it
is found in \SGAIII.

As the gauge group in classical Yang-Mills theory is given by a
Lie-group, we have to introduce the notion of group functors. In
regard to Yang-Mills theory it is particularly important that the
maps $\Ad$ and $\ad$ have algebraic geometric analogues.

\section{The Zariski tangent space}\label{1101}

\begin{Def}\label{1102}
    \begin{enumerate}
    \item
    Let $X$ be a scheme and $x\in X$. Let $\m_x$ be the maximal
    ideal of $\co_{X,x}$ and $k(x)=\co_{X,x}/\m_x$ be the residue
    field. Then $\m_x/\m_x^2=\m_x \otimes_{\co_{X,x}}k(x)$ is in
    a natural way a $k(x)$-vector space. Its dual $(\m_x/\m_x^2)
    ^\vee$ is called the {\it Zariski tangent space} to $X$ at $x$.
    We denote it by $T_{X,x}$.
    \item
    Let $f:X\to Y$ be a morphism of schemes, let $x\in X, \; y=f(x)$.
    Then $f_x^\#:\co_{Y,y}\to \co_{X,x}$ canonically induces a
    $k(x)$-linear map
    \begin{align*}
        T_{f,x}:T_{X,x}\to T_{Y,y}\otimes_{k(y)} k(x)
    \end{align*}
    the {\it tangent map} of $f$ at $x$.
    \end{enumerate}
\end{Def}
If $X \to S$ is a smooth morphism, then $X$ is regular, i.e.
dim~$\co_{X,x}=$dim$_{k(x)} T_{X,x}$ for all $ x\in X$. Let us see
that we are reduced to the ordinary differential geometric notions
in the standard situation of classical physics. \\  \\
\underline{physical interpretation:} Let {$S=$ Spec $\br$} and let
{$X=V(f) \hookrightarrow \ba_S^n $ be smooth over $S$ (e.g.
$f=\sum_{i=1}^n X_i^2 -1 $, i.e. $X(\real)=(n-1)$-sphere).} {Let
$x\in X(\br)$ be a closed point}. Then each irreducible component
of $X$ has the same dimension (\Liu, 2.5.26). So we may assume
that $X$ is irreducible. Then we get:
\begin{eqnarray*}
    \mbox{dim} \, \co_{X,x} = \mbox{dim} \, X = \mbox{dim} \, \ba_{\br}^n -1
    = \underbrace{\mbox{dim Spec}\,\br}_{=0} + n-1=n-1.
\end{eqnarray*}
The first equality is due to \Liu, Cor. 2.5.24, the second one is
due to \Liu, Cor. 2.5.26, and finally the third equality follows
from \Liu, Cor. 2.5.17. Therefore, $\mbox{dim}_\br \, T_{X,x} =
n-1$. Furthermore, if we write the polynomial $f$ in the form
$f=f(T_1, \ldots, T_n)$, one has due to \Liu, Prop. 4.2.5
\begin{align*}
    T_{X,x}= \left\{ (t_1, \ldots t_n) \in \real^n \ \big|
    \ \sum_{i=1}^n \frac{\partial f}{\partial T_i}(x) t_i = 0
    \right\}.
\end{align*}
Thus $T_{X,x}$ is really the tangent space in the classical sense.
\\ \\
Important is the following theorem which justifies the notion of
physical points (see Definition \ref{0057}). Let us stipulate that
we denote the dual of a vector space $V$ by $V^{\vee}$.
\begin{Theorem}\label{1103}
Let $f:X\to S$ be a smooth morphism, $s\in S$ and $x\in X_s$
closed. Then
\begin{align*}
    (\Omega_{X/S}^1 \otimes_{\co_{X,x}} k(x))^\vee = T_{X_s,x} \oplus
    (\Omega_{k(x)/k(s)}^1)^\vee.
\end{align*}
Especially for physical points we get
\begin{align*}
(\Omega_{X/S}^1 \otimes_{\co_{X,x}} k(x))^\vee = T_{X_s,x}.
\end{align*}
\end{Theorem}

\begin{proof}
    \Liu, Ex. 6.2.5
\end{proof}
{\bf Remark:}~ By the preceding theorem we establish the desired
duality of tangent and co-tangent vectors at physical points.

\begin{Def}\label{1104}
    \begin{enumerate}
    \item Let $(A, \maxi)$ be a regular noetherian local ring of dimension $d$.
    Any system of generators of $\maxi$ with $d$ elements is coordinate system for $A$.
    \item Especially if $X\to S$ is smooth, $\co_{X,x}$ is a
    regular noetherian local ring. A local {\it coordinate system} at $x\in X$ is a coordinate system for
    $\co_{X,x}$.
    In particular, the residue class of a local coordinate system at
    $x\in X$ gives rise to a basis of the tangent space $T_{X,x}$ and
    conversely by Nakayama's lemma.
    \end{enumerate}
\end{Def}
\section{The tangent bundle}\label{1110}

Let us now summarize the construction of the tangent bundle
following \SGAIII. We will see that, in physical points, this
definition reduces to the Zariski tangent space. But it is
designed in such a way that it also preserves all classical
properties of the tangent bundle (like the duality between tangent
and co-tangent vectors one is used to from differential geometry)
in non-physical points, too. Therefore we will use this notion for
our purposes.

\begin{Def}\label{1111}
    Let $S$ be a scheme and $\cm$ a quasi-coherent $\co_S$-module. Let
    $D_{\co_S}(\cm)$ denote the quasi-coherent $\co_S$-algebra
    $\co_S\oplus\cm$ (where $\cm$ is considered as $\co_S$-algebra via
    the definition $\cm\cdot\cm=0)$. Then
    \begin{align*}
        I_S(\cm):=\underline{\mbox{\rm Spec}} \; D_{\co_S}(\cm).
    \end{align*}
    In particular we define:
    \begin{align*}
        D_{\co_S}&:=D_{\co_S}(\co_S) \\
        I_S&:=I_S(\co_S).
    \end{align*}
\end{Def}

\begin{Def}\label{1112}
    Let $S$ be a scheme. For all $\co_S$-modules $\cf$ we define the functors $V(\cf)$ and $W(\cf)$ upon $(Sch)/S$
    \footnote{$(Sch)$ denotes the category of schemes, and
    $(Sch)/S$ denotes the category of $S$-schemes, i.e. the objects are schemes $X$ which come with a unique
    morphisms $X \to S$} by:
    \begin{eqnarray*}
    & & V(\cf)(S'):=\mbox{\rm Hom}_{\co_{S'}}(\cf\otimes\co_{S'},\co_{S'}) \\
    & & W(\cf)(S'):=\Gamma(S',\cf\otimes\co_{S'}).
    \end{eqnarray*}
\end{Def}

\begin{Def}\label{1113}
    Let $S$ be a scheme and $X$ an $S$-functor. We say that $X$ verifies
    $(E)$ relative to $S$, if for all $S'\to S$ and for all
    free $\co_{S'}$-modules of finite type $\cm$ and $\cn$ the commutative diagram
    \begin{align*}
        \xymatrix @-0.2pc @dr {  X(I_{S'}(\cm\oplus\cn)) \ar[d] \ar[r] &  X(I_{S'}(\cn)) \ar[d] \\
        X(I_{S'}(\cm)) \ar[r]  & X(S')}
    \end{align*}
    obtained by applying first $I_{S'}$ and then $X$ to the canonical
    commutative diagram
    \begin{align*}
        \xymatrix @-0.2pc @dr {  \cm\oplus\cn \ar[d] \ar[r] &  \cn \ar[d] \\
        \cm \ar[r]  & 0}
    \end{align*}
    is cartesian. (Recall that $I_{S'}(0) \cong S'$.)
\end{Def}
{\bf Remark:} If $X$ is an $S$-scheme, it verifies condition
$(E)$. In the situation of Definition \ref{1113} we will sometimes
simply say: $X/S$ verifies $(E)$.

\begin{Def}\label{1114}
    Let $\underline{\co}$ denote the functor
    \begin{align*}
        S\leadsto \Gamma(S,\co_S)
    \end{align*}
    provided with its structure of an
    $\widehat{(Sch)}$-ring.\footnote{If $\mathfrak{C}$ denotes a
    category, then $\widehat{\mathfrak{C}}$ denotes the category of
    contravariant functors on $\mathfrak{C}$.} In particular $\underline{\co}$ is
    represented by the scheme {\rm Spec}~$\bz[T]=\ba_\bz^1$. This
    induces by base change over $\bz$ an affine $S$-ring which we
    denote by $\underline{\co}_S$.
\end{Def}
Now we are prepared to introduce the tangent bundle.
\subsection{The tangent bundle as a functor.}\label{1114z}
\begin{Def}\label{1115}
    Let $S$ be a scheme and $\cm$ an $\co_S$-module of finite type.
    Let $f:X\to S$ be a scheme over $S$. The relative tangent bundle
    of $X$ over $S$ relative to $\cm$ is the
    $S$-functor\footnote{For all $S$-functors $\cf$, $\cg$ and for
    all $S$-morphisms $S' \to S$ one has got by definition
    \begin{align*}
        \GHom_S (\cf,\cg)(S'):=\Hom_{S'}(\cf_{S'},\cg_{S'}),
    \end{align*}
    where $\cf_{S'}$ (resp. $\cg_{S'}$) denotes the restriction of $\cf$ (resp.
    $\cg$) from $(Sch)/S$ to $(Sch)/S'$.
    }
    \begin{align*}
        T_{X/S}(\cm):=\GHom_S (I_S(\cm),X).
    \end{align*}
    The relative tangent bundle of $X$ over $S$ is the $S$-functor
    \begin{align*}
        T_{X/S}:=T_{X/S}(\co_S)=\GHom_S(I_S,X).
    \end{align*}
\end{Def}
$\cm\leadsto T_{X/S}(\cm)$ is a covariant functor from the
category of free $\co_S$-modules of finite type to the category of
$S$-functors. In particular, we get two canonical morphisms of
$S$-functors:
\begin{tabbing}
 \= 1.) structure morphism: \quad \= $T_{X/S}(\cm) \to X$ \\
 \> 2.) zero section:             \> $X \to T_{X/S}(\cm)$
\end{tabbing}
induced by the canonical morphism $0\to\cm$ and $\cm\to 0$
(recall: $T_{X/S}(0)\cong X$).

\begin{Satz}\label{1116}
    Let $I_S:=I_S(\co_S)$. Then there is an isomorphism of $X$-functors:
    \begin{align*}
        \GHom_S(I_S,X)\cong V(\Omega_{X/S}^1).
    \end{align*}
\end{Satz}

\begin{proof}
 Let $X'\stackrel{x'}{\longrightarrow} X$ be an object of
 $(Sch)/X$. Then
\begin{eqnarray*}
 \GHom_S(I_S,X)(X') & & \stackrellow{\mbox{\scriptsize in}\, (Sch)/X}{=} X(\ce_{\co_{X'}})^{-1}(x') \\
 & &  \quad \ \; = \mbox{Hom}_{\co_{X'}} ((x')^\ast\Omega_{X/S}^1, \co_{X'})
    \quad \text{\footnotesize{by \SGAIII, Expose II, Prop. 2.2}} \\
 & & \quad \ \; = V(\Omega_{X/S}^1)(X') \quad \text{\footnotesize{by Definition \ref{1112}}}.
\end{eqnarray*}
In particular, $\cm\leadsto T_{X/S}(\cm)$ is already a covariant
functor
\begin{eqnarray*}
 \mbox{(free} \; \co_S-\mbox{modules of finite type)} & \to & (X-\mbox{functors}) \\
 \cm & \leadsto & T_{X/S}(\cm).
\end{eqnarray*}
\end{proof}

\begin{Def}\label{1117}
    Let $u\in X(S)=\Gamma(X/S)$. The {\it (relative) tangent
    space of $X$ over $S$ at $u$ relative to $\cm$} is the $S$-functor
    \begin{align*}
        L_{X/S}^u(\cm),
    \end{align*}
    given by the inverse image of the $X$-functor $T_{X/S}(\cm)$ under
    the morphism $u:S\to X$; i.e. we have a pull-back diagram
    \begin{align*}
        \xymatrix @-0.2pc {  L_{X/S}^u(\cm) \ar[d] \ar[r] &  T_{X/S}(\cm) \ar[d] \\
        \quad S \quad \ar[r]  & \quad X \quad }
    \end{align*}
    The (relative) tangent space of $X$ over $S$ at $u$ is the
    $S$-functor
    \begin{align*}
        L_{X/S}^u:=L_{X/S}^u(\co_S).
    \end{align*}
\end{Def}

\begin{Satz}\label{1118}
    The functors $T_{X/S}(\cm)$ and $L_{X/S}^u(\cm)$ are functorial in
    $X$: If $f:X\to X'$ is an $S$-morphism, then there are commutative
    diagrams:
    \begin{align*}
        \xymatrix @-0.2pc {  T_{X/S}(\cm) \ar[d] \ar[r]^{T(f)} &  T_{X'/S}(\cm) \ar[d]  &
         L_{X/S}^u(\cm) \ar[dr] \ar[rr]^{T(f)} & &  L_{X'/S}^{f\circ u}(\cm) \ar[dl]\\
        \quad X \quad \ar[r]^f  & \quad X' \quad  & &  S  }
    \end{align*}
\end{Satz}

\begin{proof}
    \SGAIII, Expose II, Prop. 3.7
\end{proof}
\begin{Rem}\label{1119}
    If $f:X\to X'$ is \'etale, the above square is cartesian. In
    general there is a morphism of $X$-functors:\footnote{Let
    $X \to X'$ be a morphism and let ${\cal F}$ be a $X'$-functor. Recall that we
    denote the restriction of ${\cal F}$ to an $X$-functor
    by ${\cal F}_X$.}
    \begin{align*}
        T_{X/S}(\cm) \to (T_{X'/S}(\cm))_X.
    \end{align*}
    The morphism $T_{X/S}(\cm) \to (T_{X'/S}(\cm))_X$ (resp.
    $L_{X/S}^u(\cm)\to L_{X'/S}^{f\circ u}(\cm)$) is a morphism of
    $\co_X$-modules (resp. $\co_S$-modules) if $X/S$ and $X'/S$ verify
    $(E)$.
\end{Rem}

\begin{Satz}\label{1120}
    Let $X$ and $Y$ be two functors above $S$. Then there are
    isomorphisms which are functorial in $\cm$.
    \begin{eqnarray*}
        & & T_{X\times_S Y/S}(\cm) \stackrel{\sim}{\longrightarrow} T_{X/S}(\cm) \times_S \, T_{Y/S}(\cm) \\
        & & L_{X\times_S Y/S}^{(u,v)} (\cm) \stackrel{\sim}{\longrightarrow} L_{X/S}^u(\cm) \times_S L_{Y/S}^v (\cm).
    \end{eqnarray*}
\end{Satz}

\begin{proof}
    \SGAIII, Expose II, Prop. 3.8
\end{proof}
\begin{Satz}\label{1121}
    If $X/S$ and $Y/S$ verify $(E)$, then $X\times_S Y/S$ verifies
    $(E)$, too, and the above isomorphisms are compatible with the
    module-structures.
\end{Satz}

\begin{proof}
    \SGAIII, Expose II, Prop. 3.8.bis
\end{proof}
As $\GHom$ commutes with base change, we get

\begin{Satz}\label{1122}
    The formation of $T_{X/S}(\cm)$ and $L_{X/S}^u(\cm)$ commutes with
    base change: Let $S'\to S$. Then there are isomorphisms functorial
    in $\cm$:
    \begin{eqnarray*}
        & & T_{X_{S'}/S'}(\cm\otimes\co_{S'}) \stackrel{\sim}{\longrightarrow} [T_{X/S}(\cm)]_{S'} \\
        & & L_{X_{S'}/S'} (\cm\otimes\co_{S'}) \stackrel{\sim}{\longrightarrow} [L_{X/S}^u (\cm)]_{S'},
        \quad \text{where $u':=u(s)$.}
    \end{eqnarray*}
\end{Satz}

\begin{proof}
    \SGAIII, Expose II, Prop. 3.4
\end{proof}
\subsection{The tangent bundle as a scheme.}\label{1122a}
Let $R$ be a ring, $M$ a $R$-module. Consider the symmetric
algebra $ S(M):=S_R(M)$ defined by the following universal
property: For all commutative $R$-algebras $A$ there exist a
commutative diagram:
\begin{align*}
    \xymatrix @-0.2pc {  M \ar[d] \ar[r]^{T(f)} & S(M) \ar@{-->}[dl]^{\exists !}\\
    A}
\end{align*}
i.e.:
\begin{center}
    {$ \quad \quad \Hom_R(S(M),A) \cong \Hom_{R-\text{lin}}(M,A)$.}
\end{center}
Let now $(X,\co_X)$ be a scheme and let $\ce$ be a $\co_X$-module
over $X$. Then an open subset $U\subset X$ induces a
$\Gamma(U,\co_X)$-module $S(\Gamma(U,\ce))$, and thus a presheaf
of algebras
\begin{align*}
    U\leadsto S(\Gamma(U,\ce)).
\end{align*}
Let $S(\ce):=S_{\co_X}(\ce)$ be the sheaf associated to this
presheaf. Then $S(\ce)$ has the following universal property: For
all $\co_X$-algebras $\ca$ and for all homomorphisms $\ce\to\ca$
$\co_X$-modules there exists a commutative diagram
\begin{align*}
    \xymatrix @-0.2pc {  \ce \ar[d] \ar[r]^{} & S(\ce) \ar@{-->}[dl]^{\exists !}\\
    \ca}
\end{align*}
$S$ is a functor $\{\co_X$-module sheaves over $X\}\to
\{\co_X$-algebras over $X\}$. It has the following properties:
\begin{itemize}
    \item[a)] $S(\ce)_x=S(\ce_x) \quad \text{for all} \; x\in X$;
    \item[b)] $S(\ce\oplus\cf)=S(\ce)\otimes_{\co_X} S(\cf)$;
    \item[c)] $S_{\co_X}(\co_X)=\co_X[T]:=\co_X\otimes_\bz \bz[T]$;
    \item[d)] $S(f^\ast \cf)=f^\ast S(\cf)$ for all morphisms $f:(X,\co_X)\to(Y,\co_Y)$ ;
    \item[e)] $S(\tilde{M})=\widetilde{S(M)}$ if $X=\Spec A $ is affine and if $\ce=\tilde{M}$ is associated to an
              $A$-module $M$.
\end{itemize}

\begin{Def}\label{1122e}
    Let $S$ be a scheme and $\ce$ a quasi-coherent $\co_S$-module.
    Then
    \begin{align*}
        \bv(\ce):=\underline{\mbox{\rm Spec}} \, S_{\co_S}(\ce)
    \end{align*}
    is called the \emph{vector bundle} associated to $\ce$.
\end{Def}
Let now $f:X\to S$ be a morphism of schemes. Due to the universal
property of $S_{\co_S}(\ce)$ there are canonical bijections:
\begin{eqnarray*}
    \Hom_S(X,\bv(\ce))
    & = & \GHom_{\co_S}(S(\ce),f_\ast\co_X) \\
    & = & \GHom_{\co_S-\text{lin}}(\ce,f_\ast \co_X) \\
    & = & \GHom_{\co_X-\text{lin}}(f^*\ce, \co_X),
\end{eqnarray*}
where the last bijection is due to Proposition \ref{1325}. By
means of these equations, we deduce the following theorem.

\begin{Theorem}\label{1122f}
The sheaf of $S$-sections of $\bv(\ce)$ is in canonical bijection
with the dual $\ce^\vee:=\GHom_{\co_S}(\ce,\co_S)$ of $\ce$.
\end{Theorem}
Especially, if $X=\{\xi\}$ is the spectrum of a field $K$, the
structure-morphism $f:X\to S$ corresponds to a monomorphism
$k(s)\hookrightarrow K$, where $s:=f(\xi)$. The set of all
$S$-morphisms $\{\xi\}\to\bv(\ce)$ are the points of $\bv(\ce)$
with values in the field extension $K$ of $k(s)$ (and in
particular they are elements of the fibre $\pi^{-1}(s)$, where
$\pi: \bv(\ce)\to S$ is the structure-morphism). By the above,
this set is in bijection with the set of
$\co_X$-module-homomorphisms $f^\ast\ce\to\co_X=K$. Now,
\begin{align*}
    f^\ast \ce=(f^\ast\ce)_\xi = \ce_s
    \otimes_{\co_{S,s}}K=\ce_s/\m_s\ce_s \otimes_{k(s)} K.
\end{align*}
Thus we deduce the following one-to-one correspondence between
sets:
\begin{eqnarray*}
    \left\{ \begin{array}{cccc}
    \{\xi\} & \longrightarrow & \bv(\ce) \\[1ex]
    \downarrow & & \downarrow \\[1ex]
    \{s\} & \hookrightarrow & S \end{array} \right\} \;
    \stackrel{1:1}{\longleftrightarrow} & &
    \Hom_{K} \left(\ce_s/\m_s \ce_s \otimes_{k(s)} K,K\right) \\
    & = & \Hom_{k(s)} \left(\ce_s/\m_s \ce_s,k(s)\right) \otimes_{k(s)} K  \footnotesize{\text{ \, \
    (if $\ce_s/\m_s \ce_s$ or $K$ is}} \\
    & & \footnotesize{\text{\quad \quad \quad \quad \quad \quad \quad \quad \quad
    \quad  \quad \quad \quad \quad \quad \quad \quad \ \, finite-dim. over $k(s)$)}}  \\
    & = & (\ce_s/\m_s \ce_s)^\vee \otimes_{k(s)} K.
\end{eqnarray*}

\begin{Satz}\label{1123}
    The association  $\ce \rightsquigarrow \bv(\ce)$ is a
    contravariant functor
    \begin{align*}
        \bv: \mbox{(quasi-coherent} \; \co_S\mbox{-modules)} \; \to \;
        \mbox{(affine} \; S\mbox{-schemes)}.
    \end{align*}
    It has got the following properties:
    \begin{enumerate}
        \item
        $\bv(\ce)$ is of finite type over $S$ if $\ce$ is an $\co_S$-module of finite
        type.
        \item
        $\bv(\ce\oplus\cf)= \bv(\ce) \times_S \bv(\cf)$.
        \item
        $\bv(g^\ast \ce)=\bv(\ce) \times_S S'$ for all morphisms $g:S'\to S$.
        \item
        $\bv(\cf) \hookrightarrow \; \bv(\ce)$ is a closed
        immersion if $\ce\to\cf$ is a surjective morphism of quasi-coherent $\co_X$-modules.
    \end{enumerate}
\end{Satz}

\begin{proof}
    \EGAz, Prop. 1.7.11
\end{proof}
Now we use this general construction in order to define the
tangent bundle. We will see that sections of the tangent bundle
are indeed linked with derivations and actually  provide a
generalization of the differential geometric notion of
vector-fields.

\begin{Def}\label{1124}
    Let $f:X\to S$ be a morphism of schemes. Let $\cf$ be a
    $\co_X$-module. A homomorphism of sheaves of additive groups
    \begin{align*}
        \cd: \co_X\to\cf
    \end{align*}
    is called an $S$-derivation of $\co_X$ into $\cf$ if and only if:
    \begin{enumerate}
        \item[a)]
        For all open subsets $V\subset X$ and each pair $(t_1,t_2)$ of sections of $\co_X$ above $V$ one has:
        \begin{align*}
            \cd(t_1\otimes t_s)=t_1\cdot\cd(t_2)+\cd(t_1)\cdot t_2.
        \end{align*}

        \item[b)]
        For all open subsets $V\subset X$, each section $t$ of $\co_X$ above
        $V$ and each section $s$ of $\co_S$ above an open subset $U\subset S$
        such that $V\subset f^{-1}(U)$, one has
        \begin{align*}
            \cd(s|_V\cdot t)=s|_V\cdot \cd(t).
        \end{align*}
    \end{enumerate}
\end{Def}

\begin{Rem}\label{1125}
    \begin{enumerate}
        \item
        In the  situation above $\cd:\co_X\to\cf$ is an $S$-derivation if
        and only if for all $x\in X$ the homomorphism of additive groups
        $\cd_x:\co_{X,x}\to\cf_x$  is a $\co_{S,f(x)}$-derivation in the ordinary sense.
        \item
        The $S$-derivations of $\co_X$ into $\cf$ form a $\Gamma(X,\co_X)$-module
        $\GDer_S(\co_X,\cf)$.
        \item
        Let $\cf=\co_X$. An $S$-derivation of $\co_X$ into
        itself is simply called an $S$-derivation of $\co_X$.
    \end{enumerate}
\end{Rem}

\begin{Satz}\label{1126}
    Let $f: X\to S$ be a morphism of schemes and $d_{X/S}:\co_X \to
    \Omega_{X/S}^1$ the canonical $S$-derivation (see our exposition following Remark \ref{0013}). Then for all
    $\co_X$-modules $\cf$ there is a canonical isomorphism of $\Gamma(X,\co_X)$-modules:
    \begin{eqnarray*}
    \GHom_{\co_X} (\Omega_{X/S}^1,\cf) & \tilde{\longrightarrow} & \GDer_S(\co_X,\cf) \\
    u & \mapsto & u\circ d_{X/S}
    \end{eqnarray*}
\end{Satz}

\begin{proof}
    We may check the statement on stalks, but then the statement reduces to
    the defining universal property of $\Omega_{X/S,x}^1$, $x\in X$, by Remark \ref{1125} a).
\end{proof}
{\bf Remark:} Due to the pointwise characterization of
$S$-derivations, the presheaf
\begin{align*}
    U\leadsto \Der_S(\co_U, \cf|_U)
\end{align*} is already a sheaf.
This $\co_X$-module
\begin{align*}
    \GDer_S(\co_X,\cf)
\end{align*}
is called the sheaf of \emph{$S$-derivations} of $\co_X$ into
$\cf$. Thus we may restate the above proposition as follows:

\begin{Cor}\label{1127}
    For all $\co_X$-modules $\cf$ the homomorphism of $\co_X$-modules
    \begin{eqnarray*}
        \GHom_{\co_X} (\Omega_{X/S}^1,\cf) & \longrightarrow & \GDer_S(\co_X,\cf) \\
        u & \mapsto & u_d:=u\circ d_{X/S}
    \end{eqnarray*}
    is bijective.
\end{Cor}

\begin{Def}\label{1128}
    The dual ${\cal T}_{X/S}$ of the $\co_X$-module $\Omega_{X/S}^1$
    is called the tangent sheaf of $X$ relative to $S$. By
    Corollary \ref{1127} we can write:
    \begin{align*}
        {\cal T}_{X/S}:=\GHom_{\co_S}(\Omega_{X/S}^1,\co_X)=\GDer_S(\co_X,\co_X).
    \end{align*}
\end{Def}
Therefore we see how the tangent bundle must be defined.

\begin{Def}\label{1129}
    The tangent bundle $T_{X/S}$ of $X$ relative to $S$ is
    \begin{align*}
        T_{X/S}:=\bv(\Omega_{X/S}^1).
    \end{align*}
    This is well-defined, because the sheaf of relative differential forms $\Omega_{X/S}^1$ is a
    quasi-coherent $\co_X$-module.\footnote{It is admissible to
    use here the same label $T_{X/S}$ for the tangent-bundle as in
    Definition \ref{1115}, because due to Proposition \ref{1132} both notions coincide if the functor $X$
    is representable by a scheme.}
\end{Def}

\begin{Rem}\label{1130}
    Recalling Theorem \ref{1122f}, we see that there is a canonical bijection
    \begin{align*}
        \Gamma(T_{X/S}/X) \tilde{\longrightarrow} \;
        \Hom_{\co_X}(\Omega_{X/S}^1,\co_X) \, = \, \Gamma(X,{\cal T}_{X/S}).
    \end{align*}
    In this bijection, which is compatible with restrictions, we
    may  replace $X$ by an open subset $U\subset X$. Thus we conclude:
    There is an isomorphism between the tangent sheaf ${\cal T}_{X/S}$ of $X$
    relative $S$ and the sheaf of germs of $X$-sections of the fibre
    bundle $T_{X/S}$ of $X$ relative $S$. This means that
    vector fields are indeed given by sections of the tangent
    bundle as we are used to from differential geometry.
\end{Rem}
But let us finally prove that this notion of vector fields
actually reduces to the classical differential geometric notion.
We do so by considering the tangent space at physical points (see
Definition \ref{0057}). Consider  the special case of a
commutative diagram
\begin{align*}
    \xymatrix{\{\xi\} \ar@{^{(}->}[r] \ar[d] &  T_{X/S} \ar[d]\\
    \{x\} \ar@{^{(}->}[r] & X}
\end{align*}
where $x \in X$ and $\xi \in T_{X/S}$ are points such that
$k(x)=k(\xi)$. Let $T_{X/S}(x)$ denote the set of these diagrams;
i.e. $T_{X/S}(s)$ is the set of $x$-valued points of $T_{X/S}$. By
our exposition preceding Proposition \ref{1123}, we already know
that:
\begin{eqnarray*}
    T_{X/S}(x) & = & \Hom_{k(x)} (\Omega_{X/S}^1 \otimes k(x), \, k(x)) \\
\end{eqnarray*}
On the other hand we saw in Theorem \ref{1103} that there is also
a bijection
\begin{align*}
    T_{X_s,x}=\Hom_{k(x)} (\Omega_{X/S}^1 \otimes k(x),k(x))
\end{align*}
in terms of the Zariski tangent space at $x$, if $f:X\to S$ is
smooth and if $x\in X$ is a physical point. This yields indeed
$T_{X/S}(x)=T_{X_s,x}$ in physical situations.

\begin{Def}\label{1131}
    $T_{X/S}(x)$ is called the tangent-space of $X$ at $x$ relative to $S$.
\end{Def}
Finally let us remark that this notion of the tangent bundle and
of the tangent-space in terms of schemes as given above is a
special case of the notion in terms of functors which was
introduced  in the preceding subsection \ref{1114z}. Both notions
coincide if the considered functors are representable by schemes:

\begin{Satz}\label{1132}
    Let $X$ be an $S$-functor which is representable by an $S$-scheme $X \to
    S$. Let $T_{X/S}$ (resp. $T_{X/S}(\cm)$) denote the relative tangent bundle in the
    sense of Definition \ref{1115}.
    Then $T_{X/S}(\cm)$ and $L_{X/S}^u(\cm)$ are representable. In particular, we can write:
    \begin{eqnarray*}
        T_{X/S} & = & \bv (\Omega_{X/S}^1) \\
        L_{X/S}^u & = & \bv (u^\ast \Omega_{X/S}^1).
    \end{eqnarray*}
\end{Satz}

\begin{proof}
    \SGAIII, Expose II, Prop.3.3
\end{proof}

\begin{Rem}\label{1133}
Within the setting of Proposition \ref{1132} let $u$ be the
canonical inclusion $u:\{x\}\hookrightarrow X$ corresponding to a
point $x \in X$. Then there is a bijection
\begin{align*}
L_{X/S}^u(\{x\})=T_{X/S}(x).
\end{align*}
In particular, it is justified to label $L_{X/S}^u$ as tangent
space at $u$ as we did in Definition \ref{1117}.
\end{Rem}
Using the properties of fibre bundles, which are stated in
Proposition \ref{1123}, we conclude as follows.
\begin{Rem}\label{1134}
    \begin{enumerate}
        \item
        Let $f: X\to Y$ be an $S$-morphism of schemes and let $f^\ast
        \Omega_{Y/S}^1\to \Omega_{X/S}^1$ be the induced morphism
        pull-back morphism (see Definition \ref{0014}). Then, due to the identity
        $\bv(f^\ast \Omega_{Y/S}^1)=\bv (\Omega_{X/S}^1)\times_Y X$,
        the pull-back morphism gives rise to a canonical morphism induced by $f$:
        \begin{align*}
            T_{X/S}(f): T_{X/S} \to T_{Y/S} \times_Y X.
        \end{align*}
        If $g: Y\to Z$ is a second $S$-morphism, we get
        \begin{align*}
            T_{X/S} (g\circ f)=(T_{Y/S}(g) \times \mbox{id}_X)\circ T_{X/S}
            (f).
        \end{align*}

        \item For all base changes $S'\to S$ we get an isomorphism:
        \begin{align*}
            T_{X'/S'} \tilde{\longrightarrow} \; T_{X/S} \times_S S' = T_{X/S}
            \times_X \, X',
        \end{align*}
        where $X':=X \times_S S'$. If $x'\in X'$ lies over $x \in X$, we
        get
        \begin{align*}
            T_{X'/S'}(x')=T_{X/S}(x) \otimes_{k(x)} k(x').
        \end{align*}

        \item
        Let $f:X\to Y$ be an $S$-morphism, let $x \in X$ and
        $y:=f(x)$. Then
        \begin{align*}
            f^\ast \Omega_{Y/S}^1 \otimes_{\co_{X,x}} k(x) = \Omega_{Y/S}^1 \otimes_{\co_{Y,y}} \co_{X,x}
            \otimes_{\co_{X,x}} k(x)    = (\Omega_{Y/S}^1 \otimes_{\co_{Y,y}} k(y)) \otimes_{k(y)}
            k(x).
        \end{align*}
        If $\Omega_{Y/S}^1$ is a $\co_Y$-module of finite type (which will
        be the case in the situations we are going to consider), we have
        an isomorphism
        \begin{align*}
            \Hom_{k(x)} (f^\ast \Omega_{Y/S}^1 \otimes k(x),
            k(x))=T_{Y/S}(y) \otimes_{k(y)} k(x),
        \end{align*}
        because $k(y)\to k(x)$ is flat. Thus the pull-back of differential
        forms $f^\ast \Omega_{Y/S}^1 \to \Omega_{X/S}^1$ gives rise to a
        homomorphism of $k(x)$-vector spaces
        \begin{align*}
            T_{f,x}: T_{X/S}(x)\to T_{Y/S}(y) \otimes_{k(y)} k(x),
        \end{align*}
        called the \emph{tangent map} of $f$ at $x$. At physical points this definition coincides
        with the former Definition  \ref{1102} b).
    \end{enumerate}
\end{Rem}
As the sheaf $\GHom$ occurs in a canonical way if we consider
vector bundles (like the relative tangent-bundle), we finish with
a collection of some properties of $\GHom$ which might be useful.

\begin{Satz}\label{1135}
    Let $(X,\co_X)$ be a ringed space and let
    $\ce$ be a locally free $\co_X$-module of finite rank. We define
    the dual of $\ce$, denoted ${\ce}^{\vee}$, to be the sheaf
    of $\GHom_{\co_X} (\ce,\co_X)$. Then the following assertions
    are true:
    \begin{enumerate}
        \item[a)]
        $({\ce}^{\vee})^\vee \cong \ce$.

        \item[b)]
        For any $\co_X$-module $\cf$
        \begin{align*}
            \GHom_{\co_X} (\ce,\cf) \cong {\ce}^{\vee} \otimes \cf.
        \end{align*}

        \item[c)]
        For any $\co_X$-modules $\cf, \cg$
        \begin{align*}
            {\Hom}_{\co_X} (\ce\otimes\cf,\cg) \cong {\Hom}_{\co_X}
            (\cf,\GHom_{\co_X}(\ce,\cg)).
        \end{align*}

        \item[d)] (Projection Formula):
        If $f:(X,\co_X)\to (Y,\co_Y)$ is a morphism of ringed spaces, if
        $\cf$ is an $\co_X$-module, and if $\ce$ is a locally free
        $\co_Y$-module of finite rank, then there is a natural isomorphism
        \begin{align*}
            f_\ast(\cf \otimes_{\co_X} f^\ast\ce)\cong f_\ast \cf
            \otimes_{\co_Y} \ce.
        \end{align*}

        \item[e)]
        Let $X$ be a noetherian scheme, let $\cf$ be a coherent sheaf
        on $X$, let $\cg$ be any $\co_X$-module and let $x\in X$ be a point. Then we have
        \begin{align*}
            \GExt^i(\cf,\cg)_x \cong \Ext_{\co_{X,x}}^i (\cf_x,\cg_x)
        \end{align*}
        for all $i\ge 0$, where the right-hand side is $\Ext$ over the
        local ring $\co_{X,x}$.
    \end{enumerate}
\end{Satz}

\begin{proof}
    \Hart, Chap. II, Ex. 5.1 and \Hart, Chap. III, Prop. 6.8
\end{proof}

\begin{Satz}\label{1136}
    Let $X=\Spec A$ an affine scheme, and let $\cf, \cg$ be
    $\co_X$-modules. Then the canonical map
    \begin{align*}
        \Hom_{\co_X}(\cf,\cg) \to \Hom_{A}(\cf(X),\cg(X))
    \end{align*}
    is bijective if $\cf$ is quasi-coherent.
\end{Satz}

\begin{proof}
    \Liu, Ex. 5.1.5
\end{proof}

\begin{Satz}\label{1137}
    Let $X$ be a scheme and let $\ce$ be a locally free $\co_X$-module.
    Then the sheaf $\GHom$ commutes with pull-back, i.e.:
    \begin{align*}
        \alpha^* \ce^{\vee} \cong (\alpha^* \ce)^{\vee}
    \end{align*}  for all morphisms $\alpha: Y \to X$. Especially,
    if $p:X \to S$ is a smooth $S$-scheme, if $Y=S$, if $\alpha:S \hookrightarrow
    X$ is a section of $p$ and if $\ce=\Omega_{X/S}^1$, this
    isomorphism is canonical.
\end{Satz}

\begin{proof}
    If $\alpha$ is flat, one knows that the canonical morphism
    \begin{align*}
        \alpha^*  \ce^{\vee} \stackrel{\sim}\longrightarrow
        (\alpha^* \ce)^{\vee}, \qquad \varphi \otimes a \mapsto a \cdot
        \widetilde{\varphi},
    \end{align*}
    $\widetilde{\varphi}(e \otimes a):= \varphi(e) \otimes a$,
    is an isomorphism (see e.g. \Liu, Ex. 1.2.8). Also if $p$ is not flat, one can
    construct an explicit inverse of the canonical morphism above.
    In our physical situation $p:X \to S$ will be a smooth $S$-group
    scheme (see Definition \ref{1204}), $\ce$ will be the module of differential forms
    $\Omega^1_{X/S}$, and $\alpha:S \hookrightarrow X$ will be a
    section of $p$. So we will only perform the proof in this case, because
    then no explicit construction is necessary. We may simply apply the general theory: Recalling
    that $\alpha = \tau_{\alpha} \circ \varepsilon$, where
    $\varepsilon$ is the unit section and where
    $\tau_{\alpha}$ is the isomorphism given by left-translation
    with $\alpha$ (see subsection \ref{1240}), and evoking
    Proposition \ref{1241a}, we get canonical isomorphisms
    \begin{align*}
        p^*(\alpha^* \Omega_{X/S}^1)^{\vee}
        \cong p^*(\varepsilon^* \tau_{\alpha}^* \Omega_{X/S}^1)^{\vee}
        \cong p^*(\varepsilon^* \Omega_{X/S}^1)^{\vee}
        \cong (p^* \varepsilon^* \Omega_{X/S}^1)^{\vee}
        \cong (\Omega_{X/S}^1)^{\vee},
    \end{align*}
    because $p$ is flat due to smoothness. Applying the pull-back
    functor $\alpha^*$ and using the fact that $\alpha^* p^* = (p \circ \alpha)^* =
    \text{id}^*$ we are done.
\end{proof}
\section{Differential $p$-forms and the exterior
differential}\label{1150}

We already introduced the sheaf $\Omega_{X/S}^1$ of differential
$1$-forms in the beginning of section \ref{0011s}. In the previous
section \ref{1110} on the tangent-bundle, we recognized
differential forms as sections of the tangent-bundle: If we denote
by ${\cal
T}_{X/S}:=\left(\Omega_{X/S}^1\right)^{\vee}:=\GHom_{\co_X}(\Omega_{X/S}^1,\co_X)$
the dual of $\Omega_{X/S}^1$ (i.e. the sheaf of vector fields),
then in accordance with Definition \ref{1129} we define the
tangent-bundle $T_{X/S}$ and the cotangent-bundle $T_{X/S}^*$ as
follows:
\begin{equation*}
    T_{X/S}:=\bv(\Omega_{X/S}^1) \quad \quad \quad
    T_{X/S}^*:=\bv({\cal T}_{X/S}). \quad
\end{equation*}
Vector fields and differential forms may be identified with
sections of the respective bundles (see Theorem \ref{1122f}):
\begin{equation*}
    {\cal T}_{X/S}(X)= \Gamma(T_{X/S}/X) \quad \quad
    \Omega_{X/S}^1(X)=\Gamma(T_{X/S}^*/X)
\end{equation*}
As already exposed in section \ref{0200g}, a global section
$\omega$ of some tensorpower $\Omega_{X/S}^{\otimes n}$ of
$\Omega_{X/S}^1$ may be interpreted as a ``multilinear'' morphism
\begin{align*}
    \omega: T_{X/S} \times_X \ldots \times_X T_{X/S} & \to
    \Affin_X^1.
\end{align*}
This morphism is multilinear in two regards:
\begin{enumerate}
    \item[\underline{a) pointwise multilinearity:}]
    Let $S=\Spec R$ be
    an affine scheme, and let $X$ be a smooth and separated
    $S$-scheme. Consider an $S$-valued point $\alpha: S \to X$ of $X$.
    Physically we interpret $\alpha \in X(S)$ as an ``adelic'' point of
    \sto. First pulling back $\omega$ via $\alpha$ (i.e.
    performing a base change of $\omega$ with $\alpha$) we get a
    morphism $\alpha^* \omega$. Recalling the identity $T_{X/S}
    \times_X S = \bv(\alpha^*\Omega_{X/S}^1)$ (Proposition \ref{1123} c)), then
    applying the global section functor $\Gamma(\, \cdot \, ,S)$ to
    $\alpha^* \omega$, using the universal property $\Hom_X(X',Y
    \times_X Z) \cong \Hom_X(X',Y) \times \Hom_X(X',Z)$ of fibre
    products and finally evoking Theorem \ref{1122f} in order to provide the
    identity $\Gamma(\bv(\alpha^* \Omega_{X/S}^1)/S)=(\alpha^*
    \Omega_{X/S}^1)^{\vee}(S)$, we arrive at a morphism
    \begin{align*}
        \omega (\alpha): (\alpha^* \Omega_{X/S}^1)^{\vee}(S) \times
        \ldots \times (\alpha^* \Omega_{X/S}^1)^{\vee}(S)
        \longrightarrow R.
    \end{align*}
    This morphism is $R$-multilinear. Due to Proposition \ref{1137} and Proposition \ref{1136} there are isomorphisms
    \begin{align*}
        \left( \alpha^* \ct_{X/S} \right)(S)
        = (\alpha^* \Omega_{X/S}^1)^{\vee}(S)
        = \Hom_R\left((\alpha^* \Omega_{X/S}^1)(S),R\right)
    \end{align*}
    where $(\alpha^* \Omega_{X/S}^1)(S)$  is  the cotangent-space,
    and where $\left( \alpha^* \ct_{X/S}
    \right)(S)$ is the tangent-space at the ``adelic'' \st point $\alpha$. Thus
    $\omega(\alpha)$ (i.e. the evaluation of the differential form
    $\omega$ at the ``adelic'' \st point $\alpha$) behaves as one would
    expect: It is a multilinear form which maps a tuple of
    tangent-vectors at $\alpha$ to numbers $R$.

    \item[\underline{b) global multilinearity:}] $\omega$ does not
    only induce a multilinear form pointwise, but also on the global
    sections of the tangent-bundle: In order to see this apply the
    global section functor $\Gamma(\, \cdot \, ,X)$ to $\omega$.
    Recalling the bijection $\Hom_Y(Y,X \times_S Y)=\Hom_S(Y,X)$ for
    $S$-schemes $X$ and $Y$, we obtain in particular
    $\ba_X^1(X):=\Hom_X(X,\ba_{X}^1)$ $=\Hom_{\Spec \bz}(X,\ba_{\Spec \bz}^1)$. The
    latter set equals $\Hom_{\bz}(\bz[T],\co_X(X))=\co_X(X)$ due to
    the following lemma.
    \begin{Lemma}\label{1150e}
        Let $Y$ be an affine scheme. For any scheme $X$, the canonical
        map
        \begin{align*}
            \Hom(X,Y) \longrightarrow \Hom(\co_X(Y), \co_X(X))
        \end{align*}
        is bijective.
    \end{Lemma}

    \begin{proof}
        \Liu, Prop. 2.3.25
    \end{proof}

    Finally, making once again use of the universal property $\Hom_X(X',Y
    \times_X Z) \cong \Hom_X(X',Y) \times \Hom_X(X',Z)$ of fibre
    products and recalling the identification ${\cal T}_{X/S}(X)=
    \Gamma(T_{X/S}/X)= \Hom_X(X,T_{X/S})$ of vector fields with global
    sections of the tangent-bundle, we see that $\omega$ gives rise to
    a $\co_X(X)$-multilinear map:
    \begin{align*}
        \omega: {\cal T}_{X/S}(X) \times \ldots \times {\cal T}_{X/S}(X) \longrightarrow \co_X(X)
    \end{align*}
    All in all, $\omega$ indeed maps a tuple of vector fields to
    functions and is linear in each component.
\end{enumerate}
Thus we may think of differential forms in completely the same way
as we are used to from differential geometry. Using this
interpretation of differential forms as multilinear forms on
vector fields, we may state the following definitions in an
elegant way.

\begin{Def}\label{1151}
    \begin{enumerate}
    \item
    Let $f:X\to S$ be a morphism of schemes. The sheaf $\Omega_{X/S}^p$
    of {\it differential $p$-forms of $X$ relative $S$} is the
    $p$-th exterior product of the $\co_X$-module $\Omega_{X/S}^1$:
    \begin{align*}
        \Omega_{X/S}^p:= \bigwedge_{i=1}^p (\Omega_{X/S}^1).
    \end{align*}

    \item
    The \emph{exterior product} is a map $\wedge: \Omega_{X/S}^{k}
    \oplus
    \Omega_{X/S}^{l} \to \Omega_{X/S}^{k+l}$ defined  as follows:
    \begin{align*}
        (\alpha \wedge \beta)(\vecv_1,\dots,\vecv_{k+l}) = \frac{1}{k!l!}
        \sum_{\pi \in \mathfrak{S}_{k+l}} \sigma_\pi \alpha(\vecv_{\pi(1)},\dots
        \vecv_{\pi(k)}) \beta(\vecv_{\pi(k+1)},\dots ,\vecv_{\pi(k+l)})\;.
    \end{align*}
    Thereby $\mathfrak{S}_n$ denotes the group of permutations of the numbers $1,\dots,
    n$, and $\sigma_\pi$ is the sign of the permutation, i.e. the number of
    transpositions which transfer $1,2,\dots,n$ into
    $\pi(1),\pi(2),\dots,\pi(n)$.
    \end{enumerate}
\end{Def}
The exterior product has the following properties:
\begin{itemize}
    \item
    $(f_1 \alpha_1+f_2 \alpha_2) \wedge \beta = f_1 \alpha_1
    \wedge \beta + f_2 \alpha_2 \wedge \beta$,\qquad \ $f_1,f_2 \in
    \co_X$

    \item
    $\alpha \wedge (\beta \wedge \gamma)= (\alpha \wedge \beta)
    \wedge \gamma$

    \item
    $\alpha \wedge \beta=(-1)^{kl} \beta \wedge \alpha$,\qquad
    \quad \qquad \qquad \qquad \quad \quad \
    $\alpha \in  \Omega_{X/S}^{ k}$, $\beta \in \Omega_{X/S}^{l}$.
\end{itemize}

\begin{Rem}\label{1152}
    From these properties we conclude that there is a
    graded algebra of differential forms in algebraic
    geometry, too. More precisely, we may state:
    \begin{enumerate}
    \item
    $\Omega_{X/S}^0 = \co_X$ and $\Omega_{X/S}^p=0$ for $p<0$.

    \item
    The $\Omega_{X/S}^p$ are the homogeneous components of
    (the graded) exterior algebra of $\Omega_{X/S}^1$:
    \begin{align*}
        \Omega_{X/S}^\bullet:=\bigwedge (\Omega_{X/S}^1)=:
        \bigoplus_{p\in\bz} \Omega_{X/S}^p.
    \end{align*}

    \item
    $\Omega_{X/S}^\bullet$ is a graded, quasi-coherent,
    anti-commutative $\co_X$-algebra via the exterior product . For open affine $U\subset X$ one has
    \begin{align*}
        \Gamma(U,\Omega_{X/S}^\bullet)=\bigwedge \Gamma (U,\Omega_{X/S}^1)
    \end{align*}
    where $\Gamma(U,\Omega_{X/S}^1)$ is considered as a
    $\Gamma(U,\co_X)$-module.

    \item
    $\Omega_{X/S}^p=\left(\Omega_{B/R}^p\right)^\sim:= \left(\bigwedge\limits_{i=1}^p \Omega_{B/R}^1\right)^\sim$
    if $X=\Spec B$ and $S= \Spec R$ are affine.
    \end{enumerate}
\end{Rem}

\begin{Def}\label{1153}
The \emph{interior product} between a vector field $\vecv \in
\ct_{X/S}$ and a differential $k$-Form $\alpha \in
\Omega_{X/S}^k$, $k\geq 1$, is a differential $(k-1)$-form
$i_{\vecv} \alpha \in \Omega_{X/S}^{k-1} $ defined as follows:
\begin{align*}
    (i_{\vecv} \alpha)(\vecv_1,\dots,\vecv_{k-1}) :=
    \alpha(\vecv,\vecv_1,\dots,\vecv_{k-1})\;,
\end{align*} where $\vecv_i \in
\ct_{X/S}$.
\end{Def}
Applying the above construction of the exterior product to the
sheaf $\ct_{X/S}$ of vector fields instead of the sheaf
$\Omega_{X/S}^1$ of co-vector fields, we may consider the exterior
product
\begin{align*}
    \ct_{X/S}^k:=\bigwedge\limits_{i=1}^k \ct_{X/S}.
\end{align*}
In particular $\ct_{X/S}^0 =\co_X$, $\ct_{X/S}^1 =\ct_{X/S}$ and
$\ct_{X/S}^k =0$ if $k < 0$.

\begin{Def}\label{1154}
A \emph{$k$-vector field} is a global section of $\ct_{X/S}^k $.
\end{Def}
Then we may define a $\co_X$-bilinear morphism $\langle \cdot ,
\cdot \rangle: \ct_{X/S}^k \oplus \Omega_{X/S}^k  \longrightarrow
\co_X$ by means of $\langle \vecv_1 \wedge \dots \wedge
\vecv_k,\alpha\rangle := k! \ \alpha(\vecv_1,\dots,\vecv_k)$.

\begin{Def}\label{1155}
    The $\Verjungung$ of a differential $(k+l)$-form $\omega \in
    \Omega_{X/S}^{k+l}$ with a $k$-vector field $\vecv \in \ct_{X/S}^k$
    is the $\co_X$-bilinear morphism
    \begin{align*}
        \rfloor: \ct_{X/S}^k \oplus \Omega_{X/S}^{k+l} \longrightarrow
        \Omega_{X/S}^l, \quad (\vecv,\omega) \mapsto \vecv \rfloor \omega
    \end{align*}
    defined as follows: $\left(\vecv \rfloor \omega \right)(\vecv_1,
    \ldots, \vecv_l) := \langle \vecv \wedge \vecv_1 \wedge \ldots
    \wedge \vecv_l, \omega \rangle$.
\end{Def}
The $\Verjungung$ has the following properties:
\begin{itemize}
    \item
    $(f_1 X_1+f_2X_2)\rfloor \alpha = f_1 (X_1 \rfloor \alpha) +
    f_2 (X_2\rfloor \alpha)$,

    \item
    $X\rfloor (f_1 \alpha_1 + f_2 \alpha_2) = f_1 (X \rfloor
    \alpha) + f_2 (X\rfloor \alpha_2)$.
\end{itemize}
A further very important operation is the \emph{exterior
differential}.

\begin{Theorem}\label{1156}
    There exists a unique endomorphism $d:\Omega_{X/S}^\bullet \to
    \Omega_{X/S}^\bullet$ of sheaves of additive groups, the so called \emph{exterior
    differential}, such that:
    \begin{enumerate}
        \item[(i)]
        $d\circ d=0$.

        \item[(ii)]
        For all open $U\subset X$ and for all
        $f\in\Gamma(U,\co_X)$ one has
        \begin{align*}
        df=d_{X/S} f.
        \end{align*}

        \item[(iii)]
        For all open $U\subset X$, all integers $p,q\in\bz$ and for all pairs of sections
        $\omega'_p\in\Gamma(U,\Omega_{X/S}^p), \;
        \omega''_q\in\Gamma(U,\Omega_{X/S}^q)$ we have:
        \begin{align*}
            d(\omega'_p\wedge\omega''_q)=(d\omega'_p)\wedge\omega''_q+(-1)^p
            \omega'_p\wedge (d\omega''_q).
        \end{align*}
    \end{enumerate}
\end{Theorem}

\begin{proof}
    \EGAvv, Thm. 16.6.2
\end{proof}
Especially $d$ is local, because it is a morphism of sheaves; i.e.
if $\alpha|_U=\beta|_U$ for some sections $\alpha, \beta$ of
$\co_X$ and an open subset $U \subset X$, then
\begin{align*}
    (d\alpha)|_U=d(\alpha|_U)=d(\beta|_U)=(d\beta)|_U.
\end{align*}

\begin{Cor}\label{1157}
    Let $\omega \in \Omega_{X/S}^k$ and $\vecv_0, \vecv_1, \ldots, \vecv_k \in
    \ct_{X/S}$. Then the exterior differential has the following
    form:
    \begin{align*}
        (d \omega)(\vecv_0,\vecv_1,\ldots,\vecv_k)
        &:= \, \  \sum_{i=0}^k \ \ \, (-1)^i \ \vecv_i ( \omega(\vecv_0,\dots,\check{\vecv_i},\dots,\vecv_k))
        \\
        &  + \sum_{0 \leq i < j \leq k} (-1)^{i+j}
        \omega([\vecv_i,\vecv_j],\vecv_0,\dots,\check{\vecv_i},\dots,\check{\vecv_j},\dots,\vecv_k)\;,
    \end{align*}
    where $\check{\vecv_i}$ means that this vector is
    omitted.\footnote{
    In order to give sense to the expression
    $\vecv_i ( \omega(\vecv_0,\dots,\check{\vecv_i},\dots,\vecv_k))$,
    let us point out that in this situation $\vecv_i$ is  considered as
    an $S$-derivation of $\co_X$ by means of Corollary \ref{1127}.} The
    commutator $[\cdot,\cdot]$ of vector fields is defined in
    Definition \ref{1158}.
\end{Cor}

\begin{proof}
    One has to check that the three conditions of Theorem \ref{1156}
    are fulfilled.
    In order to show $(ii)$ let $f \in \co_X$ and $\vecv \in \ct_{X/S}$. Then by
    definition $(df)(\vecv):=\vecv(f)$, where on the right hand side $\vecv$ is considered as a
    derivation by means of Corollary \ref{1127}. Thus $(df)(\vecv)=\vecv(d_{X/S} f)$ if $\vecv$ is now
    considered as a dual differential from. Finally the canonical
    isomorphism $\ce \to (\ce^{\vee})^{\vee}$ provides an equality
    $\vecv(d_{X/S}f)= (d_{X/S} f)(\vecv)$ . This proves $(ii)$, because
    $\vecv$ is arbitrary.
    A longer algebraic manipulation (which is known from differential geometry)
    shows that $d$ fulfills the conditions $(i)$ and $(iii)$ of Definition \ref{1153}, too.
\end{proof}

\begin{Def}\label{1158}
    Let $X \to S$ be an $S$-scheme, let $\vecv, \vecw \in
    \Gamma(T_{X/S}/X)$ be vector fields, and let $f \in \co_X$.
    Let us introduce the following abbreviation $\vecv_d:=\vecv \circ d$.
    From the algebraic properties of the derivations $\vecv_d$
    and $\vecw_d$ it follows that
    \begin{align*}
        [\vecv,\vecw](f)
        := (\vecv_d)\left((\vecw_d)(f)\right)- (\vecw_d)\left((\vecv_d)(f)\right)
    \end{align*}
    is a derivation, too.\footnote{
        $[\vecv,\vecw](f g)
        = (\vecv_d)\left((\vecw_d)(fg)\right)-
        (\vecw_d)\left((\vecv_d)(fg)\right)$

        $ \qquad \quad = (\vecv_d)\left((\vecw_d)(f) \cdot g + f \cdot (\vecw_d)(g)  \right)
            - (\vecw_d)\left((\vecv_d)(f) \cdot g + f \cdot (\vecv_d)(g)
            \right)$

        $ \qquad \quad = (\vecv_d)((\vecw_d)(f)) \cdot g + (\vecw_d)(f) \cdot
            (\vecv_d)(g) +
            (\vecv_d)(f)(\vecw_d)(g) + f \cdot
            (\vecv_d)((\vecw_d)(g))$

             $\qquad \quad \quad \ -
            (\vecw_d)((\vecv_d)(f)) \cdot g - (\vecv_d)(f) \cdot
            (\vecw_d)(g) -
            (\vecw_d)(f)(\vecv_d)(g) - f \cdot
            (\vecw_d)((\vecv_d)(g))$

        $\qquad \quad = [\vecv,\vecw](f) \cdot g + f \cdot [\vecv,\vecw](g) \qquad \text{for all $f,g \in \co_X$}$.
    }
    The corresponding vector field
    $[\vecv,\vecw] \in \Gamma(T_{X/S}/X)$ is called the \emph{commutator
    of $\vecv$ and $\vecw$}.
\end{Def}
Let us conclude with the definition of the Hodge-star operator.
For this purpose recall that the metric yields an isomorphism
$g:{\cal T}_{X/S} \to ({\cal T}_{X/S})^{\vee} = \Omega_{X/S}^1$
(Definition \ref{0201}). In particular we get isomorphisms
\begin{align*}
    \Lambda^k g^{-1}: \Omega_{X/S}^k \stackrel{\sim}\longrightarrow
    {\cal T}_{X/S}^k
\end{align*}
for all $k \in \Natural$. By Theorem \ref{1244} there exists a
volume form $v_g$ in physical situations.

\begin{Def}\label{1159}
    Let $X \to S$ be of relative dimension $n$ and assume that
    there exists a volume form $v_g$ on $X$.
    Let $\alpha \in \Omega^k_{X/S}$ be a differential $k$-form.
    Then the operator $*: \Omega^k_{X/S} \to \Omega^{n-k}_{X/S}$,
    \begin{align*}
        * \, \alpha:= (\Lambda^k g^{-1})(\alpha) \rfloor v_g
    \end{align*}
    is called the \emph{Hodge-star operator}.
\end{Def}
\section{Group Schemes}\label{1200}

\subsection{Definition of group schemes}\label{1200a}

Let $\Ccat$ be a category; for example, let $\Ccat$ be the
category $(Sch/S)$ of schemes over a fixed scheme $S$. Each object
$X\in \Ccat$ gives rise to its functor of points
\begin{align*}
    h_X:\Ccat \to \left( Sets \right)
\end{align*}
which associates to any $T\in \Ccat$ the set
\begin{align*}
    h_X(T):=X(T):={\Hom} (T,X)
\end{align*}
of $T$-valued points of $X$. Each morphism $X\to X'$ in $\Ccat$
induces a morphism $h_X\to h_{X'}$ of functors by the composition
of morphisms in $\Ccat$. In this way one gets a covariant functor
\begin{align*}
    f:\Ccat \to {\Hom} \left( \Ccat^0, {\left( Sets \right)} \right)
\end{align*} of $\Ccat$ to the category of covariant functors from
$\Ccat^0$ (the dual of $\Ccat$) to the category of sets; the
category $\Hom(\Ccat^0$,(Sets)) is denoted by $\hat{\Ccat}$; it is
called the category of contravariant functors from $\Ccat$ to
(Sets).

\begin{Satz}\label{1201}
    The functor $h: \Ccat \to\hat{\Ccat}$ is fully faithful; i.e.,
    for any two objects $X, \; X'\in \Ccat$, the canonical map
    \begin{align*}
        {\Hom}_{\Ccat}(X,X')\to {\Hom}_{\hat{\Ccat}}(h_X,h_{X'})
    \end{align*} is bijective. More generally, for all objects $X\in
    \Ccat$ and $\cf \in \hat{\Ccat}$, there is a canonical bijection
    \begin{align*}
        \cf(X) \stackrel{\sim}{\longrightarrow} {\Hom}_{\hat{\Ccat}}(h_X,\cf)
    \end{align*}
    mapping $u\in\cf(X)$ to the morphism $h_X\to\cf$ which to a
    $T$-valued point $g\in h_X(T)$, where $T$ is an object of $\Ccat$,
    associates the element $\cf(g)(u)\in\cf(T)$. The bijection
    coincides with the above one if $\cf=h_{X'}$ and is functorial in
    $X$ and $\cf$ in the sense that $\cf\mapsto$
    $\Hom_{\hat{\Ccat}}(h(\cdot),\cf)$ defines an isomorphism
    $\hat{\Ccat}\to\hat{\Ccat}$.
\end{Satz}
In particular, if a functor $\cf\in$ $\Hom(\Ccat^0$,(Sets)) is
isomorphic to a functor $h_X$, then $X$ is uniquely determined by
$\cf$ up to an isomorphism in the category $\Ccat$. In this case,
the functor $f$ is said to be {\it representable}. Thus
Proposition \ref{1201} says that the functor $h$ defines an
equivalence between the category $\Ccat$ and the full subcategory
of $\Hom(\Ccat^0$,(Sets)) consisting of all representable
functors.

In order to define group objects in the category $\Ccat$, it is
necessary to introduce the notion of a law of composition on an
object $X$ of $\Ccat$. By the latter we mean a functorial morphism
\begin{align*}
    \gamma: h_X\times h_X\to h_X.
\end{align*}
Thus, a law of composition on $X$ consists of a collection of maps
\begin{align*}
    \gamma_T: h_X(T)\times h_X(T)\to h_X(T)
\end{align*}
(laws of composition on the sets of $T$-valued points of $X$)
where $T$ varies over the objects in $\Ccat$. The functoriality of
$\gamma$ means that all maps $\gamma_T$ are compatible with
canonical maps between points of $X$, i.e., for any morphism $u:
T'\to T$ in $\Ccat$, the diagram
\begin{align*}
    \xymatrix @-0.2pc {  h_X(T)\times h_X(T) \ar[r]^{ \quad \quad \gamma_T}
    \ar[d]_{h_X(u)\times h_X(u)}  & h_X(T) \ar[d]^{h_X(u)} \\
    h_X(T')\times h_X(T') \ar[r]^{\quad \quad \gamma_{T'}}   &  h_X(T')}
\end{align*}
is commutative. If the law of composition has the property that
$h_X(T)$ is a group under $\gamma_T$ for all $T$, then $\gamma$
defines on $h_X$ the structure of a group functor, i.e., of a
contravariant functor from $C$ to the category of groups. In this
case, $\gamma$ is called a group law of $X$.

\begin{Def}\label{1202}
    A group object in $\Ccat$ is an object $X$ together with a law
    of composition $\gamma: h_X\times h_X\to h_X$ which is a group law.
\end{Def}
It follows that a group object in $\Ccat$ is equivalent to a group
functor which, as a functor to the category of sets, is
representable.

When dealing with group objects, it is convenient to know that the
category in question contains direct products and a final object,
say $S$. The latter means that, for each object $T$ of $\Ccat$,
there is a unique morphism $T\to S$. So, in the following, assume
that $\Ccat$ is of this type, and consider a groups object $X$ of
$\Ccat$ with group law $\gamma$. Then, since the product $X\times
X$ exists in $\Ccat$ and since the functor $h:\Ccat \to$
Hom$(\Ccat^0$,(Sets)) commutes with direct products, the law of
composition $\gamma: h_X\times h_X\to h_X$ corresponds to a
morphism $m: X\times X\to X$, as is seen by using Proposition
\ref{1201}. Furthermore, the injection of the unit element into
each group $h_X(T)$ yields a natural transformation from $h_S$ to
$h_X$, hence it corresponds to a morphism
\begin{align*}
    \ve: S\to X,
\end{align*} called the {\it unit section of} $X$ which is a
section of the unique morphism $X\to S$. Finally, the formation of
the inverse in each $h_X(T)$ defines a natural transformation
$h_X\to h_X$ and hence a morphism
\begin{align*}
    \iota: X\to X,
\end{align*} called the {\it inverse map on} $X$.
The group axioms which are satisfied by the groups $h_X(T)$, and
hence by the functor $h_X$, correspond to certain properties of
the maps $m,\ve$ and $\iota$. Namely, the following diagrams are
commutative:

\begin{enumerate}
    \item[a)] {\it associativity}
    \begin{align*}
        \xymatrix @!C {  X\times X\times X \ar[r]^{ \ \ \ \ m\times id_X}
        \ar[d]_{id_X\times m}  & X\times X \ar[d]^{m} \\
        X\times X \ar[r]^{m}   & X }
    \end{align*}

    \item[b)] {\it existence of a left-identity}
    \begin{align*}
        \xymatrix @!C {X \ar[r]^{(p,id_X)} \ar[drr]_{id_X} &  S \times X \ar[r]^{ \ve \times id_X}
        & X\times X \ar[d]^{m} \\
        &  & X }
    \end{align*}
    where $p:X\to S$ is the morphism from $X$ to the final object $S$.

    \item[c)] {\it existence of a left-inverse}
    \begin{align*}
        \xymatrix @!C {  X \ar[r]^{(\iota, id_X) \quad }
        \ar[d]_{p}  & X\times X \ar[d]^{m} \\
        S \ar[r]^{\ve}   & X }
    \end{align*}

    \item[d)] {\it commutativity} (only if all groups $h_X(T)$ are commutative)
    \begin{align*}
        \xymatrix @!C {X\times X \ar[r]^{\tau} \ar[dr]_{m} &  X \times X \ar[d]^{m} \\
        &  X }
    \end{align*}
    where $\tau$ commutes the factors.
\end{enumerate}
Note that a left-identity is also a right-identity and that a
left-inverse is also a right-inverse. It is clear that once we
have an object $X$ and morphisms $m$, $\ve$ and $\iota$ with the
above properties, we can construct a group object in the given
category from these data, and furthermore that group objects in
$\Ccat$ and data $(X,m,\ve,\iota)$ correspond bijectively to each
other.

\begin{Satz}\label{1203}
    The group objects in a category $\Ccat$ correspond one-to-one
    to data $(X,m,\ve,\iota)$ where $X$ is an object of $\Ccat$ and where
    \begin{align*}
        m:X\times X\to X, \quad \ve:S\to X, \quad \iota: X\to X
    \end{align*}
    are morphisms in $\Ccat$ such that the diagrams $a), b), c)$ above
    are commutative. Furthermore, a group object in $\Ccat$ is
    commutative if and only if, in addition, the corresponding diagram
    $d)$ is commutative.
\end{Satz}
In the following we restrict ourselves to the category $(Sch/S)$
of $S$-schemes where $S$ is a fixed base scheme. Then the direct
product in $(Sch/S)$ is given by the fibred product of schemes
over $S$, and the $S$-scheme $S$ is a final object in $(Sch/S)$.

\begin{Def}\label{1204}
    An $S$-group scheme is a group object in the category of $S$-schemes $(Sch/S)$.
\end{Def}
Due to Proposition \ref{1203}, an $S$-group scheme $G$ can be
viewed as an $S$-scheme $X$ together with appropriate morphisms
$m,\; \ve$ and $\iota$. When no confusion about the group
structure is possible, we will not mention these morphisms
explicitly. In particular, in our notation we will make no
difference between the group object $G$ and the associated
representing scheme $X$. Also we want to pint out that there exist
group functors on $(Sch/S)$ which are not representable and thus
do not correspond to $S$-group schemes.

It follows immediately from Definition \ref{1204} that the
technique of base change can be applied to group schemes. Thus,
for any base change $S'\to S$, one obtains from an $S$-group
scheme $G$ an $S'$-group scheme $G_{S'}:=G\times_S S'$.

Let us look at some examples of $S$-group schemes. We start with
the classical groups $\bg_a$ (the additive group), $\bg_m$ (the
multiplicative group), $GL_n$ (the general linear group), and
$PGL_n$ (the projective general linear group). In terms of group
functors, these groups are defined as follows. For any $S$-scheme
$T$ set
\begin{eqnarray*}
    \bg_a(T)  & := & \mbox{the additive group} \; \co_T(T) \\
    \bg_m(T)  & := & \mbox{the group of units in} \; \co_T(T) \\
    GL_n(T)   & := & \mbox{the group of} \; \co_T(T)\mbox{-linear automorphisms of} \; (\co_T(T))^n \\
    PGL_n(T)  & := & \mbox{Aut}_T(\bp(\co_T^n)).
\end{eqnarray*}
All these group functors are representable by affine schemes over
$\bz$. Working over $S:=$ Spec $\bz$, the additive group is
represented by the scheme $ X:=\mbox{Spec} \; \bz[\zeta] $
($\zeta$ is an indeterminate), where the group law $m: X\times
X\to X$ corresponds to the algebra homomorphism
\begin{align*}
    \bz[\zeta]\to\bz[\zeta]\otimes_\bz \bz[\zeta], \quad \zeta\mapsto
    \zeta\otimes 1+1\otimes\zeta.
\end{align*}
Similarly, for $\bg_m$, the representing object is $\Spec
\bz[\zeta,\zeta^{-1}]$ with the group law given by
$\zeta\mapsto\zeta\otimes\zeta$.
\subsection{The Lie-algebra as a functor}\label{1220}

Let $G$ be a group functor over $S$ and denote by $\ve: S\to G$
the unit section. Recall from Definition \ref{1117} that the
$\underline{\co}_S$-module $L_{G/S}^\ve(\cm)$ denotes the tangent
space of $G$ over $S$ at $\ve$. Then we introduce the following
notation:

\begin{Def}\label{1221}
    The $\underline{\co}_S$-module {\rm Lie}$(G/S,\cm):=L_{G/S}^\ve(\cm)$ is
    called the Lie-algebra of $G$ relative $\cm$. In particular,
    \begin{align*}
    \Liealg:={\Lie}(G/S):= {\Lie}(G/S,\co_S)
    \end{align*}
    is called the \emph{Lie-algebra} of $G$.
\end{Def}
By Proposition \ref{1120}, $  T_{G/S}(\cm)$ and $\Lie(G/S,\cm)$
carry a group structure over $S$ which is induced by the group
structure on $G$. Furthermore, by definition of
$L_{G/S}^\ve(\cm)$, there is a cartesian diagram
\begin{align*}
    \xymatrix  {  \Lie(G/S,\cm)  \ar[r]^{\ \ i}
    \ar[d]_{}  & T_{G/S}(\cm) \ar[d]^{p} \\
     \quad S \quad  \ar[r]^{\ve}   & \quad  G \quad  }
\end{align*}
and we recognize Lie$(G/S,\cm)$ as the kernel of the canonical
projection $p$. In particular, we get morphisms of groups
\begin{align*}
    \mbox{Lie}(G/S,\cm) \stackrel{i}{\longrightarrow} T_{G/S}(\cm) \;
    \stackrel{p}{\longrightarrow \atop
    \stackrellow{s}{\longleftarrow}} \; G,
\end{align*}
where $s$ is a section of $p$. Due to the following Proposition
\ref{1222}, $T_{G/S}(\cm)$ is the semi-direct product of $G$ and
$\Lie(G/S,\cm)$.

\begin{Satz}\label{1222}
    Let $\Ccat:=(Sch/S)$ be the category of schemes over $S$. Let $f:
    W\to G$ be a morphism of $\hat{\Ccat}$-groups, and let $H(S):=ker
    \, f(S)$. Let $u: G\to W$ be a morphism of $\hat{\Ccat}$-groups
    which is a section of $f$ (in particular a monomorphism). Then $W$
    is the semi-direct product of $H$ and $G$ with an operation of $G$
    on $H$ given by the interior automorphism
    \begin{align*}
        (g,h) \mapsto \; {{\Int}} (u(g))\cdot h := u(g) \cdot h
        \cdot u(g)^{-1}
    \end{align*}
    for all $g\in G(S)$, $h\in H(S)$ and $S \in$ \emph{Ob}$\,
    \Ccat$.
\end{Satz}

\begin{proof}
    \SGAIII, Expose I, Prop. 2.3.7
\end{proof}
The corresponding operation of $G$ on Lie$(G/S,\cm)$ is denoted by
$\Ad$ and is called the \emph{adjoint representation} of $G$. By
definition we have got
\begin{align*}
    \Ad(x)X= i^{-1} (s(x)i(X)s(x)^{-1}) \quad \text{for all $x\in G(S')$ and $X \in \Lie(G/S,\cm)(S')$}.
\end{align*}
If $G$ and $H$ are two group functors over $S$ and if $f:G\to H$
is a group homomorphism, then (by means of functoriality) we
arrive at a commutative diagram with exact rows:
\begin{align*}
    \xymatrix{ 1 \ar[r] & \Lie(G/S,\cm) \ar[r] \ar[d]^{\Lie(f)}&
    T_{G/S}(\cm) \ar[r]\ar[d]^{T(f)}& G \ar[r]\ar[d]^f & 1 \\
    1 \ar[r] & \Lie(H/S,\cm) \ar[r] & T_{H/S}(\cm) \ar[r] & H \ar[r] & 1}
\end{align*}
\begin{Satz}\label{1223}
    Let $g\in G(S)$. Then the adjoint representation $\Ad$ of $G$ can
    be written as follows:
    \begin{align*}
        \emph{\Ad}(g)=\mbox{\rm Lie} (Int (g)).
    \end{align*}
\end{Satz}

\begin{proof}
 Ad$(g) X \stackrellow{\mbox{\scriptsize def}}{=} i^{-1} (Int (g)i(X))=\mbox{Lie} (Int(g))(X)$.
\end{proof}
{\bf Remark:} Recalling Definition \ref{1113} let us assume that
$G/S$ verifies $(E)$. This is for example true if $G$ is
representable by an $S$-group scheme. Then Ad$(g)$ respects the
$\underline{\co}_S$-module structure of Lie$(G/S,\cm)$, i.e. in
this case Ad is a linear representation of $G$ in the
$\underline{\co}_S$-module Lie$(G/S,\cm)$:
\begin{align*}
    \mbox{Ad}: G\to \mbox{Aut}_{\underline{\co}_S-lin}
    (\mbox{Lie}(G/S,\cm)).
\end{align*}
As one is used to from differential geometry, one may identify the
Lie-algebra with left-invariant vector fields.
\begin{Satz}\label{1224}
    There is an isomorphism
    \begin{eqnarray*}
        \begin{array}{ccc}
            \Hom(G,\mbox{Lie}(G/S,\cm)) & \stackrel{\sim}{\longrightarrow} & \Gamma(T_{G/S}(\cm)/G), \\[1ex]
            \qquad \qquad \qquad f & \mapsto & S_f \qquad
        \end{array}
    \end{eqnarray*}
    where   $S_f(g):=i(f(g))\cdot s(g)$  \ for all $ g\in G(S')$ and for
    all $S'\to S$.
\end{Satz}

\begin{proof}
    see \SGAIII, page 61
\end{proof}
If $h$ is an automorphism of functors over $S$, to each section
$t$ of $T_{G/S}(\cm)$ can be associated the unique section $h(t)$
making the following diagram commutative:

\begin{align*}
    \xymatrix  { G   \ar[r]^{t \ \ \quad } \ar[d]_{h}  & T_{G/S}(\cm) \ar[d]^{T(h)} \\
    G   \ar[r]^{h(t) \quad \ \ }   &  T_{G/S}(\cm)  }
\end{align*}
In particular, we can choose $h$ to be the right-translation with
$x\in G(S)$:
\begin{align*}
    h(g):=t_x(g):=g\cdot x \quad \text{$g\in G(S')$ and $S'\to S$}.
\end{align*}
Then the isomorphism of Proposition \ref{1224} is compatible with
right-translations. In particular those sections, which are
invariant under right translations, are mapped to constant
morphisms of $G$ to $\Lie(G/S,\cm)$, i.e. to morphisms which
factorize over $S$.

\begin{Satz}\label{1225}
 The map
\begin{eqnarray*}
    \mbox{\rm Lie} (G/S,\cm)(S) & \to & \Gamma(T_{G/S}(\cm)/G) \\
    X & \mapsto & (x\mapsto X\cdot x),
\end{eqnarray*}
is an isomorphism of $\, \Lie(G/S,\cm)(S)$ to those sections of
$\, \Gamma (T_{G/S}(\cm)/G)$ which are invariant under
right-translation.
\end{Satz}

\begin{proof}
    \SGAIII, Expose II, Prop. 4.1.2
\end{proof}
\subsection{The maps Ad and ad}\label{1230}

\begin{Def}\label{1231}
    A $\underline{\co}_S$-module $\cf$ is called \emph{good}, if the canonical
    morphism
    \begin{align*}
        \cf \otimes_{\underline{\co}_S} T_{\underline{\co}_S/S}(\cm)\to T_{\cf/S}(\cm)
    \end{align*}
    is an isomorphism.
\end{Def}

\begin{Def}\label{1232}
    A group functor $G$ over $S$ is called {\it good}, if it verifies
    $(E)$ (see Definition \ref{1113}) and if $\Lie(G/S)$ is a good $\underline{\co}_S$-module.
\end{Def}
{\bf Remark:} Let us remark that a group functor $G$ over $S$
which is representable by a scheme $G$ is good. This will be the
case in the physical situations we are going to consider. For more
details on ``good $\underline{\co}_S$-modules'' see e.g. \SGAIII,
Expose II, Chap. 4. We are just interested in some theorems about
good modules which show that many classical, differential
geometric results concerning the Lie-algebra of Lie-groups carry
over to algebraic geometry.

\begin{Satz}\label{1233}
    If $\cf$ is a good $\underline{\co}_S$-module, the $\underline{\co}_S$-module
    structure of $\Lie(\cf/S,\cm)$ is induced by that of $\cf$.
\end{Satz}

\begin{proof}
    \SGAIII, Expose II, Cor. of Def. 4.4
\end{proof}

\begin{Satz}\label{1234}
    If $\cf$ is a good $\underline{\co}_S$-module, there is a functorial
    isomorphism
    \begin{align*}
        \mbox{\rm Lie(Aut}_{\underline{\co}_S-lin}(\cf)/S,\cm)
        \stackrel{\sim}{\longrightarrow} \Hom_{\underline{\co}_S-lin} (\cf,\mbox{\rm
        Lie}(\cf/S,\cm)).
    \end{align*}
    In particular:
    \begin{align*}
        \mbox{\rm Lie(Aut}_{\underline{\co}_S-lin}(\cf)/S)
        \stackrel{\sim}{\longrightarrow} \mbox{\rm End}_{\underline{\co}_S-lin}(\cf).
    \end{align*}
\end{Satz}

\begin{proof}
    \SGAIII, Expose II, Prop. 4.6
\end{proof}
Let now $G$ be a good $S$-group-scheme. Then $\Lie(G/S)$ is a good
$\underline{\co}_S$-module. We already introduced the linear
representation Ad of $G$ in the $\underline{\co}_S$-module
Lie$(G/S)$:
\begin{eqnarray*}
    \Ad: G & \longrightarrow & \Aut_{\underline{\co}_S-lin}(\Lie(G/S)) \\
    g & \mapsto & \Lie(Int(g)).
\end{eqnarray*}
Applying the functor Lie and using Proposition \ref{1234}, we
arrive at the $\underline{\co}_S$-module morphism ad:
\begin{align*}
    \mbox{ad: Lie}(G/S) \longrightarrow
    \mbox{End}_{\underline{\co}_S-lin}(\mbox{Lie}(G/S))
\end{align*}
which may also be considered as a bilinear morphism
\begin{align*}
    \mbox{Lie}(G/S) \times_S \mbox{Lie}(G/S)  \longrightarrow
    \mbox{Lie}(G/S),
    \quad  (x,y)   \mapsto  [x,y]:= \mbox{ad}(x)(y)
\end{align*}
where $x,y\in \Lie(G/S)(S') = \Lie(G_{S'}/S')(S')$.

\begin{Satz}\label{1235}
    Let $G$ be a representable group functor over $S$ (or more generally
    a group functor which is a subfunctor of a representable
    group over $S$). Then $\Lie(G/S)$ is a $\underline{\co}_S$-Lie-algebra, i.e.:
    \begin{itemize}
        \item[(i)] $[x,x]=0$;
        \item[(ii)] $[x,y]+[y,x]=0$;
        \item[(iii)] $[x,[y,z]]+[y,[z,x]]+[z,[x,y]]=0$.
    \end{itemize}
\end{Satz}

\begin{proof}
    \SGAIII, Expose II, pages 68-72
\end{proof}
\subsection{The Lie-algebra and invariant differential
forms}\label{1240}

In subsection \ref{1220} we stated several results about the
Lie-algebra of an $S$-group scheme $G$. In analogy to differential
geometry, Proposition \ref{1224} states the one-to-one
correspondence  of elements of the Lie-Algebra
$\mbox{Lie}(G/S,\cm)(S)$ and left-invariant global sections of the
tangent-bundle $\Gamma(G,T_{G/S}(\cm))$.

In this paragraph we will recall these statements from the dual
point of view, i.e. we will  consider differential forms instead
of vector fields. In order to this, we first introduce the notion
of translations on an $S$-group scheme $G$ for a given $T$-valued
point $g:T\to G$, i.e. an $S$-morphism from an $S$-scheme $T$ to
$G$.

Throughout this section, let $G$ be a group scheme over a fixed
scheme $S$. First we want to introduce the notion of translations
on $G$. In order to do this, consider a $T$-valued point
\begin{align*}
    g: T\to G
\end{align*}
of $G$; i.e., an $S$-morphism from an $S$-scheme $T$ to $G$. Then
$g$ gives rise to the $T$-valued point
\begin{align*}
    g_T:=(g,\mbox{id}_T): T\to G_T:= G\times_S T
\end{align*}
of the $T$-scheme $G_T:=G \times_S T$. If $p_1: G_T\to G$ denotes
the first projection, we have $g=p_1\circ g_T$. In the special
case where $T:=G$ and where $g:=$ id$_G$ is the so called
universal point of $G$, the morphism $g_T$ equals the diagonal
morphism $\Delta$ of $G$. For any other $T$-valued point $g$ of
$G$, the morphism $g_T$ is obtained from $\Delta$ by performing
the base change $g: T\to G$.

As usual, let $m:G\times_S G\to G$ be the group law of $G$ and
write $m_T$ for its extension when a base change $T\to S$ is
applied to $G$. Then, for any $T$-valued point $g$ of $G$, we
define the left translation by
\begin{align*}
    \tau_g: G_T \stackrel{\sim}{\longrightarrow} T \times_T G_T
    \stackrel{g_T\times \mbox{\scriptsize id}}{\longrightarrow} G_T
    \times_T G_T \stackrel{m_T}{\longrightarrow} G_T
\end{align*}
and the right translation by
\begin{align*}
    \tau'_g: G_T \stackrel{\sim}{\longrightarrow} G_T \times_T T
    \stackrel{\mbox{\scriptsize id}\times g_T}{\longrightarrow} G_T
    \times_T G_T \stackrel{m_T}{\longrightarrow} G_T.
\end{align*}
Both morphisms are isomorphisms. Quite often we will drop the
index $T$ and characterize the map $\tau_g$ by writing
\begin{align*}
    \tau_g: G\to G, \quad x\mapsto gx;
\end{align*}
the same procedure will be applied for $\tau'_g$ and for similar
morphisms. In the special case where $T:=G$ and $g:=$~id$_G$ is
the universal point, $\tau_g$ is the so-called {\it universal left
translation,} namely the morphism
\begin{align*}
    \Phi: T\times_S G\to T \times_S G, \quad (x,y)\mapsto (x,xy).
\end{align*}
Similarly, $\tau'_g$ gives rise to the {\it universal right
translation}
\begin{align*}
    \Psi: G\times_S T\to G \times_S T, \quad (x,y)\mapsto (xy,y).
\end{align*}
Each left translation by a $T$-valued point $g:T\to G$ is obtained
from the universal left translation $\Phi$ by performing the base
change $g: T\to G$; in a similar way one can proceed with right
translations.

Now let us consider the sheaf $\Omega_{G/S}^i$ of relative
differential forms of some degree $i\ge 0$ on $G$; it is defined
as the $i$-th exterior power of $\Omega_{G/S}^1$ (see Definition
\ref{1151}). For any $S$-scheme $T$ and any $T$-valued point $g\in
G(T)$, the left translation $\tau_g: G_T\to G_T$ gives rise to an
isomorphism\footnote{This isomorphism is the canonical pull-back
morphism induced by $\tau_g$. In the special case of differential
forms it was introduced in Definition \ref{0014}; for the general
case of forms of higher degree see Definition \ref{1329}.}
\begin{align*}
\tau_g^\ast \Omega_{G_T/T}^i \stackrel{\sim}{\longrightarrow}
\Omega_{G_T/T}^i.
\end{align*}
A global section $\omega$ in $\Omega_{G/S}^i$ is called {\it
left-invariant} if $\tau_g^\ast \omega_T = \omega_T$ in
$\Omega_{G_T/T}^i$ for all $g\in G(T)$ and all $T$, where
$\omega_T$ is
 the pull-back of $\omega$ with respect to the projection $p_1: G_T\to G$.
 Using right translations $\tau'_t$, one defines {\it
right-invariant} differential forms in the same way. Since each
translation on the group scheme $G_T$ is obtained by base change
from the universal translation, it is clear that one has to check
the invariance under translations only for the universal
translation. Generally, in connection with translations, we will
drop the index $T$ and write $\omega$ instead of $\omega_T$ if no
confusion is possible. Within this context, let us state the
following two remarks:

\begin{enumerate}
    \item
    All morphisms $g_T: T\to G_T$ for a given $g: T\to G$
    are induced by base change from $\Delta: G\to G \times_S G$.

    \item
    Consider two global sections $\omega$ and $\omega'$ of a
    sheaf $F$ on $G$, let $F_T$ denote the pull-back of $F$ to $G_T$. Then:
    \begin{align*}
        \omega=\omega' \  \Leftrightarrow \ \ g_T^\ast \omega_T=g_T^\ast
        \omega'_T \in g_T^\ast F_T.
    \end{align*}
    Similarly two sheaves are isomorphic if their restriction to each
    $T$-valued point of $G$ are isomorphic.
\end{enumerate}
\begin{Satz}\label{1241}
    Let $G$ be an $S$-group scheme with unit section $\ve: S\to G$.
    Then, for each $\omega_0\in \Gamma(S,\ve^\ast \Omega_{G/S}^i)$,
    there exists a unique left-invariant differential form
    $\omega\in\Gamma(G,\Omega_{G/S}^i)$ such that $\ve^\ast \omega=\omega_0$
    in $\ve^\ast\Omega_{G/S}^i$. The same is true for right-invariant differential forms.
\end{Satz}

\begin{proof}
    Let us shortly indicate, how the left invariant differential form $\omega$
    is obtained from $\omega_0$. For details of the prove, we
    refer the reader to \BLR, Prop 4.2/1. Due to the
    uniqueness assertion, the problem is local in $S$
    (because we may glue). Furthermore
    it suffices to consider the case $i=1$. Thus we may assume
    that $\omega_0$ lifts to a section $\omega'$ of
    $\Omega_{G/S}^1$ which is defined over a neighborhood $U$ of the unit section.
    This is due to the fact that $(\ve^*\Omega_{G/S}^1)(S) =(\Omega_{G/S}^1)(U) \otimes_{\co_G(U)}
    \co_S(S)$ if $S$ and $U\subset G$ are affine,  and that
    $p^*:\co_S(S) \hookrightarrow \co_G(U)$ is injective,
    because $p:G\to S$ has the section $\ve$. Then the
    decomposition $p_1^*\Omega_{G/S}^1 \oplus p_2^*\Omega_{G/S}^1
    \stackrel{\sim}\longrightarrow \Omega_{G \times_S G/S}^1$ of
    Proposition \ref{0018} gives a decomposition $m^* \omega' = \omega_1 \oplus
    \omega_2$ over $m^{-1}(U)$, where $m: G \times_S G \to G$ is
    the multiplication. If $\delta: G \to G \times_S G$, $x \mapsto
    (x,x^{-1})$, denotes the twisted diagonal morphism, $m^*
    \omega'$ is defined in a neighborhood of the image of
    $\delta$ so that $\delta^*\omega_2$ gives rise to a global
    section $\omega$ of $\Omega_{G/S}^1$. Then $\omega$ is the
    searched left-invariant differential form with $\ve^* \omega=
    \omega_0$ in $\ve^* \Omega_{G/S}^1$.

    In particular the association $\omega_0 \mapsto \omega$ is
    $\co_S(S)$-linear.
\end{proof}

\begin{Satz}\label{1241a}
    There are canonical isomorphisms
    \begin{align*}
        p^\ast \ve^\ast \Omega_{G/S}^i \stackrel{\sim}{\longrightarrow}
        \Omega_{G/S}^i \quad \text{for all } i\in\bn
    \end{align*}
    which are obtained by extending sections in $\ve^\ast
    \Omega_{G/S}^i$ to left-invariant sections in $\Omega_{G/S}^i$.
    Similar isomorphisms are obtained by using right-invariant
    differential forms.
\end{Satz}

\begin{proof}
    \BLR, Prop 4.2/2
\end{proof}
As a direct consequence we obtain:

\begin{Cor}\label{1242}
    Let $G$ be a smooth $S$-group scheme of relative dimension
    $n$, and assume  that $\ve^\ast \Omega_{G/S}^i$ is free. Then $\Omega_{G/S}^i$
    is free $\co_S$-module generated by
    ${n \choose i}$ left-invariant differential forms of degree $i$.
    The same is true for right-invariant differential forms.
\end{Cor}
In physical applications, \st is given by the set of $S$-sections
$X(S)$ of a smooth, separated algebraic space $X \to S$ with
certain further properties (see Definition \ref{5022} for a
complete list of properties of $X$). In special cases, $X$ is even
a group object and thus carries a canonical $CPT$-operation (see
Definition \ref{5024}). Then, all results on group schemes, which
are summarized above, are valid for $X$. The preceding Corollary
\ref{1242} is the basis for the proof that there exists a volume
form $\vol$ in $X$. This result is essential, because we need to
make use of a volume form within our physical purposes of a
Yang-Mills theory over rings.

\begin{Def}\label{1243}
    Let $X \to S$ be a smooth $S$-scheme of relative dimension
    $n$ (in particular $\Omega_{X/S}^n$ is a line-bundle). Let us assume that there exists a global section
    $\vol \in \Omega_{X/S}^n(X)$ which generates $\Omega_{X/S}^n$.
    Then $\vol$ is called a \emph{volume form} on $X$.
\end{Def}

\begin{Theorem}\label{1244}
    Let $S=\Spec R$ be the spectrum of a Dedekind ring $R$ which is a principal ideal domain.
    Let $X \to S$ be a smooth $S$-group scheme of relative dimension $n$.
    Then the sheaves $\Omega_{X/S}^i$ are free $\co_X$-modules
    generated by left-invariant differential forms of degree $i$. In particular there exists a
    volume form on $X$.
\end{Theorem}

\begin{proof}
    By Corollory \ref{1242} it suffices to show that $\ve^\ast
    \Omega_{X/S}^i$ is  a free $\co_S$-module. We already know that $\ve^\ast
    \Omega_{X/S}^i$ is a locally free
    $\co_S$-module, because it is the pull-back of the locally free $\co_X$-module  $\Omega_{X/S}^i$.
    Due to the  well-known fact that vector bundles over a scheme $S$
    and locally free $\co_S$-modules are in one-to-one
    correspondence, it suffices to show that (up to isomorphism) there are only trivial
    vector bundles over $S$; i.e. it suffices to show that $K_0(R) \cong \Ganz$.
    But the ideal class group ${\text{Cl}}(R)$ of $R$ is the trivial group $\{1\}$ if and only
    if $R$ is a principal ideal domain. Thus we are done by the following
    result which may be found in nearly every book on algebraic
    $K$-theory.
\end{proof}

\begin{Satz}\label{1245}
    Let $R$ be a Dedekind ring and let $\emph{\text{Cl}}(R)$ denote the ideal class group of $R$.
    Then there is an isomorphism of groups:
    \begin{align*}
        K_0(R) \cong \Ganz \, \oplus \, \, \emph{\text{Cl}}(R)
    \end{align*}
\end{Satz}
\begin{Cor}\label{1246}
    Let $X \to S$ be a {\SR} (see Definition \ref{5024}). Then there
    exists a volume form $\vol$ on $X$.
\end{Cor}
As we are finally interested in physical applications, let us from
now on assume that $G$ is a smooth $S$-group scheme of relative
dimension $n$, and that there is a left-invariant differential
form $\omega \in\Omega_{G/S}^n(G)$ generating $\Omega_{G/S}^n$ as
an $\co_G$-module. Performing a base change with the canonical
morphism $G_K \to G$ (where $K$ is the field of fractions of $S$),
$\omega$ is pulled back to a left-invariant differential form
$\omega_K \in \Omega_{G_K/K}^n(G_K)$ generating
$\Omega_{G_K/K}^n$. Interpreting the generic fibre as classical
limit (as explained in section \ref{0052}), we may interpret
$\omega_K$ as the classical limit of $\omega$. Let us finish this
section with the prove that the volume form $\omega_K$ is
invariant with respect to translations if $G$ describes \st
itself. By Theorem \ref{0123} the latter means that the generic
fibre $G_K$ of $G$ is given by an abelian variety. In particular,
$G_K$ is bounded by Proposition \ref{0130}, i.e. $G_K(K^{sh})$ is
bounded. More precisely, we claim that $\omega_K$ is not only
invariant with respect to left-translations, but that furthermore
$|{\tau'}_g^* \omega_K| = |\omega_K|$ for all right-translations
by physical points $g\in G_K(K^{sh})$ and for all valuation
$|\cdot|$ (i.e. prime spots $\prim$) corresponding to closed
points $s \in S$. Before stating the proof of this translation
invariance, let us first recall the definition of the interior
automorphism which was already considered in subsection
\ref{1220}.

\begin{Def}\label{1247}
    For an arbitrary $T$-valued point $g$ of $G$ we define the
    interior automorphism
    \begin{align*}
        \intg:=\tau_g\circ
        \tau'_{g^{-1}}: G\longrightarrow G, \quad  x \mapsto g x g^{-1}
    \end{align*}
    induced by $g$.
\end{Def}
Now our claim is a direct consequence of the following two
propositions.

\begin{Satz}\label{1248}
    There exists a unique group homomorphism
    $\chi:G\to\bg_m $ (a so called character on $G$), such that
    \begin{align*}
        \intg^\ast \omega=\tau_{g^{-1}}^{'\ast} \omega=\chi(g) \omega
    \end{align*}
    for each $T$-valued point $g$ of $G$.
\end{Satz}

\begin{proof}
    \BLR, Prop. 4.2/4
\end{proof}

\begin{Satz}\label{1249}
    Let $G_K$ be a smooth $K$-group of relative dimension $n$, and
    assume that $G_K(K)$ (resp. $G_K(K^{sh})$) is bounded in
    $G_K$. Then the character $\chi$ considered in Proposition \ref{1248}
    satisfies $|\chi(g)|=1$ for each $g \in G_K(K^{sh})$ (resp. each $g \in
    G_K(K^{sh})$).
\end{Satz}

\begin{proof}
   \BLR, Prop. 4.2/5
\end{proof}
\section{Faithfully flat descent}\label{1380}

Let $p:S' \to S$ be a morphism of schemes and consider the functor
$\cf \to p^* \cf$ which associates to each quasi-coherent
$\co_S$-module $\cf$ its pull-back under $p$. Then, in its
simplest form, the problem of descent relative to $p:S' \to S$ is
to characterize the image of this functor. But descent may also be
viewed as a natural generalization of a patching problem. The
procedure of solution is as follows. Set $S'':= S' \times_S S'$,
and let $p_i:S'' \to S'$ be the projection onto the $i$-th factor
$(i=1,2)$. For any quasi-coherent $\co_{S'}$-module $\cf'$, call
an $S''$-isomorphism $\varphi: p_1^* \cf' \to p_2^* \cf' $ a
\emph{covering datum} of  $\cf'$. Then the pairs $(\cf',\varphi)$
of quasi-coherent $\co_{S'}$-modules with covering data form a
category in a natural way. A morphism between two such objects
$(\cf',\varphi)$ and $(\cg',\psi)$ consists of an
$\co_{S'}$-module-homomorphism $f: \cf' \to \cg'$ which is
compatible with the covering data $\varphi$ and $\psi$; thereby we
mean that the diagram
\begin{align*}
    \xymatrix{p_1^* \cf' \ar[r]^{\varphi} \ar[d]_{p_1^*f} & p_2^* \cf' \ar[d]^{p_2^* f} \\
p_1^* \cg' \ar[r]^{\psi}  & p_2^* \cg' }
\end{align*}
is commutative.

Starting with a quasi-coherent $\co_{S}$-module $\cf$, we have a
natural covering datum on $p^* \cf$ which consists of the
canonical isomorphism
\begin{align*}
    p_1^*(p^* \cf) \cong (p \circ p_1)^* \cf = (p \circ p_2)^* \cf \cong p_2^*(p^* \cf).
\end{align*}
So we can interpret the functor $\cf \rightsquigarrow p^* \cf$ as
a functor into the category of quasi-coherent $\co_{S'}$-modules
with covering data. It is this functor which will be of interest
in the following. It turns out that this functor is fully faithful
if $p:S' \to S$ is faithfully flat and quasi-compact (see e.g.
\BLR, Prop. 6.1/1). Furthermore, it is an equivalence of
categories if, instead of covering data, we consider descent data.
A descent datum is a special covering datum which satisfies a
certain cocycle condition. In order to introduce them, set
$S''':=S' \times_S S' \times_S S'$,  and let $p_{ij}:S''' \to S''$
be the projection onto the factors with indices $i$ and $j$ for
$i<j$; $i,j=1,2,3$. In order that a quasi-coherent
$\co_{S'}$-module $\cf'$ with covering datum $\varphi: p_1^* \cf'
\to p_2^* \cf'$ belongs to the essential image of the functor $\cf
\rightsquigarrow p^* \cf$, it is necessary that the diagram
\begin{align*}
    \xymatrix{ p_{12}^*p_1^* \cf' \ar@{=}[d] \ar[r]^{p_{12}^* \varphi \quad \quad \ \,}
    & p_{12}^*p_2^* \cf' = p_{23}^*p_1^* \cf'  \ar[r]^{ \quad \ \, \quad p_{23}^* \varphi}  & p_{23}^*p_2^* \cf' \ar@{=}[d]  \\
    p_{13}^*p_1^* \cf'  \ar[rr]^{p_{13}^* \varphi} &  &p_{13}^*p_2^*
    \cf' }
\end{align*}
is commutative; the unspecified identities are the canonical ones.
Namely, if $\cf'$ is the pull-back under $p$ of a quasi-coherent
$\co_{S}$-module and if $\varphi$  is the natural covering datum
on $\cf'$, then the diagram is commutative, because all occurring
isomorphisms are the identical ones. The commutativity of the
diagram is referred to as the \emph{cocycle condition} for
$\varphi$; in short, we can write it as
\begin{align*}
    p_{13}^*\varphi = p_{23}^*\varphi \circ  p_{12}^*\varphi.
\end{align*}
It corresponds to the usual cocyle condition on triple overlaps
when a global object is to be constructed by gluing local parts. A
covering datum $\varphi$ on $\cf'$ which satisfies the cocycle
condition is called a \emph{descent datum} on $\cf'$. The descent
datum is called \emph{effective} if   the pair $(\cf', \varphi)$
is isomorphic to the pull-back $p^* \cf$ of a quasi-coherent
$S$-module $\cf$ where, on $p^* \cf$, we consider the canonical
descent datum. Also we want to mention that the notions of
covering and descent datum are compatible with base change over
$S$. Now we are ready to state the desired result on the descent
of quasi-coherent $S'$-modules due to Grothendieck.
\begin{Theorem}\label{1381}
Let $p:S' \to S$ be faithfully flat and quasi-compact. Then the
functor $\cf  \rightsquigarrow p^* \cf $, which goes from
quasi-coherent $\co_{S}$-modules to quasi-coherent
$\co_{S'}$-modules with descent data, is an equivalence of
categories.
 \end{Theorem}

\begin{proof}
    \BLR, Thm. 6.1/4
\end{proof}

\begin{Rem}\label{1382}
    (Etale coverings). Consider a quasi-separated scheme $S$ and a
    finite \'etale covering $(S_i \to S)_{i \in I}$ of $S$ (see
    Definition \ref{1306}). Let $S':= \coprod_{i\in I} S_i$ be the disjoint
    union of the $S_i$, and let $p:S' \to S$ be the canonical
    projection. Note that $p$ is faithfully flat and quasi-compact. A
    quasi-coherent $\co_{S'}$-module $\cf'$ may be thought of as a
    family of $\co_{S_i}$-modules $\cf_i$. Under what conditions does
    $\cf'$ descent to a quasi-coherent $\co_{S}$-module $\cf$; i.e.
    under what conditions can one glue the $\cf_i$ in order to obtain
    a quasi-coherent $\co_{S}$-module $\cf$ from them? By Theorem \ref{1381}
    we need a descent datum for $\cf'$ with respect to $p:S' \to S$.
    Such a datum consists of an isomorphism $\varphi: p_1^* \cf'
    \stackrel{\sim}\to p_2^* \cf'$ satisfying the cocycle condition,
    where $p_1$ and $p_2$ are the projections from $S''$ to $S'$. In
    our case, we have
    \begin{align*}
        S''= S' \times_S S' = \coprod_{i,j \in I} S_{ij},
    \end{align*}
    where $S_{ij}:=S_i \times_S S_j$. Thus the isomorphism $\varphi$
    consists of a family of isomorphisms
    \begin{align*}
        \varphi_{ij}: \cf_i \mid_{S_{ij}} \stackrel{\sim}\longrightarrow  \cf_j \mid_{S_{ij}}
    \end{align*}
    satisfying the cocycle condition, namely, the condition that
    \begin{align*}
    \varphi_{ik} \mid_{S_{ijk}}  = \varphi_{jk}\mid_{S_{ijk}}
    \circ \,
    \varphi_{ij} \mid_{S_{ijk}}
    \end{align*}
    for all $i,j,k \in I$, where $S_{ijk}:=S_i \times_S S_j \times_S
    S_k$. Thereby, $\cf_i \mid_{S_{ij}}$ denotes the pull-back
    $p_{i}^* \cf_i$  of $\cf_i$ under the canonical projection
    $p_{i}:S_i \times_S S_j \to S_i $, etc.

    So the descent datum $\varphi$ in the form of the above family of
    isomorphism  is the patching datum for the $\co_{S_i}$-modules
    $\cf_i$ with respect to \'etale topology on $S$.
\end{Rem}
Keeping the morphism $S' \to S$, one may also study the problem of
when an $S'$-scheme $X'$ descends to an $S$-scheme $X$. The
general setting will be the same as in the case of quasi-coherent
modules, and the definitions we have given can easily be adapted
to the new situation. For example, a \emph{descent datum on an
$S'$-scheme $X'$} is an $S''$-isomorphism
\begin{align*}
    \phi : p_1^* X' \to p_2^* X'
\end{align*}
which satisfies the cocycle condition; $p_i^* X'$ is  the scheme
obtained from $X'$ by applying the base change $p_i:S'' \to S'$.
Again there is a canonical functor $X \rightsquigarrow p^*X$ from
$S$-schemes to $S'$-schemes with descent data.

\begin{Theorem}\label{1383}
    Let $p:S' \to S$ be faithfully flat and quasi-compact. Then
    the functor $X \rightsquigarrow p^*X$ from
    $S$-schemes to $S'$-schemes with descent data is fully
    faithful.
\end{Theorem}

\begin{proof}
    \BLR, Thm. 6.1/6
\end{proof}

\end{document}